\documentclass[11pt,letterpaper,oneside,openany]{book}
\title{Partonic transport model application to heavy flavor}
\author{Weiyao Ke}
\usepackage[utf8]{inputenc}
\usepackage[T1]{fontenc}
\usepackage{lmodern}
\usepackage{bbold}
\usepackage{wrapfig}
\makeatletter
  \newenvironment{abstract}{%
      \titlepage
      \null\vfil
      \@beginparpenalty\@lowpenalty
      \begin{center}%
        \Huge\sffamily\color{offblack}Abstract
      \end{center}}%
     {\par\vfil\null\endtitlepage}
     \vspace{2ex}
\makeatother

\usepackage{mathtools}
\usepackage[titletoc]{appendix}
\usepackage{amsmath}
\usepackage{amssymb}
\usepackage{amsfonts}
\usepackage{tabularx}
\usepackage{booktabs}
\usepackage{color}
\usepackage{xcolor}
\usepackage{multirow}
\usepackage{verbatim}
\usepackage[inline]{enumitem}
\usepackage{slashed}

\usepackage{graphicx}
\graphicspath{{figures/}{figures/LBT/}{figures/Feynman/}{figures/Lido/}{figures/New-analysis/}{figures/Citing/}}

\usepackage{booktabs}
\usepackage{enumerate}
\usepackage{tikz}
\usetikzlibrary{arrows,backgrounds,calc,matrix,shapes}

\usepackage{listings}
\definecolor{deepred}{RGB}{146,11,36}
\definecolor{offblack}{RGB}{35,31,32}
\usepackage[pdfusetitle,colorlinks=true,allcolors=deepred,bookmarksdepth=2,
linktoc=section,bookmarksopen=false]{hyperref}

\usepackage{titlesec}
\titleformat{\chapter}[display]{\Huge\sffamily\color{offblack}\filcenter
}{\thechapter}{1ex}{}
\titleformat*{\section}{\Large\bfseries\color{deepred}}
\titleformat*{\subsection}{\large\bfseries\color{deepred}}
\titleformat*{\subsubsection}{\bfseries\color{deepred}}
\titleformat*{\paragraph}{\bfseries\color{offblack}}
\titlespacing{\paragraph}{0pt}{*1.4}{*2.5}

\usepackage[font=small,labelfont={bf,color=deepred},labelsep=quad]{caption}

\usepackage{lettrine}

\usepackage{lipsum}

\newcommand{\sqrts}{\sqrt{s_\mathrm{NN}}}

\newcommand{\trento}{T\raisebox{-0.5ex}{R}ENTo}

\newcommand{\paddedhline}{\noalign{\smallskip}\hline\noalign{\smallskip}}
\newcommand{\dnchdy}{dN_\text{ch}/d\eta}

\newcommand{\x}{\mathbf x}

\newcommand{\ppi}{\frac{\partial}{\partial p_i}}

\newcommand{\ppl}{\frac{\partial}{\partial p_l}}
\newcommand{\Kpara}{\kappa_{\|}}
\newcommand{\Kperp}{\kappa_{\perp}}
\newcommand{\Ppara}{\hat{P}^{\|}}
\newcommand{\Pperp}{\hat{P}^{\perp}}

\DeclareFontFamily{OMS}{oasy}{\skewchar\font48 }
\DeclareFontShape{OMS}{oasy}{m}{n}{%
         <-5.5> oasy5     <5.5-6.5> oasy6
      <6.5-7.5> oasy7     <7.5-8.5> oasy8
      <8.5-9.5> oasy9     <9.5->  oasy10
      }{}
\DeclareFontShape{OMS}{oasy}{b}{n}{%
       <-6> oabsy5
      <6-8> oabsy7
      <8->  oabsy10
      }{}
\DeclareSymbolFont{oasy}{OMS}{oasy}{m}{n}
\SetSymbolFont{oasy}{bold}{OMS}{oasy}{b}{n}

\DeclareMathSymbol{\smallleftarrow}     {\mathrel}{oasy}{"20}
\DeclareMathSymbol{\smallrightarrow}    {\mathrel}{oasy}{"21}
\DeclareMathSymbol{\smallleftrightarrow}{\mathrel}{oasy}{"24}
\newcommand{\tensor}[1]{\overset{\scriptscriptstyle\smallleftrightarrow}{#1}}

\DeclareMathOperator{\diag}{diag}

\begin{document}

\frontmatter

\makeatletter

\begin{titlepage}
  \centering
  \sffamily
  \vspace*{.25\textheight}
  \Huge\@title \\
  \vspace{.05\textheight}
  \LARGE\@author \\
  \vspace{.05\textheight}
  \Large Ph.D.\ dissertation \\[.25ex]
  Advisor: Steffen A.\ Bass \\[.25ex]
  Department of Physics, Duke University \\
  \vspace{.05\textheight}
  \today
\end{titlepage}
\makeatother


\titlespacing{\chapter}{0pt}{-45pt}{30pt}
\setcounter{tocdepth}{1}
\titlespacing{\chapter}{0pt}{50pt}{40pt}

\begin{abstract}
Heavy-flavor particles are excellent probes of the properties of the hot and dense nuclear medium created in the relativistic heavy-ion collisions.
Heavy-flavor transport coefficients in the quark-gluon plasma (QGP) stage of the collisions are particularly interesting, as they contain important information on the strong interaction at finite temperatures.
Studying the heavy-flavor evolution in a dynamically evolving medium requires a comprehensive multi-stage modeling approach of both the medium and the probes, with an accurate implementation of the physical ingredients to be tested.
For this purpose, I have developed a new partonic transport model (Linear-Boltzmann-plus-Diffusion-Transport-Model) LIDO and applied it to heavy quark propagation inside a QGP.
The model has an improved implementation of parton in-medium bremsstrahlung and a flexible treatment of the probe-medium interactions, combining both large angle scatterings and diffusion processes.
The model is then coupled to a high-energy event-generator, a hydrodynamic medium evolution and a hadronic transport model.
Finally, applying a Bayesian analysis, I extract the heavy quark transport coefficients from a model-to-data comparison.
The results, with uncertainty quantification, are found to be consistent with earlier extraction of the light-quark transport coefficients at high momentum and with first-principle calculations of the heavy flavor diffusion constant at low momentum.
\end{abstract}

\mainmatter
\pagenumbering{arabic}
\tableofcontents

\chapter{Introduction}
\label{chapter:introduction}
The fundamental theory of strong interaction -- Quantum chromodynamics (QCD) -- describes a rich set of phenomena, from low-energy vibrations of atomic nuclei to the production of energetic jets of particles with over trillion electron volts of energy on the Large Hadron Collider (LHC).
Its force binds over 99\% of the mass in the visible universe, yet its complexity has made it a subject very hard to approach both experimentally and theoretically.
To understand its dynamics, people look for simplified scenarios to do experiments and examine our understandings.
One such limit is the high energy limit, where the coupling constant of the strong interaction $\alpha_s$ becomes relatively small, known as the asymptotic freedom \cite{Gross:1973id,Politzer:1973fx}. 
Theoretical tools such as perturbative QCD (pQCD) can be applied.
Observations from high energy collisions have confirmed the success of perturbation theory \cite{RevModPhys.59.465}.
Another interesting aspect is to understand physical systems in the ``many-particle'' (thermodynamic) limit.
Instead of exciting a few fundamental particles and observing their evolution, one deposits a huge amount of energy into a tiny region, for example, by colliding heavy nuclei that excites a medium with thousands of particles. 
In this limit, one is more interested in the collective dynamics of the strong force.
Features such as the structure of the equation-of-state (EoS), the medium transport coefficients, and the medium stopping power (opacity), etc, are also fundamental properties of the strong interaction.
Applying tools from many-body physics such as the finite temperature field theory, kinetic transport theory, and hydrodynamics, many facets of QCD have been revealed from data taken at particle colliders.

I shall briefly review the basic concepts of QCD and our current understanding of nuclear matter, in particular, the quark-gluon plasma (QGP). 
Then, key experimental discoveries made by studying relativistic heavy-ion collisions are reviewed and I will show how different probes can be used to characterize the different transport properties of the QGP through a model-to-data comparison.
This dissertation is an example of such practice: I developed a heavy-flavor transport model and combined it with an advanced statistical method to reverse-engineer the heavy quark transport coefficient from experimental data.

\section{Quantum chromodynamics and nuclear matter}
QCD describes the interaction of objects that carries ``color'' charges.
Quarks (fermions) and gluons (bosons) are the elementary degrees-of-freedom (DoF). 
The QCD Lagrangian (with one flavor of quark) is,
\begin{eqnarray}
\mathcal{L} = \bar{\psi_i} \left(i\gamma_\mu D^\mu_{ij} -m \delta_{ij} \right)\psi_j - \frac{1}{4}G_{\mu\nu}^a G^{\mu\nu,a},
\end{eqnarray}
where $\psi_i$ is the Dirac spinor of the quark field with color $i=1,\cdots, N_c$. 
There are three types of color charges $N_c = 3$ in the physical world.
\begin{eqnarray}
D_{ij}^\mu = \partial^\mu - i g T_{ij}^a A^{\mu, a}
\end{eqnarray}
is the covariant derivative, containing the interaction between quark field and the gluon field with coupling strength $g$.
Here $T_{ij}^a$ are the generators of the SU(3) group in the fundamental representation in the color space and they satisfy the commutation relation,
\begin{eqnarray}
[T^a, T^b] = i f^{abc} T^c
\end{eqnarray}
where $f^{abc}$ are known as the structure constants of SU(3).
The field tensor of the gluon field with color $a$ is,
\begin{eqnarray}
G^{\mu\nu,a} = \partial^\mu A^{\nu, a} - \partial^\nu A^{\mu, a} + g f^{abc} A^{\mu,b}A^{\mu,c}. \label{eq:gluon-L}
\end{eqnarray}
The gluon field transforms as the adjoint representation of SU(3) and $a$ index runs from 1 to $N_c^2 - 1$.
The first term in equation \ref{eq:gluon-L} is the kinetic term, and the second term is the gluon field self-interaction (also with strength $g$), which is a unique feature of the non-Abelian gauge field.

\subsection{Asymptotic freedom and confinement}
Due to quantum fluctuations, the effective coupling strength $g$ changes with the energy scale of a process. 
The rate of change of $g$ with respect to the scale parameter is called the $\beta$-function,
\begin{eqnarray}
\frac{\partial g}{\partial \ln\mu} = \beta(g),
\end{eqnarray}
which can be evaluated as a perturbation series of $g$ at weak coupling.
At leading order, the QCD $\beta$ function with number of colors $N_c$ and $n_f$ flavors of quarks is,
\begin{eqnarray}
\beta(g) = - \left( \frac{11}{3}N_c - \frac{2}{3}n_f \right) \frac{g^3}{16\pi^2}.
\end{eqnarray}
This $\beta$-function is negative for QCD ($N_c=3$) using realistic number of quark flavors $n_f = 2\cdots 6$, meaning the effective coupling constant decreases with increasing energy scale.
This property is known as the asymptotic freedom of QCD because the interaction becomes small at asymptotically high energy, which also makes possible the use of perturbation theory in such limit.

Often, the strong coupling constant is defined as $\alpha_s = g^2/4\pi$.
Using the leading order $\beta$-function, its scale dependence is
\begin{eqnarray}
    \alpha_s(Q^2) = \frac{4\pi}{\left(\frac{11}{3}N_c - \frac{2}{3}n_f\right)\ln\left(\frac{Q^2}{\Lambda^2}\right)}.
\end{eqnarray}
The integration constant has been absorbed into the QCD scale parameter $\Lambda$.
Therefore, at least in perturbation theory, $\Lambda$ becomes the only parameter of QCD. 
Its value is determined by anchoring $\alpha_s(\mu)$ to an experimental measurement at a fixed scale, for example, at the scale that equals to the $Z$ boson mass where $\alpha_s(M_z) = 0.1185$.
The leading order $\Lambda$ is then around $200$ MeV.

The decrease of $\alpha_s(Q)$ is logarithmically slow at high energy, but it rises quickly when $Q$ approaches $\Lambda$ from above.
Even before reaching this scale, the coupling constant is already too large for a reliable perturbative calculation.
Near the $\Lambda$ scale, QCD enters the non-perturbative region.
Nowadays, the only reliable {\it ab initio} theoretical tool for solving non-perturbative QCD is lattice field theory, where the QCD Lagrangian is discretized on a finite lattice and studied on a computer.

At long distances ($l \gtrsim 1/\Lambda$), only hadrons exist as color-neutral bound states of quarks and gluons.
The fact that color is not directly observable at large distances is known as ``color confinement'' of QCD. 
To pull a quark out of the hadron, the color field becomes so strong that eventually more quark-antiquark pairs populate the space in-between the pulled quark and the remnant and form new hadrons.
Depending on its valence quarks\footnote{ quarks that carry the net quantum number of the hadron} content, hadrons are generally categorized into baryons and mesons.
Baryons have three valence quarks or anti-quarks, such as neutrons and protons.
Mesons have a valence quark and an anti-quark, such as pions and kaons.
Hadrons are also populated with sea-quarks and gluons that are constantly produced and annihilated as quantum fluctuations.
The momentum of a hadron is mostly carried by the valence quarks.
Sea quarks and gluons together share the remaining fraction of the total momentum, but their abundance at high energy is very important to the particle production in relativistic hadron / heavy-ion collisions.

\subsection{The phase-diagram of the QCD matter}
\begin{figure}
 
    \centering
    \includegraphics[width=.8\textwidth]{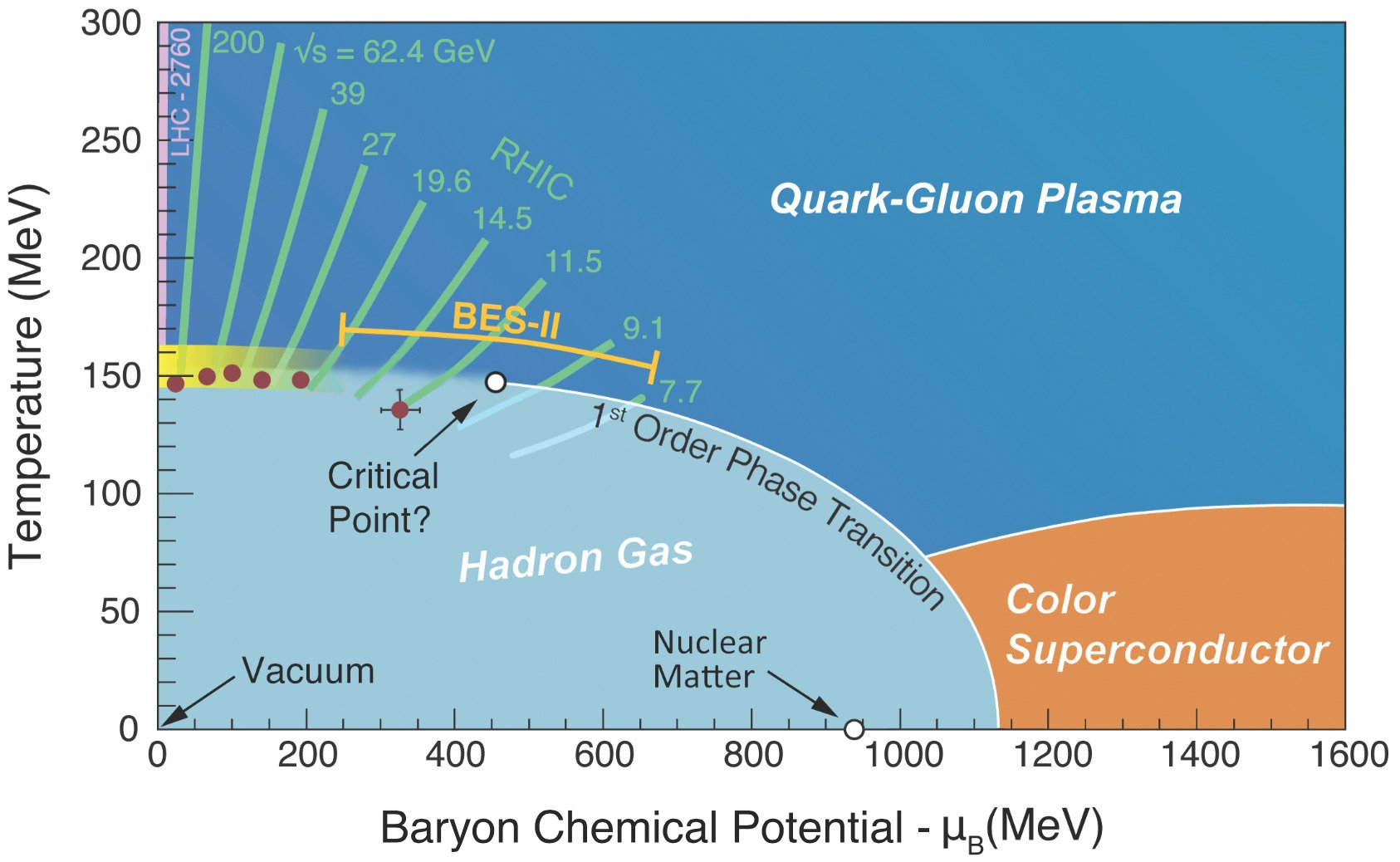}
    \caption[The (partly projected) phase-diagram of the nuclear matter from]{The (partly projected) phase-diagram of the nuclear matter from reference \cite{Geesaman:2015fha} with the red dots from lattice calculation \cite{PhysRevLett.109.192302}. The horizontal variable is Baryon chemical potential $\mu_B$ and the vertical variable is temperature $T$. The red and green trajectories indicates the reachable regions by the heavy-ion program at the LHC and RHIC.}
    \label{fig:phase-diagram}
\end{figure}

At zero temperature ($T$), protons and neutrons form bound states of atomic nuclei that are the building blocks of the ordinary matter.
One can define the baryon chemical potential $\mu_b$, which for ordinary matter is around $1$ GeV, close to the proton mass.
The region of ordinary nuclei is denoted as the white dot on the (partly conjectured) phase diagram in figure \ref{fig:phase-diagram} \cite{Geesaman:2015fha}.
If one increases the system temperature, nucleons start to escape from the nuclear potential, and thermal collisions, resonance formation and decay may create other hadrons.
This system is known as the hadron gas (the cyan region in figure \ref{fig:phase-diagram}).

\begin{figure}
 
    \centering
    \includegraphics[width=.8\textwidth]{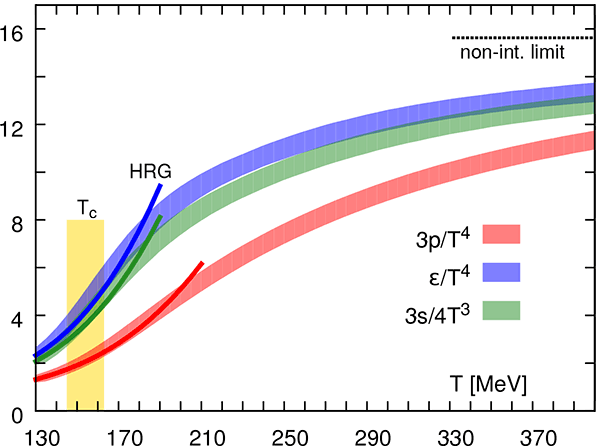}
    \caption[The lattice equation of state for 2+1 flavor QCD taken from]{The lattice equation of state for 2+1 flavor QCD taken from reference \cite{Bazavov:2014pvz}. The rescaled pressure, energy density and entropy density as functions of temperature at zero chemical potential are shown as red, blue, and green bands. The dashed lines denotes non-interaction (Stephan-Boltzmann) limit, and the solid lines show the expected values from a hadron resonance gas. The yellow band is the region of pseudo-critical temperature.}
    \label{fig:qcd_eos}
\end{figure}

Because QCD has asymptotic freedom at high energy and confinement occurs at a low energy scale, a so-called deconfinement phase-transition exists when the temperature crosses the QCD non-perturbative scale.
At asymptotically high temperature, the weakening of the coupling should lead to the transition from the color confined hadronic matter to a system of deconfined quarks and gluons, termed the quark-gluon plasma (QGP). 
First principle lattice QCD calculations have studied this transition at zero baryon chemical potential with 2+1 flavors (up, down plus strange quark).
Figure \ref{fig:qcd_eos} quotes the equation of state computed by the HotQCD Collaboration \cite{Bazavov:2014pvz}.
It shows the pressure $P$, energy density ($\epsilon$) and entropy density ($s$) of the system.
These thermodynamic quantities are scaled by powers of temperature so that the ratio can be loosely related to the effective number of degrees-of-freedom of the system.
The dashed line on the up-right corner denotes the Stefan-Boltzmann limit of non-interacting gas of quarks and gluons.
The effective number of DoF converges to the expectation from a hadron resonance gas model (solid lines) at low temperature and rapidly increases to a value closer to the Stefan-Boltzmann limit in a narrow temperature window.
This observation suggests a release of the quark and gluon in the system at high temperature.
More dedicated studies indicate that this is not a real phase-transition at $\mu_b = 0$ and refer it as a ``cross-over'' phase transition, where the thermodynamic quantities smoothly across this region of phase-diagram.
Nevertheless, a pseudo critical temperature can be defined using chiral condensate and susceptibility and is found to be $T_c \approx 154 \pm 9 $ MeV corresponding to 1.5 trillion Kelvin \cite{Bazavov:2011nk}.

With a finite baryon chemical potential, the lattice approach runs into the fermion sign problem, though recent studies have been pushing the realm of lattice QCD into regions of small $\mu/T$  \cite{Gunther:2016vcp,Bazavov:2017dus}.
Effective models studies have suggested the existence of a first-order phase transition at large $\mu_B/T$ and one may refer to \cite{Fukushima:2010bq} for a review.
If true, the first-order coexistence line must end at a point on the phase-diagram at lower $\mu_b$, beyond which the phase-transition is of the cross-over type.
Such a point, called the critical-end-point (CEP), has attracted significant interests from both the theoretical and the experimental community.

It is believed that the QCD high-temperature phase-transition occurred in the early universe around microseconds after ``the Big Bang'' when its temperature drops down to the QCD scale.
In ``nowadays'' universe, compact stars are ``celestial laboratories'' to test the QCD equation-of-state in the high density and low-temperature region, providing crucial physical input for simulating the recently discovered gravitational wave emission from neutron star mergers \cite{TheLIGOScientific:2017qsa}.
In laboratories, we create hot and dense nuclear matter by colliding heavy nuclei at ultra-relativistic high energies.
While the created matter is transient and tiny compared to the cosmic nuclear matter, we can study not only thermodynamic properties but also essential dynamical properties of QCD in these experiments.

\section{Phenomenology of relativistic heavy-ion collision}
Relativistic heavy-ion collisions are currently the only tool to access the high energy density QCD medium in a laboratory.
Since 2000, the Relativistic Heavy-ion Collider (RHIC) at the Brookhaven National Laboratory (BNL) has been colliding gold nuclei at 200 GeV. 
The Large Hadron Collider (LHC) started its heavy-ion programs later, colliding lead nuclei at 2.76 TeV and 5.02 TeV.
Since then, evidence has been pointing to the existence of a new state-of-matter: the strongly coupled quark-gluon plasma (sQGP).

In this section, I shall introduce useful concepts and terminology used in heavy-ion collision physics.
Then I will review a few essential experimental observables and how they can help us understand the properties of the sQGP.

\subsection{Kinematics}
In ultra relativistic collisions, it is advantageous to use a new set of coordinates, related to the Cartesian coordinates by,
\begin{eqnarray}
x_\perp &=& x_\perp\\
\tau &=& \sqrt{t^2 - z^2}\\
\eta_s &=& \frac{1}{2}\ln\frac{t+z}{t-z}
\end{eqnarray}
where the $z$ direction aligns with the beam direction.
$\tau$ is called the ``proper time'' and $\eta_s$ is called the space-time rapidity.
One advantage of using this set of coordinates is that $\tau$ and $\eta_s$ transform much simpler than $t$ and $z$ under a Lorentz boost in the beam direction with velocity $\beta_z$,
\begin{eqnarray}
\tau' &=& \tau,\\
\eta_s' &=& \eta_s + \frac{1}{2}\ln\frac{1+\beta_z}{1-\beta_z}
\end{eqnarray}
Similarly, the four momentum $p^\mu$ is parametrized as 
\begin{eqnarray}
p_x &=& p_T\cos\phi\\
p_y &=& p_T\sin\phi\\
m_T &=& \sqrt{m^2 + p_T^2}\\
y &=& \frac{1}{2}\ln\frac{E+p_z}{E-p_z}.
\end{eqnarray}
$p_T$ is transverse momentum relative to the beam ($z$) direction, $\phi$ is the azimuth angle of particle emission. 
$m_T$ is referred as the transverse mass, and $y$ is the rapidity of a particle.
Besides, pseudorapidity is often used in experiments,
\begin{equation}
\eta = \frac{1}{2}\ln\frac{|p|+p_z}{|p|-p_z} = \frac{1}{2}\ln\frac{1+\cos\theta}{1-\cos\theta}
\end{equation}
It has the merits that it is directly related to the polar angle  $\theta$ of particle emission.
When the transverse mass is small compared to $p_z$, the pseudorapidity is also a good proxy of rapidity.

\subsection{Nuclear collision geometry}
Nuclei are extended objects.
The radius of heavy nuclei scales like $A^{1/3}$, where  $A$ is the atomic number; therefore, the collision geometry plays a far more critical role than it is in the proton-proton collision.
In the center-of-mass frame,  nuclei ``shrink'' in the $z$ direction by the factor $\gamma = (1-v^2)^{-1/2} = E/M$ due to Lorentz contraction.
$\gamma$ is about $100$ for gold nuclei at top RHIC energy and is larger than $2500$ for lead nuclei at the LHC.
As a result, the approaching nuclei takes a very short time to penetrate each other $t_L = 2R/\gamma$, while dynamics in the transverse direction can only propagate within a causal circle of $r < t_L$ that is much smaller than the nuclear radius.

\paragraph{Impact-parameter and centrality} Defining the impact parameter $\vec{b}$ as the transverse separation between the centers-of-mass of the two approaching nuclei, the initial deposition of the energy largely depends on $\vec{b}$.
The collision geometry is a useful handle to study QGP dynamics; however, it is impossible to control $b$ directly in high energy experiments.
What is used as an approximate geometry indicator is the so-called ``centrality''.
Centrality is defined in different ways (detector response, multiplicity or transverse energy) and with different kinematic cuts, but the idea is that the nuclear collision geometry strongly correlates with the particle production activity.
It is reasonable to anticipate that the average number of charged particles produced or the total transverse energy deposited within a particular detector's acceptance is higher if the collision is more central (small impact parameter), and is lower for a peripheral collision (large impact parameter).
Of course,  as fluctuations smear out the exact one-to-one correspondence between centrality and impact-parameter.
Correctly accounting for these fluctuations is particularly important for small collision systems, such as proton-lead and deuteron-gold collisions.

\paragraph{Centrality selection} Experimentally, one sorts a minimum-biased (a minimum set of event triggers) sample of recorded events according to the centrality definition, e.g., multiplicity. 
Then the events are binned by percentile.
For example,  the top 0--5\% highest multiplicity events are associated with the centrality class 0--5\%. 
The one uses a model to deduce the collision geometry of a specific centrality class.
Usually, the model is one of many variants of the Glauber model \cite{Miller:2007ri}, which we shall explain it in detail in section \ref{chapter:simulation}).
It computes the number of binary nucleon-nucleon ($N_{\textrm{bin}}$) collisions and number of participant nucleons ($N_{\textrm{part}}$, nucleons that suffers at least one binary collisions) at a given impact parameter.
Experimentally, $N_{\textrm{part}}$ is often used as the centrality estimator of the model as it is roughly proportional to the bulk particle production; while the cross-section of hard processes that involves large momentum transfer $Q \gg \Lambda$ scales like the $N_{\textrm{bin}}$.
While this correspondence can be model dependent, the uncertainty can be quantified, and model predictions can be validated by studying the production of colorless probes such as photon and weak-boson \cite{Afanasiev:2012dg,Chatrchyan:2011ua,Aad:2012ew,Aad:2015lcb,Adam:2015lda}.
In particular, recent measurement of $Z$ boson production in Pb-Pb collisions at $\sqrt{s}=5.02$ TeV has reached a very high precision to constrain collision geometry models in the future \cite{ATLAS:2017zkv}.

\begin{figure}
 
\centering
\includegraphics[width=.8\textwidth]{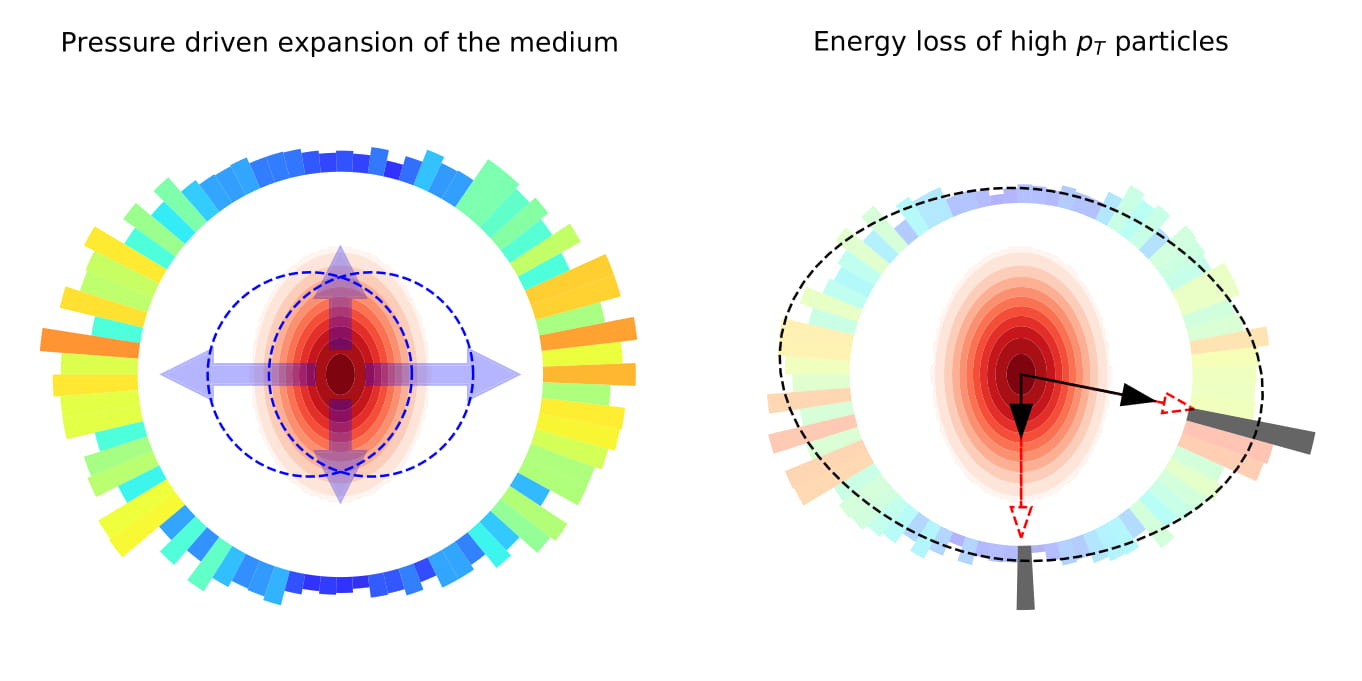}
\caption[Illustration of the relation between final state particles and]{Illustration of the relation between final-state particles and the formation of a dense medium and its geometry. Left: blue circles indicate the nuclei separated by a finite impact-parameter. An off-central collision creates an almond-shaped hot medium (red contour) in the overlapped region. The medium undergoes pressure-driven expansion and creates a $\cos(2\phi)$ like modulation in the final particle distribution (denoted as a colored histogram). Right: hard particles travel through the medium loses energy. On average, their final momenta (black arrow) are smaller than the original one (red arrow), and hard partons moving along the long-axis lose more energy than those moving along the short-axis direction.}
\label{fig:v2_demo}
\end{figure}

\subsection{Particle production at low-$p_T$ and collective flow} 
Immediately after the nuclei pass through each other at $t\sim 2R/\gamma$, a huge amount of energy is deposited, and entropy is produced, creating a fireball in the middle while the nuclear remnants recede.
This highly-excited fireball of fields undergoes complex dynamics and cools down rapidly due to its longitudinal and transverse expansion.
Eventually, the system hadronizes, and the hadrons can have further interactions and may decay into other hadrons, photons, and lepton that are measured by the detectors.

Particles produced in relativistic nuclear collisions distribute across a wide (pseudo)rapidity range, and have steep falling transverse momentum spectra \cite{Adler:2003cb,Abelev:2012hxa,ALICE:2015kda}.
The majority of the particles are soft hadrons with relatively small transverse momenta $p_T \lesssim 3$ GeV and their creation is a consequence of final-state interactions.
One of the most striking discoveries from the RHIC and the LHC heavy-ion programs is that these soft particles display a strong collectivity and the patterns are describable by relativistic viscous hydrodynamic-based models to a high precision \cite{Dusling:2007gi,Song:2008si,Schenke:2010nt,Petersen:2008dd,Niemi:2015qia,Bernhard:2016tnd,Bernhard:2018hnz}.
This success of the hydrodynamic model reveals the strongly coupled nature of the matter produced with a temperature of several times $T_c$, and it has been given the name strongly coupled quark-gluon plasma (sQGP).
An sQGP is in contrast to a weakly coupled gas of quarks and gluons that would not exhibit any collectivity.

\begin{figure}
 
\centering
\includegraphics[width=.6\textwidth]{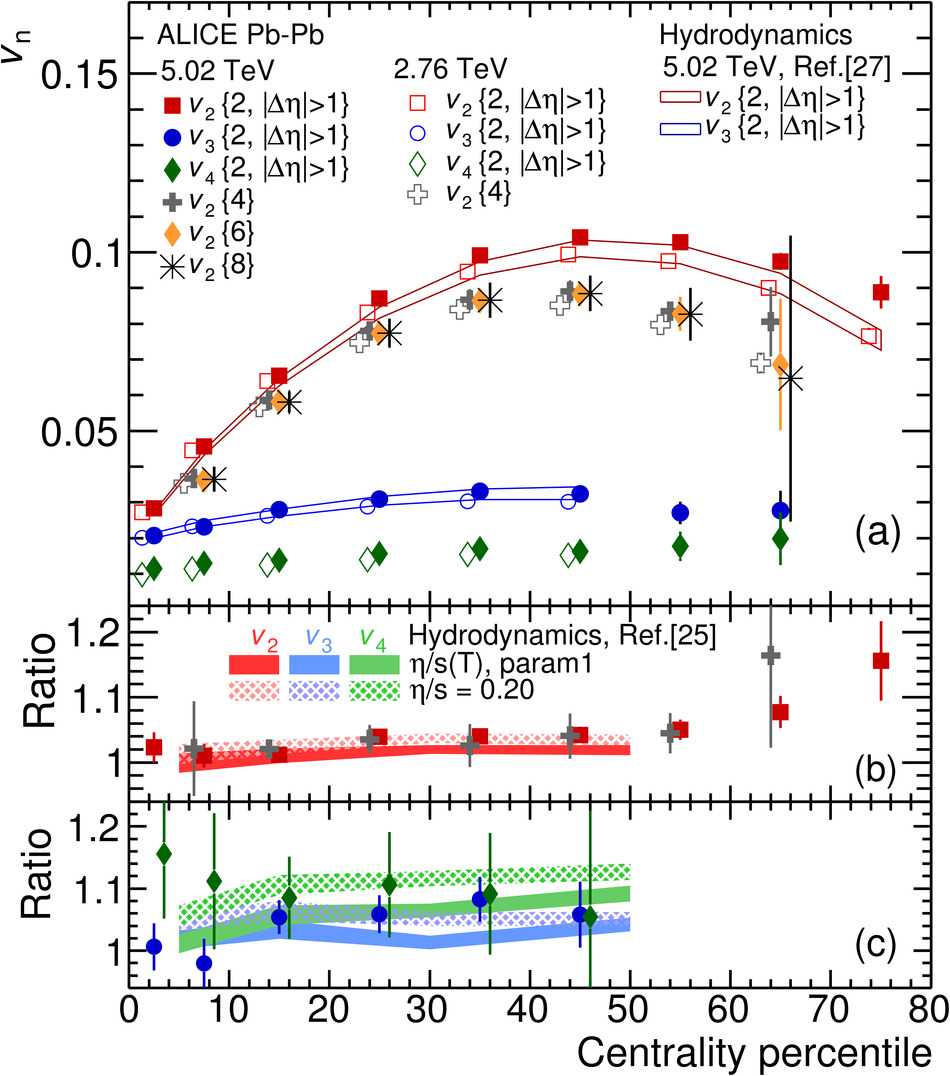}
\caption[The momentum anisotropy coefficient $v_2$ estimated from]{The momentum anisotropy coefficient $v_2$ estimated from multi-particle correlation for charged particle from the ALICE collaboration \cite{ALICE:2011ab,Adam:2016izf}.
The hydrodynamic-based calculations \cite{Niemi:2015voa,Noronha-Hostler:2015uye} (lines) very well explains the centrality dependence of the second (red), third (blue) and fourth (green) order coefficients at different beam energies.}
\label{fig:intro:vn}
\end{figure}

One manifestation of collectivity is the momentum-space anisotropy or collective flow of the bulk medium.
One decomposes the charged-particle spectra into a Fourier series of the azimuth angle,
\begin{eqnarray}
E\frac{dN}{p_T dp_T d\phi dy} = \frac{1}{2\pi}\frac{dN}{p_T dp_T dy}\left(1 + 2\sum_{n=1}^{\infty}v_n(p_T)\cos\left[n(\phi-\Psi_n)\right]\right).
\end{eqnarray}
The first term is an averaged yield, and subsequent terms in the sum encode the angular dependence. 
The $n=1$ term is the center-of-mass momentum of the distribution.
From $n=2$, $v_n$s are momentum anisotropy coefficients of $\cos({n\phi})$ modulations.
If the particle production is simply an independent sum of elementary nucleon binary collisions, then the anisotropy would be zero after the averaging process. 
However, experiments observe surprisingly large elliptic flow ($v_2$), triangular flow ($v_3$) ,and higher order $v_n$ at both RHIC and LHC in nuclear collisions.
Figure \ref{fig:intro:vn} shows the variation of the $v_n$ as function of centrality from ALICE measurements \cite{ALICE:2011ab,Adam:2016izf}.
The $v_2$ coefficient first increases from central to mid-central collisions and slightly decreases at peripheral collisions, while $v_3$, $v_4$ signals are smaller and vary slower with centrality.

In the hydrodynamic picture, the initial fireball in a non-central collision has a finite spatial eccentricity $\epsilon_n$ (please refer to the left of figure \ref{fig:v2_demo} for an illustration),
\begin{eqnarray}
\epsilon_n e^{i n\Phi_n} = \frac{\int dx_\perp^2 r^2 e^{i n\phi} \epsilon(x_\perp)}{\int dx_\perp^2 r^2 \epsilon(x_\perp)}
\end{eqnarray}
The energy density $\epsilon$ is higher in the middle and lower at the boundary, so a hydrodynamic pressure builds up and drives the transverse expansion of the fireball.
Because the pressure gradient is stronger in the short axis-direction and the long-axis direction, the matter flows in an anisotropic way, creating the observed momentum space second-order anisotropy $v_2$.
The existence of higher-order flow harmonics and non-zero $v_2$ in the most central collisions can be explained by nuclear configuration fluctuations. 
For example, randomized nucleon positions create all orders of eccentricity $\epsilon_n$.
In short, a hydrodynamic expansion transfers initial geometry eccentricities $\epsilon_n$ into final state momentum anisotropy $v_n$ of the particle.

\paragraph{Extracting the QGP transport coefficients}
A relativistic ideal hydrodynamic model assumes an infinitely strong interaction that the medium always stays in local thermal equilibrium.
A more sophisticated treatment is relativistic viscous hydrodynamics, which accounts for deviations from the local thermal equilibrium due to large gradients in the expansion.
Responses of the hydrodynamic evolution to the gradients are characterized by the shear viscosity $\eta$ and bulk viscosity $\zeta$.
The QGP shear viscosity and bulk viscosity are of fundamental importance. 
The shear viscosity to entropy ratio $\eta/s$ is an indicator of the strong/weak coupling nature of the QGP. 
The bulk viscosity to entropy ratio $\zeta/s$ is directly related to the scale-invariance breaking of QCD.

Dynamical quantities such as viscosity are extremely hard to compute from a first-principle approach, so currently, the determination of these numbers and their temperature dependence requires phenomenological extraction from experiments \cite{Muronga:2004sf, Chaudhuri:2006jd, Romatschke:2007mq, Dusling:2007gi, Song:2007ux, Luzum:2008cw}.
The flow harmonics $v_n$ are particularly sensitive to the viscous effects, as a finite viscosity dampens the development of anisotropic flows, reducing the transition efficiency from $\epsilon_n$s to $v_n$s.
Global comparisons of the state-of-the-art medium modeling to a collection of soft observables have corroborated the need of a small $\eta/s$ that is likely to be slowly increasing with temperature and a non-vanishing $\zeta/s$.

\subsection{Probing sQGP using hard probes}
Very occasionally, an initial collision involving large momentum transfer creates high-$p_T$ particles in the system ($p_T\gtrsim 10$ GeV) that referred to as ``hard'' particles.
By uncertainty principle, they can only be produced at the beginning of the nuclear collision on a time scale $\delta t \sim 1/p_T$, then they pass through and interact with the surrounding bulk medium.
Hard particles serve as self-generated probes of the system.
Due to asymptotic freedom, the initial production of hard processes can be computed in the perturbative framework, granting a good theoretical control of its initial state.
The final-state interaction with the medium then modifies the initial production and leaves fingerprints of the medium on the hard probes.

\paragraph{Jet and jet quenching}
Initial hard partons (gluons, quarks) undergoes complex QCD dynamics, radiating more partons which then hadronize into a collimated bunch of hadrons and decay products.
The final collection of particles observed by the detector is classified as jets.
In proton-proton collisions, perturbative calculations and Monte-Carlo simulations explain the production cross-section of the jet and leading hadrons (the hardest hadron in the jet) reasonably well.
In nuclear collisions, the initial parton and its radiative daughter partons interact with the medium, causing energy loss and triggering medium-induced radiation.
As a result, one expects the jet / leading parton yield at high-$p_T$ being reduced compared to the reference in proton-proton collisions (please refer to the right of figure \ref{fig:v2_demo} for an illustration).
Focusing on the difference caused by medium effects, the reference has to cancel out a na\"ive difference that rises simply because there are more effective nucleon-nucleon collisions in a nuclear collision.
One defines the so-called ``nuclear modification factor'',
\begin{eqnarray}
R_{AA}(y, p_T) = \frac{\frac{dN_{AA}}{dy dp_T}}{\langle N_{\textrm{coll}}\rangle \frac{dN_{pp}}{dy dp_T}} = \frac{\frac{dN_{AA}}{dy dp_T}}{\langle T_{AA} \rangle \frac{d\sigma_{pp}}{dy dp_T}}.
\end{eqnarray}
It is the ``average-$N_{\textrm{bin}}$-normalized'' ratio between the yield in AA collisions and pp collisions.
The number of binary collisions is sometimes replaced by the average nuclear overlapping function $\langle N_{\textrm{coll}}\rangle = \langle T_{AA} \rangle /\sigma_{pp}^{\textrm{inel}}$, and the yield is replaced by the inclusive cross-section for the proton-proton collision $\frac{dN_{pp}}{dy dp_T}\rightarrow \frac{d\sigma_{pp}}{dy dp_T}$.
These two expressions are equivalent.
The ratio is expected to be unity if there is no medium effect, though we do remind the readers that $N_{\textrm{coll}}$ is not a directly observed quantity and has to be estimated in a model-dependent way.

At both RHIC and LHC, colored probes as measured by the $R_{AA}$ of leading hadrons and jets are found to be significantly below unity in nuclear collisions, while the $R_{AA}$ of color neutral probe such as $Z$-boson is consistent with unity \cite{Adare:2008qa,Chatrchyan:2011ua,Afanasiev:2012dg,Aad:2012ew,Aad:2015lcb,Adam:2015lda,ATLAS:2017zkv}.
These discoveries indicate the creation of a color-deconfined medium that strongly modifies the hard parton propagation.
The interaction strength between the hard parton and the medium is theoretically quantified as the jet transport coefficient $\hat{q}$. It is defined as the momentum broadening per unit path-length in the direction transverse to the direction of motion,
\begin{eqnarray}
\hat{q} = \frac{d\langle p_\perp^2 \rangle}{dL}
\end{eqnarray}
The jet transport coefficient is another quantity of fundamental interest  in heavy-ion collisions, and there has been a great effort in both first principle computation and phenomenological extraction \cite{Wang:1994fx,Zakharov:1996fv,Baier:1996sk,Zakharov:1997uu,Arnold:2002zm,Gyulassy:2003mc,Kovner:2003zj,Jeon:2003gi,CasalderreySolana:2007pr,Djordjevic:2008iz,Bass:2008rv,Schenke:2009gb,Majumder:2009zu,Majumder:2010qh,Armesto:2011ht,Zapp:2011ya,Ovanesyan:2011xy,Kang:2014xsa,Cao:2016gvr,Kauder:2018cdt,Cao:2017zih}.
The hard parton / jet probes the medium at different scales at different stages of its evolution, and more sophisticated observables and theoretical tools are being constructed to answer more microscopic questions such as the effective degrees of freedom of the strongly coupled QGP.
Future experimental upgrades might provide the precision to look into this problems \cite{ATLAS-Collaboration:2012iwa,Abelevetal:2014dna,STAR:upgrade-hf,Adare:2015kwa,CMS:2017dec}.

\paragraph{Heavy-flavor probes}
Heavy flavors are the primary focus of this work.
Heavy quarks have masses well above the QCD non-perturbative scale.
The charm quark ($M=1.3$ GeV) and the bottom quark (M=$4.2$ GeV) are the most relevant ones for the present study.
The reason that the top quark ($M = 173$ GeV) is out of our discussion is due to its extremely short lifetime ($\sim 5\times 10^{-25} \approx 0.15$  fm/$c$ in its rest frame), so it barely interacts with the QGP before it decays into, predominantly, bottom quarks.
Though there has also been a proposal that takes advantage of this short lifetime to probe the temporal structure of the QGP evolution \cite{Apolinario:2017sob}, we only focus on the charm and bottom flavor in this work.

\begin{figure}
 
\centering
\includegraphics[width=.7\textwidth]{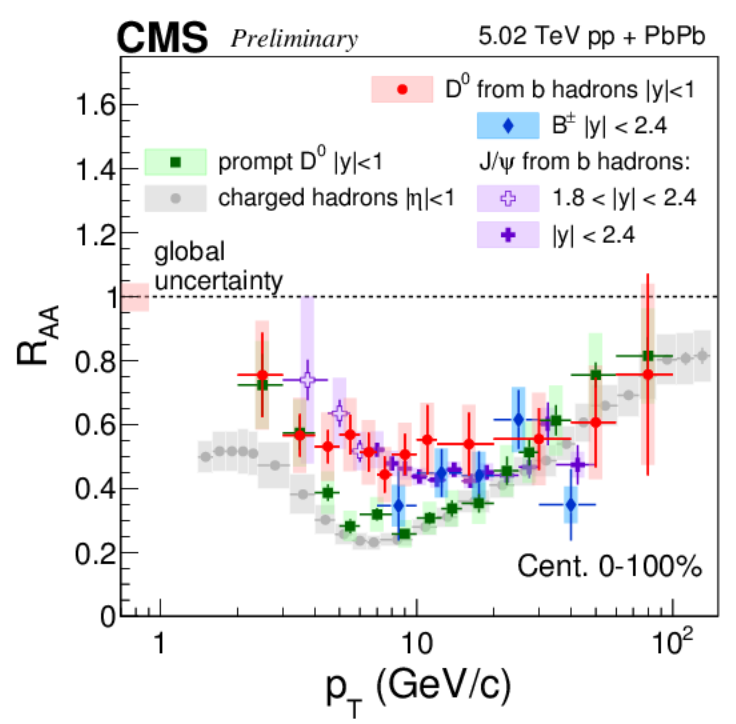}
\caption[The nuclear modification factor $R_{AA}(p_T)$ of charged hadron (gray),]{The nuclear modification factor $R_{AA}(p_T)$ of charged hadron (gray), D meson (green), B meson (blue), $b$-decayed D meson (red), and $b$-decayed J$/\Psi$ (purple) measured by the CMS collaboration \cite{Khachatryan:2016odn,Sirunyan:2017isk,Sirunyan:2017xss,Sirunyan:2017oug} (figure from Matthew Nguyen).}
\label{fig:intro:Raa}
\end{figure}

A large mass guarantees a negligible thermal production contribution, at least for the present top LHC beam energies (there are estimates that thermal production can play a role for the future FCC collider \cite{Zhou:2016wbo}).
Therefore, heavy flavors, regardless of $p_T$, are almost always created in initial hard processes.
Moreover, the tiny population of heavy flavors in the collision also suppresses the chances that they annihilate/recombine with their anti-particles.
Certainty heavy mesons have a long lifetime such that their decay vertices are outside of the fireball and can be resolved by experiments.
Therefore, the number of heavy-flavor particles is almost conserved from the beginning to the end of the entire medium evolution, including both the QGP phase and the hadronic phase.

\begin{figure}
 
\centering
\includegraphics[width=\textwidth]{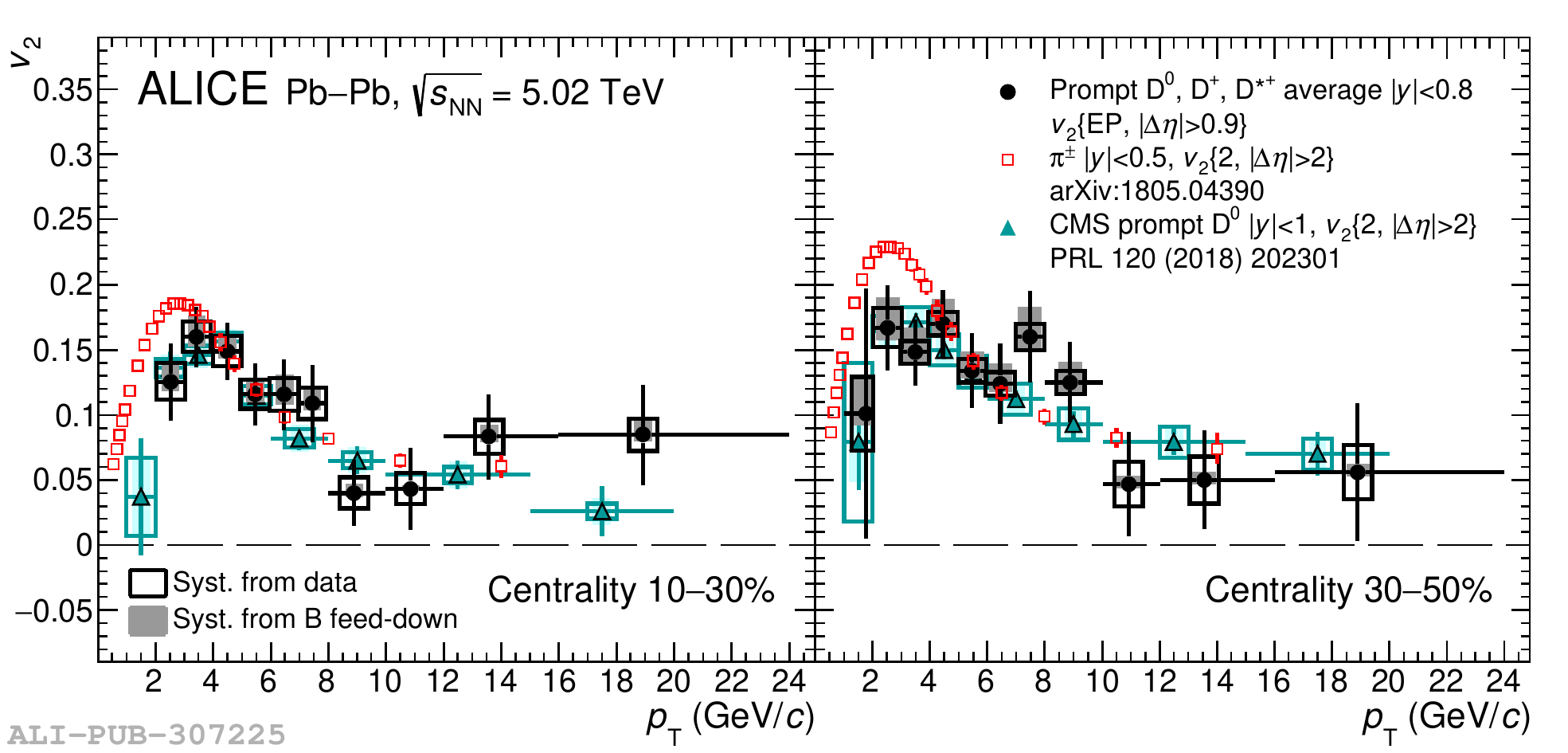}
\caption[The charged pion (red) and D meson momentum anisotropy]{The charged pion (red) and D meson momentum anisotropy measured by the ALICE Collaboration (black) and the CMS Collaboration (green). Left and right panels show the results for 10-30\% centrality and 30-50\% centrality respectively.}
\label{fig:intro:D-vn}
\end{figure}

Heavy flavors are of physical interests in many ways.
Mass effects are less important at very high $p_T$ , and heavy flavors are ideal flavor-tagged probes for jet energy loss studies.
At intermediate $p_T$, the mass effect is expected to suppress the medium-induced radiation, which dominates the energy loss of light quark.
There may also be a competition between collisional and radiative energy loss.
Experimentally, these differences lead to a hierarchy in the amount of nuclear suppression depending on the parton masses.
For example, figure \ref{fig:intro:Raa} from the CMS collaboration shows a collection of $R_{AA}$ measurements, for charged (mostly light flavor) hadron, prompt D meson (charmed meson), prompt B meson (bottom meson) and D and J/$\Psi$ meson from $b$ quark decays.
All the $R_{AA}$ tends to collapse onto the same trend at very high-$p_T$, while for $p_T$ range around 5 to 20 GeV, despite the large uncertainty, there is a suggestive hierarchy of $R_{AA}(\textrm{light}) < R_{AA}(\textrm{charm}) < R_{AA}(\textrm{bottom})$.
It would be interesting to study whether a theoretical framework can explain this difference quantitatively and distinguish different energy loss mechanisms.

For low-$p_T$ heavy quark, collisional process dominates over radiative process, and the description of the heavy-flavor dynamics under the influence of medium ``kicks'' reduces to a succinct diffusion equation \cite{Moore:2004tg}.
A spatial diffusion constant $D_s$ controls the close-to-equilibrium dynamics, and can be related to the momentum diffusion by $D_s = 4T^2/\hat{q}(p\rightarrow 0)$.
The large inertia $M\gg T$ delays its relaxation time $\tau_{\textrm{th}} \propto M/T D_s$ and one expects to find a different degree of thermalization for light, charmed, and bottom hadrons during the final lifetime of the created QGP fireball \cite{PhysRevD.37.2484,Moore:2004tg,Riek:2010fk,Cao:2013ita}.
For instance, take a look at the low $p_T$ momentum anisotropy shown in figure \ref{fig:intro:D-vn} \cite{Acharya:2017qps,Sirunyan:2017plt},
the large $v_2$ of the charged pions below $3$ GeV is due to the collective expansion.
D mesons also catch up a significant amount of flow \footnote{ As a remark, the finite $v_2$ of D meson at high $p_T$ is not directly related to the flow phenomena, but as a result of anisotropic energy loss in a spatially eccentric medium}, though still less than the pion.
To explains this amount of D meson $v_2$, phenomenological studies suggests a $D_s$ close to the lattice QCD calculations \cite{He:2012df,Cao:2013ita,Xu:2017obm,Banerjee:2011ra,Ding:2012sp,Francis:2015daa}, while leading order weakly coupled result \cite{Moore:2004tg} is inadequate . 
This finding suggests the importance of non-perturbative phenomenon in understanding the coupling between low momentum heavy quark and the medium.

Finally, the unique flavors of heavy quarks help to tag specific processes of interest,
for example, the study of recombination hadronization mechanism, strangeness enhancement, and implementation of selection bias on the quark / gluon-initiated jet ratio.

\subsection{Transport modeling of hard probes}
The understanding of jets and heavy-flavor in heavy-ion collisions needs a comprehensive non-equilibrium modeling effort.
Such a model includes initial production and evolution of hard partons, partonic propagation in the QGP, and eventually hadronic interaction.
Transport equations are convenient tools to couple the microscopic hard probes propagation to the macroscopic medium evolution, though one has to be very careful with the multiple scales of the problem.
For example, the hardest scale is the initial parton transverse momentum $p_T$.
The probe-medium interactions happen at typical scales around temperature $T$ and the screening mass $m_D \sim gT$, while the medium-induced radiation occurs at a momentum broadening scale of order $\hat{q} t$.
Eventually, hadronization happens at scale $\Lambda$.
Meanwhile, the medium evolution has a another set of (time) scales, the hydrodynamization time $\sim 1$ fm/$c$, the finite-size and life-time of the QGP fireball $\sim O(10)$  fm/$c$, and the expansion time scale $\tau_{\textrm{ex}}$.
Usually, separation of scales allows significant simplification to theory calculations.
For example, the Boltzmann transport equation requires a separation between the mean-free-path and the coherence time of the scattering.
However, in realistic event simulations, treating regions of overlapping scales seems inevitable.
We shall develop in this thesis a new transport model to account for a few of these overlapping scale issues.

\paragraph{Heavy quark transport models}
In early days, the heavy quark measurements at RHIC energy did not extend to very high-$p_T$, and the radiative energy loss is not as important as elastic ones.
Therefore, early studies relied on non-equilibrium dynamics using a pure diffusion model \cite{Moore:2004tg,vanHees:2007me}. 
Later, a radiation-improved diffusion model was developed \cite{Cao:2013ita} to include high-$p_T$ radiation processes, and has been applied to the first Bayesian extraction of the heavy quark transport coefficients \cite{Xu:2017obm}.
Apart from the diffusion-based models, Boltzmann and linearized Boltzmann models, including both elastic and inelastic interactions, has also been developed \cite{Scardina:2017ipo,Cao:2017hhk,Ke:2018tsh}.
Regarding the physical inputs, the Boltzmann-equation-based models take a weakly-coupled picture; the diffusion-based model is more flexible since the transport coefficient can be computed in both weakly coupled or strongly coupled approaches.
Using different models, the extracted transport parameters $D_s$, $\hat{q}$ have notable differences \cite{Rapp:2018qla,PhysRevC.99.014902,Cao:2018ews}.
The major sources that lead to these differences are:
\begin{itemize}
\item Inclusion of radiative energy loss.
\item Assumptions on the medium: close-to-equilibrium hydrodynamic medium, or non-equilibrium medium from a full partonic Boltzmann equation.
\item Weakly coupled approach versus strongly coupled approach.
\item More subtle differences such as Langevin versus Boltzmann dynamics, and the detailed treatment of the radiative processes.
\end{itemize}
To make progress, improved theoretical calculations and more accurate modeling are important; on the other hand, more subtle differences can be treated as intrinsic modeling uncertainty so that the extracted transport parameters are not over-interpreted by putting finite theoretical uncertainty bands on the extracted $D_s$ and $\hat{q}$. 

\subsection{Understanding QGP as a parameter inference problem}
All the interesting dynamics of heavy-ion collisions last for $O(10) $ fm/$c$, while we can only observe the collision remnants by detectors meters away.
Therefore, the determination of any intrinsic properties of the QGP is essentially  a parameter inference problem:
given measurements, models, and parameters of QGP such as $\eta/s, \hat{q}$, and what the favored range of parameters to explain the data is.
Finally, by comparing the theoretical expectations and the phenomenological constraints of these parameters, one can evaluate the theoretical assumptions, which provide new information on the QGP.

This inversion from observables to parameters is not as simple as it appears, because of the following difficulties,
\begin{itemize}
\item The dynamical models are complex and computationally expensive.
\item Models take multiple parameters or unknown functions that have infinitely many degrees of freedom.
\item Global comparison to many experimental observables.
\item The quantification of uncertainty: including experimental uncertainty, model uncertainty, and theoretical uncertainty.
\end{itemize}
A Bayes analysis for model parameter calibration solves most of these issues, and its key ingredients will be explained in chapter \ref{chapter:bayes}.
One remaining issue is the theoretical uncertainty, which is hard to quantify.
Our solution regarding the theoretical uncertainty is two-fold.
First, if there exist several theoretical assumptions without compelling reasons to disfavor either of them, then this difference should be propagated into the extracted model parameters.
Including these uncertainties will undoubtedly decrease the constraining power on the transport parameters, but it prevents biasing the estimated number from imposing an assumption that is too strong.
Second, existing theoretical calculations are often worked out in certain idealized limits, while a dynamical modeling approach is much more complicated and may not strictly follow the underlying theory.
Such differences between theory and model implementations obscure the interpretation of the extracted parameters.
Therefore, as an essential practice for dynamical modeling, the model should be able to calibrate to theoretical calculations at least in those idealized limits, and then be generalized to the more complex scenarios.
We devoted chapter \ref{chapter:transport} to improve the accuracy of hard parton transport model.

\section{Outline of this Thesis}
In this thesis, I focus on the extraction of the heavy quark transport coefficients from experiments using a newly developed transport model for hard parton propagation in a QGP.
In chapter \ref{chapter:simulation}, I introduce a hydrodynamic-based model for medium evolution. 
As an application of this simulation framework, I review my earlier project on reverse engineering a three-dimensional initial entropy deposition of the heavy-ion collision from experimental data. 
In chapter \ref{chapter:transport}, we develop the transport model for hard parton (including heavy flavor) propagation inside the QGP.
This model interpolates a small-angle diffusion picture and a large-angel scattering picture of the probe-medium interaction.
I show the limitation of the semi-classical transport approach in implementing parton branching processes (radiation) at high energy and demonstrate how to modify the semi-classical approach to treat it properly.
Chapter \ref{chapter:coupling} builds a comprehensive simulation workflow that couples the initial production and in-medium transport of heavy flavors to the bulk medium evolution.
Benchmark calculations with simple guesses of parameters are compared to the experimental measurements.
Chapter \ref{chapter:bayes} is a brief description of the Bayes methodology of model parameter calibration.
Applying the Bayes method, in chapter \ref{chapter:results}, a systematic model-to-data comparison is performed, extracting the heavy flavor transport properties.
Finally, chapter \ref{chapter:conclusion} summarizes the work and discuss possible future improvements.

\chapter{Bulk-medium evolution and initial conditions}
\label{chapter:simulation}
This chapter introduces a hydrodynamic-based model for the medium evolution in heavy-ion collisions.
I will also review my project on applying this framework to the reverse engineering of the initial three-dimensional entropy deposition from experimental observables \cite{PhysRevC.96.044912}.

\section{A Hydrodynamics-based dynamical modeling}
\subsection{Relativistic hydrodynamic}
Relativistic hydrodynamics plays a central role in the modeling of heavy-ion collisions.
It is relativistic as the flow velocity of the QGP fireball expansion can reach a significant fraction of the speed of light.
It is a macroscopic description that propagates the energy-momentum tensor of the system without detailed knowledge of the microscopic interaction.
The first set of equations comes from energy-momentum conservation, which should always be satisfied,
\begin{eqnarray}\label{eq:hydro:0-4}
\partial_\mu T^{\mu\nu} = 0.
\end{eqnarray}
$T^{\mu\nu}$ is the energy momentum tensor and $\partial_\mu = \partial/\partial x^\mu$. 
Here we have chosen the metric as $g^{\mu\nu} = \diag\{1, -1, -1, -1\}$.

\paragraph{Ideal hydrodynamics} Ideal hydrodynamics assumes that the system relaxes to local thermal equilibrium much faster compared to other time scales. Then, $T^{\mu\nu}$ can be specified by given only the energy density $e$, pressure $P$, and flow velocity $u^\mu$ of a fluid element,
\begin{eqnarray}
\partial_\mu T^{\mu\nu} = e u^\mu u^\nu - P (g^{\mu\nu}-u^{\mu\nu}).
\end{eqnarray}
Boosting into the co-moving frame of the fluid element, $T^{\mu\nu}$ reduces to the diagonal form $T^{\mu\nu} = \diag\{e, -P, -P, -P\}$.

There are five unknowns $e, P, u_x, u_y, u_z$ ($u_t$ is determined by $u^2 = 1$), but the conservation laws \ref{eq:hydro:0-4} only provide four equations.
A fifth equation is the equation-of-state (EoS) $P = P(e)$, relating pressure and energy density in the thermal equilibrium, which completes the ideal-hydrodynamic equations.
Lattice QCD calculations have determined the 2+1 flavor QCD EoS to high precision.
Though it is not a prior that the lattice QCD EoS computed in an infinite, static matter is the right choice for describing a transient system with large gradients, using the lattice input does result in reasonable agreement with the data.
Moreover, there has been a study that tries to constrain the form of the EoS from experimental data and the ``calibrated'' EoS is very close the Lattice calculation \cite{Pratt:2015zsa}.

\paragraph{Viscous hydrodynamics and QCD transport coefficients}
A physical relaxation rate is always finite, and the system can be driven out of local thermal equilibrium by large gradients of the fast-expanding fireball.
Relativistic viscous hydrodynamics takes into account these off-equilibrium effects by including viscous corrections.
The energy-momentum tensor deviates from the ideal one by a bulk viscous pressure $\Pi$, and a shear viscous tensor $\pi^{\mu\nu}$,
\begin{eqnarray}
T^{\mu\nu} = u^\mu u^\nu e - (g^{\mu\nu}- u^\mu u^\nu) (P+\Pi) + \pi^{\mu\nu}.
\end{eqnarray}
The $\Pi$ and $\pi^{\mu\nu}$ then respond to the finite gradients of the fluid field.
To first order in the gradient, they are given by the constitutive relations,
\begin{eqnarray}
\pi^{\mu\nu} &=& 2\eta\sigma^{\mu\nu},\\
\Pi &=& -\zeta\theta,
\end{eqnarray}
and the hydrodynamic equations are the relativistic version of the Naiver-Stokes equations.
Here, $\sigma^{\mu\nu} = \partial^{\langle \mu} u^{\nu\rangle}, \theta = \partial\cdot u$ are the fluid shear stress and expansion rate.
The proportionality constants are known as the QCD shear ($\eta$) and bulk ($\zeta$) viscosity, encoding dynamical information of the QCD medium. 
Their dimensionless ratio to the entropy density $\eta/s$ and $\zeta/s$ are highly-important indicators of the interaction strength and the scale-violation of the QCD matter.
A significant effort is underway to either compute these quantities from first principles or effective models or extract these numbers from experiments \cite{Bernhard:2015hxa,Bernhard:2016tnd,Auvinen:2017fjw,Bernhard:2018hnz,Novak:2013bqa}.

However, it has been shown that one has to go to the second order in the gradient expansion to render the viscous correction compatible with special relativity \cite{ISRAEL1976310}.
Meanwhile, $\pi$ and $\Pi$ become dynamical quantities that relax to the Naiver-Stokes limit.
\begin{eqnarray}
\tau_\pi \dot{\pi}^{\langle\mu\nu\rangle}+\pi^{\mu\nu} &=& 2\eta\sigma^{\mu\nu}- \delta_{\pi\pi}\pi^{\mu\nu}\theta + \phi_7 \pi_{\alpha}^{\langle\mu}\pi^{\nu\rangle\alpha}\\ 
\nonumber
&& -\tau_{\pi\pi}\pi_{\alpha}^{\langle\mu}\sigma^{\nu\rangle\alpha} + \lambda_{\pi\Pi}\Pi\sigma^{\mu\nu},
\\
\tau_{\Pi}\dot{\Pi} + \Pi &=& -\zeta\theta - \delta_{\Pi\Pi}\Pi\theta + \lambda_{\Pi\pi}\pi^{\mu\nu}\sigma_{\mu\nu}.
\end{eqnarray}
The $\tau_\pi, \tau_\Pi, \delta, \phi, \lambda$ are known as second order transport coefficients.
This complicated set of equations together with the conservation law and the EoS forms the viscous hydrodynamic equations.
Nowadays, well tested numerical packages have been developed to solve these equations in the context of heavy-ion collisions \cite{Song:2007ux,Shen:2014vra,Schenke:2010nt,Karpenko:2013wva}.

\paragraph{Boost-invariance approximation and beyond}
In general, the hydrodynamic equations have to be solved in 3+1 dimensions.
But a reduction to a 2+1 dimension is possible, if a boost-invariant symmetry is assumed near mid-rapidity \cite{Miller:2007ri, Drescher:2006ca, Schenke:2012wb, Niemi:2015qia, Moreland:2014oya, Chatterjee:2015aja}.
Bjorken first proposed this symmetry in \cite{PhysRevD.27.140}.
A ``boost-invariance'' symmetry assumes that the system at different space-time rapidity behaves the same up to a longitudinal boost.
Then, the 2+1 dimensional solution obtained at one space-time rapidity ($\eta_s = 0$) can be boosted to get the solution at other $\eta_s$.
As a result, particle emissions from such a fireball does not strongly depend on $\eta_s$.

The $\eta_s$-dependence cannot be detected but is related to the emitted particle rapidity/pseudorapidity for a reason below.
Suppose that the Lorentz contracted nuclei interact at $z=0$ and produce excitations that free-stream in the longitudinal direction, then for each of these excitations,
\begin{equation}
  \frac{z}{t} = \frac{p_z}{E}.
\end{equation}
This approximation provides the equivalence of $\eta_s$ and $y$ at early stages of the collision,
\begin{equation}
  \eta_s = \frac{1}{2}\log\frac{t+z}{t-z} \sim y = \frac{1}{2}\log\frac{E+p_z}{E-p_z}.
\end{equation}
Assume further that these initial excitations have small masses, then the rapidity can be approximated by pseudorapidity as $E\approx |p|$.
Since the following hydrodynamics is boost-invariant, then emitted particle pseudorapidity should be flat.
Experimentally, the event-averaged rapidity-distribution of charged particles $dN_{\textrm{ch}}/dy$ in both proton-proton and symmetric nuclei-nuclei collisions at the RHIC and energy falls off at large rapidity but has a central plateau at least within $|y|<2$.

However, since $dN_{\textrm{ch}}/dy$ are event-averaged quantities, it being flat within $|y|<2$ does not rule out event-by-event particle production fluctuation which breaks the boost-invariance in each event, and these fluctuations can be different at different transverse locations.
Moreover, asymmetric nuclear collision such as p-Pb, p-Au, d-Au, He-Au, etc clearly break the boost-invariance even on an event averaged level 
\cite{PhysRevLett.94.032301,Abelev:2014mda, Aad:2014lta, Aad:2013fja, CMS:2012qk, Chatrchyan:2013nka, Khachatryan:2015waa, Khachatryan:2015oea, Khachatryan:2016ibd, Adare:2014keg, Adare:2015cpn, Adare:2018toe}.
Therefore, the study of longitudinal fluctuation related observables or the search for hydrodynamic behavior in small collision systems requires one to go beyond the boost-invariance set-up.

\subsection{Particularization and microscopic transport}
The longitudinal and transverse expansion cools down the system temperature, and the density and relaxation rate also drop rapidly.
Eventually, the relaxation time is too long for the hydrodynamic approach to be applicable.
At this point, it is proper to switch to a microscopic transport model description.
This switching is usually performed near or below the pseudo-critical temperature $T_{sw} \lesssim T_c$ so that the energy-momentum tensor can be particularized as an ensemble of hadrons, whose interactions are well known.
If this matching is performed well above $T_c$, then we have to deal with the problem of modeling quark/gluon dynamics and hadronization in the strongly-coupled regime near $T_c$.
Of course, whether hydrodynamics with lattice EoS and a hadronic transport model are both valid in the vicinity of $T_{sw}$ is another question.

The hydrodynamic $T^{\mu\nu}$ is usually particlized at a constant-energy-density or a constant-temperature hypersurface using the Cooper-Frye prescription \cite{PhysRevD.10.186}.
The number of specie ``a'' particles emitted with momentum $p$ from a hypersurface element $i$ is computed from,
\begin{eqnarray}
\Delta N_i^a(p) = \frac{g^a f^a(p) dp^3}{(2\pi)^3}  \frac{p^{\mu}}{E} \Delta \sigma_{i,\mu}.
\end{eqnarray}
$f^a(p)$ is the phase space density, $\Delta \sigma_{i,\mu}$ are the four components of the surface element, with unit of a three-dimensional volume. 
$g^a$ is the degeneracy of the specie.
This distribution function should include both a equilibirum part and a viscous correction, $f = f_0 + \delta f$.
There are more than one way to construct the viscous correction $\delta f$  from $e, P, n, \Pi$, and $\pi^{\mu\nu}$ based on different assumptions of the non-equilibrium processes.
In this work, we use a non-additive $\delta f$ correction that has been developed in \cite{Pratt:2010jt,Pratt:2014vja} and implemented by \cite{Bernhard:2018hnz}.
Please refer to these references for the original formulation and numerical implementation details.

The particlized hadronic system is then solved by the Ultra-relativistic Quantum Molecular Dynamics (UrQMD) model \cite{Bass:1998ca,Bleicher:1999xi} until the system is dilute enough and interaction ceases, called the point of kinetic freezeout. 
The UrQMD model includes processes such as resonance decays, elastic and inelastic scatterings, and string formations and fragmentations.

\subsection{Pre-equilibrium stage}
At very early times of the collision, the system is off equilibrium. 
However, viscous hydrodynamic assumes a closeness to the local thermal equilibrium \footnote{ There are also recent efforts in understanding the effectiveness of hydrodynamics outside of its traditional range of applicability \cite{PhysRevLett.115.072501,Romatschke:2017vte,Strickland:2019jut}).}
A successful prediction using an early onset of hydrodynamic evolution at $\tau_0\lesssim 1$fm/$c$ suggests a fast hydrodynamization, though the mechanism is still under debate. 
There are different models of the pre-equilibrium stage available\footnote{  Such a pre-equilibrium stage was not included in my study of the 3D initial condition, and the hydrodynamics starts at $\tau_0 = 0.6$ fm/$c$ assuming local thermal equilibrium. For the later study of the heavy flavor transport, I used a 2+1D collision-less Boltzmann equation implemented by \cite{Bernhard:2018hnz}}, including solving the classical Yang-Mills equation \cite{Schenke:2012wb,Schenke:2016ksl}, applying partonic transport models \cite{PhysRevC.97.034915}, a collision-less Boltzmann equation (free-streaming) plus a Landau matching \cite{Liu:2015nwa}, and the linear response method of the effective kinetic approach \cite{Kurkela:2018wud}.

Here we briefly introduce the free-streaming model \cite{Liu:2015nwa}.
The initial energy density ($\tau = 0^+$) at mid-rapidity is thought to be carried by massless partons that propagate at the speed of light in the transverse direction. 
The initial distribution function is put into a factorized form $f(x_\perp, p_\perp, \tau=0) = n(x, \tau=0) \times dN/dp_\perp^2$.
The momentum distribution $dN/dp_\perp^2$ does not evolve with time as collisions are neglected, while the spatial density propagates as
\begin{eqnarray}
n(\vec{x}, \tau) = \int n(\vec{x}', \tau=0) \delta^{(2)}(\vec{x} - \vec{x}'- \tau) d\vec{x}'^2
\end{eqnarray}
Then, the model assumes a sudden hydrodynamization at time $\tau_{\textrm{hydro}}$, where the free-streamed
\begin{eqnarray}
T^{\mu\nu}(x_\perp, \tau_{\textrm{hydro}}) = \int f(x_\perp, p_\perp, \tau=0) \frac{p^\mu p^\nu}{E} dp^3
\end{eqnarray}
is used for initializing the hydrodynamic equations by the Landau matching procedure \cite{Liu:2015nwa}.

\section{Initial condition model}
Unlike the dynamical models that are governed by a few equations with a few parameters, the initial condition model parametrizes many more unknowns.
For different initial condition models, these unknowns can be the initial color density of the nuclear wave function, the effective size of a nucleon, the form factor of the nucleon-nucleon inelastic cross-section, and the amount of fluctuations in particle production/energy deposition, et cetera.
There are two classes of initial condition models:
\begin{itemize}
\item Models that take into account the particle-production dynamics. Such as mini-jet production, strings productions and flux-tubes, coarse-grained hadronic transport models \cite{Wang:1991hta, Zhang:1999bd, Werner:2010aa, Bozek:2015bna, Petersen:2008dd}, and color-glass condensate (CGC) effective-field-theory-based models \cite{Schenke:2016ksl, Dumitru:2011wq, Hirano:2012kj}.
\item Parametric models. Models provide macroscopic initial conditions without a dynamical component. Such as the Monte-Carlo Glauber models and its extensions \cite{Bozek:2015bha, Rybczynski:2013yba}, and the \trento\ initial condition model to be explained \cite{Moreland:2014oya}.
\end{itemize}
This section introduces the original boost-invariant \trento\ model and my work that extends the model to include longitudinal structure and fluctuations.

\subsection{The original (boost-invariant) \trento\ model}
The original \trento\ model is proposed as a flexible ansatz to investigate a family of entropy/energy deposition behaviors at mid-rapidity.
First, the impact parameter $\vec{b}_{AB}$ between the two colliding nuclei $A$ and $B$ is sampled at random.
Then, nucleon positions inside each nucleus are sampled according to the Woods-Saxon distribution\footnote{   The Woods-Saxon distribution is intended for heavy nuclei, for light nuclei such as Deuteron, Helium, Oxygen, the few-body wave function or pre-tabulated nuclear configurations are used for sampling},
\begin{eqnarray}
\frac{df_N}{r^2 dr d\phi d\cos\theta} = \frac{1}{\exp\{\frac{r-R(1+\beta_2 Y_{20}(\theta)+\beta_4 Y_{40}(\theta))}{a}\}+1}
\end{eqnarray}
including the quadrupole and hexadecapole deformation of certain nuclei.
The randomized nucleon position is critical to explain the odd order of flow harmonics observed in experiments \cite{Alver:2010gr}.

The collision between the two nuclei is determined at the nucleon level. 
Every nucleon pair $\{i, j\}$ with $i$ from nucleus $A$ and $j$ from nucleus $B$ has a certain probability of colliding inelastically.
This probability at given nucleon-nucleon impact-parameter $\vec{b}_{ij} = \vec{b}_{AB} + \vec{x}_{i, \perp} -  \vec{x}_{j, \perp}$ is parametrized as,
\begin{eqnarray}
P(b_{ij}; \sigma) = 1 - \exp\left[-\sigma T_{pp}(b_{ij})\right],
\label{dsigma_db}
\end{eqnarray}
where $T_{pp}(b)$ is the overlap function between the $z$-integrated density of the nucleon $T_p$,
\begin{eqnarray}
T_{pp}(b) = \int d\vec{x}_\perp^2 T_p(\vec{x}_\perp-\vec{b}/2) T_p(\vec{x}_\perp+\vec{b}/2).
\end{eqnarray}
Assuming a 3D Gaussian shaped nucleon with width parameter $w$, $T_p$ is
\begin{eqnarray}
T_p(\vec{x}_\perp^2) = \frac{1}{2\pi w^2} \exp\left(-\frac{\vec{x}^2}{2w^2}\right)
\end{eqnarray}
and $w$ is treated as a tunable parameter.
The $\sigma$ parameter in equation \ref{dsigma_db} is the nucleon opacity parameter with units of an area.
It is determined by fitting the $b$-integrated collision probability to the experimental measured proton-proton inelastic cross-section at a given beam energy,
\begin{eqnarray}
\sigma_{pp}^\text{inel}(\sqrt{s}) = \int d\vec{b}^2 P(b; \sigma(\sqrt{s})).
\end{eqnarray}
We apply the probabilistic collision criteria of equation \ref{dsigma_db} to each pair of nucleons and sample the binary collisions. 
Nucleons that suffer at least one inelastic collision are called participants, and the total number of binary inelastic collisions is denoted as $N_{\textrm{bin}}$.
A minimum-biased event sample in the \trento\ model includes all events that have at least one binary collision.

The above procedure is similar to the that of a Monte-Carlo Glauber model in determining the nuclear inelastic cross-section \cite{Miller:2007ri}.
Some experimental versions of the Glauber model use a black disk proton instead of a Gaussian profile.
We found that this difference results in little change in the centrality dependence of $N_{\textrm{part}}$, but can significantly affect $N_{\textrm{bin}}$ and may affect the computation of the hard probe nuclear modification factor.

The novel component of \trento\ is an ansatz that maps the participants to the energy/entropy density deposited at mid-rapidity.
Defining the participant densities as a sum of the participants' thickness function,
\begin{equation}
T_{A, B}(\x) = \sum_{i\in \textrm{Parts}_{A, B}} w_i\, T_p(\x - \x_i).
\end{equation}
The summation goes over all participants in nuclei $A$ and $B$, and each contribution is modulated by a fluctuating weight $w_i$ that follows a $\Gamma$-distribution.
The fluctuation has a unit mean and variance $1/k$, and $k$ is a free parameter.
This weight accounts for the large multiplicity fluctuation in minimum biased proton-proton collisions.
The entropy/energy density deposition at mid-rapidity is assumed to be a function that depends only on local values of $T_A$ and $T_B$,
\begin{eqnarray}
\frac{dS(\vec{x})}{d\eta dx_\perp^2} \textrm{ or } \frac{dE(\vec{x})}{d\eta dx_\perp^2} = f(T_A(\vec{x}), T_B(\vec{x})).
\end{eqnarray}
This simplification is possible because at $\tau=0^+$, causality requires that the entropy production at one location cannot be correlated with the information at a different transverse location. 
Also, the partons that contribute to the bulk low-$p_T$ particle production at high $\sqrt{s}$ are predominately low energy gluons whose longitudinal wavelength is longer than the contracted proton radius in the $z$-direction; therefore, the entropy production should not be sensitive to the details of how the participants are aligned but only its $z$-integrated density.
\trento\ parametrizes this mapping from $T_A$ and $T_B$ to the energy / entropy deposition using a ``generalized mean'' ansatz,
\begin{eqnarray}
f(T_A, T_B; p) = \left(\frac{T_A^p + T_B^p}{2}\right)^{1/p}.
\end{eqnarray}
$p$ is another tunable parameter.
Some special values of $p$ reduce the ansatz to well known averaging procedures as shown in table \ref{tab:trento-p}.
\begin{table}
\centering
\caption{\trento\ $p$-parameter}\label{tab:trento-p}
\begin{tabular}{ccc}
\hline
$p\in \mathbb{R}$ & $f(x, y)$ & Entropy / energy production\\
\hline
$-\infty$ & $\min\{x, y\}$ &  dominated by the thinner target\\
$-1$ & $2xy/(x+y)$ &   the harmonic mean scaling\\
$0$ & $\sqrt{xy}$ &  the geometric mean scaling\\
$1$ & $(x+y)/2$ &  the arithmetic mean (participant) scaling\\
$+\infty$ & $\max\{x, y\}$ &  dominated by the thicker target \\
\hline
\end{tabular}
\end{table}
This way the model is able to parametrize a class of entropy/energy production schemes and includes certain type of initial condition uncertainty.
Through a global model-to-data comparison, this $p$ parameter has been calibrated to be very close to 0, suggesting the data favors a mid-rapidity entropy/energy deposition that scales as $\sqrt{T_A T_B}$.
A similar scaling is also noticed in the EKRT initial condition model based on pQCD particle production and saturation physics \cite{Eskola:1999fc}.

\subsection{Parametrizing the longitudinal dependence in \trento}
The \trento\ model has been very successful describing observables at mid-rapidity \cite{Moreland:2014oya,Bernhard:2016tnd,Moreland:2018gsh}.
My work focuses on extending the parametrization to a rapidity-dependent initial condition and seeks a reverse-engineered 3D entropy production from rapidity-dependent observables \cite{PhysRevC.96.044912}.

Boost-invariance can be broken if the initial participant density is asymmetric ($T_A(\mathbf{x}_\perp) \neq T_B(\mathbf{x}_\perp)$); in addition, dynamical fluctuations can generate asymmetry as well.
We listed what contributions are included and what are not in the model
\begin{itemize}
\item In asymmetric collisions like p-A and non-central A-A, the local thickness functions are imbalanced $T_A \neq T_B$.
\item Even in central $A$-$A$ collision, nuclear / nucleon configuration fluctuations also contribute to the asymmetry. It is included as the randomized nucleon position fluctuation and the $\gamma$-fluctuation of the nucleon thickness function.
\item This model does not include dynamical fluctuation, either initial flow in the $z$-direction.
\end{itemize}
Therefore, the asymmetry in the extended \trento\ model only comes from the imbalance between $T_A$ and $T_B$.
Take the following decomposition of $s(\x, \eta_s)$ at the hydrodynamic starting time $\tau_0$,
\begin{equation}
  s(\x, \eta_s)\vert_{\tau=\tau_0} \propto f(T_A(\x), T_B(\x)) \times g(T_A(\x), T_B(\x), \eta_s).
  \label{factorized}
\end{equation}
$f$ is the entropy production at mid-rapidity as explain above.
The function $g$ parametrizes the rapidity-dependence and is always normalized such that $g(\eta_s=0)=1$ so that it reduces to the original model at mid-rapidity.
We parametrize the $g$ function in terms of rapidity and then transformed to the space-time rapidity.
\begin{eqnarray}
g(\x, \eta) &=& g(y; T_A(\x), T_B(\x)) \frac{J \cosh \eta_s}{\sqrt{1 + J^2 \sinh^2 \eta_s}},
\label{jacobian}
\end{eqnarray}
where the species-dependent factor $J$ is replaced with an effective value $J \approx \langle p_T \rangle / \langle m_T \rangle$.
To relate the asymmetry of $g(y)$ to the difference of $T_A, T_B$, we parametrize the $y$-cumulants of $g$ as functions of $T_A$ and $T_B$.
The $g(y)$ function is then reconstruct using its first few cumulants (mean $\mu$, standard deviation $\sigma$, and skewness $\gamma$) by,
\begin{eqnarray}
g(\x, y) &=& \mathcal{F}^{-1}\{\tilde{g}(\x, k)\}, \\
\log \tilde{g} &=&  i \mu k - \frac{1}{2}\sigma^2 k^2 - \frac{1}{6} i \gamma \sigma^3 k^3  e^{-\frac{1}{2}\sigma^2 k^2} + \cdots \label{eq:cgf}.
\end{eqnarray}
where for the skewness term, we have included an exponential factor that systematically includes higher order cumulants to regulate the behavior of the function at large $y$.
Numerically, we find that within the range of $|y| < 3.3\sigma$, the reconstructed function has good behavior and remains positive definite.
The mean, standard deviation (width), and the skewness are parametrized as follows,
\begin{itemize}
\item For the mean parameter, we assume it is proportional to the center-of-mass rapidity of the local participant density $\mu = \mu_0\eta_\text{cm}$,
\begin{equation}
  \eta_\text{cm}=\frac{1}{2} \log \left[\frac{T_A e^{y_b}+T_Be^{-y_b}}{T_A e^{-y_b}+T_B e^{y_b}}\right]
\end{equation}
where $y_b$ is the beam rapidity.
\item For the standard deviation, currently we leave it as a global parameter independent of the transverse location $\sigma = \sigma_0$, but only a function of the center-of-mass energy.
\item For the skewness, there is not a preferred form, so we tested two parametrizations. 
Also, in the end, we will check whether the 3D initial condition extracted from data is sensitive to the different parametrizations schemes.
These two choices are termed ``relative skewness'' and ``absolute skewness''.
\begin{itemize}
\item The ``relative skewness'' parametrization assume a skewness proportional to the relative difference of $T_A$ and $T_B$,
\begin{equation}
  \mathcal{A}(T_A, T_B) = \gamma_r\frac{T_A - T_B}{T_A + T_B},
\end{equation}
\item The ``absolute skewness'' parametrization assume a skewness proportional to the absolute difference of $T_A$ and $T_B$,
\begin{equation}
  \mathcal{A}(T_A, T_B) = \gamma_a \frac{T_A - T_B}{T_0}.
\end{equation}
And we put the unit for thickness function $T_0=1$~fm$^{-2}$ to restore $\gamma$ as a dimensionless parameter.
\end{itemize}
\end{itemize}
We have summarized the two parametrizations in table \ref{tab:parametrization}, and $\mu_0$, $\sigma_0$, $\gamma_r$ or $\gamma_a$, along with the effective Jacobian $J$ are the four additional parameters introduced for the three-dimensional extended \trento model.
\begin{table}
\centering
\caption{Rapidity cumulants parametrizations}\label{tab:parametrization}
\begin{tabular}{lccc}
\paddedhline
Model & mean $\mu$ & std.\ $\sigma$ & skewness $\gamma$ \\
\paddedhline \noalign{\smallskip}
Relative  & $\frac{1}{2} \mu_0 \ln\left(\frac{T_A e^{y_b}+T_B e^{-y_b}}{T_A e^{-y_b} + T_B e^{y_b}}\right)$ & $\sigma_0$ & $\gamma_r \dfrac{T_A - T_B}{T_A + T_B}$ \smallskip\\
Absolute & $\frac{1}{2} \mu_0 \ln\left(\frac{T_A e^{y_b}+T_B e^{-y_b}}{T_A e^{-y_b} + T_B e^{y_b}}\right)$  & $\sigma_0$ & $\gamma_a (T_A - T_B)/T_0$\smallskip\\
\paddedhline 
\end{tabular}
\end{table}

\begin{figure}
 
\centering
\includegraphics[width=.8\textwidth]{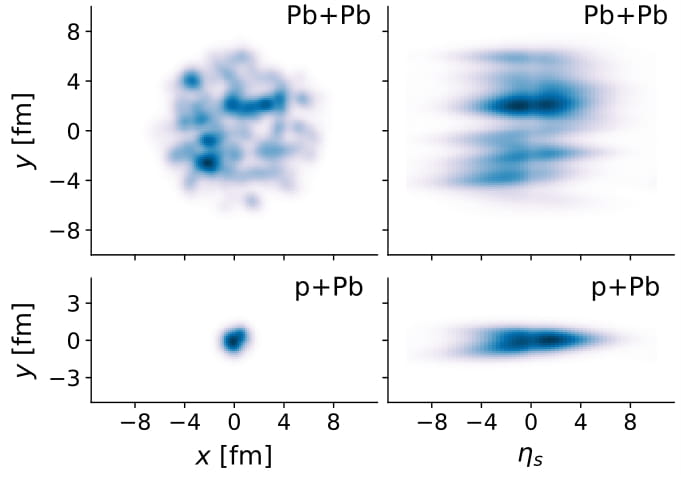}
\caption[Two sample events for Pb-Pb collision (top row) and p-Pb]{Two sample events for Pb-Pb collision (top row) and p-Pb collision (bottom row). The left column and the right column slice the initial entropy density at middle rapidity and the $x=0$ plane respectively. The relative skewness parametrization is used with $\mu_0=1$, $\sigma_0=3$ and $\gamma_r=6$.}
\label{fig:3d-example}
\end{figure}

In figure \ref{fig:3d-example}, we show two sample events from a Pb-Pb collision (top plots) and a p-Pb collision (bottom plots) generated by \trento.
The 3D initial entropy densities are sliced at mid-rapidity $\eta_s=0$ (left plots) and at the $x=0$ plane (right plots).
The mid-rapidity results are identical to the one predicted in the original \trento\ model.
The model is capable of generating fluctuating longitudinal structures that are local in the transverse plane, and breaking the boost-invariance both locally and globally.
For the $p$-Pb event, there is a clear structure that one hot spot extends into the proton going side $\eta_s >0$, while the participant clusters from the lead nuclei push the entropy production into the lead going side $\eta_s <0$.

\section{Reverse engineering the 3D initial condition}
In the final section of this chapter, I apply the hydrodynamic-based simulation framework and the flexible \trento-3D initial condition model to reverse engineer the 3D entropy deposition at the onset of hydrodynamics at LHC energies.

Experimentally, one can only measure rapidity-dependent observables on an event averaged level, which already integrates the contribution of particle production over the transverse plane; while our parametrization in the \trento\ model only involves local functions of the participant density function.
So it is a nontrivial task to infer the functional form of local entropy production $s(\x, \eta_s)$ from these ``global'' measurements.
The statistical technique for the reverse-engineering is the Bayes analysis explained in chapter \ref{chapter:bayes}.

\subsection{Sensitive observables to the initial entropy deposition}
\paragraph{The single particle spectra}
The most direct observable is the charged particle pseudorapidity density $\dnchdy$ measured for different collision systems and centralities.
The ALICE collaboration and the ATLAS collaboration has measured this quantity for both the Pb-Pb system ($-3.5<\eta<5.0$) and the p-Pb system ($|\eta| < 2.7$).
$\dnchdy$ can very well constrain the global rapidity profile and the centrality dependence of the mode, while the limitation being that it is less sensitive to the amount of longitudinal fluctuations.

\paragraph{Two particle pseudorapidity correlation}
A good probe of event-by-event longitudinal fluctuations is the two-particle pseudorapidity correlation observable $C(\eta_1, \eta_2)$,
\begin{eqnarray}
C(\eta_1, \eta_2) = \frac{ \left\langle N(\eta_1)N(\eta_2) \right\rangle}{\left\langle N(\eta_1)\right\rangle\left\langle N(\eta_2) \right\rangle}
\end{eqnarray}
The long range part of $C(\eta_1, \eta_2)$ is sensitive to the initial state.
It is because the correlation between particles separated by a large rapidity gap at proper time $\tau$ can only come from proper times before $\tau e^{-|\eta_1-\eta_2|/2}$ due to causality.
For example, if two particles separated by four units of rapidity are emitted at a constant $\tau = 8$ fm/$c$ hydrodynamic freeze-out hypersurface and neglecting the long-range correlation from the hadronic cascade, then any correlation must have come from before the proper time $ 8 e^{-2}\approx 1$ fm/$c$.

To see how $C(\eta_1, \eta_2)$ is related to the longitudinal fluctuation of the particle production, decompose $\dnchdy$ for each event in a finite pseudorapidity window $[-Y, Y]$ using the normalized Legendre polynomials basis \cite{Bzdak:2012tp, Jia:2015jga, ATLAS:2015kla},
\begin{eqnarray}
\frac{dN}{d\eta} &=& \biggl\langle\frac{dN}{d\eta}\biggr\rangle \biggl[1 + \sum_{n=0}^\infty a_n T_n\left(\frac{\eta}{Y}\right) \biggr],\\
T_n(x) &=& \sqrt{n + \frac{1}{2}} P_n(x)
\end{eqnarray}
Where $\langle\frac{dN}{d\eta}\rangle$ is the reference multiplicity at mid-rapidity for a certain centrality.
$a_0$ is the total multiplicity fluctuation and $a_1$ controls how the multiplicity distribution is tilted in rapidity in each event and so on.
Define the normalized event-wise distribution $R(\eta) = dN/d\eta /\langle dN/d\eta\rangle$, then the correlator is,
\begin{eqnarray}
C(\eta_1, \eta_2) &=& \left\langle R(\eta_1) R(\eta_2)\right\rangle \\
&=& 1 + \sum_{m, n}\langle a_m a_n\rangle  T_{mn}(\eta_1, \eta_2),\\
T_{mn}(\eta_1, \eta_2) &=& \frac{T_n(\eta_1)T_m(\eta_2) + T_m(\eta_1)T_n(\eta_2)}{2}.
\end{eqnarray}
Therefore, the two-particle correlation $C(\eta_1, \eta_2)$ measures the variance of these $a_n$ coefficients, and $\langle a_m a_n\rangle$ can be extracted by projecting $C(\eta_1, \eta_2)$ on to the basis function $T_{mn}$.

Combinations like $\langle a_0 a_n\rangle$ simply reflect how the event-wise rapidity fluctuation correlates with the multiplicity fluctuation. 
These contributions are canceled to first order by another normalization to define $C_N$,
\begin{eqnarray}
 C_N(\eta_1, \eta_2) &=& \frac{C(\eta_1, \eta_2)}{C_1(\eta_1)C_2(\eta_2)},\\
C_{1,2}(\eta_{1,2}) &=& \int_{-Y}^{Y}C(\eta_1, \eta_2)\frac{d\eta_{2,1}}{2Y}.
\end{eqnarray}
Finally, $C_N$ is directly related to the rapidity fluctuation itself,
\begin{eqnarray}
C_N(\eta_1, \eta_2) \sim 1 + \frac{3}{2}\langle a_1 ^2 \rangle \frac{\eta_1\eta_2}{Y^2} + \cdots.
\end{eqnarray}
We make use of the $\langle a_1 ^2 \rangle$ measurements to constrain the maginitude of linearly-tilting fluctuation in the model.

In additional to the initial condition fluctuation, short range correlation also contribute to the variance of $a_1$ \cite{Denicol:2015bnf}.
The UrQMD hadronic afterburner captures certain types of short-range correlations coming from resonance decay and collisions, but jet-like correlations are hard to include.

\subsection{Calibration of the 3D initial condition parameters}
The degrees of freedom of the initial condition model are,
\begin{itemize}[itemsep=0pt]
  \item[1--2.] Two normalization factors for Pb-Pb and p-Pb collisions at $\sqrts=2.76$~TeV and 5.02~TeV beam energies. They are not fully independent as $N_{\textrm{p+Pb}} > N_{\textrm{Pb-Pb}}$ is always imposed in the parameter sweep.
  \item[3.] The mid-rapidity entropy deposition parameter $p$.
  \item[4.] The $\Gamma$-fluctuation variance parameter $1/k$.
  \item[5.] The Gaussian nucleon width $w$, which determines the initial state granularity.
  \item[6--8.] Three coefficients that modulate the local rapidity distribution's shift $\mu_0$, width $\sigma_0$, and skewness $\gamma_r$ or $\gamma_a$,
  \item[9.] An average Jacobian $J$ for the conversion from rapidity to pseudorapidity.
\end{itemize}
The range of the parameters is shown in \ref{tab:trento:parameters}.
One may notice that we did not use different width parameters for the rapidity distribution $\sigma_0$ for Pb-Pb and p-Pb collisions because the beam rapidity changes less than $8\%$ from $2.76$ TeV to $5.02$ TeV.

\begin{table}
\centering
\caption{Three-dimensional initial condition parameters}
\label{tab:trento:parameters}.
\begin{tabular}{lll}
      Parameter & Description	& Range \\
      \paddedhline
      $N_{\textrm{p-Pb}}$    & Overall p-Pb normalization      & 140.0--190.0 \\
      $N_{\textrm{Pb-Pb}}$   & Overall Pb-Pb normalization     & 150.0--200.0  \\
      $p$	                   & Generalized mean parameter      & -0.3--0.3 (with a prior)  \\
      $k$	                   & Multiplicity fluct.\ shape      & 1.0--5.0  \\
      $w$	                   & Gaussian nucleon width     & 0.4--0.6  \\
      $\mu_0$                & Rapidity shift mean coeff.\     & 0.0--1.0  \\
      $\sigma_0$             & Rapidity width std.\ coeff.\    & 2.0--4.0  \\
      \multirow{2}{*}{$\gamma_0$}             & \multirow{2}{*}{Rapidity skewness coeff.\ }      & 0.0--10.0 (rel) \\
                  &        & 0.0--3.6 (abs)  \\
      $J$	                   & Pseudorapidity Jacobian param.  & 0.6--0.9
\end{tabular}  
\end{table}

The dynamical model consists of a 3+1D relativistic hydrodynamics and a hadronic afterburner.
The equation-of-state interpolates the state-of-the-art lattice-QCD EoS \cite{Bazavov:2014pvz} at high temperature and zero baryon density to a hadron resonance gas EoS at low temperature \cite{Moreland:2015dvc}.
The energy density at which the hydrodynamic energy momentum tensor is particlized into hadrons is $\epsilon_{sw} = 0.322$~GeV/fm$^3$ corresponding to a switching temperature close to the pseudo-critical temperature $T_{sw} \sim T_c = 0.154$~GeV).
As a remark, the relativistic hydrodynamics code vHLLE \cite{Karpenko:2013wva} includes viscous corrections, but we used its ideal mode in the parameter extraction.
The reason for this is that the parameter optimization process requires running the model on hundreds of different parameter sets.
For each parameter set, thousands of minimum-biased events needed to be generated to control the statistical uncertainty, especially for the correlation observables. 
The full 3+1D viscous hydrodynamics is extremely time-consuming; therefore, we choose to run the hydrodynamic model in its ideal mode and on a rather coarse grid.
The justification is that the rapidity distribution of the multiplicity and normalized two-particle correlations is less sensitive to the viscosity.
In the end, as a validation of this procedure, we will be using a set of high-likelihood parameters and run the dynamical model with the full viscous hydrodynamic model to see if it describes other observables such as the anisotropic flows, and event-plane decorrelations.

\begin{figure}
 
\includegraphics[width=\textwidth]{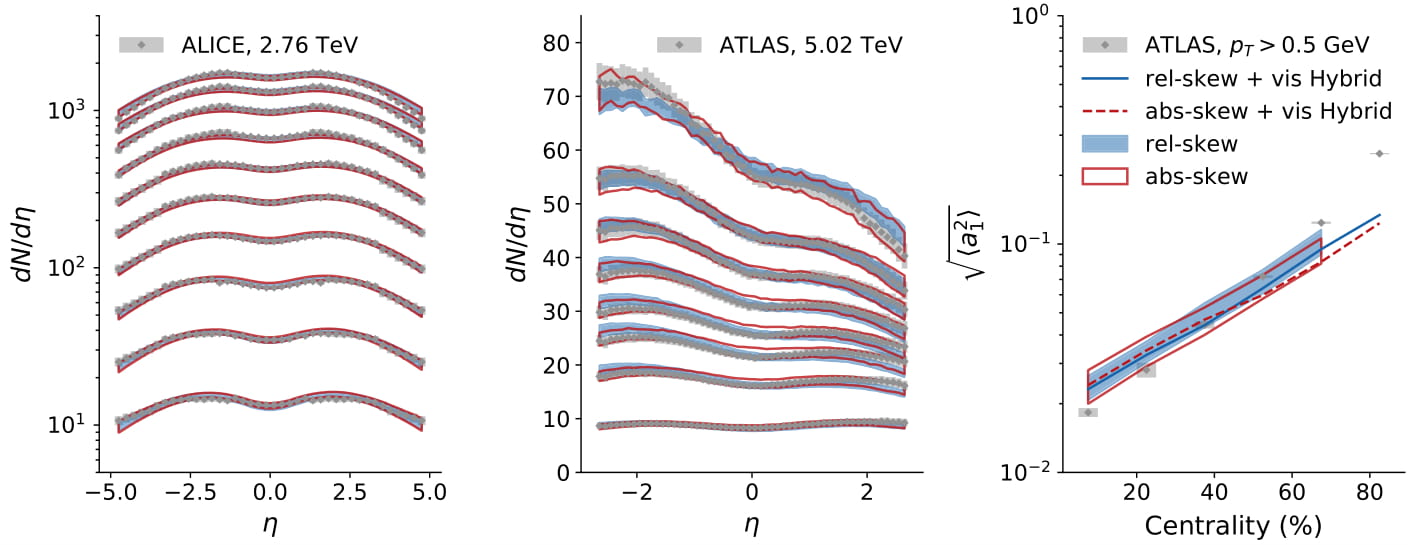}
\caption[The posterior distribution of the observables after the fitting]{The posterior distribution of the observables after the fitting process. Blue and red stands for the results calibrated using the ``relative skewness'' and the ``absolute skewness'' ansatz respectively.
From the left to the right, the subplots show charged-particle pseudorapidity density in Pb-Pb collision, in p-Pb collision, and the root-mean-square (rms) of the $a_1$ coefficient in the two-particle pseudorapidity correlation decomposition. The lines in the rightmost subplot also show that the rms $a_1$ is not very sensitive to the viscous effect in the hydrodynamic evolution.}
\label{fig:trento:post_obs}
\end{figure}

Four thousand Pb-Pb and ten thousand p-Pb events are generated at $100$ sets of parameter values; then the Bayesian analysis makes inference on the probability distribution of the parameters by comparing to measurements.
After the calibration, the $\dnchdy$ and $a_1$ fluctuations as function of centrality are compared to measurements in figure \ref{fig:trento:post_obs}.
The single-particle distribution can be well reproduced by the calibrated initial condition plus dynamical evolution, while the correlation observables can be described up to 50\% centrality. 
For more peripheral Pb-Pb collisions, the hydrodynamic model significantly underestimates the $a_1$ fluctuation.
We understand this as a consequence of inadequate modeling of the short-range correlation for peripheral collisions, as they are more important in low multiplicity events ($\dnchdy\sim 250$ and $N_{\textrm{part}} \sim 75$ at 50\% centrality).
The authors of \cite{ATLAS:2015kla} compared the measurements to the HIJING event generator that is based on mini-jet production.
They found that these jet-like correlations agree well with the $a_1$ fluctuation for peripheral collisions with $N_{\textrm{part}} \lesssim 80$, while it overshoots the data for more central collisions.
The hydrodynamic model and mini-jet production model provide a complementary picture to fully understand the rms $a_1$: at larger centrality, mini-jet production dominates the fluctuation of the $\dnchdy$, while as multiplicity increases and the final state interactions are frequent, the event-by-event asymmetry in the single-particle distribution dominates the rms $a_1$.

\begin{figure}
 
\centering
\includegraphics[width=\textwidth]{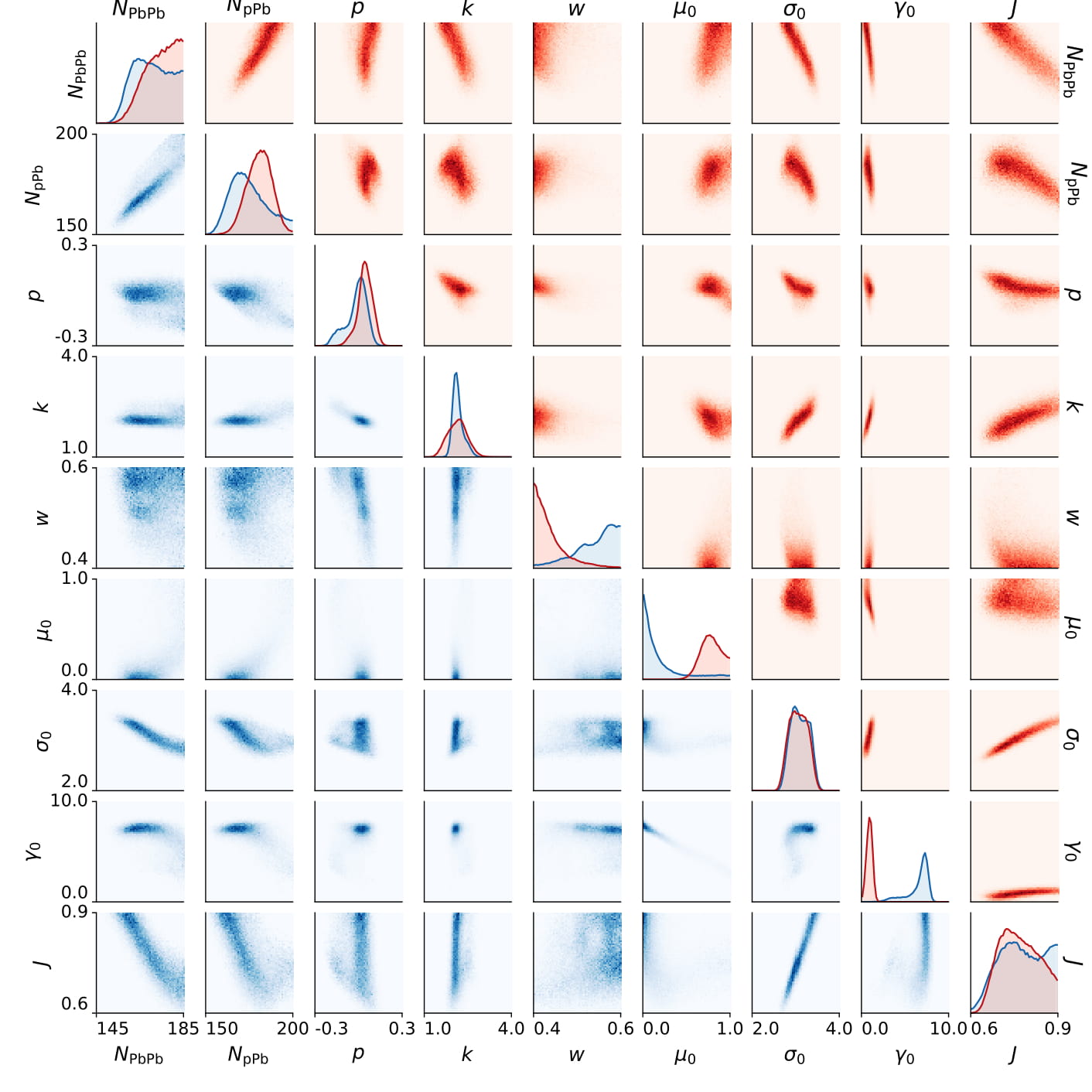}
\caption[The posterior probability distribution of the model parameters,]{The posterior probability distribution of the model parameters, the colors distinguish the use of ``absolute skewness'' (red) and ``relative skewness'' ansatz. The diagonals are single-parameter distribution, and the off-diagonals are two-parameter joint distributions.}
\label{fig:trento:posterior}
\end{figure}

Regarding the performance of different parametrizations of the skewness, the ``relative'' skewness ansatz does better in reproducing the uprising trend of rms $a_1$, while the absolute skewness ansatz better describes the large $\dnchdy$ asymmetry in the top 1\% p-Pb collisions.
However, there is no substantial evidence to favor any of them over the other.
We will see later that this is because the two parametrizations, though take different forms, actually behave similar in terms of $ds/d\eta(\eta; T_A, T_B)$, for typical values of $T_A$ and $T_B$ of heavy nuclei.

\begin{figure}
 
\centering
\includegraphics[width=.8\textwidth]{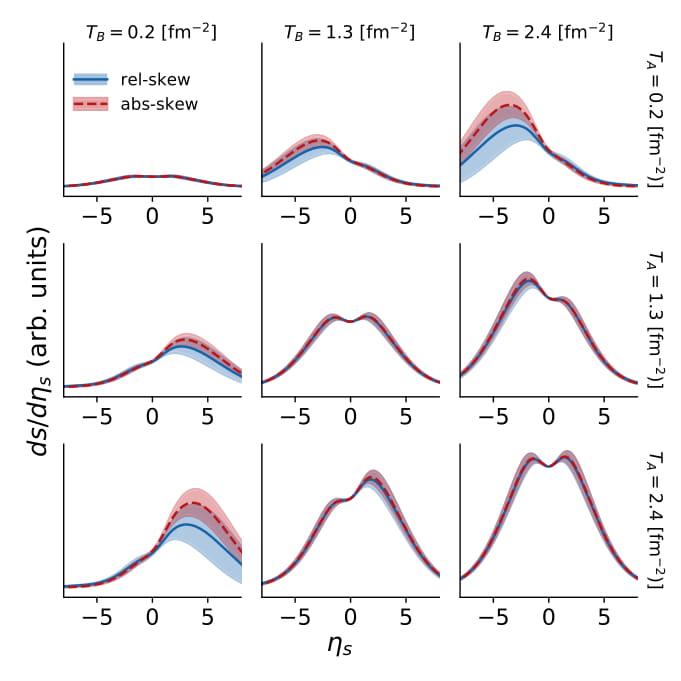}
\caption[The constrained functional form of the three-dimensional initial]{The constrained functional form of the three-dimensional initial entropy deposition $s(T_A(\mathbf{x}_\perp), T_B(\mathbf{x}_\perp), \eta_s)$. $T_A$ and $T_B$ is varied from $0.2$ to $2.4$ fm${}^{-2}$. Colors distinguish the results from using ``absolute skewness'' (red) and ``relative skewness'' ansatz.}
\label{fig:trento:post_dsdy}
\end{figure}

The distribution of the calibrated parameters is shown in figure \ref{fig:trento:posterior}.
The red and blue lines and color map correspond to results using the ``absolute'' and ``relative'' skewness respectively.
The normalization parameters $N_{\textrm{PbPb}}$ and $N_{\textrm{pPb}}$, mid-rapidity entropy deposition parameter $p$, nucleon thickness function fluctuation parameter $k$, the width of the rapidity distribution $\sigma_0$ and the effective Jacobian $J$ have similar probability distributions between the two parametrizations.
However, the distribution of the asymmetry related parameters $\mu_0$ and $\gamma_0$, and nucleon width $w$ are very different.
The reason is that the Bayesian calibration looks for the high-likelihood region of the parameter space to explain the data, and the two different parametrizations can achieve this same goal by optimizing the parameter combinations differently.
For the ``relative'' skewness one, the optimized parameters have a small shift of the mean $\mu$ and a large skewness $\gamma$; for the ``absolute'' skewness one, they correspond to the region with a large $\mu$ but vanishing $\gamma$.
Such result does not mean that there are ``two models'' explaining the same data, because both of them are simple parametrization of $ds/d\eta(\eta; T_A, T_B)$ with infinitely many degrees of freedom. 
Instead, we should treat them as an estimation of the systematic uncertainty in extracting the function form of the initial 3D entropy deposition $ds/d\eta(\eta; T_A(\x), T_B(\x))$.
It is more instructive to see the probability distribution of $ds/d\eta$ it self, given different values of $T_A(\x)$ and $T_B(\x)$.
In Fig.~\ref{fig:trento:post_dsdy}, we sample the calibrated parameter distributions and use them to draw the entropy density as a function of rapidity at different values of $T_A$ and $T_B$. 
Again, the red and blue colors represent ``absolute'' and ``relative'' skewness parametrizations; the lines show the median prediction and the bands are one standard deviation uncertainties. 
In each row and each column, $T_A$ and $T_B$ varies from $0.2~\text{fm}^{-2}$ to $2.4~\text{fm}^{-2}$. 
For references, the value of the thickness function at the center of a Gaussian proton with nucleon width $0.5$ fm is about $0.6~\text{fm}^{-2}$ and is $0.2~\text{fm}^{-2}$  at 1.5 width away from its center; while the maximum nuclear thickness function of a Pb nucleus is about $2.2~\text{fm}^{-2}$.
Therefore, the chosen range of $T_{A,B}$ already cover the typical ranges and combinations for entropy production in a realistic heavy-ion collision.
One observes that the functional form of the $ds/dy$ between the two parametrizations agrees within one standard deviation, with the discrepancy increasing as the asymmetry increases.
Indeed, with sufficiently different $T_A$ and $T_B$ combinations, the two results will have totally different predictions; however, within the physical range of the thickness function, the two methods converge onto a similar behavior,
We conclude that by applying the hydrodynamic-based model and a flexible 3D initial condition, the form of the local rapidity distribution as a function of participant densities can be reverse engineered using single-particle pseudorapidity distributions and two-particle pseudorapidity correlations.

\subsection{Prediction with the calibrated 3D initial condition}
The calibrated initial condition is useful in predicting other rapidity-dependent observables.
Especially, since we have only used multiplicity observables $\dnchdy$ and rms $a_1$ in the calibration, a prediction of azimuthal anisotropy related observables would provide non-trivial validation of the initial condition.
We choose a set of high-likelihood parameters shown in table \ref{tab:chosen_parameters}.
The selected validation/prediction observables are pseudorapidity dependent harmonic flows, event-plane decorrelations, and the symmetric cumulants, which quantify the correlation between different flow harmonics.

\begin{table}
\centering
\caption{A high likelihood parameter set}
\begin{tabular}{lll}
\hline
Parameter & rel-skew	& abs-skew \\
\hline
$N_{\textrm{Pb-Pb}}$   & 150.0     & 154.0  \\
$p$	    & 0.0      & 0.0  \\
$k$	    & 2.0     & 2.0  \\
$w$	    & 0.59     & 0.42  \\
$\mu_0$   & 0.0     & 0.75  \\
$\sigma_0$ & 2.9    & 2.9  \\
$\gamma_0$ & 7.3		& 1.0	\\
$J$	     & 0.75 & 0.75	\\
\hline
\end{tabular}
\label{tab:chosen_parameters}    
\end{table}

\begin{figure}
 
\centering
\includegraphics[width=.7\textwidth]{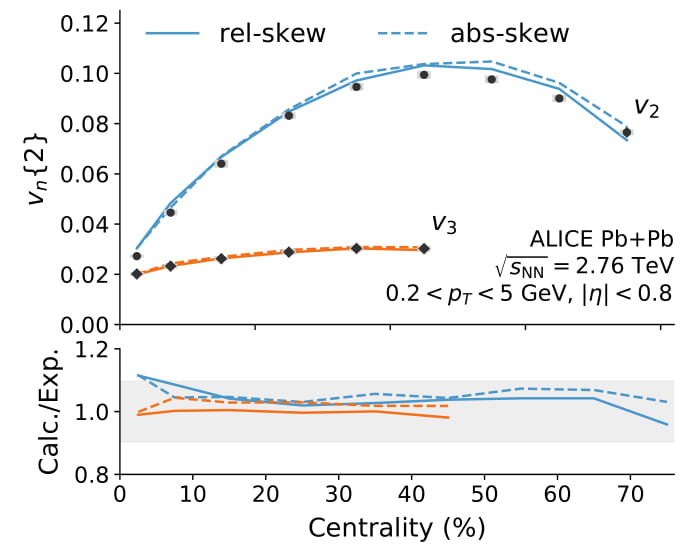}
\caption[Flow coefficients $v_2\{2\}$ (blue) and $v_3\{2\}$ (orange) are plotted as]{Flow coefficients $v_2\{2\}$ (blue) and $v_3\{2\}$ (orange) are plotted as a function of centrality. The calculation uses a 3+1D viscous hydrodynamic-based model with constant $\eta/s=0.17$ and $0.19$ for ``relative skewness'' (solid) and ``absolute skewness'' (dashed) ansatz. $\zeta/s=0$ and the switching temperature is $T_\text{sw}=154$~MeV.
The initial condition parameters are selected from the high-likelihood region of the posterior.}
\label{fig:trento:vn_cen}
\end{figure}

\paragraph{$v_n$ at mid-rapidity} The elliptic and triangular flow from two-particle correlation $v_2\{2\}$ and $v_3\{2\}$ are calculated at as functions of centrality at mid-rapidity (figure \ref{fig:trento:vn_cen}).
We can describe flow measured by ALICE \cite{Adam:2016izf} at mid-rapidity with a shear viscosity $\eta/s = 0.17$--$0.19$ close to other phenomenological studies, though the bulk viscosity is not included in this study.

\begin{figure}
 
\includegraphics[width=\textwidth]{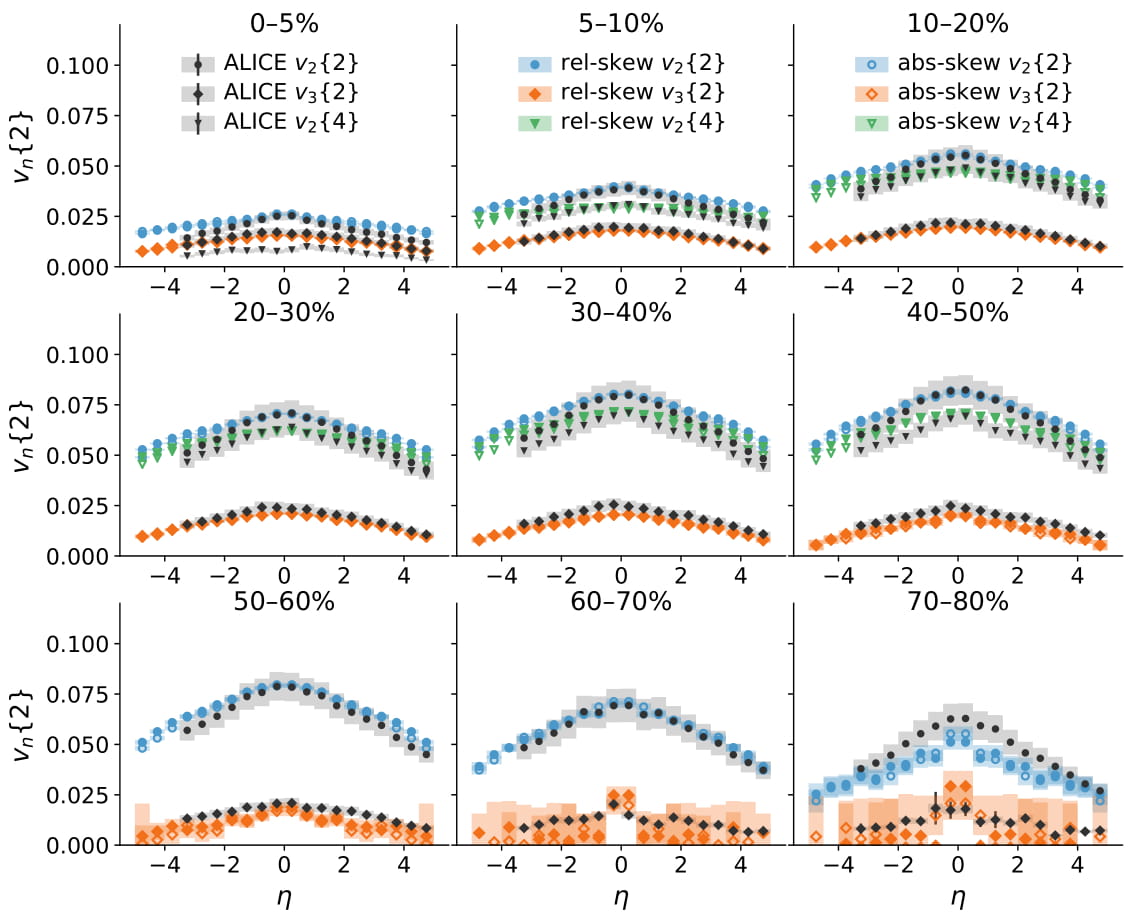}
\caption[Pseudorapidity dependent flow coefficients $v_2\{2\}$ (blue circle), ]{Pseudorapidity dependent flow coefficients $v_2\{2\}$ (blue circle), $v_3\{2\}$ (green triangle) and $v_2\{4\}$ (orange diamond).
Constant $\eta/s=0.25$ and $0.28$ are used for ``relative skewness'' (closed symbol) and ``absolute skewness'' (open symbol) ansatz. Data is taken from the ALICE Collaboration \cite{Adam:2016ows}.}
\label{fig:trento:vn_eta}
\end{figure}

\paragraph{Rapidity-dependent $v_n$} For the rapidity-dependent flow figure \ref{fig:trento:vn_eta}, we uses a larger $\eta/s=0.25$--$0.28$. 
This inconsistency is because the ALICE measurement extrapolates the $p_T$ range for the rapidity-dependent flow down to 0, while the mid-rapidity  $p_T$ cut is $0.2 < p_T < 5.0$ GeV.
It would not be a problem for the model to describe both with the same transport parameters if the $p_T$ differential flow and $p_T$ differential particle spectra were both reproduced.
However, due to the lack of a systematic tuning of the model parameters, including both shear and bulk viscosity, the current mean $p_T$ is too high.
Therefore, agreement with the $p_T$-integrated flow in one kinematic cut does not guarantee the agreement to measurements that extrapolate to $p_T = 0$.
Since our primary interest is the $\eta$-dependence of the flow, we have chosen this larger-than-usual shear viscosity to match the $v_2\{2\}(\eta=0)$ values to data.
The calculated $v_2\{2\}$, $v_3\{2\}$, and $v_2\{4\}$ gradually decrease from mid-rapidity to forward / backward rapidity, which is also the trend seen in the data.
The shape of $v_3\{2\}$ is well described; but for $v_2\{2\}$ and $v_2\{4\}$ in the region $|\eta| > 2.0$, the data decreases faster than our predictions.
There could be several reasons for this difference.
It is possible that the current initial condition model produces enough fluctuations, but an inadequate variance of the overlapping geometry as a function of space-time rapidity.
It is also showed in a study of at RHIC energies that the decreasing slope of $v_n(\eta)$ can be sensitive to the shear viscosity in the hadronic phase \cite{Denicol:2015bnf, Bozek:2010bi}.
In our model, the transport properties of the system below $T_c$ is all encoded in UrQMD and are not tunable. 
Moreover, we have assumed that hydrodynamization happens at a constant proper time hypersurface;
while it is possible that a constant entropy density hypersurface is a better criterion, and as a result, the matter at larger rapidity experiences a shorter period of pressure-driven expansion.
To systematically investigate all these effects, a future global parameter calibration including both initial condition parameters, transport parameters, and matching parameters is inevitable and is also feasible given the advances in the dynamical models and programming developments as well as more powerful computing resources.

\paragraph{Event-plane decorrelations} The event-planes are defined as the phase angle of the anisotropic flow,
\begin{eqnarray}
\Psi_n^\text{EP} = \frac{1}{n}\arctan\frac{\langle \sin n \phi \rangle}{\langle\cos n \phi \rangle}.
\end{eqnarray}
Due to longitudinal fluctuations, the event-planes separated by a rapidity gap decorrelate from each other.
This decorrelation is important for observables that involve a large rapidity gap.

\begin{figure}
 
\centering
\includegraphics[width=.8\textwidth]{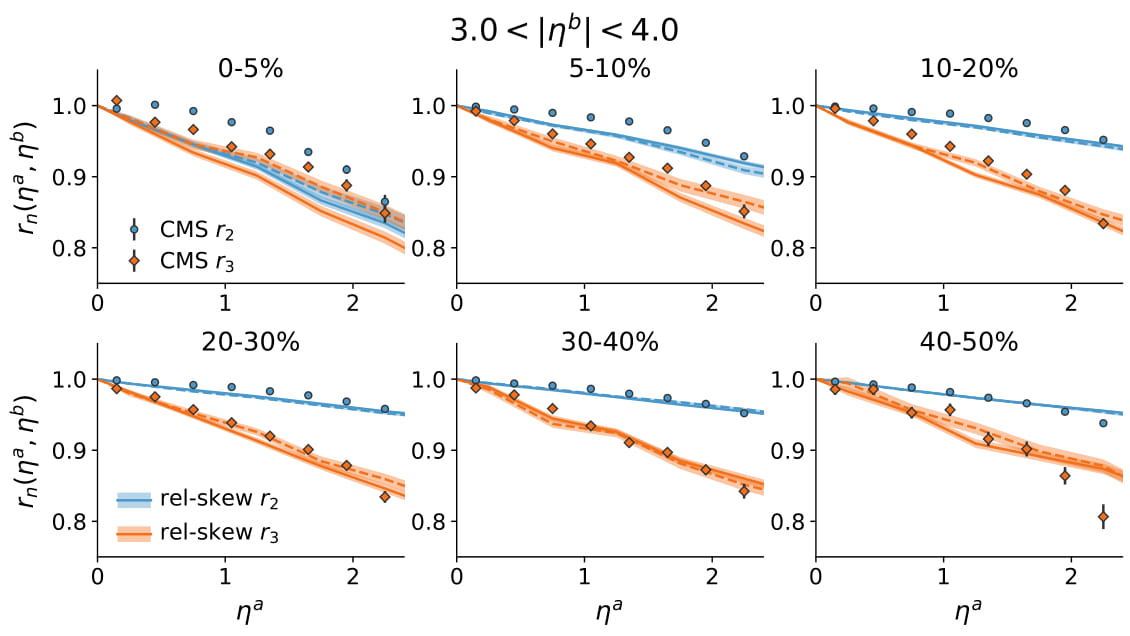}\\
\includegraphics[width=.8\textwidth]{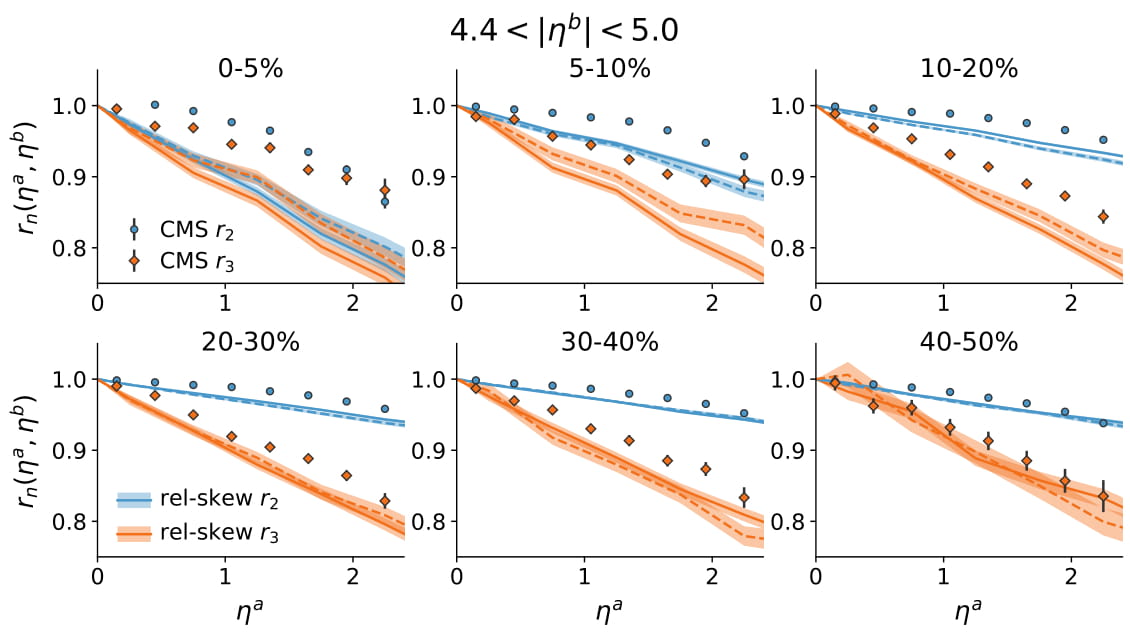}
\caption[The $n=2$ (blue) and $n=3$ (orange) event-plane decorrelations]{The $n=2$ (blue) and $n=3$ (orange) event-plane decorrelations obtained for ``relative skewness'' (lines with bands) and ``absolute skewness'' (lines with hatches) ansatz. The top and bottom plots uses different reference particle bins $3.0<|\eta_b|<4.0$ and $4.4<|\eta_b|<5.0$ respectively.}
\label{fig:trento:epd}
\end{figure}

This event-plane decorrelation has been studied using initial conditions from a 3D extended Glauber model \cite{Bozek:2015bna} and A Multi-Phase Transport (AMPT) model \cite{Jia:2014ysa, Xiao:2012uw,Pang:2015zrq}.
Gluonic Yang-Mills dynamics in three-dimensions have also been used to study decorrelation at the initial stage level \cite{Schenke:2016ksl}.
In our model, the geometry at forward and backward rapidity are dominated by the participant density of two different nuclei.
In between, the entropy production smoothly transits from one to another and so are the orientation of the energy density eccentricities that drive the flow of particles.
The CMS collaborations quantifies the decorrelation by a three-bins factorization ratio \cite{Khachatryan:2015oea},
\begin{eqnarray}
r_n(\eta^a, \eta^b) &= \frac{V_{n\Delta}(-\eta^a, \eta^b)}{V_{n\Delta}(\eta^a, \eta^b)}, \\
V_{n\Delta}(\eta^a, \eta^b) &= \langle\langle \cos(n\Delta\phi) \rangle\rangle,
\end{eqnarray}
where the double average runs over all particle pairs and all events.
One of the rapidity bins is near $\eta^b$, and the rest of the two bins are around $\pm\eta^a$.
This ratio measures the decorrelation between two event planes separated by a larger rapidity gap $\eta^a + \eta^b$ relative to the decorrelation over a smaller gap $|\eta^a - \eta^b|$.
In experiments, one would like to take the reference bin $\eta^b$ farther from $\eta_a$ to suppress the short-range correlations correlation.
However, the way we build our model is to extend the mid-rapidity entropy deposition to finite rapidity, and this extension will eventually fail at sufficiently large rapidity because the model behavior in those regions is not well constrained.
In figure \ref{fig:trento:epd}, we compare the model calculations to data with both a large $4.4<\eta^b<5.0$ (bottom) and a relative smaller $3.0 < \eta^b< 4.0$ (top).
The predicted factorization ratios decrease linearly with the increasing rapidity gap.
Using reference particles from $3.0 < \eta^b< 4.0$, the decorrelation is reproduced at larger centrality but not for central collisions.
Because the $n=2$ event-plane has a preferred direction defined by the impact parameter, while $n=3$ event-plane does not, we observe that the $n=2$ factorization ratio decorrelates slower than the $n=3$ ratio, except for the most central collision when $n=2$ event-plane is also dominated by fluctuations.
Comparing to data with reference particles from $4.4<\eta^b<5.0$, the experimental data stays similar to the previous case except for central collisions, but the prediction is completely off.
This is simply because the model fails on the details at large rapidity as we explained earlier.
Given the present comparison, we conclude that the valid range of the model should be restricted to $|\eta| < 4.0$.

\paragraph{Symmetric cumulants} Finally, we predict the symmetric cumulants (SC), which is a four-particle correlation between different orders of anisotropic flows $v_m$ and $v_n$ \cite{Niemi:2012aj,Bilandzic:2013kga}.
\begin{eqnarray}
SC(m, n) &=& \langle\langle \cos(m\phi_1+n\phi_2-m\phi_3-n\phi_4)\rangle\rangle \nonumber \\
\nonumber &-& \langle\langle\cos[m(\phi_1-\phi_2)]\rangle\rangle\langle\langle\cos[n(\phi_1-\phi_2)]\rangle\rangle \label{eq:scmn}\\
&=& \langle v_m^2 v_n^2 \rangle - \langle v_m^2\rangle\langle v_n^2\rangle.
\end{eqnarray}
It is clear from the second equation that it measures the covariance between $v_n^2$ and $v_m^2$.
One can also define the normalized symmetric cumulants (NSC),
\begin{equation}
NSC(m,n) = \frac{SC(m,n)}{\langle v_m^2\rangle\langle v_n^2\rangle}.
\end{equation}
These observables are interesting because they are robust against non-flow effects, and have been shown to be sensitive to the temperature dependence of the transport coefficients \cite{ALICE:2016kpq}.
Other analyses show that $SC(4,2)$ probes the non-linear response of the hydrodynamic evolution, while $SC(3,2)$ is more sensitive to initial conditions \cite{Zhu:2016puf}.

\begin{figure}
 
\includegraphics[width=\textwidth]{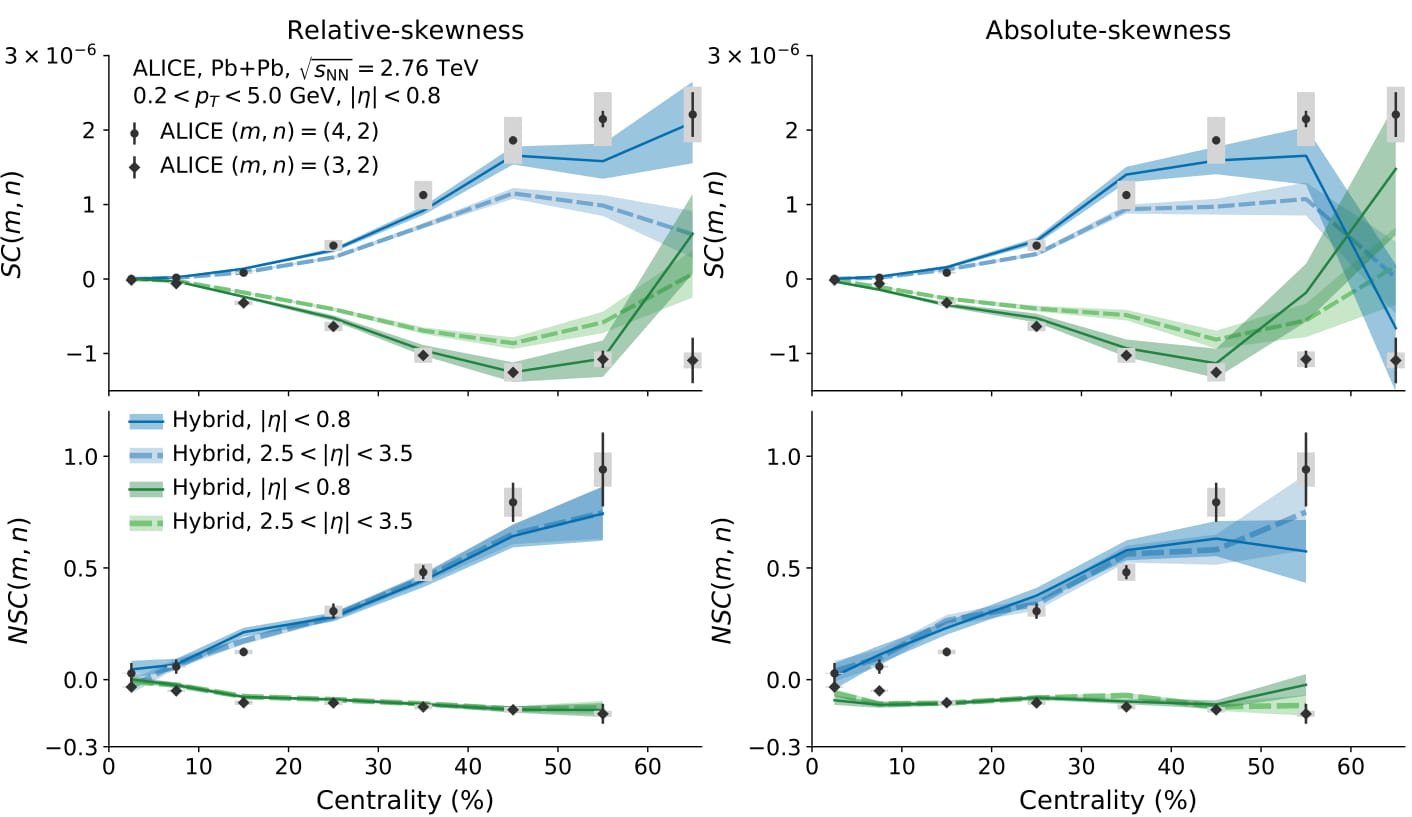}
\caption[The symmetric cumulants (top row) and normalized symmetric]{The symmetric cumulants (top row) and normalized symmetric cumulants (bottom row) obtained from the ``relative skewness'' (left column) and the ``absolute skewness'' (right column) ansatz.
$SC(4,2)$ (blue) and $SC(3,2)$ (green) are shown as functions of centrality. The experimental measurements at mid-rapidity (black symbols) are taken from \cite{ALICE:2016kpq}.
The calculations are performed at both mid-rapidity $|\eta|<0.8$ (solid lines) and forward/backward rapidity $2.5<|\eta|<3.5$ (dashed lines).
}
\label{fig:trento:smn} 
\end{figure}

The left and right of figure \ref{fig:trento:smn} show the calculations using relative- and absolute-skewness.
Symmetric cumulants $SC(4,2)$ (blue) and $SC(3,2)$ (green) are displayed in the top plots, and $NSC(4,2)$ (blue) and $NSC(3,2)$ (green) in the bottom plots.
Within each plot, black dots are ALICE measurements at mid-rapidity $|\eta|<0.8$ \cite{ALICE:2016kpq}, which should be compared to the calculation shown in solid lines.
The calculation shown in dashed lines are our predictions at forward/backward rapidity $2.5 < |\eta| < 3.5$,
\begin{eqnarray}
SC^\prime(m, n) &=& \langle\langle \cos(m\phi_1+n\phi_2-m\phi_3^\prime-n\phi_4^\prime)\rangle\rangle \\
\nonumber &-& \langle\langle\cos[m(\phi_1-\phi_2^\prime)]\rangle\rangle\langle\langle\cos[n(\phi_1-\phi_2^\prime)]\rangle\rangle, \label{eq:scmn-diff}
\end{eqnarray}
where the two of the particles (primed) are selected from the forward/backward rapidity bins, while the rest of the two still come from the mid-rapidity bin $|\eta|<0.8$.

At mid-rapidity, the calculated (normalized) symmetric cumulants reproduce the trend of the measurements, and the ``relative'' skewed initial condition does a better quantitative job than the ``absolute'' skewed model.
At forward and backward rapidity, though the magnitudes of both $SC(m,n)$ and $v_n$, $v_m$ changed, the normalized symmetric cumulants remains the same as the one at mid-rapidity.
These predictions can be checked in future measurements to put further constraints on the three-dimensional initial condition model.

As a summary of this chapter, I have introduced the hydrodynamic-based medium evolution model that is very successful in describing the bulk observables.
The sensitivity of the harmonic flows allows one to extract QCD transport coefficients using advanced statistical techniques.
The initial condition for the dynamical model is still a significant source of uncertainty in both data interpretation and parameter extraction.
The \trento\ model was developed as a flexible ansatz for mid-rapidity entropy/energy deposition so that the initial condition and transport coefficients can be calibrated simultaneously to data.
Finally, I discussed my work on extending the \trento\ initial condition to include rapidity dependence and the use of multiplicity observable to reverse-engineer the functional form of initial 3D entropy deposition.
The results can be used to predict more rapidity-dependent observables, and are useful to other studies involving a large rapidity range.
Due to the simplicity of the modeling, and the lack of global fit of all parameters including both 3D initial conditions and transport coefficients, we cannot yet answer how the inclusion of longitudinal fluctuation quantitatively affects the extracted QCD transport coefficients, but it is a critical question for future precision analysis.

\chapter{A transport model for hard partons in the QGP}
\label{chapter:transport}
Hard partons are predominately created in perturbative scatterings at the earliest stage of relativistic heavy-ion collisions.
The distribution of the hard partons gets modified by the medium, and the final distribution carries information about the medium, as well as the hard-soft interaction properties.

Among the many ways of describing the in-medium evolution of hard partons, 
the transport approach has a unique advantage. 
Here, we refer to the transport approach as a class of models that evolve the semi-classical particle distribution function of hard partons in real-time.
Transport models can often be formulated as particle-level simulations, which provide easy coupling to local properties of a dynamically evolving and fluctuating medium, and an exclusive final state.
Of course, challenges exist when applying transport models to high energy collisions.

First, there are different assumptions of the interactions between the hard partons and the medium to be made.
Two commonly assumed extremes are:
\begin{itemize}
\item[1] A weakly coupled picture: the medium consists of perturbative quasi-particles (scattering centers) whose distribution is close to local thermal equilibrium.
Hard partons scatter perturbatively with these well-separated scattering centers. A Boltzmann equation describes its dynamics.
\item[2] Diffusion picture: interactions between the medium and the hard parton are frequent and soft, and there are substantial many-body and non-perturbative effects. Such dynamics are solved using a Langevin equation, and a drag and a diffusion coefficient model the effect of these soft interactions on the hard parton. 
\end{itemize}
These two commonly used approaches are not necessarily mutually exclusive and can have an overlapping range of validity. 
For example, the effect of soft momentum exchange processes in the perturbative calculation can be very well modeled by a diffusion equation \cite{Ghiglieri:2015ala,Dai:2019hbi}.
These different assumptions on the interaction between the medium and the hard probe are primarily due to our inadequate theoretical tools in describing the QGP medium in the strongly coupled regime.
On the one hand, this becomes an uncertainty intrinsic to the transport approach, until one finds a convincing way of calculating the dynamics of the strongly coupled QGP medium from first principles.
On the other hand, experiments may be able to tell which assumption (or a combination of both) is preferred and answer the very question of how the sQGP participates in the jet-medium interactions.

A second difficulty is that a semi-classical transport equation is inadequate in treating quantum coherence.
Indeed, a quantum transition will always be bounded by the uncertainty principle: a process with momentum scale $Q$ can not be localized within a space-time extent of $1/Q$.
While in the semi-classical transport model, one always specifies a local point in space-time where the interaction takes place.
This is valid if the momentum scale $Q$ is high enough that $1/Q$ is much smaller than the resolution of the transport model, e.g., characterized by the mean-free-path in the Boltzmann equation.
However, the soft and collinear divergences of QCD bremsstrahlung (or more generally, parton branching and merging) processes generate an abundance of small-$Q$ events whose spatial extent can be much greater than the mean free path, which happens for both the vacuum parton shower and the medium-induced parton shower in certain phase space regions.
For medium-induced branching, this is the QCD analog of the Landau-Pomeranchuk-Migdal (LPM) effect \cite{PhysRev.103.1811,Wang:1994fx,Zakharov:1996fv}, and the radiation pattern is changed qualitatively.
When this happens, strictly speaking, the semi-classical transport equation is not the appropriate tool.
However, considering the advantages of the transport formulation, we want to develop a minimum set of modification to the semi-classical transport that can mimic the quantum effects of medium-induced branchings approximately.

We start by introducing a class of widely used transport equations: the (linearized) Boltzmann equation and the Langevin equation.
Then, we combine these two approaches into a hybrid one by introducing a cut-off distinguishing hard and soft momentum transfer processes and build the transport model in the incoherent limit.
After that, we provide a brief review of the theory of QCD in-medium parton branching processes at leading order, discussing its various approximations and also the numerical solutions in a simplified medium.
With these theoretical insights, the primary outcome of this chapter is developing a ``modified Boltzmann'' transport approach, treating the medium-induced parton branching with an approximate LPM effect.
Finally, the simulations of the transport model are compared to theoretical expectations to validate the implementation in different regimes.
We will show that the modified transport approach can reasonably describe the energy spectrum of the medium-induced splitting vertex for different channels $q\rightarrow q+g$, $g\rightarrow g+g$ and $g\rightarrow q+\bar{q}$.
Treatment of the heavy quark masses effect and running coupling are also investigated.
For future references, in the very end, we make comparisons between two other Monte-Carlo approaches for medium-induced radiation with the present one and comment on potential problems.

\section{The Boltzmann equation}
The Boltzmann equation evolves the particles' distribution function under the effect of localized collisions. 
By localization, it means that the time scale of the collision has to be much smaller than the mean free-path $\tau \ll \lambda$. 
Therefore, the collision probability can be evaluated using a local particle distribution function.
It also allows one to include only few-body collision processes, because the probability of interacting with an additional particle during this collision is small $P \approx \tau/\lambda \ll 1$.
At weak coupling, we will see in the next section that this is indeed the case for elastic collisions or soft and large-angle radiation. 
However, for radiation with a large formation time, the process becomes ``non-local''.
Accordingly, the Boltzmann formulation needs to be modified quite fundamentally for such processes.
In this section, we only focus on local interactions.

With two-body to two-body (elastic) and two-body to three-body (inelastic, including the reverse process) processes, the Boltzmann equation for particle specie $a$ takes the following form,
\begin{eqnarray}
\frac{\partial f^a}{\partial t} + \vec{v}\cdot\frac{\partial f^a}{\partial \vec{x}} + \frac{\partial E}{\partial \vec{x}}\cdot\frac{\partial f^{a}}{\partial \vec{p}} = - \sum_{b,c,d}\mathcal{C}_a^{a+b\leftrightarrow c+d}- \sum_{b,c,d,e}\mathcal{C}_a^{a+b\leftrightarrow c+d+e}
\end{eqnarray}
On the left hand side, the distribution function $f^{a}(t, \vec{x}, \vec{p})$ undergoes transport with velocity $\vec{v} = \partial E/\partial \vec{p}$, and a potential force $\vec{f} = -\partial E/\partial \vec{x}$.
On the right-hand side of the equation, the $2\leftrightarrow 2$ and $2\leftrightarrow 3$ collision terms are functionals of the distribution functions.
The summation of $b,c,d,e$ iterates over all other particle species including $a$.
Using the elastic process as an example and neglecting degeneracy of the internal quantum number for simplicity, the collision term can be separated into gain and loss terms,
\begin{eqnarray}
\mathcal{C} &=& \mathcal{C}_\textrm{loss} + \mathcal{C}_\textrm{gain}\\
&=& \int_{bcd} f^a(p_1)f^b(p_2)[1+\epsilon^c f^c(p_3)][1+\epsilon^d f^d(p_4)] \overline{|M|^2}(p_1, p_2; p_3, p_4) \\\nonumber
&-& \int_{bcd} f^c(p_3)f^d(p_4)[1+\epsilon^a f^a(p_1)][1+\epsilon^b f^b(p_2)] \overline{|M|^2}(p_3, p_4; p_1, p_2) \\
&=& \int_{bcd} \left\{
f^a(p_1)f^b(p_2)[1+\epsilon^c f^c(p_3)][1+\epsilon^d f^d(p_4)] \right. \label{eq:collision-term:symmetry} \\\nonumber
&& \left.- f^c(p_3)f^d(p_4)[1+\epsilon^a f^a(p_1)][1+\epsilon^b f^b(p_2)]\right\}
\overline{|M|^2},
\end{eqnarray}
where the crossing symmetry of the matrix-elements has been used in the last line of the equation ($\overline{|M|^2}(p_1^a, p_2^b; p_3^c, p_4^d) = \overline{|M|^2}(p_3^c, p_4^d; p_1^a, p_2^b) = \overline{|M|^2}$), and the phase-space integral is 
\begin{eqnarray}
\int_{bcd} = \prod_{i\in {b,c,d}}\int\frac{dp_i^3}{2E_i (2\pi)^3} (2\pi)^4 \delta^{4}(p_a+p_b - p_c-p_d).
\end{eqnarray}
The $\epsilon = 0, -1, 1$ corresponds to classical, Fermi-Dirac, and Bose-Einstein statistics depending on the nature of the particle.
The first term in the integration represents the loss of a particle of type ``$a$'' in the phase-space around the point $(x, p_1)$ due to elastic collision, and the second term represents the gain of a particle of type ``$a$'' due to the reverse process.
The symmetry in the microscopic matrix-element is essential for the kinetic equation to satisfy detailed balance: the probability to transition from one microscopic state to anther equals that of the reverse process.
The detailed balance ensures the thermal equilibrium limit of the system. 
Assume the system evolves long enough in a box of finite volume and there is no spatial variance of the distribution function.
Then the left side of equation \ref{eq:collision-term:symmetry} is zero, and the static solution has to satisfy the relation,
\begin{eqnarray}
f^a f^b (1+\epsilon f^c) (1+\epsilon f^d) = f^c f^d (1+\epsilon f^a) (1+\epsilon f^b),
\end{eqnarray}
for the entire phase-space and every combination of particle species.
Therefore, the following combination is conserved for each reaction channel.
 \begin{eqnarray}
\frac{f^a}{(1+\epsilon^a f^a)} \frac{f^b}{(1+\epsilon^b f^b)}
= \frac{f^c}{(1+\epsilon^c f^c)} \frac{f^d}{(1+\epsilon^d f^d)}
\end{eqnarray}
The available conservation quantities are the four-momentum; therefore, one solution to the previous equation is,
\begin{eqnarray}
\frac{f^a}{(1+\epsilon^a f^a)} = e^{-\beta \mu_a-\beta p\cdot u}
\end{eqnarray}
for every particle species with parameters $\mu, \beta$ and a four vector $u$ ($u^2 = 1$). 
So the static solution of the distribution is 
\begin{eqnarray}
f^a(p) = \frac{1}{ e^{\beta \mu_a+\beta p\cdot u} - \epsilon^a} \label{eq:thermal}\\
\mu_a +\mu_b = \mu_c + \mu_d \label{eq:chem}
\end{eqnarray}
The fist line is the distribution function in thermal (kinetic) equilibrium, and the second line is the requirement for reaching chemical equilibrium. 
One can identify the $\beta$ and $\mu$ parameter as the inverse temperature and chemical potential. The $u$ vector is the flow velocity of the cell as can be seen from the average velocity,
\begin{eqnarray}
\left\langle \frac{p^\mu}{M} \right\rangle = \frac{\int f(p) \frac{p^\mu}{M} dp^3}{\int f(p) dp^3} = \frac{\int f(p) \frac{p^\mu}{M} dp^3}{\int f(p) dp^3} = u^\mu
\end{eqnarray}

\subsection{The linearized Boltzmann equation and the diffusion limit}
Analytic solutions of the Boltzmann equations are almost impossible; even numerical solutions and simulations are highly non-trivial tasks.
However, under certain circumstances, a linearization of the Boltzmann equation is possible and greatly simplifies both the analytic analysis as well as the numerical implementation.

Hard particles (jet partons, heavy flavors) either have a large momentum $p\gg T$ or a large mass $M \gg T$. 
In heavy-ion collisions, the hard cross-section drops fast with the increase of $p_T$ and $m_T$, so hard partons are very rare in an actual event and the occupation number of hard partons is small $f_H \ll 1$.
Therefore, one can neglect the quantum statistics terms in the Boltzmann equation for them $1+\epsilon f_H \approx 1$.
Also, the collision terms with more than one uncorrelated hard particle in the initial state can also be neglected since these contributions are proportional to $f_H^2 \ll f_H$. 
Finally, we also assume that the response of the bulk of the particles to the hard particles is small, and shall neglect any collision terms that involve a hard parton in the Boltzmann equation for the bulk distribution function\footnote{ recently studies show that such back reactions are important for the study of full jet observables \cite{PhysRevC.99.054911}, but we only consider leading particles in this thesis}.
Under these approximations, one arrived at a set of equations that are linearized with respect to the hard parton:
\begin{eqnarray}
\frac{df_H}{dt} &=& -\mathcal{C}_H[f_H, f_{\textrm{bulk}}], \label{eq:hard-bulk-eq}\\
\frac{df_{\textrm{bulk}}}{dt} &=& -\mathcal{C}[f_{\textrm{bulk}}].
\end{eqnarray}
The collision term $\mathcal{C}_H$ is a linear operator on $f_H$.

For the medium particles, the equations are still complicated.
However, by observing that the time it takes for the low momentum bulk particles to reach local thermalization is much shorter than the relaxation time of the hard particles, a zeroth-order approximation would be using the local thermal distribution \ref{eq:thermal}.
The space-time evolution of the temperature $T$, chemical potential $\mu$ and flow velocity $u$ can be obtained from a hydrodynamic simulation.
Replacing the medium distribution function by the thermal one in equation \ref{eq:hard-bulk-eq}, one arrives at a closed and linearized equation for the hard particles.
Here we write down the equations assuming both classical statistics and the conservation of the hard parton's species, and only the elastic collision terms are shown for simplicity,
\begin{eqnarray}
\frac{df^H}{dt} &=& -\sum_{b} \int_{234} \left\{
f^H(p_1)f^b_{eq}(p_2) - f^H(p_3)f^b_{eq}(p_4)\right\}
\overline{|M|^2} \\
&=& - \int \left\{
f^H(p_1) w(p_1; p_3) - f^H(p_3) w(p_3, p_1)\right\}\frac{dp_3^3}{2E_3 (2\pi)^3}.
\end{eqnarray}
The $w(p; p')$ is the transition probability density for a particle with momentum $p$ into momentum state $p'$,
\begin{eqnarray}
w(p; p') = \sum_b\int_{24} f_{eq}^b(p_2) \overline{|M|^2}(p, p_2; p', p_4)
\end{eqnarray}

Using local thermal solutions for the bulk particles is a strong assumption. 
The degree of local thermalization in realistic events is still an open question, especially at the early stage of the heavy-ion collision. 
Moreover, whether the system can be understood in terms of quasi-particle degrees of freedom is a different question.
In an extreme weakly coupled system $g\ll 1$, one expects that the pressure and energy density can be explained using fundamental degrees of freedom: quarks and gluons, with perturbative corrections \cite{Blaizot:2000fc,Strickland:2010tm,Su:2015esa}.
However, with a large $g$ estimated from phenomenological studies, such a perturbative description may not be the most efficient way of understanding the bulk medium, and non-perturbative physics can play an essential role. 
Interpreting the medium in terms of microscopic degrees-of-freedom seem to be an unavoidable step of the Boltzmann equations; however, it is possible to ``integrate out'' the microscopic details in the soft limit of interaction into a set of transport coefficients.

\paragraph{The Fokker Planck equation}
In the soft-momentum transfer $q = p'-p$ limit  $|q| \ll |p|$, one can expand the collision term to second-order in $q$, and the linearized Boltzmann equation reduces to the Fokker-Planck type of equation,
\begin{eqnarray}
\frac{df}{dt} &=& - \int \left\{
f(p) - \left[f(p) +  \vec{q}\frac{\partial f}{\partial\vec{p}} + \frac{1}{2}\vec{q}\vec{q}\frac{\partial^2 f}{\partial\vec{p} \partial\vec{p}} \right]
\right\} w(p',p)\frac{dp_3^3}{2E_3 (2\pi)^3} \\
&=& - \int \left\{ \vec{q}\frac{\partial f}{\partial\vec{p}} - \frac{1}{2}\vec{q}\vec{q}\frac{\partial^2 f}{\partial\vec{p} \partial\vec{p}}
\right\} w(p',p)\frac{dp_3^3}{2E_3 (2\pi)^3} \\
&=&  -\eta_D(p) \frac{\partial f}{\partial\vec{p}} + \frac{1}{2}B(p)\frac{\partial^2 f}{\partial\vec{p} \partial\vec{p}},
\end{eqnarray}
where the vector function $A$ and tensor function $B$ are the first- and second-order moments of the transition rate,
\begin{eqnarray}
A_i(p) &=& \int w(p,p+q) q_i \frac{dp_3^3}{2E_3 (2\pi)^3},\\
B_{ij}(p) &=& \int w(p,p+q) q_i q_j \frac{dp_3^3}{2E_3 (2\pi)^3}.
\end{eqnarray}
One remark is that although the form of the Fokker-Planck equation can be derived as the soft limit of the linearized Boltzmann equation, its range of applicability is different from the latter.
It is because the transport coefficients are well defined in general, regardless of whether one assumes quasi-particle type microscopic dynamics.
Therefore in our model, we replace the soft sector of the Boltzmann equation with the Fokker-Planck equation so that the use of ``medium quasi-particles'' is restricted to hard momentum transfer processes.

Moments beyond second order are neglected in deriving the Fokker-Planck equation. 
This truncation is justified if the interaction is frequent enough so that within the smallest time scale that is concerned, a statistical description of the effect of many interactions in terms of the first (mean) and the second moments (variance) is adequate.
However, if the physical processes are rare, fluctuations contained in higher moments are indispensable and a diffusion equation is not a good approximation.

\paragraph{Transport coefficients and the Einstein relation} 
The $A$ and $B$ functions have to satisfy certain symmetries, as the only special direction after averaging over medium effects is the direction of motion.
Therefore,  $\vec{A} = \eta_D \vec{p}$ defines the drag coefficient  $\eta_D$; the tensor $B$ can be decomposed into a transverse part and a longitudinal part, with the respective momentum diffusion coefficients $\kappa$ and $\kappa_L$,
\begin{eqnarray}
B_{ij} = \kappa_L \frac{p_i p_j}{p^2} + \kappa \left(\delta_{ij} - \frac{p_i p_j}{p^2}\right).
\end{eqnarray}

One notices that the medium temperature does not show up explicitly in the Fokker-Planck equation,
\begin{eqnarray}
\frac{df}{dt} = \frac{\partial}{\partial p_i}\left(\eta_D p_i + \frac{1}{2}\frac{\partial}{\partial p_j} B_{ij}\right)f.
\end{eqnarray}
To guarantee the system has a thermalized solution, $A$, $\Kpara$ and $\Kperp$ are not independent.
Given a static and homogeneous medium at equilibrium with temperature $T$, $f = N\exp\left(-\beta E\right)$, the equation reduces to
\begin{eqnarray}
0 &=& \ppi(\phi p_i f)\\
\phi &=& \eta_D - \frac{\Kpara}{2TE} + \frac{\partial \Kpara}{\partial p^2} + \frac{\Kpara-\Kperp}{p^2}.
\end{eqnarray}
The Einstein relation $\phi = 0$ guarantees the existence of an equilibrium solution,
\begin{eqnarray}
\eta_D = \frac{\Kpara}{2TE} - \frac{\partial \Kpara}{\partial p^2} - \frac{\Kpara-\Kperp}{p^2}
\label{eq:ein-rel}
\end{eqnarray}
This is where the temperature shows up explicitly in the Fokker-Planck equation.

\section{Hard parton transport in the incoherent limit}
In this section, we proceed to use a local and incoherent calculation of hard parton scatterings and will defer a detailed discussion on the inclusion of the LPM effect to the next section.
The partonic processes are categorized into elastic (particle number conserving) and inelastic processes (particle number non-conserving). 
The inelastic processes are further divided into parton-splitting and parton-fusing contributions. 

\subsection{Hard/soft separation: elastic collisions}
In a quasi-particle picture of the QGP, the hard parton collides with medium partons and transfers a certain amount of four-momentum.
These processes can be computed at leading order in the weakly coupled theory, where the collision cross-section is calculated using the dressed gluon propagator inside the medium \cite{PhysRevD.44.1298},
\begin{eqnarray}
D^{\mu\nu}(\omega, k) = \frac{\delta^{\mu 0}\delta^{\nu 0}}{k^2 - \Pi_L(\omega, k)} + \frac{\hat{P}_T^{\mu\nu}}{\omega^2 - k^2 - \Pi_T(\omega, k)}
\end{eqnarray}
where $\Pi_T$ and $\Pi_L$ are the self energies for the transverse and longitudinal modes.
Due to the presence of the medium, the dressed propagator loses its Lorentz invariance and depends on the complicated functions $\Pi_T$ and $\Pi_L$.
The resulting cross-section formula will be equally complicated.
Fortunately, it has been shown recently in \cite{Ghiglieri:2015ala} that simplification is possible at leading order in rewriting the elastic processes as large-angle scattering and small-angle diffusion.
In such an approach, one chooses a scale $Q_\textrm{cut}$ with a formal range of $gT \ll Q_\textrm{cut} \ll T$.
For processes with momentum transfer to the medium larger than the cut-off  (hard modes), the medium screening effect is neglected and we use matrix-elements in the vacuum.
While for processes smaller than the cut-off (soft-mode), the propagator receives significant contributions from the screen effect.
The soft processes happen frequently and only involve small momentum transfers, satisfying the requirements of diffusion approximation.
This separation allows the following modeling of the elastic interaction between hard partons and the medium,
\begin{eqnarray}
\frac{df}{dt} = \mathcal{D}(Q_{\textrm{cut}})[f] + \mathcal{C}^{2\leftrightarrow 2}(Q_{\textrm{cut}})[f].
\end{eqnarray}
Here, particles are continuously evolved by the diffusion process with their momenta occasionally changed by large-$Q$ scatterings.
Later we will verify that the cut-off dependence in the diffusion and scattering component indeed cancels for ``physical observations'' at a sufficiently small coupling.
However, the phenomenological value of $g$ is very large so that the residue cut-off dependence may be significant. 
The advantage of the current formulation is that a diffusion process can also model certain non-perturbative effects with an additional contribution to the transport coefficient.

\paragraph{Transport coefficients for soft modes} The transverse and longitudinal transport parameters below the cut-off have been calculated in \cite{Ghiglieri:2015ala},
\begin{eqnarray}
\hat{q}_S &=& \int_0^{Q_{\textrm{cut}}^2} d\mathbf{q}^2 \frac{\alpha_s m_D^2 T}{\mathbf{q}^2 (\mathbf{q}^2+m_D^2)} = g^2 C_R T m_D^2  \ln\left(1+\frac{Q_{\textrm{cut}}^2}{m_D^2}\right).
\label{eq:qS} \\
\hat{q}_{S,L} &=& \int_0^{Q_{\textrm{cut}}^2} d\mathbf{q}^2 \frac{\alpha_s m_\infty^2 T}{\mathbf{q}^2 (\mathbf{q}^2+m_\infty^2)} = g^2 C_R T m_\infty^2  \ln\left(1+\frac{Q_{\textrm{cut}}^2}{m_\infty^2}\right)
\label{eq:qSL} 
\end{eqnarray}
$m_\infty^2 = m_D^2/2$ is the thermal mass of the gluon.  
In this chapter, we use the boldface to denote the transverse component of a four-vector.
The drag force is determined by the Einstein relation in equation \ref{eq:ein-rel},
\begin{eqnarray}
\eta_D = \frac{\hat{q}_{S,L}}{2ET} - \frac{d\hat{q}_{S,L}}{dp^2} - \frac{2\hat{q}_{S,L} - 2\hat{q}_S}{2p^2}
\end{eqnarray}

\paragraph{Scattering rate for hard modes} For the large-$Q$ $2\rightarrow 2$ scattering processes, the collision rates are computed by integrating the vacuum matrix-element, 
\begin{eqnarray}
R = \frac{d}{2E_1}\int  \frac{d^3p_2}{2E_2(2\pi)^3} f_0(p_2)2\hat{s} \int_{-\hat{s}}^{Q_{\textrm{cut}^2}}\frac{d\sigma}{d\hat{t}}d\hat{t}
\end{eqnarray}
The integration is restricted to large momentum transfers above $Q_{\textrm{cut}}$, and therefore we do not impose additional screening effects to regulate the matrix-element.
In this work, the $2\rightarrow 2$ matrix-element only includes the $\hat{t}$-channel contribution.

\subsection{Hard/soft separation: inelastic collisions}
Similarly, incoherent inelastic processes are divided into small-$Q$ diffusion induced radiation/absorption ($1\leftrightarrow 2$), and large-$Q$ $2\leftrightarrow 3$ processes.

\paragraph{Diffusion induced branching} For the incoherent diffusion-induced splitting rate, we borrow the expression from \cite{Cao:2017hhk} while stripping the time-dependent phase factor,
\begin{eqnarray}
R_{1\rightarrow 2} = \int d \mathbf{k}^2 dx \frac{\alpha_s P(x) \hat{q}_S(Q_{\textrm{cut}})}{2\pi (\mathbf{k}^2 + m_\infty^2)^2}
\end{eqnarray}
where a gluon thermal mass is added to screen the divergence.
Because these gluons are induced by processes with medium momentum transfer below the cut-off, $Q_{\textrm{cut}}$ appears in the formula.
For the reverse $2\rightarrow 1$ processes, a similar reaction rate can be written down,
\begin{eqnarray}
R_{2\rightarrow 1} = \int e^{-\beta \omega} d \mathbf{k}^2 dx \frac{\alpha_s P(x) \hat{q}_S(Q_{\textrm{cut}})}{2\pi (\mathbf{k}^2 + m_\infty^2)^2}.
\end{eqnarray}
$\omega$ is the thermal parton's energy, and $x$ is defined as the fraction of the thermal parton's energy to that of the final state hard parton.
These specific expressions are associated to the medium rest frame.

\paragraph{Large-$Q$ $2\leftrightarrow 3$ process} 
Regarding the $2\rightarrow 3$ matrix-element, in a previous study \cite{Ke:2018tsh}, we used to employ an improved version of the original Gunion-Bertsch cross-section that works under the limits $k, q \ll \sqrt{s}$ and $x q \ll k$ \cite{PhysRevD.25.746,Fochler:2013epa,Uphoff:2014hza}.
In the present study, we keep improving the matrix-elements by following the derivation in \cite{Fochler:2013epa} while relaxing the condition $x q \ll k$.
Therefore the updated matrix-elements contain the correct vacuum splitting function in the collinear limit.
We summarize the matrix-elements here and have attached a derivation in appendix \ref{app:ME},
\begin{eqnarray}
\overline{|M^2|}_{g+i\rightarrow g+g+i} &=& \overline{|M^2|}_{g+i\rightarrow g+i} P_{gg}^{g(0)}  D_{gg}^{g},\\
\overline{|M^2|}_{g+i\rightarrow q+\bar{q}+i} &=& \frac{C_F d_F}{C_A d_A}\overline{|M^2|}_{g+i\rightarrow g+i} P_{q\bar{q}}^{g(0)} D_{q\bar{q}}^{g,}\\
\overline{|M^2|}_{q+i\rightarrow q+g+i} &=& \overline{|M^2|}_{q+i\rightarrow q+i} P_{qg}^{q(0)} D_{qg}^{q}.
\end{eqnarray}
Index $i$ represents a quark, an antiquark or a gluon.
The two body matrix-elements that enter the $2\rightarrow 3$ matrix-element are always required to be the $t$-channel contribution.
$P_{bc}^{a(0)}(x)$ are vacuum splitting functions from parton $a$ to partons $b$ and $c$,
\begin{eqnarray}
P_{gg}^{g(0)}  &=& g^2  C_A\frac{1+x^4+(1-x)^4}{x(1-x)},\\
P_{qg}^{q(0)} &=& g^2  C_F\frac{1+(1-x)^4}{x},\\
P_{q\bar{q}}^{g(0)} &=& g^2  \frac{N_f}{2}\left(x^2+(1-x)^4\right).
\end{eqnarray}
The $D_{bc}^{a}$ contains the interference structure,
\begin{eqnarray}
D_{qq}^{g} &=& 
C_A(\mathbf{a}-\mathbf{b})^2 + C_A(\mathbf{a}-\mathbf{b})^2 \\\nonumber
&-& C_A (\mathbf{a}-\mathbf{b})\cdot (\mathbf{a}-\mathbf{c}),
\\
D_{q\bar{q}}^{g} &=& 
C_F(\mathbf{a}-\mathbf{b})^2 + C_F(\mathbf{a}-\mathbf{b})^2 \\\nonumber
&-& (2C_F-C_A) (\mathbf{a}-\mathbf{b})\cdot (\mathbf{a}-\mathbf{b}),
\\
D_{qg}^{q} &=& 
C_F(\mathbf{c}-\mathbf{a})^2 + C_F(\mathbf{c}-\mathbf{b})^2 \\\nonumber
&-& (2C_F-C_A) (\mathbf{c}-\mathbf{a})\cdot (\mathbf{c}-\mathbf{b}).
\end{eqnarray}
with the vectors given by
\begin{eqnarray}
\mathbf{a} = \frac{\mathbf{k} - x\mathbf{q}}{(\mathbf{k} - x\mathbf{q})^2},
\mathbf{b} = \frac{\mathbf{k} - \mathbf{q}}{(\mathbf{k} - \mathbf{q})^2},
\mathbf{c} =  \frac{\mathbf{k}}{\mathbf{k}^2}.
\end{eqnarray}

\subsection{The final incoherent transport equation and Monte-Carlo technique}
Combining all these processes, we summarize the incoherent linearized-Boltzmann plus Langevin equation:
\begin{eqnarray}
\frac{df}{dt} = \mathcal{D}[f] + \mathcal{C}_{1\leftrightarrow 2}[f] + \mathcal{C}_{2\leftrightarrow 2}[f] + \mathcal{C}_{2\leftrightarrow 3}[f].
\end{eqnarray}
The distribution function undergoes soft diffusion and diffusion induced-radiation. 
Hard collisions with the medium are included as $2\leftrightarrow 2$ and $2\leftrightarrow 3$ collision terms.
The next section is devoted to the inclusion of the LPM effect to such an incoherent transport equation.
We now discuss the numerical techniques for simulating the above equation.

The Monte Carlo method starts from representing the distribution function by an ensemble of particle states,
\begin{eqnarray}
f(t,x,p) \approx \sum_{i} \delta^3(x-x_i(t)) \delta^3(p-p_i(t))
\end{eqnarray}
For linearized transport equations, it is sufficient to consider the dynamics of one such particle.
Within a short time step $\Delta t$, a particle undergoes scattering with a certain probability.
In between subsequent collisions, the particle propagates with drag force and the random thermal force.

\paragraph{Order of operation} In the presence of two types of operation: collision $\mathcal{C}[f_Q]$ and diffusion $\mathcal{D}[f_Q]$:
\begin{eqnarray}
\nonumber
  \frac{df}{dt}  &=& 
\left( \mathcal{\hat{C}} + \mathcal{\hat{D}} \right) f_Q.
\end{eqnarray}
In principle, the order of operations on the particle should matter.
However, a different choice of ordering only results in an $O(\Delta t^2)$ change in the updated distribution function.
It is evident with the formal solution of the equation,
\begin{eqnarray}
\nonumber
f_Q(x,p) &=& \exp\left\{ \int_{x'}^x \gamma u \cdot dx \left( \mathcal{\hat{C}} + \mathcal{\hat{D}} \right) \right\} f_Q(x',p)\\
&\approx & e^{\Delta t \hat{C}}e^{\Delta t \hat{D}} f_Q(x', p) + \mathcal{O}(\Delta t^2)
\end{eqnarray}

\paragraph{The diffusion solver}
The Fokker Planck equation can be solved as an ensemble of particles governed by Langevin dynamics.
The Langevin equation in the post-point discretization scheme is \cite{He:2013zua},
\begin{eqnarray}
\Delta \vec{x}_i &=& \frac{p}{E}\Delta t\\
\Delta \vec{p}_i &=& -\Gamma \vec{p}_i \Delta t + \sqrt{\tensor{B}(p+\Delta p) \Delta t  }\vec{\xi}
\end{eqnarray}
$\Gamma$ is the Langevin drag term, and $\vec{\xi}$ is a unit-variance Gaussian random force.
$\tensor{B} = \Ppara \Kpara + \Pperp \Kperp$ are the diffusion coefficients in the tensor form.
$\Ppara$ and $\Pperp$ project any vector into the direction parallel and perpendicular to the direction of motion.

The diffusion coefficients are directly related to the one in the Fokker Planck equation $\Kpara = \hat{q}_L$, and $\Kperp = \hat{q}/2$.
While the relation between the drag coefficient in the Fokker Planck equation $\eta_D$ and the drag force $\Gamma$ in the Langevin equation is discretization scheme dependent.
In the post-step scheme, this relation is \cite{He:2013zua},
\begin{eqnarray}
p_j \Gamma  = p_jA + \left(\sqrt{\Kpara}\Ppara_{lk} + \sqrt{\Kperp}\Pperp_{lk}\right) \ppl \left( \sqrt{\Kpara}\Ppara_{kj} + \sqrt{\Kperp}\Pperp_{kh} \right).
\end{eqnarray}
and reduces to,
\begin{eqnarray}
\Gamma &=& \eta_D + \frac{d \Kpara}{dp^2} + \frac{2\sqrt{\Kpara\Kperp} - 2\Kperp}{p^2} \\
 &=& \frac{\Kpara}{2TE} - \frac{1}{p^2}\left( \sqrt{\Kpara} - \sqrt{\Kperp} \right)^2.
\end{eqnarray}
The Einstein relation between $\eta_D$ and diffusion coefficient is used in the last step.

\paragraph{The scattering solver}
For two-body scattering, neglecting quantum statistics, the collision rate in the rest frame of the medium is,
\begin{eqnarray}
R_a(p) = \sum_{b,c,d}\frac{1}{2E_a}\int \frac{dp_b^3}{(2\pi)^3 2E_b} f_0(p_b) \int d\Phi_m |M^2|_{ab\rightarrow cd}
\end{eqnarray}
A similar expression can also be obtained for $2\leftrightarrow 3$ processes.
For a short amount of time $\Delta t$, the probability to have no collision is $P_{0} = \exp(-\Delta t R)$.
The number of multiple independent collisions satisfies a Poisson distribution with mean $N = \Delta t R$. 
For a particle-based simulation, one always needs to ensure that $\Delta t$ is small enough $(\Delta t \ll 1/R)$ so that effectively there is at most one collision happening within the time step.
Once a collision is sampled to happen, the full final state can be obtained by further sampling each scattering channel and the momentum phase-space differential rates.

The multi-dimensional phase-space sampling is performed sequentially for the initial state and final state phase-space.
For $2\rightarrow 2$ and $2\rightarrow 3$ body processes, we rewrite the integrated rate in the fluid cell rest frame as,
\begin{eqnarray}
R_{2m}(E_1, T) &=& \frac{d}{\nu} \frac{1}{2E_1}\int \frac{e^{-\beta E_2}dp_2^3}{(2\pi)^32E_2} 
\int d\Phi_m\overline{|M|^2}.
\end{eqnarray}
If vacuum matrix-elements are used, the nested integration is a Lorentz invariant quantity, and we can choose to calculate it in the center-of-mass frame of the two-body collision, 
\begin{eqnarray}
\int d\Phi_m\overline{|M_{22}|^2} &=& 2E_12E_2v_{\textrm{rel}}\sigma \nonumber \\
 &=& 2(s-M^2)\sigma_{\textrm{CoM}}^{22}(\sqrt{s}, T)\nonumber \\
  &=& F_{2m}(\sqrt{s}, T)
\end{eqnarray}
where $\sigma$ is the cross-section of the process.
In practice, we tabulate the values of the integrated rates and cross-sections. 
The sampling of initial state $p_2$ determines the center-of-mass energy of the process $s = (p_1+p_2)^2 = 2(E_1 E_2 - p_1p_2 \cos\theta_{12})$.
Subsequently, we sample the momentum-transfer $q$-differential cross-section with $\sqrt{s}, T$ as inputs, and reconstruct the final states given the initial state and $q$.

The sampling of the $3\rightarrow 2$ body process is more difficult due to the larger number of parameters to specify the initial state kinematics,
\begin{eqnarray}
R_{32}(E_1, T) = \frac{d}{\nu} \int \frac{e^{-\beta E_2}dp_2^3}{(2\pi)^32E_2} \frac{e^{-\beta k}dk^3}{(2\pi)^32k}
\int d\Phi_2\overline{|M|^2}.
\end{eqnarray}
The Lorentz invariant nested integral is a function of the initial state 3-body kinematics and temperature,
\begin{eqnarray}
\int d\Phi_2\overline{|M|^2} = F_{32}(\sqrt{s}, \sqrt{s_{12}}, \sqrt{s_{1k}}, T),
\end{eqnarray}
where $s = (p_1+p_2+k)^2$ is the center of mass energy, $s_{12} = (p_1+p_2)^2$ and $s_{1k} = (p_1+k)^2$.
It requires a four-dimensional table for the value of $F_{32}$ and a five-dimensional initial state sampling.
The tabulation of a high-dimensional rate and cross-section tables is manageable if a proper approximating function $A(x, y, \cdots)$ is proposed that captures the limiting behavior of the target function $T(x, y, \cdots)$.
Then, tabulating the ratio of $T/A$ would be extremely efficient and accurate with a moderate-size table.

The sequential sampling breaks the original $m+n$ body phase-space sampling into two lower-dimensional samplings.
However, as one should notice, the prerequisite is that we rewrite the integration over the final state momentum of the matrix-element into a Lorentz invariant form; therefore, the table only depends on the Mandelstam variables and temperature.
This appealing feature is broken by the inclusion of either
quantum statistics or in-medium propagators in the matrix-elements. 
Because quantum statistics introduces factors like $1\pm f(p\cdot u)$ to the final momentum integral and the in-medium propagator is not Lorentz invariant, resulting in a $F_{nm}$ that depends on the relative velocity between the collision system and the medium rest frame.
The result is a significant increase in the dimensionality and complexity of the problem. 
Fortunately, the separation of hard and soft modes allows one to use vacuum matrix-elements at large momentum transfer and absorbs the reference frame dependence into the diffusion equation, which is much easier to solve.

\section{The Landau-Pomeranchuk-Migdal effect: theory}
\begin{figure}

\centering
\includegraphics[width=.8\textwidth]{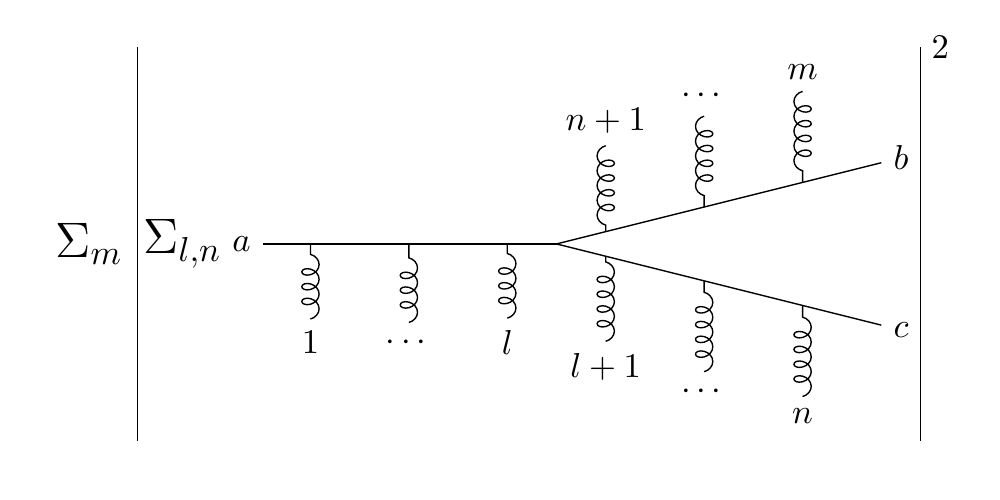}
\caption[A schematic demonstration of the physics of in-medium parton]{A schematic demonstration of the physics of in-medium parton bremsstrahlung. The hard parton ``$a$'' splits into two hard partons ``$b$'' and ``$c$''. The hard partons (quarks or gluon) are denoted as straight lines. The hard system constantly interacts with the medium through gluon exchanges, denoted as the loop lines. One sums the squared amplitudes ($|\cdots|^2$) over an arbitrary number ($m$) of interactions with the medium. Each amplitude has to include all the interference contribution.
Finally, one does an ensemble average over the medium configurations.}
\label{fig:split}
\end{figure}
In the last section, all the processes are treated as instantaneous; however, a process takes a finite amount of time for its final states to lose coherence. 
For elastic collisions, this time is $1/m_D \sim 1/gT$ which is still short compared to the mean-free-path $\lambda \sim 1/g^2 T$, provided a sufficiently small $g$.
For inelastic process, the light-cone energy difference between the initial and final states is,
\begin{eqnarray}
\delta E = \frac{\mathbf{k}^2}{2k} + \frac{\mathbf{p}^2}{2p} - \frac{\mathbf{p'}^2}{2{p'}} = \frac{ [(1-x)\mathbf{k} - x\mathbf{p}]^2}{2x(1-x)E}
\end{eqnarray}
By the uncertainty principle, the coherence time for such a transition is on the order of $\tau_f = 1/\delta E$, termed the ``formation time'' of the radiation. 
In this region of phase space $\mathbf{k}^2, \mathbf{p}^2 < g^2x(1-x)ET$, the average number of collisions during the radiation becomes a relatively large number $N = \tau_f/\lambda >1$, so the picture of induced-radiation from independent scattering centers breaks down.
It has been shown that these multiple scatterings should be resummed \cite{Zakharov:1996fv,Zakharov:1997uu,Baier:1996kr} (please refer to figure \ref{fig:split} for a schematic demonstration), and the leading order resummed radiation probability is,
\begin{eqnarray}
\frac{dP^{a}_{bc}}{d\omega} &=& \frac{\alpha_s P^{0,a}_{bc}(x)}{x^2(1-x)^2 E^2}\mathfrak{Re}\int_0^\infty dt_1 \int_{t_1}^{\infty} dt_2\\\nonumber &&\nabla_{\mathbf{b}_1} \cdot\nabla_{\mathbf{b}_2} \left\{G(t_2, \mathbf{b}_2; t_1, \mathbf{b}_1) - G_0(t_2, \mathbf{b}_2; t_1, \mathbf{b}_1) \right\}|_{\mathbf{b}_1, \mathbf{b}_2 \rightarrow 0}
\label{eq:theory-dR}
\end{eqnarray}
$P^{0,a}_{bc}(x)$ is the vacuum splitting function, and $x$ is the energy fraction carried by the daughter $b$.
Inside the double-time integral, $G$ is the propagator of the following Hamiltonian for the transverse dynamics of the splitting system,
\begin{eqnarray}
\hat{H} &=& \frac{-\nabla^2_{\mathbf{b}} + m^2_\textrm{eff}}{2x(1-x)E} - i \Gamma_3(\mathbf{b})\\
\Gamma_3(\mathbf{b}) &=& \frac{C_a-C_b+C_c}{2}\Gamma_2(\mathbf{b}) + \frac{C_a-C_c+C_b}{2}\Gamma_2(x\mathbf{b}) \\\nonumber
&&+ \frac{C_b+C_c-C_a}{2}\Gamma_2((1-x)\mathbf{b}),
\end{eqnarray}
and $G_0$ is the free propagator. 
The variable $\mathbf{b}$ is the Fourier transformation dual of the transverse momentum, and is usually referred to as the impact-parameter (not to be confused with the one used in the nuclear collision geometry).
$m^2_\textrm{eff}$ is a combination of both parton bare masses and thermal masses.
Finally, the interaction $\Gamma_3(\mathbf{b})$ encodes the transverse broadening of the three body system $a\rightarrow b+c$ \cite{Zakharov:1997uu}.
It has three two-body contributions $\Gamma_2(\mathbf{b})$.
The vacuum piece $G_0$ is subtracted from $G$ so that this formula only computes the medium-induced radiation.
The two gradient operators at time $t_1$ and $t_2$ come from the action of the radiation vertices, meaning this transition receives coherence contribution from $t_1$ to $t_2$.
If one neglects the mass term and rewrites $G$ as $G_0 -i G_0\Gamma_3 G$, the equation simplifies to 
\begin{eqnarray}
\nonumber
\frac{dP^{a}_{bc}}{d\omega} &=& \frac{g^2 P^{0,a}_{bc}(x)}{\pi}\mathfrak{Re}\int_0^\infty dt_1 \int_{t_1}^{\infty} dt_2\\ &&\frac{1}{(2x(1-x) E)^2}\nabla_{\mathbf{b}_1} \cdot\nabla_{\mathbf{b}_2} [G_0(-i\Gamma_3) G](t_2, \mathbf{b}_2;t_1, \mathbf{b}_1)|_{\mathbf{b}_1, \mathbf{b}_2 \rightarrow 0},\label{eq:theory-dR-1}\\
&=& \frac{g^2 P^{0,a}_{bc}(x)}{\pi}\mathfrak{Re}\int_0^\infty dt_1 \int_{t_1}^{\infty} dt_2 F(t_2; t_1).
\label{eq:theory-dR-2}
\end{eqnarray}
$F(t_2, t_1)$ is a short notation for the term to be integrated.

The interaction potential $\Gamma_2(\mathbf{b})$ depends on the assumption of the probe-medium interaction.
For example, in a weakly coupled theory, a compact result is obtained at leading order \cite{Aurenche:2002pd},
\begin{eqnarray}
\Gamma_2(\mathbf{b}) = \frac{1}{\pi}\int \frac{d\mathbf{q}^2}{(2\pi)^2} \frac{g^2 T m_D^2 (1-e^{i\mathbf{b}\cdot\mathbf{q}})}{\mathbf{q}^2(\mathbf{q}^2+m_D^2)}
\end{eqnarray}

There are two systematic ways to investigate equation \ref{eq:theory-dR}.
In a method called the opacity expansion \cite{Wiedemann:2000za,Gyulassy:1999zd}, one solves the propagator in a perturbation series expanding in terms of the number of interactions $\Gamma_3$, or opacity $N \sim L/\lambda$.
Another approach works in the limit of large number of collisions $N\gg 1$ and expands in terms of $1/\ln(N)$.
In this limit, consider soft interactions (small-$q$) and approximate the collision kernel by a harmonic oscillator  $\Gamma_2(b) \approx \frac{1}{4}\hat{q}b^2$, the propagator can be solved analytically \cite{Baier:1996kr,Baier:1998yf,Baier:1996sk},  known as the leading-log ($1/\ln(N)$) approximation.
Taking the residue potential $\Gamma(b) - \frac{1}{4}\hat{q}b^2$ as a perturbation, improvements at the next-to-leading-log level has also been investigated in \cite{Arnold:2008zu,Mehtar-Tani:2019tvy}.

Though the leading-order calculation has this compact form in equation \ref{eq:theory-dR}, it is not trivial to include its effect (even approximately) in the semi-classical Boltzmann simulation.
One can see this problem by observing that it requires a finite time interval of $t_1 \rightarrow t_2$ to compute the splitting rate at time $t_1$, while the Boltzmann equation only has a single time variable. 
We will devote the next section to an approximated solution to this problem.
For the rest of this section, we shall elaborate the details of the current understanding in the opacity expansion and harmonic oscillator (deep-LPM) regime, which significantly facilitates the discussion of the next section.

\subsection{Large medium}
For a large and static medium that approaches the infinite medium limit, 
further simplification is possible. 
The problem becomes a ``static'' one, and a branching rate $\Gamma$ can be defined as the branching probability per unit time.
This limit is known as the AMY equation \cite{Arnold:2002ja,Arnold:2002zm,Arnold:2003zc},
\begin{eqnarray}\label{eq:AMY-1}
\nonumber
\frac{d\Gamma_{a\rightarrow bc}}{dx} &=& \frac{1}{2E\nu_a} \frac{\alpha_s d_a P_{a\rightarrow bc}(x)}{x^2(1-x)^2}\int\frac{d^2\mathbf{k}}{(2\pi)^2}\mathbf{k}\cdot \mathfrak{Re} \mathbf{F}
\end{eqnarray}
where we have dropped the Bose enhancement and the Pauli blocking factors of the outgoing partons from the original formula.
The vector valued wave-function $\mathbf{F}(\mathbf{h}; p, x)$ satisfies the following integral equation \cite{Arnold:2002ja},
\begin{eqnarray}\label{eq:AMY-2}
\nonumber
2\mathbf{k} &=& i\frac{\mathbf{F}(\mathbf{k})}{\tau_f(k)}  + g^2 \mathcal{C}_3[\mathbf{F}]
\end{eqnarray} 
$\mathbf{k}$, and $\tau_f(k)$ is the transverse scale and the formation time of the branching. $\mathcal{C}_3$ is the $\Gamma_3$ operator in the momentum representation,
\begin{eqnarray}
\mathcal{C}_3[f] &=& \int_{\bf q} \mathcal{A}(\mathbf{q}^2)
\left\{  \frac{C_b+C_c-C_a}{2}\left(f_{\bf p}-f_{{\bf p}-{\bf q}}\right) \right.\\\nonumber
&& + \left. \frac{C_a+C_c-C_b}{2}\left(f_{\bf p}-f_{{\bf p}+x{\bf q}}\right) + \frac{C_a+C_b-C_c}{2}\left(f_{\bf p}-f_{{\bf p}+(1-x){\bf q}}\right)\right\}\\
\mathcal{A}(\mathbf{q}^2) &=& \frac{g^2 m_D^2 T}{\mathbf{q}^2\left(m_D^2+\mathbf{q}^2\right)}
\end{eqnarray}
The exact solution can be solved numerically.
Here, we investigate this formula in two extreme regimes: the incoherent limit (Bethe-Heitler regime) and the deep-LPM regime.

{\bf The Bethe-Heitler regime}: The quantum interference can be neglected if the formation time is sufficiently short.
In such cases, the amplitude under the double-time integral has a delta-function like time structure, and the transition probability has a nice interpretation of integrating the localized branching rate over a single time variable.
However, the kinematic range for short formation times is very limited. 
The condition $\tau_f \ll \lambda$ translates to $\omega \ll T$.
For such case, one may solve for $F$ by treating $F/\tau_f$ as a large quantity \cite{Ghiglieri:2015ala}, then, the leading equations are
\begin{eqnarray}
2\mathbf{k}\tau_f(k) &=& - \mathfrak{Im} \mathbf{F} \\
\mathfrak{Re} \mathbf{F} &=& -g^2 \tau_f(k)\mathcal{C}_3[\mathfrak{Im} \mathbf{F}] 
\end{eqnarray}
Take a quark splitting into a quark and a gluon as an example and neglecting the thermal masses, the resulting rate is then proportional to 
\begin{eqnarray}
R &\propto& 2g^2 \int d \mathbf{k}^2 \int d \mathbf{q}^2 \mathcal{A}(\mathbf{q}^2) \left\{
\frac{C_A}{2} \frac{\mathbf{k}}{\mathbf{k}^2}\cdot\left[\frac{\mathbf{k}}{\mathbf{k}^2}-\frac{\mathbf{k}-\mathbf{q}}{(\mathbf{k}-\mathbf{q})^2}\right] \right.\\\nonumber
&&+\left. \frac{2C_F-C_A}{2} \frac{\mathbf{k}}{\mathbf{k}^2}\cdot\left[\frac{\mathbf{k}}{\mathbf{k}^2}-\frac{\mathbf{k}+x\mathbf{q}}{(\mathbf{k}+x\mathbf{q})^2}\right]
+\frac{C_A}{2} \frac{\mathbf{k}}{\mathbf{k}^2}\cdot\left[\frac{\mathbf{k}}{\mathbf{k}^2}-\frac{\mathbf{k}+(1-x)\mathbf{q}}{(\mathbf{k}+(1-x)\mathbf{q})^2}\right]
\right\}.
\end{eqnarray}
Though this expression looks very different from the cross-section formula that we used in the incoherent rate of the Boltzmann equation, they are equivalent upon the integration of $d\mathbf{k}^2$. 
We provide a detailed explanation of this connection between the Bethe-Heitler approximation of the AMY rate equation and the incoherent rate computed with $2\rightarrow 3$ cross-section in appendix \ref{app:ME}.

In the high energy limit $E\gg T \gg \omega$ so that $x\ll 1$, this rate can be approximated by its $x\rightarrow 0$ limit, 
\begin{eqnarray}
\frac{dR}{dx} &=& \frac{4 E\alpha_s d_a P(x)}{\nu_a} \int \frac{d\mathbf{k}^2}{(2\pi)^2} \int \frac{d\mathbf{q}^2 \mathcal{A}(\mathbf{q}^2)}{(2\pi)^2}  \sum_{\pm}
\frac{C_A}{2} \frac{\mathbf{k}}{\mathbf{k}^2}\cdot\left[\frac{\mathbf{k}}{\mathbf{k}^2}-\frac{\mathbf{k}\pm\mathbf{q}}{(\mathbf{k}\pm\mathbf{q})^2}\right] \\
&=& \frac{2 C_A E\alpha_s d_a P(x)}{\nu_a} \int \frac{d\mathbf{k}^2}{(2\pi)^2} \int \frac{d\mathbf{q}^2\mathcal{A}(\mathbf{q}^2)}{(2\pi)^2} 
\left[\frac{\mathbf{k}}{\mathbf{k}^2}-\frac{(\mathbf{k}-\mathbf{q})}{(\mathbf{k}-\mathbf{q})^2}\right]^2
\end{eqnarray}
where in the second step, the $k$ integration of the term with the ``$+$'' sign has been shifted to an integration over $k-q$ to render the expression into the complete square form.
This form is known as the Gunion-Bertsch approximation \cite{PhysRevD.25.746} of inelastic $2\rightarrow 3$ scattering, whose improved form \cite{Fochler:2013epa,Uphoff:2014hza} has been employed in existing full Boltzmann simulations of the partonic transport equation \cite{Xu:2004mz,Uphoff:2010sh}.
To understand the physical meaning of the above expression, we can proceed to integrate and regulate the soft divergence with a screening mass whenever needed.
Eventually we have,
\begin{eqnarray}
\frac{dR}{dx} \propto \alpha_s P(x) \times \alpha_s C_A T \propto \frac{\alpha_s P(x)}{\lambda_g}
\label{eq:incoh-dR}
\end{eqnarray}
Where the second factor $\alpha_s C_A T$ can be interpreted as the inverse of the gluon-mean-free path. 
Now the physical meaning becomes clear: in the incoherent limit, a certain amount of radiation $\alpha_s P(x)$ is triggered every mean-free-path from interactions with the collision centers.

To summarize the Bethe-Heitler regime, the total number of branchings reduces to contributions from an incoherent sum of $2\rightarrow 3$ processes localized at time $t$.
Such contributions are easily incorporated into the Boltzmann equation with the incoherent rate.
However, the validity range for this approximation is at best $x E \sim$ a few times of the temperature.

{\bf The deep-LPM region (leading-log behavior)}: another useful approximation considers the limit $\tau_f$ being so large that many collisions contribute coherently to the branching. 
$\tau_f \gg \lambda$ corresponds to the region when the daughter parton's energy is large $\omega, E-\omega \gg T$.
As a result, the transverse momentum $k$ of the branching should be large compared to the average momentum transfer to each scattering center $q$.
In this limit, a diffusion approximation to the $\mathcal{C}_3$ operators is possible.
The finite difference between $F(k)$ and $F(k+O(q))$ is expanded in $\mathbf{q}$. 
The zeroth-order cancels and the first order $\mathbf{q}$ contribution vanishes due to the symmetric $q$ integration.
Keeping only second order terms in $\mathbf{q}$, the AMY equation is simplified to a diffusion type equation but with a complex diffusion constant and a source term \cite{Arnold:2008zu}
\begin{eqnarray}
- \frac{1}{4} \nabla^2_{\mathbf{k}}\mathbf{F} + i\frac{k^2 + m^2_{\textrm{eff}}}{2x(1-x)E\hat{q}_3}\mathbf{F} = \frac{2\mathbf{k}}{\hat{q}_3}.
\end{eqnarray}
This approximation of the original collision operator is also known as the harmonic oscillator approximation.
$\hat{q}_3$ is the effective transport parameter,
\begin{eqnarray}
\hat{q}_3(x, Q_0^2) &=& \alpha_s T m_D^2 \ln\left(1+\frac{Q_0^2}{m_D^2}\right) C_{abc}(x).\label{eq:qhat3}
\end{eqnarray}
This is obtained by doing the $\mathbf{q}$ integration of the expanded collision operator up to a cut-off scale $Q_0$, below which the small-$q$ approximation is considered to be valid.
The effective transport parameter also depends on the color structure of the splitting,
\begin{eqnarray}
C_{abc}(x) &=&  \frac{C_b+C_c-C_a}{2} + x^2 \frac{C_a+C_c-C_b}{2} \\
&&+(1-x)^2\frac{C_a+C_b-C_c}{2}
\end{eqnarray}
Taking the momentum fraction of the ``$b$'' particle to zero $x\rightarrow 0$, this color factor goes to $C_b$; similarly, $x\rightarrow 1$ which corresponds to ``$c$'' particle taking a vanishing fraction of the total energy, the color factor approaches $C_c$.
Therefore, in these extreme limits $x\rightarrow 0$ or $1$, the effect transport parameters look like the daughter with softer momentum.
With finite $x, 1-x$, the color factor becomes a combination of the colors of the whole splitting system.

Neglecting the thermal mass, the solution to the this diffusion equation can be obtained analytically \cite{Arnold:2008zu},
\begin{eqnarray}
\mathbf{F} = i 4x(1-x)E^2 \frac{\mathbf{k}}{k^2} \left[\exp\left(\frac{-i^{1/2}k^2}{\sqrt{2x(1-x)E\hat{q}_3}}\right)-1\right]
\end{eqnarray}
And the radiation rate can be obtained accordingly,
\begin{eqnarray}\label{eq:AMY-LL}
\frac{dR_{bc}^{a,\textrm{LL}}}{dx} &=& \frac{\alpha_s P_{bc}^{a(0)}}{\pi\sqrt{2}}
\sqrt{\frac{\hat{q}_3(x, Q_0^2)}{2x(1-x)E}} \propto \frac{\alpha_s P(x)}{\langle \tau_f \rangle}
\end{eqnarray}
Such a result is often referred to as the leading-log (leading in $1/\ln(N)$) solution.
An interesting scale $\sqrt{2x(1-x)E\hat{q}_3}$ shows up in this calculation which governs the typical transverse momentum of the splitting, or equivalently $\sqrt{\hat{q}_3/2x(1-x)E}$ which governs the rate at which the splitting happens.
A simple interpretation for these scales is:
during $\tau_f \sim 2x(1-x)E/\mathbf{k}^2$, many soft interactions contribute to the broadening of $\mathbf{k}^2$.
In a diffusion approximation, the variance $\langle \mathbf{k}^2 \rangle$ is linearly proportional to the diffusion constant and time, $\langle \mathbf{k}^2 \rangle \sim \hat{q}_3\tau_f$.
Combined with the expression of formation time, one arrives at the above typical transverse momentum and typical formation time.

Compared to a na\"ive ``incoherent expectation'' in equation \ref{eq:incoh-dR}, the actual radiation rate is reduced by a factor of $\lambda_{\textrm{el}}/\langle \tau_f \rangle$ on average. 
Therefore, in the deep-LPM regime, instead of triggering radiation every mean-free-path, many collision centers contribute coherently and trigger emission every $\tau_f$ which scales as $\lambda \sqrt{\omega/T}$.
Considering that this approximation only works for $\omega \gg T$, we combine this result with the Bethe-Heitler regime and summarize the radiation pattern in a large medium as,
\begin{eqnarray}
\frac{dR^a_{bc}}{dx} \sim \frac{\alpha_s P^{a(0)}_{bc}(x)}{\max\{\lambda, \tau_f\}}
\end{eqnarray}
This simple idea will be the foundation for modeling of the parton branching processes in section \ref{section:modified-transport}.

{\bf The deep-LPM region (next-to-leading-log level)}:
The previously introduced leading-log result has both a simplicity and a clear physical interpretation in explaining what happens in the deep-LPM region $\ln(N) \sim \ln(xE/T) \gg 1$.
Together with the Bethe-Heitler (incoherent) limit at $xE \lesssim T$, one can already develop a pretty good understanding in an infinite medium.

One thing that still deserves a detailed discussion is the upper bound $Q_0$ introduced in the $q$-integration in the leading-log approximation.
This cut-off scale, as a result of the small-$q$ simplification of the full model, is generally unknown and brings uncertainty to the approximation at this level.
This issue is improved at the next-to-leading-log (NLL) level by treating the large-$q$ part of the collision kernel as a perturbation to this approximation.
The authors of \cite{Arnold:2008zu} and more recently, authors of \cite{Mehtar-Tani:2019tvy} have found that a reasonable choice of $Q_0$ is the order of $\mathbf{k}^2$ itself.
A self-consistent determination of $Q_0$ is also possible by requiring a minimal contribution from the NLL correction.
The NLL result, takes a similar structure as the leading-log solution, but with the unknown $Q_0$ replaced by its NLL improved value $Q_{1}$,
\begin{eqnarray}
Q_1^2  \approx \sqrt{\omega \hat{q}} \approx \sqrt{\omega \alpha_s C_R m_D^2 T \ln\frac{Q_0^2}{m_D^2}}
\label{eq:Q1}
\end{eqnarray}
or a self-consistent determination as in \cite{Arnold:2008zu},
\begin{eqnarray}
Q_1^2 &=& \sqrt{2 x (1-x) E \alpha_s T m_D^2}\\\nonumber
&\times & \left(
\frac{C_b+C_c-C_a}{2}\ln\frac{2\xi Q_1^2}{m_D^2} + \frac{C_a+C_c-C_b}{2} x^2 \ln\frac{2\xi Q_1^2}{x^2 m_D^2} \right.\\\nonumber 
&+& \left.\frac{C_a+C_b-C_c}{2} (1-x)^2 \ln\frac{2\xi Q_1^2}{(1-x)^2 m_D^2} \right)^{1/2}
\label{eq:Q1-sf}
\end{eqnarray}
with $\xi \approx 9.1$ a constant. 
It suggests that the optimal choice of the scale is on the order of the branching's transverse momentum itself $\langle k^2 \rangle = \sqrt{2x(1-x)E\hat{q}_3}$, but with an improved logarithmic factor.
With the self-consistently determined $Q_1$, the NLL approximation is very close to the numerical solution of the full model when $\omega \gg T$ \cite{Arnold:2008zu}.

\subsection{Thin medium: opacity expansion}
For a thin and dilute medium, there are only a few effective collisions that contribute.
In such cases, systematic expansion of  $L/\lambda$ has been developed and is known as the opacity expansion \cite{Wiedemann:2000za,Gyulassy:1999zd,Djordjevic:2008iz}.
Solve the propagator with a perturbation series of the interaction potential $\Gamma_3$ and apply soft approximation $x\ll 1$. 
At leading order in the opacity,  the radiation rate is \cite{Djordjevic:2008iz},
\begin{eqnarray}
\frac{dR}{dx} \propto g^2 P(x) C_A \int \frac{d\mathbf{q}^2 d\mathbf{k}^2}{(2\pi)^4} \frac{g^2 T m_D^2}{\mathbf{q}^2(\mathbf{q}^2 + m_D^2)} \frac{\mathbf{k}\cdot\mathbf{q}}{(\mathbf{k}-\mathbf{q})^2 k^2} \left[1-\cos\left(\frac{(\mathbf{k}-\mathbf{q})^2 t}{2x E}\right)\right]
\end{eqnarray}
It has a notable time-dependent $1-\cos(\omega t)$ modulation due to interference from the production point at $t=0$ to the first interaction with medium at time $t$.
Therefore, there is a finite-size effect for the radiation spectrum in a thin medium and the associated energy loss of the leading parton.
The finite-size effect is critical for phenomenological studies because the QGP fireball from nuclear collisions is far from an ``infinite'' medium.

\subsection{Numerical solution for a general case}
Going beyond the above approximation and investigating how the different limiting regimes are connected, one resort to numerical approach.
We follow the approach described in \cite{CaronHuot:2010bp} to solve the propagator $G$ in momentum space.
Neglecting the thermal mass term, the momentum space representation of the splitting rate is,
\begin{eqnarray}
\frac{dR^{a}_{bc}(t)}{dx} = \frac{P^{a(0)}_{bc}(x)}{2\pi x(1-x)E} \mathfrak{Re} \int \frac{d\mathbf{p}^2}{(2\pi)^2} \int_0^t d\tau e^{\frac{-i\tau}{\tau_f(p)}} \mathbf{p}\cdot \mathbf{\Psi}(\mathbf{p}, \tau)
\end{eqnarray}
with the time evolution of the vector-valued wave function solved in the interaction picture with the initial condition,
\begin{eqnarray}
\frac{\partial \mathbf{\Psi}}{\partial \tau} &=& - e^{\frac{-i\tau}{\tau_f(p)}} \mathcal{C}_3\left[e^{\frac{i\tau}{\tau_f(p)}}\mathbf{\psi}(\mathbf{p}, \tau)\right]\\
\mathbf{\Psi}(\mathbf{p}, \tau=0) &=& \mathcal{C}_3\left[\frac{i\mathbf{p}}{p^2+m^2_{\textrm{eff}}}\right]
\end{eqnarray}
The $\mathcal{C}_3$ operation involves a finite difference and two-dimensional integration over the transverse momentum $p$. 
Fortunately, the integration over the azimuths angle of $p$ can be performed analytically at least for the leading order collision kernel with a fixed coupling constant.
Reparametrizing the vector function into a vector part times a rotational invariant function $\mathbf{\Psi} = \mathbf{p}/(p^2+m^2_{\textrm{eff}})^2 \Phi(p^2)$, the evolution equation for the scalar function $\Phi$ is,
\begin{eqnarray}
&&\frac{\partial \Phi(p^2)}{\partial \tau} = - \alpha_s T \sum_n c_n \int dq^2 \left\{\frac{\Phi(p^2)}{|p^2-q^2|} - \frac{\Phi(p^2)}{\sqrt{(p^2+q^2+m_n^2)^2 - 4p^2q^2}}\right. \\\nonumber
&&\left.- e^{i\tau\frac{p^2-q^2}{2x(1-x)E}}\frac{\Phi(q^2)}{2p^2}\frac{(p^2+m^2_{\textrm{eff}})^2}{(q^2+m^2_{\textrm{eff}})^2} \left[\frac{p^2+q^2}{|p^2-q^2|} - \frac{p^2+q^2+m_n^2}{\sqrt{(p^2+q^2+m_n^2)^2 - 4p^2q^2}}\right]\right\}\\
&&\Phi(\tau=0)= i\sum_n c_n \int dq^2 \left\{\frac{p^2+m^2_{\textrm{eff}}}{|p^2-q^2|} - \frac{p^2+m^2_{\textrm{eff}}}{\sqrt{(p^2+q^2+m_n^2)^2 - 4p^2q^2}}\right. \\\nonumber
&&\left.-\frac{(p^2+m^2_{\textrm{eff}})^2}{2p^2(q^2+m^2_{\textrm{eff}})} \left[\frac{p^2+q^2}{|p^2-q^2|} - \frac{p^2+q^2+m_n^2}{\sqrt{(p^2+q^2+m_n^2)^2 - 4p^2q^2}}\right]\right\}.
\end{eqnarray}
Here, the summation goes over the different pieces of the three-body collision kernel, where the original integration variable $\mathbf{q}$ has been shifted to $\mathbf{p}-\mathbf{q}$, $\mathbf{p}+x\mathbf{q}$, and $\mathbf{p}+(1-x) \mathbf{q}$ accordingly.
The color factors $c_n$ are $c_1 = (C_b+C_c-C_a)/2$, $c_2 = x^2(C_a+C_c-C_b)/2$, $c_3 = (1-x)^2(C_a+C_b-C_c)/2$, and $m_n^2$ terms are $m_1^2 = m_D^2$, $m_2^2 = x^2 m_D^2$, $m_3^2 = (1-x)^2 m_D^2$.
The azimuthal integration over $\phi_q$ has been performed, and $q^2$  integrates from zero to infinity.
One may notice that one of the denominators $|q^2-p^2|$ can vanish.
However, as $q^2$ tends to $p^2$, the subtracted term in the second line approaches the cancels the divergence in the first line, and therefore leaves the function finite.
Now, the problem is reduced to an initial value problem of a 1+1 D first order differential-integral equation and can be solved quite efficiently using finite difference and numerical quadrature methods.

\subsection{Mass effect in medium-induced branching}
For radiation in the vacuum, the heavy quark mass is a natural regulator for the collinear divergence,
\begin{eqnarray}
\frac{dP}{dx d\mathbf{k}^2} = \frac{\alpha_s P(x) \mathbf{k}^2}{(\mathbf{k}^2 + x^2 M^2)^2} = \frac{\alpha_s P(x)}{\mathbf{k}^2}\left(\frac{\theta^2}{\theta^2 + \theta_M^2}\right)^2
\end{eqnarray}
Compared to light quark, the radiation off a heavy quark is suppressed within a typical angle $\theta_M = M/E$.
This is often referred to as the ``dead-cone'' (mass) effect \cite{Dokshitzer_1991}.

Inside a medium, the situation is more complicated \cite{Dokshitzer:2001zm,Armesto:2003jh,Abir:2012pu,Zhang:2003wk}. 
The mass not only changes the propagator, but also shorten the formation time
\begin{eqnarray}
\tau_f = \frac{2x(1-x)E}{\mathbf{k}^2 + m_{\textrm{eff}}^{'2}}, m_{\textrm{eff}}^{'2} = (1-x)m_\infty^2 + x^2 M^2
\end{eqnarray}
These two competing features together contribute to the mass correction.

In principle, the formula discussed in this section also applies to heavy quarks once the effective mass $m_{\textrm{eff}}^2$ is replaced by $m_{\textrm{eff}}^{'2}$, though corrections like $M/p$ are dropped.
When $M/p \gtrsim g$, further corrections may be significant, but this is also the region when elastic energy loss starts to dominate the radiative energy loss for heavy quark \cite{Moore:2004tg}.

\subsection{Treating multiple emissions}
The formula that has been discussed in this section only computes the probability of single bremsstrahlung.
In reality, the averaged number of emissions obtained with this formula can be greater than one, so one has to find a strategy to include multiple emissions.
There have been two primary approaches used to resum multiple emissions.
The first one is the modified DGLAP evolution approach for high virtuality partons \cite{Wang:2002pk,Cao:2017qpx}.
The single medium-induced emission probability is added to the vacuum splitting function, and then one applies the DGLAP evolution as in proton-proton collisions but using the so-called modified splitting function.
At low virtuality, the parton's in-medium dynamics is more conveniently described as a time evolution. 
The rate equation is used to generate multiple emissions over time \cite{Arnold:2002zm,Jeon:2003gi,Schenke:2009gb}.
The problem that naturally arises is how to interface these two techniques in a realistic event, where an initial highly virtual parton transits to an in-medium transport parton.
We shall discuss our tentative solution in the next chapter.

\section{A modified transport model for the LPM effect}
\label{section:modified-transport}
In this section, we shall investigate the approximated inclusion of the LPM effect in particle-based Boltzmann transport simulation, termed ``a modified Boltzmann transport''.
The approach is designed to work in a large medium and interpolates the deep-LPM region and the Bethe-Heitler region, with certain finite-size behaviors.
We also discuss the inclusion of the running coupling effect and mass effect for the study of heavy-flavor.

\begin{figure}

\centering
\includegraphics[width=.8\textwidth]{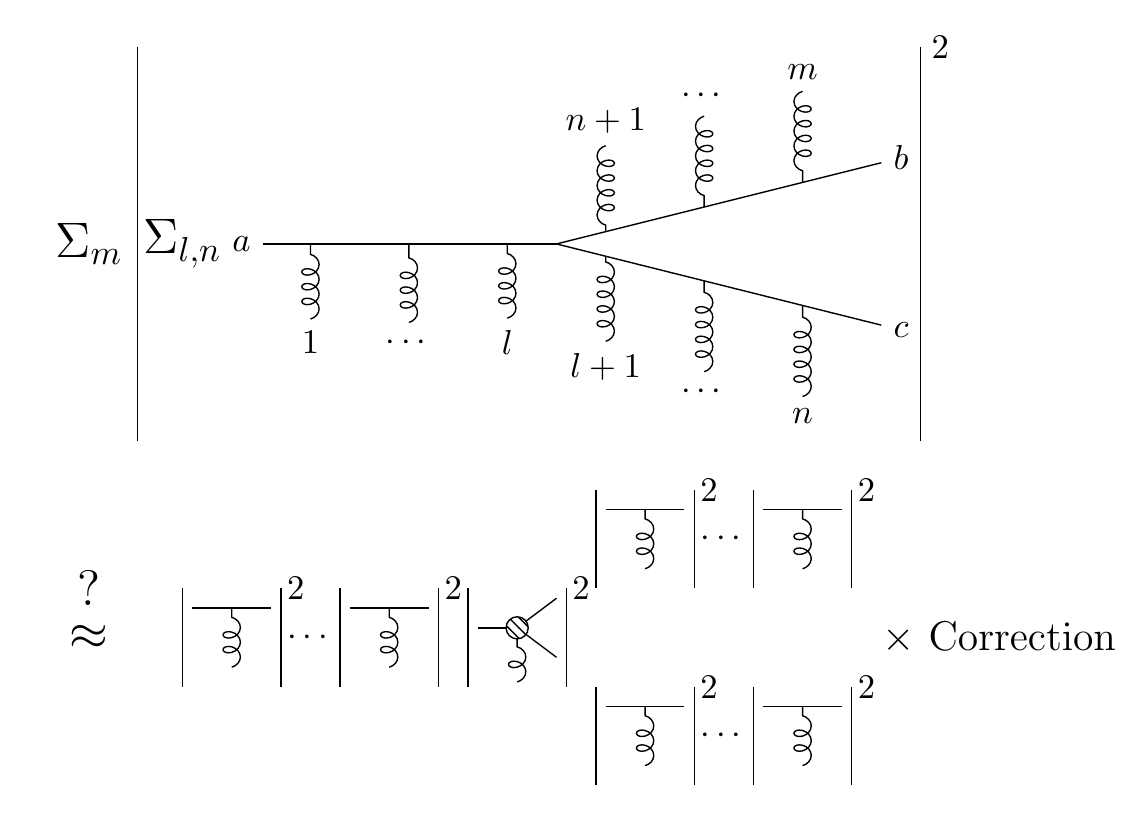}
\caption[A schematic demonstration of the ``modified transport'' approach.]{A schematic demonstration of the ``modified transport'' approach. We seek for an approximation to the coherent process by an incoherent transport simulation with a multiplicative correction to the branching probability.}
\label{fig:split-approx}
\end{figure}
\subsection{Modifying single particle evolution}
To see how to approximate the branching using a modification to the Boltzmann equation, we first go back to the formula for the single-emission rate introduced in equation \ref{eq:theory-dR-2},
\begin{eqnarray}
\frac{dP^{a}_{bc}}{dx} &=& \int_0^\infty dt \frac{g^2 P(x)}{\pi} \int_t^\infty dt'  F(t', t).
\label{eq:full-theory}
\end{eqnarray}
We have denote the function under the integration by $F(t', t)$ for convenience. 
Though only the time argument is written explicitly, $F(t', t)$ actually depends on $x, E$ and the complete medium information along the trajectory of the hard parton: $T(t)$, $u^\mu(t)$, and the coupling strength $g(t)$.

If one expands $F(t',t)$ from equation \ref{eq:theory-dR-1} in a series of products of collision operator and free propagators, then the series contains contributions from processes with an arbitrary number of multiple-interaction with the medium.
Though multiple collisions are also present in the Boltzmann simulation, the difference is that the Boltzmann multiple collisions are independent of the branching processes; therefore, they only broaden the relative transverse momentum {\it without} changing the branching probability.
From the leading-log approximation, we see that the branching probability should be reduced by a factor $\sim \lambda/\tau_f$.
The modified transport approach follows this simple observation and is summarized as follows (please also refers to figure \ref{fig:split-approx} for a schematic demonstration).
\begin{itemize}
\item[1.] Assume an incoherent branching process is generated at $t=t_0$. Do not treat the daughter partons as immediately independent.
\item[2.] Both mother and daughter partons receive elastic broadening from interacting with the medium, which also changes the formation time of the branching.
\item[3.] Evolve the branching system until $t-t_0 > \tau_f$. Then, reject this branching process with an acceptance probability that is proportional to $\lambda/\tau_f$, which corrects for the fact that these multiple scatterings should contribute coherently.
\item[4.] Branching partons for those accepted processes are treated as independent objects from this point; rejected partons are discarded without causing any physical effect.
\end{itemize}

Now we shall explain this scheme in detail.
Formally, this method can be understood as replacing the $F(t',t)$ by an ensemble average over $N$ copies of the branching systems using the following ansatz,
\begin{eqnarray}
F(t', t) \rightarrow \frac{1}{N}\sum_{i=1}^N \frac{b}{\tau_i(t)} \delta(t-t'- a \tau_i(t)).
\end{eqnarray}
Each copy ``$i$'' evolves independently and its formation time can be computed at any given time from the momentum space information,
\begin{eqnarray}
\tau_f(t) = \frac{2x(1-x) E}{\mathbf{k}^2(t)}.
\end{eqnarray}
It is a function of time because elastic interactions changes the transverse momentum over time.
The $\delta$-function requires that a parton branching that starts at time $t'$  only forms at a latter time $t+a\tau_f$.
The inverse time scale $b/\tau_f$ contains the scaling of the rate: a certain amount of radiation is induced every formation time.
Such a particle-based representation of the two-point function $F(t', t)$, is indeed a crude ansatz, and its validity has to be tested later.
The dimensional numbers $a$ and $b$ shall be determined later when matching the prediction of this ansatz to the leading-log and next-to-leading-log calculations introduced in the previous section.

Plug this ansatz for $F(t, t')$ into the branching probability,
\begin{eqnarray}
\frac{dP^{a}_{bc}}{dx} &=& \frac{1}{N}\sum_i \int_0^\infty dt \frac{g^2 P(x)}{\pi} \frac{b}{\tau_i} \\  
 &=& \frac{1}{N}\sum_i \int_0^\infty dt \frac{g^2 P(x)}{\pi \tilde{\lambda}} \frac{b \tilde{\lambda}}{\tau_i(t=t'+\tau_i)}.\\
  &=& \frac{1}{N}\sum_i \int_0^\infty dt \frac{dR_{\textrm{incoh}}}{dx} \frac{b \tilde{\lambda}}{\tau_i(t=t'+\tau_i)}.
\end{eqnarray}
In the second line, we divided and multiplied back an effective mean-free-path $\tilde{\lambda} = m_D^2/\hat{q}_g$.
In doing so, the first factor is interpreted as the incoherent branching rate $R_{\textrm{incoh}}$, while the second factor is simply the acceptance factor for incoherent branching samples we introduced before.
The formation time can be determined self-consistently for each branching copy as it is evolved under the influence of elastic broadening.
It is determined at the time when
\begin{eqnarray}
t - t' < \tau_f(t). 
\end{eqnarray}
This iterative approach for determine $\tau_f$ was first developed and implemented by \cite{Zapp:2011ya}.
In the deep-LPM region where the number of rescattering is large, such a procedure reproduces the expected scaling of the average formation time $\left\langle\tau_f\right\rangle \propto \sqrt{\omega/\hat{q}}$.
This approach also generalizes to a medium with a varying temperature profile because the multiple collisions are performed along the trajectory of the probe.
In cases where the formation time is short so that the acceptance probability is bigger than unity, the acceptance is set to one and the incoherent rate recovers the Bethe-Heitler results.
Therefore, this approach naturally provides an interpolation of the deep-LPM regime for energetic branchings and the Bethe-Heitler regime for soft branching in a large medium.

\paragraph{Determination of the $a$ parameter} Now we will determine the form of parameters $a$ and $b$ with guidance from the theory in the deep-LPM region.
In the leading-log formula, the average inverse formation time is $\langle\tau_f^{-1}\rangle \sim \sqrt{\hat{q}_3 / 2x(1-x)E}$. 
One notice that the effective $\hat{q}_3$ is different from the $\hat{q}$ of the daughter parton 
$\hat{q}_3$ is related to the gluon $\hat{q}$ by the process- and $x$-dependent factor $C_{abc}$ that has been defined before.
For this reason, we chose the $a$ parameter to be the color combination for each branching channel.
\begin{eqnarray}
a \rightarrow a_{abc}(x) = \frac{C_b}{C_{abc}(x)}
\end{eqnarray}

\paragraph{Determination of the $b$ parameter} From the previous theory discussion, we know that there is a logarithmic ambiguity in the cut-off scale $Q_0$ in $\hat{q}_3$, which can be determined at the NLL level to be the same order as the branching transverse momentum.
We need to address what the $Q_0$ scale is in the Boltzmann simulation and how to improve on that.
Because the large-$Q$ part of the elastic rescattering also uses vacuum two-body matrix-elements, the upper bound of the momentum transfer integration is cut-off by the center-of-mass energy $\sqrt{s}$ of each independent collision,
\begin{eqnarray}
s = (p_1 + p_2)^2 = 2E_1 E_2 (1-\cos(\theta))
\end{eqnarray}
where $p_1$ and $p_2$ are the four momenta of the hard parton and the medium parton.
Since at high energy, the cross-section evolves slowly with $\sqrt{s}$, we can define the average $\sqrt{s}$ by simply averaging $p_2$ over the thermal distribution,
\begin{eqnarray}
2p_1\frac{ \int p_2^3 e^{-p_2/T}(1-\cos\theta) dp_2 d\cos \theta }{\int p_2^2 e^{-p_2/T} dp_2 d\cos \theta} =  6ET.
\end{eqnarray}
Therefore, the average $Q_0^2$ from the independent transport simulation is $6ET$, compared to the NLL choice of $\sqrt{\hat{q} \omega}$
The predictions from such a simulation would systematically deviate from theory predictions in a logarithmic manner, varying energy, temperature and coupling constant.
To use the correct scale, we define a scale-dependent acceptance probability to correct the na\"ive choice of $Q_0 \sim \sqrt{6ET}$ with a $b$ parameter,
\begin{eqnarray}
b &=& 0.75\sqrt{\frac{\ln(\hat{Q}_1^2 )}{\ln(\hat{Q}_0^2 )}}.
\label{eq:NLL-b}
\end{eqnarray}
with $\hat{Q}_1^2$ and $\hat{Q}_0^2$ given by,
\begin{eqnarray}
\hat{Q}_1^2 &=& 1 + \frac{\sqrt{2x(1-x)E\hat{q}}}{m_D^2} \approx 1 + \frac{\tau_f}{\tilde{\lambda}}\\
\hat{Q}_0^2 &=& 1 + \frac{6ET}{m_D^2}
\end{eqnarray}
The $0.75$ is a constant determined when the simulation is tuned to theoretical calculations in the next section, and it will be the same throughout the entire work.
This logarithmic ambiguity traces back to the cut-off $Q_0$ imposed on the large-$q^2$ perturbative tail of $\hat{t}$-channel matrix-element;
therefore, if one assumes the absence of such a tail\footnote{ For example, non-perturbative physics motivated coupling between the hard parton and the medium}, one should drop this logarithmic part in the $b-$parameter.

\subsection{Implementing mass effect}
To apply the aforementioned approach to study heavy flavor, we require
the limit that the parton energy is large compared to the heavy quark mass.
Considering that heavy quarks introduce a mass correction to the Fermion propagator, a {\it na\"ive } change is to include the mass effect in both the formation time and also the few-body matrix-elements,
\begin{eqnarray}
\tau_f = \frac{2x(1-x)E}{\mathbf{k}^2} \rightarrow \frac{2x(1-x)E}{\mathbf{k}^2 + x^2 M^2}
\end{eqnarray}
and 
\begin{eqnarray}
\overline{|M|^2}_{2\leftrightarrow 2}(m=0) \rightarrow \overline{|M|^2}_{2\leftrightarrow 2}(m=M)\\
\overline{|M|^2}_{2\leftrightarrow 3}(m=0) \rightarrow \overline{|M|^2}_{2\leftrightarrow 3}(m=M)
\end{eqnarray}
For elastic scatterings, this replacement using the massive version of the two-body matrix-element is justified because subsequent elastic collisions are incoherent in the weak coupling limit.
For inelastic scatterings, again, the problem arises from the coherence over multiple scattering centers.
At high energy, a heavy quark acquires an average transverse momentum $\hat{q} \tau_f$ larger than the typical transverse momentum of the few body matrix-element $\overline{|M|^2}_{2\leftrightarrow 3}$.
As a result, the mass-effect should be less important compared to the scale $\hat{q} \tau_f$ than comparing to the transverse momentum acquired from a single collision center. 
To solve this problem in the simulation, we choose to use the dead-cone approximation for the radiation from a heavy quark.
The $2\rightarrow 3$ and $1\rightarrow 2$ branching of the heavy quark is sampled from the massless calculation, while the formation time is determined using the massive formula.
The key change is that the dead-cone factor modifies the acceptance probability,
\begin{eqnarray}
\frac{b\lambda}{\tau_f} \rightarrow \frac{b\lambda}{\tau_f} \left(\frac{\mathbf{k}^2}{\mathbf{k}^2+x^2M^2}\right)^2.
\end{eqnarray}
Note that the $\mathbf{k}$ here is the branching transverse momentum after the elastic broadening, and on average $\langle \mathbf{k}^2 \rangle = \langle \mathbf{k}_0^2 \rangle + \langle\tau_f\hat{q}\rangle$, where $\langle \mathbf{k}_0^2 \rangle$ is the average transverse momentum sampled from the $2\rightarrow 3$ matrix-element.
One may question the accuracy of approximating the massive version of the complicated multiple scattering matrix-element using a dead-cone approximation.
We will compare the radiation spectrum from the heavy quark to the exact solution for heavy quark in the next section.

\subsection{Implementing the running of $\alpha_s$}
There are two places in the transport model where the running of the strong coupling constant is relevant:
the coupling between the hard parton and the medium, and the coupling constant for the branching vertices.
These two processes often happen at different scales.

For elastic interactions, the scale would be the $\hat{t}$-channel momentum transfer, the typical scale is on the order of the screening mass $|\hat{t}| \sim m_D^2$.  
Using leading order running of $\alpha_s$ $n_f = 3$ and $\Lambda = 0.2$ GeV, 
\begin{eqnarray}
\alpha_s(Q^2) = \frac{4\pi}{9\ln\left(Q^2/\Lambda^2\right)}
\end{eqnarray}
the coupling constant will blow up with the scale getting close to the non-perturbative scale $\Lambda^2$, and applying a leading-order perturbative calculation to such regions is problematic. 
In a medium, we introduce a minimum scale in the running coupling, proportional to the temperature $Q_{\textrm{med}} = \mu \pi T$, to regulate the leading order running formula.
Of course, regulating $\alpha_s$ to a finite value using such a medium scale does not necessarily improve the accuracy of the calculation in this temperature range. 
For example, $\alpha_s(2\pi T)$ ranges from $0.28$ to $0.45$ ($g \sim 1.9$--$2.4$) for temperature decreasing from $T=0.4$ GeV to $T_c$, which are extremely large values, considering that the next-to-leading-order correction to the probe-medium correction is $O(g)$  
\footnote{ As a remark, in the final model-to-data comparison, we try to parametrize the non-perturbative contribution by a diffusion processes in our model to prevent the attempt to explain the coupling to sQGP in a pure perturbative framework.}.
Following \cite{Arnold:2008zu}, the elastic collision couplings are evaluated at the $\hat{t}$-channel momentum transfer $\mathbf{q}^2$.
This involve both the $\alpha_s$ in the large-$q$ matrix-element ($2\leftrightarrow 2$, and the elastic matrix-element factorized in $2\leftrightarrow 3$) as well as the $\alpha_s$ in the soft transport coefficients $\hat{q}_S$ and $\hat{q}_{L, S}$.

Unlike the coupling between hard parton and the medium, the scale for the splitting process is much harder than the screening mass due to transverse momentum broadening. For example, in a static medium,  $\mathbf{k}^2$ scales like $\sqrt{2x(1-x)E\hat{q}} \sim m_D^2 \sqrt{\omega/T}$.
Therefore for splitting where both the daughter patrons are hard $xE, (1-x)E \gg T$, the running of the splitting vertex coupling is under better control than the probe-medium coupling.
The running of the splitting vertex is included in the theory by changing the $\hat{q}_3$ in the NLL formula to its running version  \cite{Arnold:2008zu},
\begin{eqnarray}
\hat{q}_3^{\textrm{running}} \approx \frac{4\pi}{9}\left(g^2(m_D^2) - g^2(Q_0^2)\right) 1.27 T^3 C_{abc}(x),
\label{eq:q3running}
\end{eqnarray}
and then evaluate the splitting $\alpha_s$ around an averaged scale (note that $\mathbf{k}^2$ in the simulation fluctuates a lot),
\begin{eqnarray}
\langle \mathbf{k}^2\rangle \sim m_D^2 \sqrt{E/T\ln(E/T)}
\label{eq:runscale}
\end{eqnarray}  
For transport simulations, the running of splitting vertex requires a two-step implementation. 
First, the $\alpha_s$ for the splitting vertex in the few body matrix-element is evaluated at $\mathbf{k}^2$.
Next, at the end of the elastic broadening for each splitting processes, the acceptance probability is multiplied by a running coupling factor
\begin{eqnarray}
p_{\textrm{running}} = p\times \frac{\alpha_s(\mathbf{k}_{t_0+\tau_f}^2)}{\alpha_s(\mathbf{k}_{t_0}^2)}.
\end{eqnarray}
Where $\mathbf{k}_{t_0}^2$ is the transverse momentum when the splitting is generated from the few-body processes, while $\mathbf{k}_{t_0+\tau_f}^2$ is the final transverse momentum including the elastic broadening, and is on average greater than $\mathbf{k}_{t_0}^2$.

\section{Validating the modified transport approach}
\label{section:valiate_lido}
In this section, we compare the simulation of the ``modified Boltzmann transport'' to theoretical calculations introduced in the previous sections for different parton energy, coupling constants, and medium temperatures that are relevant for phenomenological applications.
Such a model validation is crucial as it tells us whether the model is a good proxy of the underlying theory and quantifies the theoretical uncertainty when applying the model to phenomenological studies and transport parameter extraction.

We first compare the splitting rate $dR/d\omega$ that comes out of the modified Boltzmann simulation to the NLL approximation in the infinite medium limit.
Then, we apply the model to a finite and expanding medium, outside of the region where this approach is designed.
Nevertheless, the model achieves a good qualitative agreement with the theoretical calculation of the finite size effect
In the end, we validate the implementation of the heavy quark mass effect.

\subsection{In a large and static medium}
\begin{figure}

\centering{
\includegraphics[width=.8\columnwidth]{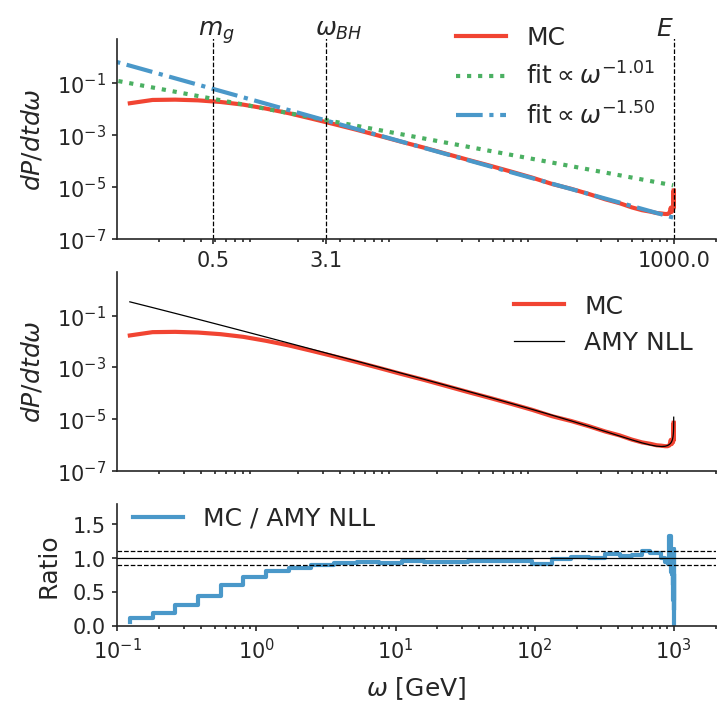}
}
\caption[The $q\rightarrow q+g$ splitting rate simulated in an infinite box with]{The $q\rightarrow q+g$ splitting rate simulated in an infinite box with $T=0.5$ GeV. The quark energy is $E=1$ TeV. $\alpha_s = 0.1$. In the top plot, the spectrum $dR/d\omega$ (red line) is fitted to a power function $\omega^\lambda$ in different gluon energy regions. The green dashed line is fitted in the Bethe-Heitler region $\omega < \omega_{BH}\approx 2\pi T$; the blue dash-dotted line is fitted in the LPM region $\omega > \omega_{BH}$. The middle plot compares to the simulation to NLL solution to the AMY equation, and their ratio is shown in the bottom plot.}
\label{fig:spectrum}
\end{figure}

In practice, to define a Monte-Carlo transport simulation in an infinite medium limit and an eikonal limit of parton propagation, an ensemble of partons of a certain species is initialized at a fixed energy $E_0$ and will be let to propagate in the ``$+z$'' direction.
Each time when a parton scatters elastically or splits, its splitting kinematics ($\omega, \mathbf{k}, t_0, \tau_f$) is recorded, then the mother parton's energy is reset back to its initial value (a test in the eikonal limit).
For elastic re-scatterings in the implementation of the LPM effect, the parton's energy is re-scaled back to the value before scatterings without changing its direction.
The system is evolved for a sufficiently long time $t_{\max}$, and only branchings that takes place within $[t_{\min}, t_{\max}]$ are analyzed to focus on the infinite time behavior of the simulation.

We start with the result for $q\rightarrow q+g$ channel shown in 
figure \ref{fig:spectrum}.
It displays the differential rate $dR/d\omega$ for a 1 TeV quark propagating through a medium of $T=0.5$ $GeV$ with coupling constant $\alpha_s = 0.1$.
The switching scale parameter takes a default value $Q_{\textrm{cut}}^2 = 4 m_D^2$
The vertical axis is the differential branching rate $dR/d\omega$, and the horizontal axis is the energy of the final state gluon $\omega$.
To better understand our result, we have put three ``landmark'' energy scales in the upper plot, which are the initial parton energy $E$, an estimate of the Bethe-Heitler energy $\omega_{\textrm{BH}}\sim\hat{q}_g \lambda_g^2 \sim 2\pi T$, and the screening mass $m_g = m_D/\sqrt{2}$.
In the LPM regime $\omega_{\textrm{BH}} < \omega$, the spectrum falls off as a power law with fitted exponent $-1.50$ (the blue dash-dotted line), and in the Bethe-Heitler regime above the screening mass $m_g < \omega < \omega_{\textrm{BH}}$, the fitted power law exponent is close to $-1$ (the green dotted line).
These exponents are in good agreement with the theoretical expectation that $dR_{\textrm{BH}}/d\omega \propto \omega^{-1}$ and $dR_{\textrm{LPM}}/d\omega \propto \omega^{-3/2}$ from equations \ref{eq:incoh-dR} and \ref{eq:AMY-LL}.
The screening mass regulates the soft divergence of the spectrum below $m_g$.
One may notice a tiny increase of the spectrum when $\omega \rightarrow E$, this is a region where the gluon takes a larger fraction of the initial quark's energy.

In the middle plot, we compare the NLL solution directly to the simulated results.
As a remark, we have tuned the prefactor in the $b$-parameter to be $0.75$  by comparing to this theory prediction at $\alpha_s=0.1, E=1 \textrm{ TeV}, T = 0.5 \textrm{ GeV}$ for the $q\rightarrow q+g$ channel.
For the rest of the comparison with different coupling, parton energy, temperature, and channels, this parameter will {\bf not} be further tuned.
The simulation agrees with the NLL solution very well when $\omega \gg \omega_{\textrm{BH}}$ where the formula is valid.
The bottom plot shows the ratio between the simulation and the theory, and it achieves a level of $\pm 10\%$ agreement in the deep-LPM region.

\begin{figure}

\centering{
\includegraphics[width=.8\columnwidth]{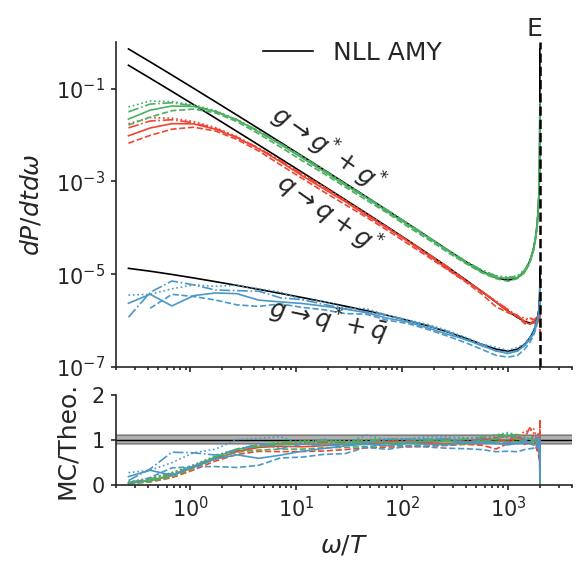}
}
\caption[Top plot shows the rates of different channels $q\rightarrow q+g^*$,]{Top plot shows the rates of different channels $q\rightarrow q+g^*$, $g\rightarrow g+g^*$, and $g\rightarrow q^* + \bar{q}$ are plotted as functions of the daughter (labeled by ``${}^*$'') parton energy. The mother parton has a energy $E=1$ TeV. The medium temperature is $T=0.5$ GeV. $\alpha_s = 0.1$. The simulations (colored lines) are compared to the NLL solutions (black lines). Different line shapes use $Q_{\textrm{cut}}^2 = 2m_D^2$ (dashed), $4m_D^2$ (solid), $8m_D^2$ (dash-dotted), and $16 m_D^2$ (dotted).
The ratios between the simulated results and theory are shown in the bottom plot.}
\label{fig:channel_rate}
\end{figure}

Next, we compare the simulation with all the three channels in figure \ref{fig:channel_rate}.
The setup is the same as the figure \ref{fig:spectrum}.
The red, green and blue lines correspond to the differential branching rate of processes $q\rightarrow q+g^*$, $g\rightarrow g^*+g^*$ and $g\rightarrow q^*+\bar{q}$; the thin back lines are the NLL solution to the AMY equation.
The ``${}^*$'' sign denotes the final-state partons whose energy  $\omega$ are recorded.
For the case of two final state gluons, both are taken into account in the simulation as they are identical particles.
We have discussed the feature for the $q\rightarrow q+g$ process in the previous paragraph. 
The spectral shape of the $g\rightarrow g+g$ process is very similar to the quark splitting channel in the range $\omega \ll E$, with a higher value.
The rate is symmetric with respect to $\omega = E/2$ due to its symmetric final states (though it is hard to tell from this double-log plot), so at large $\omega$, the rate goes up again.
The spectrum of $g\rightarrow q+\bar{q}$ is also symmetric with respect to $\omega = E/2$. Though its final state consists of two different particles,  the splitting function is symmetric.
We see that the simulation achieves a good agreement with the NLL solution in the deep-LPM region $\omega/T > 10$.
Furthermore, in this plot, we vary the switching scale $Q_{\textrm{cut}}$ between the diffusive coupling and scattering-like coupling between the probes and the medium.
The choices are $Q_{\textrm{cut}}^2 = 2 m_D^2$ (dashed lines), $4 m_D^2$ (solid lines), $8 m_D^2$ (dash-dotted lines), and $16 m_D^2$ (dotted lines).
First, the results do depend $Q_{\textrm{cut}}$.
Second, varying $Q_{\textrm{cut}}$ by a factor of $8$ only results in a $\sim 10 \%$ change in the magnitude of the spectra.
In particular, the strongest $Q_{\textrm{cut}}$ dependence appears when using the two smallest choices of  $Q_{\textrm{cut}}^2 = 2 m_D^2$ and $4 m_D^2$.
Once the switching scale is well above the Debye mass, the $Q_{\textrm{cut}}$-dependence is even weaker.

\begin{figure}

\centering
\includegraphics[width=.8\columnwidth]{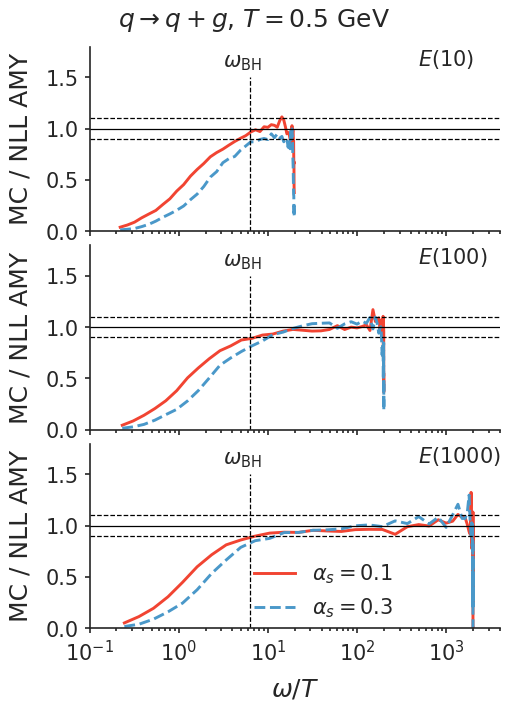}
\caption[Ratios of splitting rate $dR/\omega$ between the modified Boltzmann]{Ratios of splitting rate $dR/\omega$ between the modified Boltzmann simulation and the NLL solution for $q\rightarrow q+g$ splitting. The quark energies are $E = 10$, 100, and 100 GeV from top to the bottom plot. 
And two coupling constants are used: $\alpha_s = 0.1$ (red solid lines) and $\alpha_s = 0.3$ (blue dashed lines).
$\omega$ stands for the gluon energy.
The horizontal dashed lines denote $\pm 10\%$ deviation from unity.}
\label{fig:sys-q2qg}
\end{figure}

\begin{figure}

\centering
\includegraphics[width=.8\columnwidth]{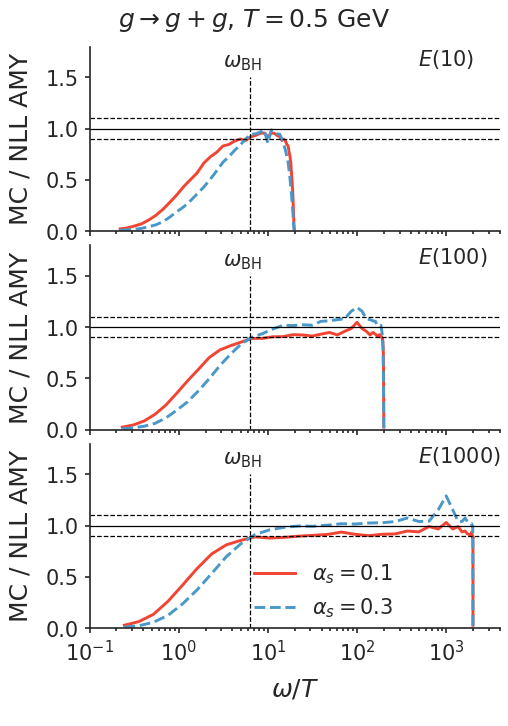}
\caption{The same as figure \ref{fig:sys-q2qg}, but $g \rightarrow g + g$.}
\label{fig:sys-g2gg}
\end{figure}

\begin{figure}

\centering
\includegraphics[width=.8\columnwidth]{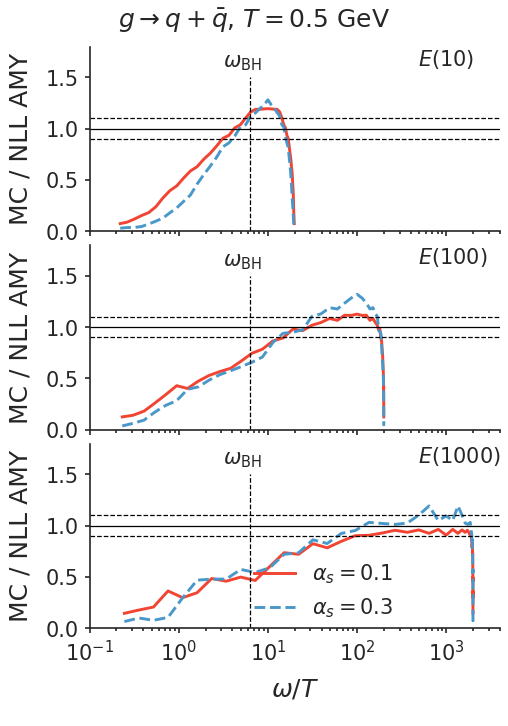}
\caption{The same as figure \ref{fig:sys-q2qg}, but $g \rightarrow q + \bar{q}$.}
\label{fig:sys-g2qqbar}
\end{figure}

Next, we test the model using different values of the coupling constant and parton energies.
We choose both a relative small coupling $\alpha_s = 0.1~(g \approx 1.12)$ and a value closer to the phenomenology coupling $\alpha_s = 0.3~(g \approx 1.94)$, and vary the energy from $10$, $10^2$, to $10^3$ GeV.
The ratios between the simulation and the NLL solutions are shown in figure \ref{fig:sys-q2qg}, \ref{fig:sys-g2gg} and \ref{fig:sys-g2qqbar}.
From these systematic comparisons,
one sees that the simulation reproduces the correct scaling in the LPM region, although this region shrinks due to the decrease of the parton energy.
In conclusion, the overall performance of the modified Boltzmann transport in describing the inelastic processes in a large medium is good and under control.
One remaining problem is that the systematic deviation for the $g\rightarrow q+\bar{q}$ channel is bigger than the other two channels, and we discuss on the cause of this in appendix \ref{app:ME}.

\begin{figure}

\centering
\includegraphics[width=.8\columnwidth]{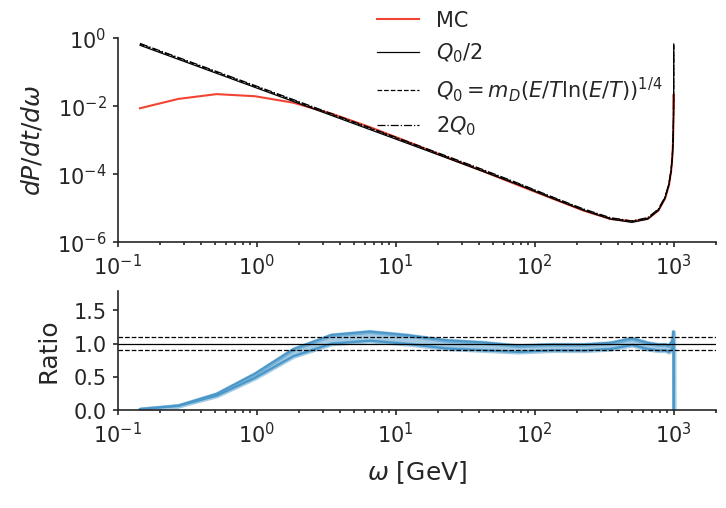}
\caption[Comparing the simulation with running coupling constant to]{Comparing the simulation with running coupling constant to the NLL solution with running $\alpha_s$ prescription.
The scale $Q$ in the theoretical formula for the effective transport parameter $\hat{q}_3$ (equation \ref{eq:q3running}) is chosen to be $1/2$, $1$ and $2$ times of $\sqrt{\langle \mathbf{k}^2 \rangle}$ in equation \ref{eq:runscale}.
The ratio is shown in the bottom plot.}
\label{fig:running}
\end{figure}

Finally, we validate the running coupling calculation in Fig. \ref{fig:running} using the $g\rightarrow g+g$ channel.
The theory curves (black lines) are obtained combining Eq. \ref{eq:AMY-LL} and Eq. \ref{eq:q3running}.
Different line styles correspond to the variation of the $Q_0$ value around an initial guess $m_D (E/T \ln(E/T) )^{1/4}$ by a factor of $2$ above and below.
For this 1 TeV parton, the scale $Q_0$ is large and the running of $\alpha_s$ is rather slow, which explains why the theory curve is not very sensitive to a factor of $4$ change in $Q_0$.
The simulation was performed using the running coupling prescription described in section \ref{section:modified-transport}.
The modified Boltzmann simulation again well describes the overall shape of the spectrum in the deep LPM region. 

\subsection{Branching in a finite / expanding medium}
We have made clear before that this approach is designed for interpolating between the Bethe-Heitler region and the deep-LPM region in a large medium, and looking at validation in the previous section; it indeed works very well.
However, the medium created in heavy-ion collisions is never in the large and static limit, its finite lifetime and spatial extension, local hot spot fluctuations and the fast radial expansion can all have a significant impact on the hard parton propagation. 
Therefore, we need to investigate how our approach would behave in a few more complicated cases including a finite medium and an expanding medium, before applying the model to phenomenological scenarios.

\paragraph{A semi-infinite medium}
Consider a semi-infinite medium with a static temperature profile,
\begin{eqnarray}
T = \begin{cases}
0 , z<0\\
T_0, z>0
\end{cases}
\end{eqnarray}
with hard partons being created at $z=0$ and propagating into the medium.
Deep inside the medium, the medium induced radiation should be asymptotically close to the calculation in an infinite medium.
At the boundary, there is a complicated interference between medium scatterings centers and the hard production vertex.
For a thin medium where the path length is short compared to the formation time, these interference terms can be worked out in the ``opacity expansion'', or by analyzing the propagator $G$ with a semi-infinite temperature profile.
This boundary effect results in a path length dependence of the medium induced branching rate that starts from zero at $t=0$ and gradually approaches the asymptotic value in a large medium.
The parton energy loss at the boundary scales quadratically with the path lenght  $\Delta E \propto L^2$ near the boundary; while it transits to $\Delta E \propto L$ deep inside the medium.

Indeed, we design the modified Boltzmann approach for the case of a large medium, but it also displays a certain finite size effect. 
Remember that the branchings in the modified transport approach take a finite amount of time, and those branchings that become independent at time $t$ are initiated by a $2\rightarrow 3$ processes from a wide range of scattering centers in the past $t' = t - \tau_f$.
Therefore, if the medium is semi-infinite, and there were no scattering centers before $t' = t-\tau_f < 0$, then the medium-induced contribution to the branchings at time $t$ are suppressed.
This reduction gets weaker and weaker when the condition $t-\tau_f > 0$ can be satisfied by more and more induced branchings and eventually, when $t\gg \langle \tau_f\rangle$, this boundary effect dies off in the simulation. 
Of course, we cannot achieve full quantitative agreement with the theory at $L \lesssim \tau_f$ since the detailed few-collision interference pattern is not implemented.
We would like to investigate if our simulation of the boundary effect can qualitatively mimic the interference physics that happens near the boundary.

\begin{figure}

\centering
\includegraphics[width=.8\columnwidth]{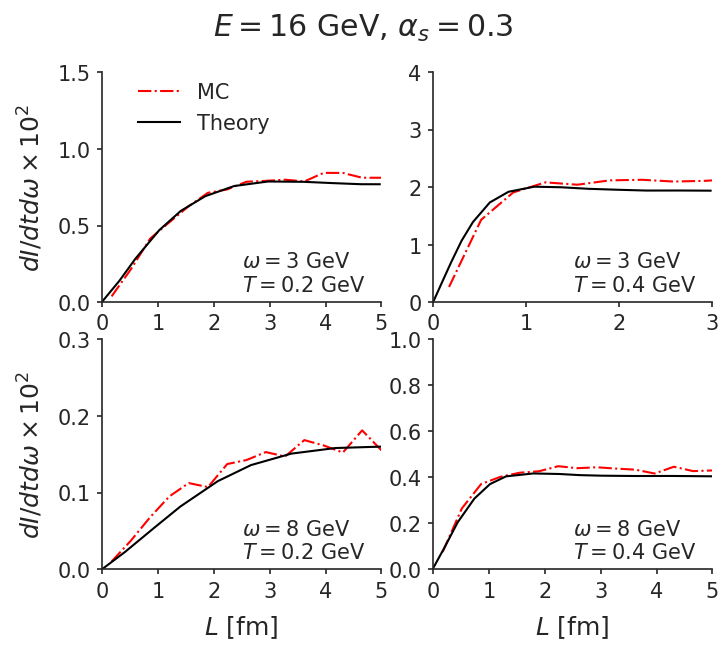}
\caption[The path-length dependence of the simulated $q\rightarrow q+g$]{The path-length dependence of the simulated $q\rightarrow q+g$ emission rate $dR/d\omega$ compared to the direction calculation of formula \ref{eq:full-theory} from \cite{CaronHuot:2010bp}. $\alpha_s = 0.3$. The quark energy is $16$ GeV. The left and right columns have medium temperature $T=0.2$ and $0.4$ GeV respectively. The top and bottom rows show the differential rate at $\omega = 3$ and $8$ GeV.}
\label{fig:spectra-L-alphas=0.3}
\end{figure}

\begin{figure}

\centering
\includegraphics[width=.8\columnwidth]{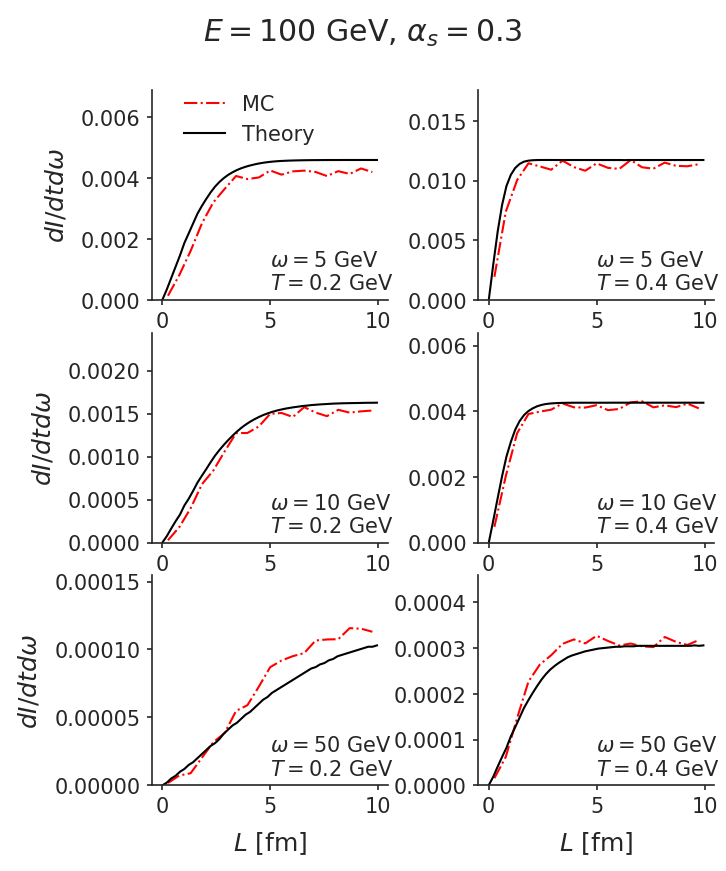}
\caption[The same as figure \ref{fig:spectra-L-alphas=0.3-E100}, but the initial quark energy is $100$ GeV,]{The same as figure \ref{fig:spectra-L-alphas=0.3-E100}, but the initial quark energy is $100$ GeV, and is plotted for gluon at 5 (top), 10 (middle) and 50 (bottom) GeV.}
\label{fig:spectra-L-alphas=0.3-E100}
\end{figure}

In Figure \ref{fig:spectra-L-alphas=0.3}, the differential rate obtained from simulation is compared to the numerical solution of the full leading-order calculation for a finite medium.
The horizontal axis is the time of travel by the hard parton (path length divided by the speed of light), and each subplot shows how the branching rate changes as a function of time with different medium temperatures ($T=0.2$ GeV on the left, $T=0.5$ GeV on the right) and for different branching parton energy ($\omega=3$ GeV at the top, $\omega=8$ GeV at the bottom).
The theory curves are taken from reference \cite{CaronHuot:2010bp} for a 16 GeV parton with coupling constant $\alpha_s = 0.3$, and the red lines are our simulation.
The theory curve first increases linearly and then turn over to a constant value in the large medium limit for $t \gg \sqrt{2x(1-x)E/\hat{q}_3}$.
The simulation, as expected, reproduces the large time limit of the rate.
Moreover, we find that the current implementation also predicts the qualitative ``turn over'' of the spectra at finite path length.
The original paper only published this calculation for a $16$ GeV quark. 
To validate if this qualitative agreement also holds at higher parton energies, we implement the numerical approach of \cite{CaronHuot:2010bp} and compute the theoretical curves for $E=100$ GeV partons.
The comparison between simulation and numerical solutions are shown in figure \ref{fig:spectra-L-alphas=0.3-E100} and again, we find qualitative agreement with the theoretical finite size effect.

\paragraph{An expanding medium}
Fast radial expansion is another important feature of the medium in heavy-ion collisions.
It causes the temperature to decrease drastically in the early stages of the expansion and introduces another time scale in which the medium temperature changes notably.
Assume a simplified power-law changing temperature profile
\begin{eqnarray}
T(\tau; \nu)^3 = T_0^3\left(\frac{\tau_0}{\tau}\right)^{2-1/\nu}.
\end{eqnarray}
The $\nu$ parameter controls the rate of expansion. 
$\nu = 1/2$ is the static medium limit, and $\nu=1$ is the Bjorken flow.
We can define the following medium expansion time, over which the transport parameter changes significantly,
\begin{eqnarray}
\tau_{\textrm{ex}} = \left(\frac{d\ln(T^3)}{d \tau} \right)^{-1} = \frac{\tau}{2-1/\nu}.
\end{eqnarray}
The larger the $\nu$ parameter is, the smaller the expansion time scale.
With $\tau_0 \sim 1$ fm/$c$, the expansion time scale can be short enough that energetic branchings already probe the changing temperature profiles within their formation time $\tau_f > \tau_{\textrm{ex}}$.
One consequence of this fast changing of temperature is that, for these branchings $\tau_f > \tau_{\textrm{ex}}$, the transition probability over a finite amount of time can not be well approximated by integrating rates that are calculated in an infinite box defined by the local temperature,
\begin{eqnarray}
\frac{dP(t_1, t_2)}{d\omega} \neq \int_{t_1}^{t_2} \frac{dR_{\infty}(T(t))}{d\omega} dt,
\end{eqnarray}
where the rate $R_\infty$ is obtained by solving the branching rate in the infinite medium setup. 
Transport models such as MARTINI and TEQUILA use this assumption \cite{Jeon:2003gi,Schenke:2009gb,Dai:2019hbi}.
Our approach takes the change of the medium temperature (and also flow velocity) into account.
It is possible as the rescattering procedure that determines amount of suppression is performed along the trajectory of the hard partons; therefore, naturally includes the effect of the cooling of the medium.
The expansion also changes the typical formation time determined by the rescattering procedure.
Recall that in a static medium the dimensionless combination that enters the leading-log formula is $t/\tau_f \sim t \sqrt{\omega/\hat{q}}$, but with a $\hat{q}$ that is decreasing with temperature.
The self-consistent determination of the formation time requires the following relation to hold on average,
\begin{eqnarray}
t_2 - t_1 &=& \frac{2x(1-x)E}{k_{\perp}^2(t_1) + \int_{t_0}^{t} \hat{q}(\tau) d\tau},\\
\hat{q}(\tau) &=& \hat{q}(\tau_0) \left(\frac{\tau_0}{\tau}\right)^{2-1/\nu}
\end{eqnarray}

\begin{figure}

\centering
\includegraphics[width=.8\columnwidth]{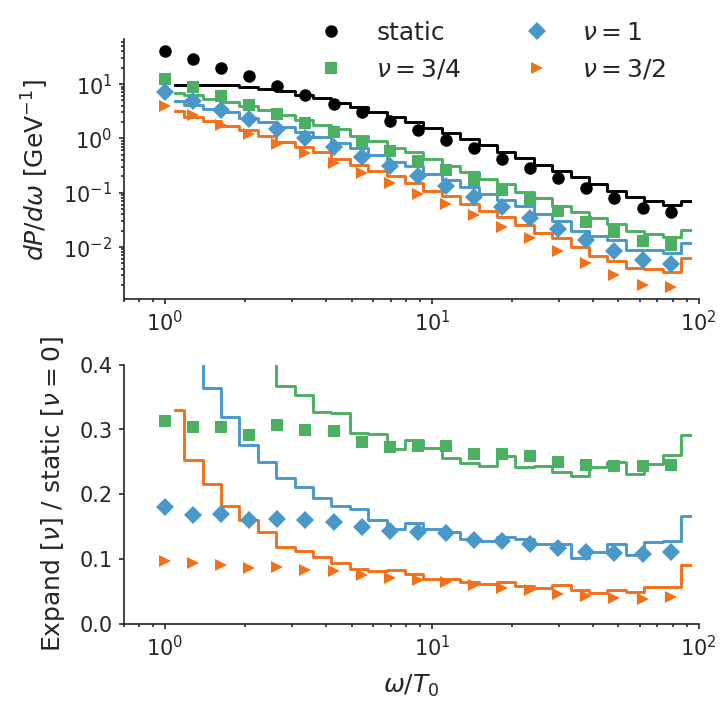}
\caption[Top plot: the simulated spectrum (diffusion plus diffusion-induced]{Top plot: the simulated spectrum (diffusion plus diffusion-induced radiation only) using the parametric medium with expansion parameters $\nu = 0$ (static, black), $3/4$ (green), $1$ (Bjorken, blue), and $3/2$ (orange). The analytic results are shown in symbols and simulations in lines. $\alpha_s=0.3$. The expansion starts at $\tau_0 = 0.2$ fm/$c$ with an initial temperature $T_0 = 1$ GeV. Bottom plot: the ratios between calculation (simulation) in an expanding medium to that in the static medium.}
\label{fig:Bjorken-BDMPS}
\end{figure}

Comparing to theoretical calculations, we utilize a result obtained in the BDMPS framework \cite{Baier:1996kr,Baier:1998yf}.
Using the power-law decreasing temperature profile, the obtained branching probability for the $q\rightarrow q+g$ splitting is \cite{Baier:1998yf},
\begin{eqnarray}
\frac{dP}{d\omega} &=& \frac{\alpha_s}{2\pi E}P_{q\rightarrow qg}(x)\mathfrak{Re}\int_{\tau_0}^{\tau_0+L}\frac{dt_f}{t_f}\int_{\tau_0}^{t_f}\frac{dt_i}{t_i} \frac{1}{\nu^2}\\
\nonumber
&& \left.\left[ I_{\nu-1}(z_i)K_{\nu-1}(z_f)-I_{\nu-1}(z_f)K_{\nu-1}(z_i)\right]^{-2}\right|_{\omega=xE}^{\omega=\infty},\\
z_{i,f} &=& 2i\nu \sqrt{\frac{\hat{q}_g(1-x+C_F/C_A x^2)}{2(1-x)\omega}} \tau_0 \left( \frac{t_{i,f}}{\tau_0}\right) ^{1/2\nu}.
\end{eqnarray}
This result recovers the static BDMPS result \cite{Baier:1996kr} when $\nu=1/2$.
One potential problem of comparing the formula to our simulation is that this BDMPS calculation works in the multiple-soft limit (leading log).
Therefore, we use only the diffusion-induced radiation in the simulation and deactivate the large-$Q$ $2\rightarrow 3$ scattering part.
Also, as mentioned before, in the absence of the perturbative tail in the collision kernel, $b=0.75$ is used without the logarithmic correcting factor in equation \ref{eq:NLL-b}.
Besides, we will not focus too much on the direct comparison between the spectra (top of figure \ref{fig:Bjorken-BDMPS}), but on the ratio between the expanding calculation/simulation over the static calculation/simulation (bottom of figure \ref{fig:Bjorken-BDMPS}).
This ratio reflects the change of the shape of the spectra due to the dropping of the temperature.

The simulation uses a medium with initial temperature $T_0=1$ GeV at $\tau_0=0.2$ fm/$c$ and lasts until $\tau = 20$ fm/$c$ using four different expansion rates $\nu = 1/2, 3/4, 1, 3/2$.
These choices correspond to a static medium, a slowly expanding medium, Bjorken flow, and a faster-than-Bjorken expansion.
We find that when $\omega \gg T_0$, the degree of change in the radiation spectra is well reproduced by the modified transport simulations.

\subsection{Comparison with a simplified solution of the transport equation}
Given the model mimic the single-emission vertex reasonably well, we now focus on the evolution of the full distribution function $f(t, \mathbf{x}, \mathbf{p})$ of hard partons. 
Though the behavior of the distribution function is far more complicated than the final-state distribution of a single-emission vertex,
surprisingly, an analytic solution exists for a simplified version of the transport equation \cite{Blaizot:2013hx,Blaizot:2015jea}.
Here we have listed the simplifications and assumptions for readers reference:
\begin{itemize}
\item Focus only on the longitudinal momentum distribution function in the high energy limit $f(t, \mathbf{x}, \mathbf{p})\rightarrow f(t, p_z \approx E)$. 
\item The initial condition consists of a single gluon $f = \delta(E-E_0)$.
\item Neglect elastic energy loss and use fixed coupling constant.
\item The system is gluonic. And retain only the singular part in the gluon splitting function $P_{gg}^g(0) = C_A(1+z^4+(1-z)^4)z^{-1}(1-z)^{-1} \rightarrow C_A z^{-1}(1-z)^{-1}$.
\item Consider the deep-LPM region ($\tau_f \gg \lambda$) in a large medium ($L\gg \tau_f$).
\item Use a leading-log picture where $\hat{q}$ is independent of the parton energy and neglect the difference between $\hat{q}$ and $\hat{q}_3$, so the formation time is $\tau_f \propto \sqrt{z(1-z)E/\hat{q}}$.
\end{itemize}
\begin{figure}
\centering

\includegraphics[width=.7\columnwidth]{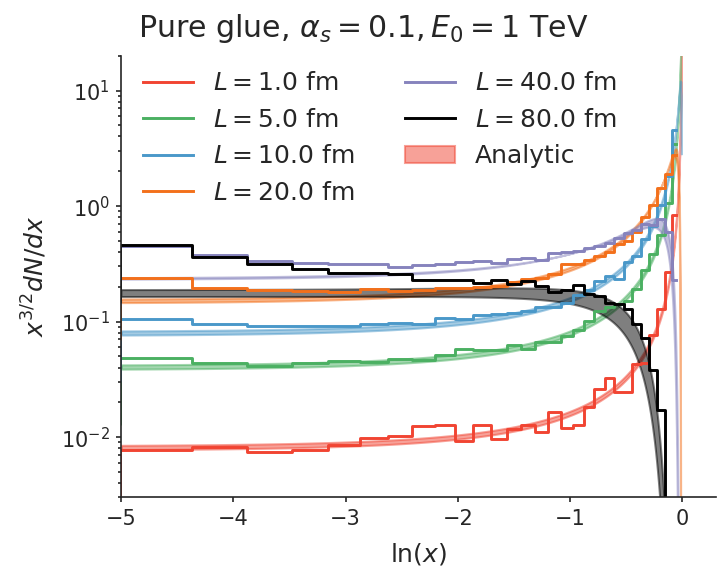}
\caption[Comparing the simulated energy distribution shared among]{Comparing the simulated energy distribution shared among the hard partons to the simplified solution in equation \ref{eq:simple-jet-solution} \cite{Blaizot:2013hx}.}
\label{fig:jet_solution}
\end{figure}
Under these assumptions, the authors of \cite{Blaizot:2013hx} write the transport equation as 
\begin{eqnarray}
\frac{\partial D(x, \tau)}{\partial \tau} = \int_x^1 \frac{dz}{z^{3/2}(1-z)^{3/2}} \sqrt{\frac{z}{x}} D\left(\frac{x}{z},\tau\right) -  \frac{D(x, \tau)}{\sqrt{x}}\int_0^1 \frac{z dz}{z^{3/2}(1-z)^{3/2}}.
\label{eq:simple-jet-eq}
\end{eqnarray}
$D(x, \tau) = E f(t, E=xE_0)$ is understood as the energy distribution with initial condition $D(x, \tau=0) = \delta(1-x)$.
$\tau$ is a rescaled dimensionless ``time'' variable,
\begin{eqnarray}
\tau = \alpha_s C_A/\pi \sqrt{\hat{q}/E_0} t.
\label{eq:simple-jet-tau}
\end{eqnarray}
The gain-term and loss-term for the energy distribution are on the right-hand side.
The solution is \cite{Blaizot:2013hx},
\begin{eqnarray}
D(x, \tau) = \frac{\tau}{x^{1/2}(1-x)^{3/2}}e^{-\frac{\pi \tau^2}{1-x}}.
\label{eq:simple-jet-solution}
\end{eqnarray}
It displays a simple $1/\sqrt{x}$ scaling at small $x$.
The Gaussian decaying factor quickly damps the initial peak of the energy distribution at $x\rightarrow 1$.

Equation \ref{eq:simple-jet-solution} requires a series of approximations; therefore, we make the following changes the simulation.
First, we use the approximated splitting function for $g\rightarrow g+g$ channel and replace $\hat{q}_3$ by $\hat{q}$ of the gluon in the transport model. 
Second, elastic broadening is essential to the implementation of the LPM effect in the model so we cannot simply turn off elastic processes completely. 
Instead, after elastic collisions/diffusion, we rescale the energy of the parton back to its value before the collision to remove the energy loss, while keeping the momentum broadening effect.
Moreover, we remove the finite-size effect in the simulation, which is done by performing the elastic broadening locally without propagating the parton forward in space-time.
These procedures take the simulation closer to the physical ingredients contained in equation \ref{eq:simple-jet-eq}.
One last issue is the determination of $\hat{q}$ used the comparison. The effective $\hat{q}$ in the model is energy dependent $\hat{q} \sim  \alpha_s C_A T m_D^2\ln(1+\sqrt{2x(1-x)\hat{q}E_0}/m_D^2)$, with $\hat{q}_0 = \alpha_s C_A T m_D^2\ln(1+6ET/m_D^2)$.
We neglect the weak $x$-dependence and use a range
\begin{eqnarray}
\ln\left(
1+\frac{\sqrt{E_0\hat{q}_0}/2}{m_D^2}
\right) < \frac{\hat{q}}{\alpha_s C_A T m_D^2} < 
\ln\left(
1+\frac{\sqrt{2 E_0\hat{q}_0}}{m_D^2}
\right)
\end{eqnarray}
in equation \ref{eq:simple-jet-tau} to quantify the uncertainty.

We initialize the system with an ensemble of $E_0=1$ TeV gluons in an infinite medium with temperature $0.5$ GeV.
This choice of parameters gives $L_c = \sqrt{E_0/\hat{q}} \approx 12$ fm;
however, since we have turned off the finite-size effect in the simulation, it is legitimate to make comparison when $L\lesssim L_c$.
In figure \ref{fig:jet_solution}, the simulated energy distribution (histograms) evolved to different path-length is compared to the simplified solutions using the two estimates $\hat{q}$ (colored bands).
Both results are multiplied by $\sqrt{x}$ to investigate the scaling behavior at small-$x$.
We find a good agreement with the analytic solution from large to moderate-small $x$ values. 
The small-$x$ part of the simulation starts to deviate from the trend at later times, e.g., the uprising tail at small-$x$ for the $L=40$ fm and $L=80$ fm cases.
One sees that the transport equation quickly builds up a power-law like energy cascade spectrum as predicted. 
The initial energy of the 1 TeV gluon gets transported along this cascade to the small-$x$ region; eventually, the initial peak near $x=1$ completely disappears, and the cascade also starts to fade.

\section{Heavy quarks and thermalization test}
\begin{figure}

\includegraphics[width=\columnwidth]{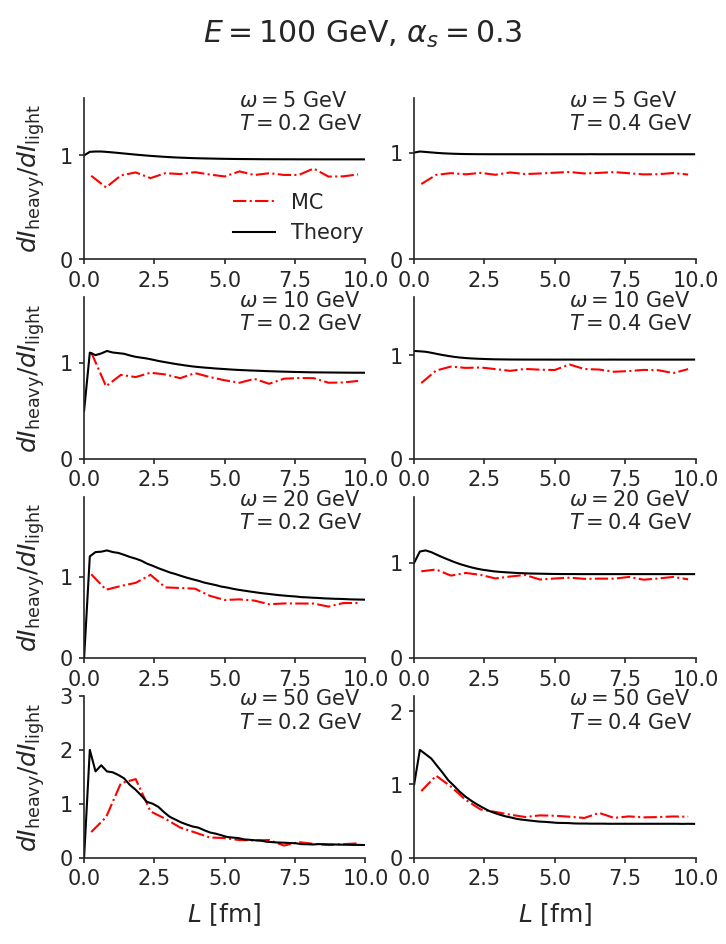}
\caption[The mass dependence of the radiation spectrum $q\rightarrow q+g$,]{The mass dependence of the radiation spectrum $q\rightarrow q+g$, presented as the ratio of bottom $dR/d\omega$ over the light quark $dR/d\omega$, as functions of path length. The left and right columns use temperatures 0.2 and 0.4 GeV. Different rows (from top to bottom) plot cases for gluon energy $\omega = 5, 10, 20, 50$ (GeV).}
\label{fig:mass}
\end{figure}
Finally, we check the model performance for heavy quarks.
The theory curves are obtained by solving the exact equation with an effective mass term,
\begin{eqnarray}
m_{\textrm{eff}}^2 = (1-x)m_g^2 + x^2 M^2
\end{eqnarray}
which includes both the thermal mass of the gluon and the current mass of the heavy quark.
We present the comparison between the simulation and the theory in terms of the ratio between the differential branching rate of the heavy quark (charm mass at 1.3 GeV, bottom mass at 4.2 GeV) and the light quark.
In figure \ref{fig:mass} for the bottom quark case, the horizontal axis is the path-length, and the vertical axis is the ratio.
Different rows have different radiated gluon energies, and different  columns have medium temperatures at $0.2$ GeV (left) and $0.4$ GeV (right) respectively.
The initial bottom quark energy is 100 GeV, and the coupling is $\alpha_s=0.3$.
We see that the dead-cone approximation agrees better with the theory calculation at larger $x$ and larger path-length.
Deviations observed at small path length are understood as the limitation of our implementation to the large medium and should be better treated by the opacity expansion. 
The deviation at small $x$ is interesting since the theory almost predicts an identical heavy quark radiation spectra as the light quark. 
This absence of a dead cone at small-$x$ is already observed in early works studying heavy quark energy loss in both the opacity expansion and the BDMPS framework \cite{Armesto:2003jh}.
It means that the treatment of the mass effect is not as simple as the dead-cone approximation and should be improved in the future.

\paragraph{Thermalization of heavy quarks}
\begin{figure}

\centering
\includegraphics[width=.8\textwidth]{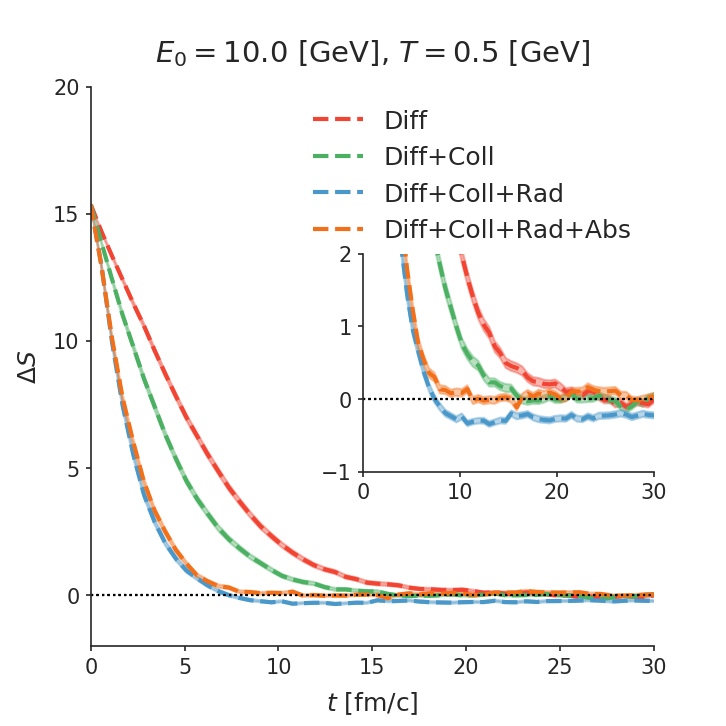}
\caption[The approach to thermal equilibrium of heavy flavor is quantified]{The approach to thermal equilibrium of heavy flavor is quantified as the change of $\Delta S$ (defined in equation \ref{eq:DeltaS}) with time. The red dotted line includes elastic processes only. The green dashed lines= further includes $2\rightarrow 3$ and $1\rightarrow 2$ processes. The solid blue line turns on the detailed balance processes $3\rightarrow 2$ and $2\rightarrow 1$.}
\label{fig:thermalization}
\end{figure}
Due to a heavy quark's large mass, it takes a longer time to thermalize, and the low-$p_T$ end of the heavy quark production in the heavy-ion collisions can carry information on non-equilibrium dynamics.
To extract the degrees of thermalization, one has to make sure the correct thermal limit is achieved in the transport model, given sufficient time.
This is trivial for large-angle elastic scatterings and diffusion processes as long as the correct Einstein relation is imposed.
The $n\rightarrow n+1$ body radiative process is approximated by an initial $2\rightarrow 3$ or $1\rightarrow 2$ process and a sequence of elastic interactions; therefore, in principle, the $n+1\rightarrow n$ absorption processes need to be treated on the same footing to restore the detailed-balance in the modified-Boltzmann equation.
It can be done but is overly complicated.
Here we argue that close to a few times of temperature, the LPM effect is not that strong and an incoherent implementation of the absorption is enough to study the bulk of particles close to thermal distribution.

We define a quantity $\Delta S$ to measure the approach to thermal distribution $f_0 = e^{-E/T}$ for an ensemble of heavy quarks,
\begin{eqnarray}
\Delta S &=& - \langle \ln f_0 \rangle - S_0 \\
 &=& - \frac{1}{N}\sum_i\ln f_0(E_i) - \frac{\int dp^3 f_0 \ln(f_0)}{\int dp^3 f_0}
 \label{eq:DeltaS}
\end{eqnarray}
Where the first term is the ensemble average of the function $-\ln f_0$, and the subtracted term is proportional to the entropy of distribution $f_0$.
Note that the quantity is zero if the ensemble is thermalized.
If the system is approaching a thermal distribution with an effective temperature such that $f = e^{-E/T'}$, then $\Delta S$ is 
\begin{eqnarray}
\Delta S = \frac{\int dp^3 E/T e^{-E/T'}}{\int dp^3 e^{-E/T'}} - S_0 = \frac{T'-T}{T}
\end{eqnarray}
which is a measure of the deviation of the effective temperature from the thermal bath temperature.

Using this definition, we plot $\Delta S$ as a function of time for 1000 heavy quarks that are initialized at $E=10$ GeV.
The temperature of the thermal bath is 0.5 GeV, and we used a fixed $\alpha_s = 0.3$.
Under the influence of diffusion (red) and diffusion plus large-angle elastic collisions (green), $\Delta S$ decreases from a large value until fluctuating around zero after 25 fm/$c$ (red) and $15$ fm/$c$ (green).
Now, adding the radiative processes (blue), $\Delta S$ reaches a value below zero, which is a false equilibrium.
Only after the balancing processes of parton absorption are also included (orange), the correct thermal equilibrium limit restores. 
We also found that the absorption process only sets in when the ensemble is close enough to the thermal distribution, as the blue line and the orange line are almost overlapping until $\Delta S$ drops to 0.3.
This is because the absorbed gluon follows the thermal distribution in the medium while phase-space for a high energy parton $E\gg T$ to absorb a low energy gluon is very limited $x<T/E$, compared to radiation processes where the value of $x$ is not restricted by the Boltzmann factor $e^{-xE/T}$.

\section{Comments on two other inelastic process implementations}
\label{section:compare_former}
I find it beneficial to discuss two other inelastic process implementations for the reader's reference. 
They are termed as the ``coherence factor'' approach and the ``blocking radiation'' approach.
I have used the previous approach in my earlier studies \cite{Ke:2018tsh}, but it is the problems I encountered in this method that later motivated the development of the ``modified Boltzmann transport'' method. 
I shall show in this section that in the deep-LPM region, the ``coherence factor'' approach still qualitatively agrees with the power counting of the LPM suppression $\lambda_{el}/\tau_f$, though it only includes the effect of one medium scattering center and the method can be logarithmically dependent on the infrared cut-off.
The ``blocking radiation'' approach, however,  does not reproduce the correction magnitude of the LPM suppression.
These two approaches, together with the ``modified Boltzmann'' approach will be compared later using the ``energy loss'' of a fixed energy quark.

\subsection{The coherence factor approach}
This approach is first implemented in the improved Langevin equation \cite{Cao:2013ita}, using the single medium-induced radiation probability from the higher-twist calculation \cite{Majumder:2009ge,Wang:2001ifa} and a prescription for multiple emissions.
The higher-twist formula of medium-induced radiation is derived for a high virtuality parton, including the interference of the hard production vertex and one medium scattering center.
The single radiation rate reads,
\begin{eqnarray}
\frac{dN_g}{dx d\mathbf{k}^2 dt} = \frac{\alpha_s P(x)\hat{q}_g}{\pi \mathbf{k}^4} 2\left(1-\cos\frac{t-t_0}{\tau_f}\right), \tau_f = \frac{2x(1-x)E}{\mathbf{k}^2}
\end{eqnarray}
Here the radiation rate is time-dependent ($t$) due to the interference with the hard production at time $t_i$. 
Note that the interference factor cancels the collinear divergence. 
The only divergence comes from soft emission $x\rightarrow 0$. 
This divergence is not a problem for computing more physical quantities such as energy loss, as gluon absorption processes will balance it.
However, an infrared cut-off $x>x_c$ has to be introduced to apply the rate equation formulation. 

The advantage is that if there is only one radiation, then the sampling of the time-dependent rate indeed reproduces the higher-twist calculation.
However, the ambiguity arises from the way it handles multiple emissions.
For example, one can compute the average number of emission by integrating this formula along the trajectory of the hard parton and then samples the fluctuating number of emissions with a Poisson distribution.
However, of course, this would assume the parton energy is not significantly changed during the process, and it is not clear how the presence of more than one scattering center would change this picture.
Here we would like to discuss another method in dealing with multiple emission using the higher-twist formula in a time evolution manner \cite{Cao:2013ita}.
The algorithm goes as follows:
\begin{itemize}
\item[1.] Choose an infrared cut-off for the gluon energy $x_c \propto T/E$, and a small enough time step $\Delta t$, so that the average number of emissions is much smaller than $1$ to suppress multiple emissions within $\Delta t$,
\begin{eqnarray}
\langle N_g \rangle = \Delta t \int_{x_c}^1 dx \int d\mathbf{k}^2 \frac{dN_g(t-t_0)}{dx d\mathbf{k}^2 dt} \ll 1.
\end{eqnarray}
\item[2.] Sample N according to a Poisson distribution with $\langle N_g \rangle$. For $\langle N_g \rangle \ll 1$, it is sufficient to sample the two leading cases of $N=0, 1$, as the probability to have more than 1 emission is negligible ($P_{N>2} = 1-e^{-\langle N_g \rangle}-e^{-\langle N_g \rangle}/\langle N_g \rangle = O(\langle N_g \rangle^2) \ll P_1 \ll P_0$).
\item[3.] If $N=0$ then propagate the parton to the $t+\Delta t$. If $N=1$, then sample the emission gluon's $x$, and $\mathbf{k}$ by the differential rate. {\it Meanwhile, $t_0$ is set to $t$}, so that the next emission's probability will be accumulated from zero again.
\item[4.] Proceed to the next time step.
\end{itemize}
We found that the key step here is resetting the clock $t_0 = t$ for the parton after every emission.
As a result, from the second emission onward, the $t-t_0$ that appears in the interference factor is measuring the time difference between two medium scattering centers.
Therefore, we will not interpret this procedure as the higher-twist rate (interference between initial hard vertex and one medium collision center) starting from the second emission;
instead, we understand it as an ansatz, from the second emission onward, to treat medium-induced radiation in a large medium, as it does not require any information on the production vertex.

Considering that it only includes one medium scattering center in the trigger of the radiation, one wonders if this approach reproduces any in-medium radiation features predicted by theory.
It is not immediately clear what this iterative procedure predicts unless one performs explicit simulations. 
To make a process, by pondering on the meaning of the ``clock resetting'' step, we find that the typical $\Delta = t-t_0$ between two emissions is a time scale within which the emission probability reaches order one,
\begin{eqnarray}
1 \sim \int_{t_0}^{t} dt\int_{x_c}^1 dx \int d\mathbf{k}^2 \frac{dN_g(t-t_0)}{dx d\mathbf{k}^2 dt}.
\end{eqnarray}
With this key observation, after a few step of algebra, we are able to isolate the qualitative features of this approach.
Taking the soft approximation $P(x) \sim 2/x$, $\tau_f\sim 2xE/\mathbf{k}^2$, and performing the time integral first, then the $\mathbf{k}$ integral with limits from $0$ to $xE$.
\begin{eqnarray}
1 &\sim& 4\alpha_s\hat{q}\Delta t \int_{x_c}^1 \frac{dx}{x} \int \frac{d\mathbf{k}^2}{\mathbf{k}^4}\left(1-\frac{\sin(\Delta t/\tau_f)}{\Delta t/\tau_f}\right)\\
&=& \alpha_s\hat{q}_g \Delta t^3 \int_{\frac{\Delta t E x_c}{2}}^{\frac{\Delta t E}{2}} 
\frac{du}{u^2} \frac{u^2 \mathrm{Si}(u) -2u + \sin(u) + u\cos(u)}{u^2}\\
&=& \frac{\alpha_s\hat{q}_g \Delta t^3}{3u^3} \left(
u^3\mathrm{Ci}(u)-3u^2\mathrm{Si}(u) \right.\\\nonumber
&&\left.\left.- u^2 \sin(u) +3u-\sin(u) - 2u\cos(u)\right)\right|_{\frac{\Delta t E x_c}{2}}^{\frac{\Delta t E}{2}} 
\end{eqnarray}
This final integral of $x$ (reparametrized by $u = xE\Delta t/2$) would have been logarithmic divergent if we had not cut it at $x_c$ at the lower bound.
The result has the following expansion at small $u$: $\frac{1}{18}(6\ln(u)+6\gamma_E - 17)$ and decays to $0$ at infinity. 
A good proxy is therefore to use the small-$u$ expansion but cut-off the upper bound of $u$ at its zero, finally
\begin{eqnarray}
1 &\sim&  \frac{\alpha_s\hat{q}\Delta t^3}{3}\ln\frac{2}{ x_c E \Delta t } \propto (g^2 T \Delta t)^3 \ln\frac{2}{ x_c E \Delta t }
\end{eqnarray}
Now, it is clear that this procedure of implementing multiple emission inside the medium resets the clock in the interference factor every $1/g^2T$ up to a certain logarithm dependence on the infrared cut-off, which is the order of the elastic collision mean-free-path.
Put this estimated $\Delta t$ back into the interference factor $2(1-\cos(\Delta t/\tau_f))$; one indeed finds a strongly suppressed radiation spectrum in regions where formation time is larger than the mean-free-path $\Delta t\sim \lambda_{el}$.

This suppression indeed mimics some property of the in-medium LPM effect but is introduced by a very different mechanism.
Recall that the LPM effect is the suppression of the single-particle emission rate through multiple collisions with the medium, without any information about how subsequent emissions are correlated.
However, the interference factor approach mimics the effect of the LPM suppression through a correlation between subsequent emissions.
This will introduce several problems: 
\begin{itemize}
\item[1.] The correlation between subsequent emissions is, in fact, physics beyond leading order and requires new types of diagrams to be computed \cite{Arnold:2015qya,Arnold:2016kek,Arnold:2016mth,Arnold:2016jnq}.
\item[2.] As we have seen, this procedure is affected by choice of the infrared cut-off. Though its dependence is very weak, it is still a dependence that we try to avoid.
\end{itemize}

\subsection{The ``blocking radiation'' approach}
Another approach we found in the literature \cite{ColemanSmith:2012vr}
is more problematic than the previous one.
In this ``blocking radiation'' approach, the splitting is also first generated through an incoherent process at time $t_0$ and then followed by the self-consistent determination of the formation time in the presence of elastic broadening. 
However, it introduces the LPM suppression by requiring that no other radiation is allowed from this radiator within the time from $t_0$ to $t_0 + \tau_f$, which again brings a correlation between subsequent emission.
While in our approach, the suppression is implemented by accepting the process with probability $\sim \lambda_{el}/\tau_f$, and more importantly, subsequent emissions are independent.

A closer investigation reveals bigger problems.
This ``blocking radiation'' approach effectively reduces every $\tau_f/\lambda_{inel}$ incoherent emission to one, resulting only in an overall reduction of the radiation spectrum without changing its shape.
Also, the suppression factor $\lambda_{inel}/\tau_f$ is different from the expected one, which is, in fact, an order of $\alpha_s$ wrong as the mean-free-path of the incoherent radiation rate contains one more power of $\alpha_s$ than $\lambda_{el}$.

\begin{figure}

\centering
\includegraphics[width=1.\textwidth]{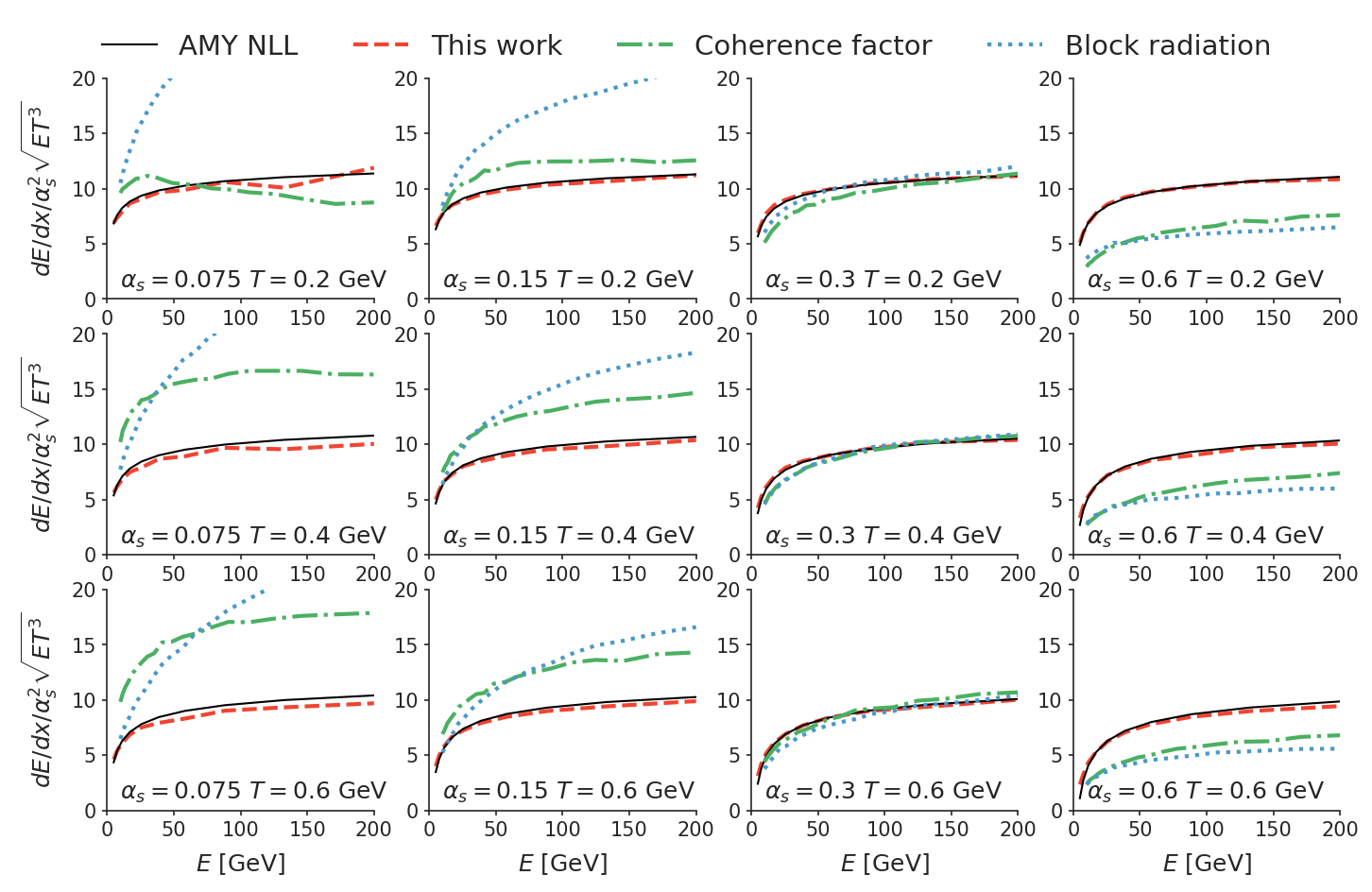}
\caption[Energy loss per unit path length $dE/dx$ as a function of energy $E$,]{Energy loss per unit path length $dE/dx$ as a function of energy $E$, temperature $T$ and coupling constant $\alpha_s$. Each column corresponds to a value of the coupling constant $\alpha_s = 0.075, 0.15, 0.3$, and $0.6$ (from left to right). Each row corresponds to a temperature of $T = 0.2, 0.4$, and $0.6$ GeV (from top to bottom). $dE/dx$ is divided by the expected scaling $\alpha_s^2 \sqrt{ET^3}$. The MC implementations in this work (red dashed lines) is compared to the ``coherence factor'' approach (green dash-dotted lines) and the ``block radiation'' approach (blue dotted lines). The analytic results are denoted as black solid lines.}
\label{fig:eloss-inf}
\end{figure}

\begin{figure}

\centering
\includegraphics[width=1.\textwidth]{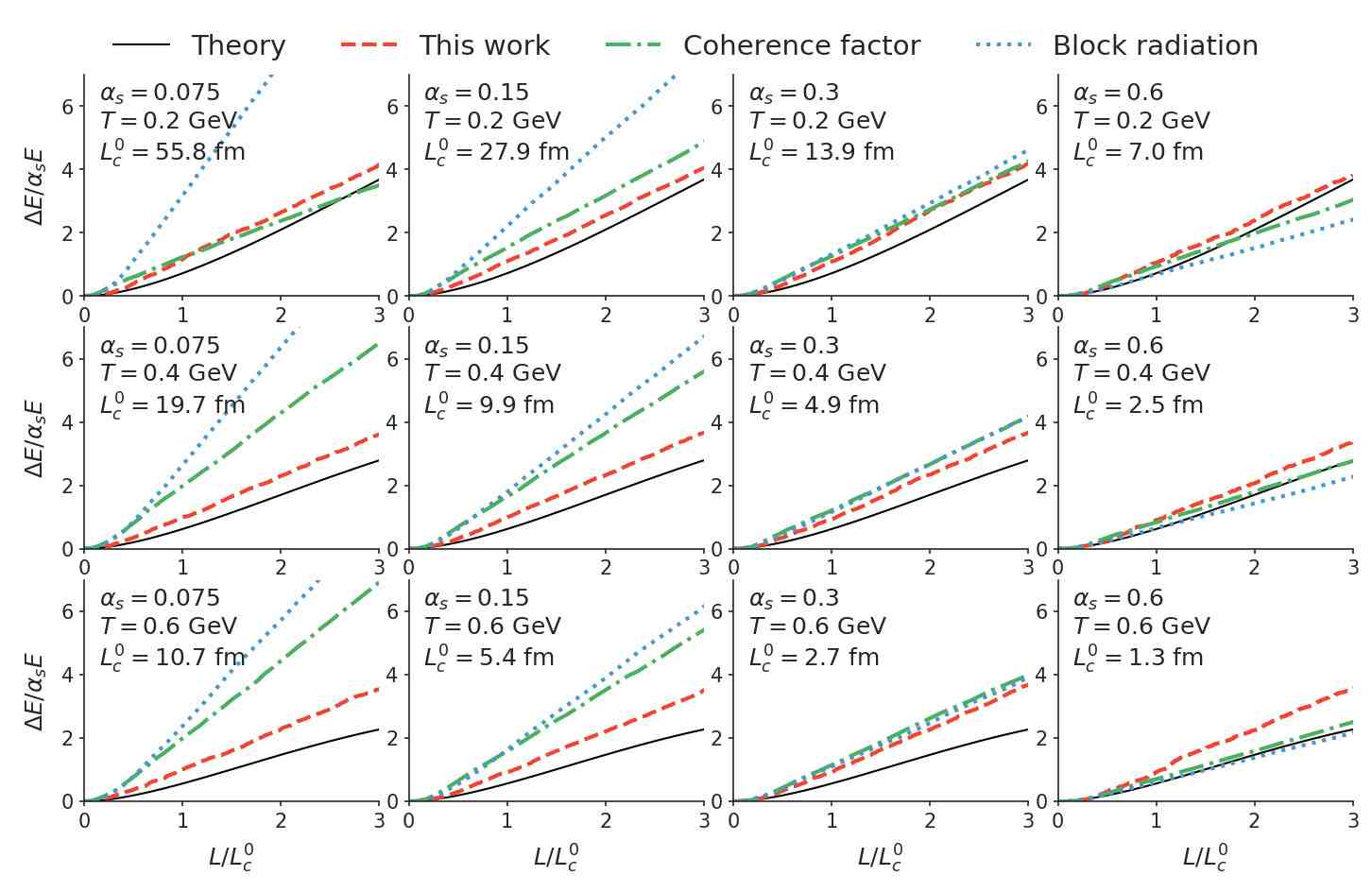}
\caption[Energy loss $\Delta E$ as a function of path length $L$, temperature $T$]{Energy loss $\Delta E$ as a function of path length $L$, temperature $T$ and coupling constant $\alpha_s$. Each column corresponds to a coupling constant of value $\alpha_s = 0.075, 0.15, 0.3$, and $0.6$ (from left to right). Each row corresponds to a temperature of value $T = 0.2, 0.4$, and $0.6$ GeV (from top to bottom). $\Delta E$ is scaled by $\alpha_s E$ and $L$ is scaled by an estimated critical path length $L_c^0 = \sqrt{E/\hat{q}_0}$, $\hat{q}_0 = C_A \alpha_s T m_D^2$. The MC implementations in this work (red dashed lines) is compared to the ``coherence factor'' approach (green dash-dotted lines) and the ``block radiation'' approach (blue dotted lines). The analytic results for a thin medium are denoted as black solid lines.}
\label{fig:eloss-ldep}
\end{figure}

\subsection{Energy loss comparison among the three approaches}
In Figure \ref{fig:eloss-inf}, we show the calculation of energy loss per unit path length $dE/dx$ of a quark in an ``infinitely large'' medium. 
Technically, $dE/dx$ is measured after an evolution time long enough ($L\gg L_c$) that finite size effects have faded away.
The results presented are normalized by $1/(\alpha_s^2 \sqrt{ET^3})$ in anticipation of the scaling $dE/dx \propto \alpha_s^2 \sqrt{ET^3}$.
For each column, we double the value of $\alpha_s$, and for each row, the temperature is increased by $0.2$ GeV. 
Within each subplot, the parton energy varies from $10$ GeV to $200$ GeV.
Different Monte Carlo implementations of the LPM effect are shown in colored lines, AMY NLL results are shown as black bands\footnote{ we only integrate $\omega$ above the Debye mass to calculate the AMY energy loss}. 
Without a surprise, the ``modified approach'' approach (red-dashed lines) reproduces the energy, temperature, and coupling constant dependence of AMY NLL energy loss very well.
The ``coherence factor'' approach (blue-dash-dotted lines) has a similar energy and temperature dependence to that of the theoretical baseline; however, it systematically deviates from the baseline for different values of the coupling constant in a logarithmic manner.
For the ``block radiation'' approaches, the deviations from the baseline regarding their $\alpha_s$-dependence are even stronger, and the energy dependence also gets worse, which is not surprising as we have discussed its problems.

Next we examine the path-length ($L$) dependence of the energy loss $\Delta E$ of a quark with an initial energy of $E = 200$ GeV in a finite medium in Figure \ref{fig:eloss-ldep}.
Again, each column uses a different coupling constant, and each row uses a different temperature. 
The path length within each subplot is varied up to four times $L_c^0$.
Here $L_c^0 = \sqrt{E/\hat{q}_0}$ with $\hat{q}_0 = C_A \alpha_s T m_D^2$ estimating the critical path length below which one expects a clear non-linear path-length dependence.
All three implementations show the non-linear increase of $\Delta E$ as a function of $L$.
The ``modified Boltzmann'' approach stays close to the theory calculations when $L<L_c^0$ for all cases, while the other two methods deviate systematically as $\alpha_s$ is varied, similar to our previous findings for the energy-loss in the infinite matter case.

\chapter{A comprehensive heavy-flavor dynamical modeling framework}
\label{chapter:coupling}
\begin{figure}
\centering
\includegraphics[width=.8\textwidth]{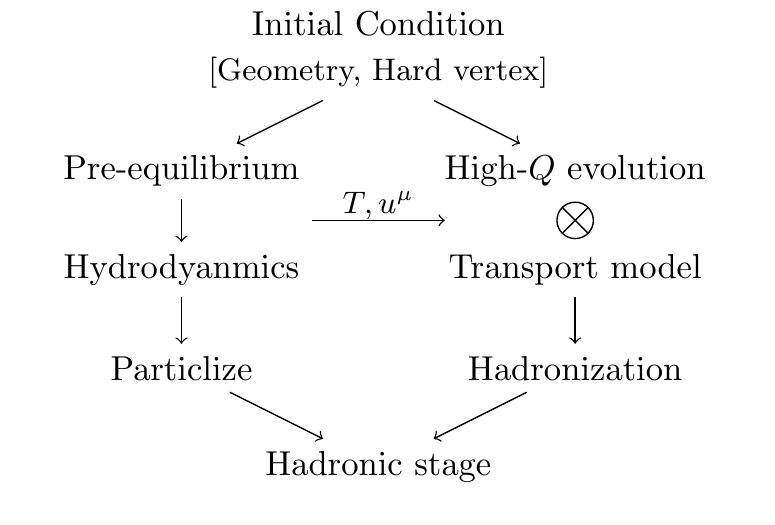}
\caption[A schematic workflow of the heavy quark simulation. The left]{A schematic workflow of the heavy quark simulation. The left branch evolves the medium in the hydrodynamic based model, providing medium information (temperature, flow velocity) to the hard probe transport in the right branch. The hard and soft hadrons are evolved in the hadronic afterburner in the last step.}
\label{fig:flowchart}
\end{figure}

The dynamical modeling framework for heavy flavor particles is summarized in the flow chart in figure \ref{fig:flowchart}.
The soft initial condition model provides both the initial energy density of the medium and the location of the hard particle production vertices in the transverse plane, while pQCD based calculations provide the momentum space distribution of the hard partons.
The left branch of this flow chart -- the hydrodynamic-based medium evolution model -- has been discussed in chapter \ref{chapter:simulation}.
We briefly review the RHS of the flow chart -- the multi-stage model for heavy-flavor evolution.
The hard production model is introduced in section \ref{section:hard}.
The initially produced partons are highly virtual and undergo the scale evolution that bring down the virtuality; eventually, at some point, this evolution will be matched to the in-medium transport calculations.
There is a complication regarding vacuum-like parton showers in a medium since certain vacuum parton branchings occupy the same space-time volume as the medium and also receive medium corrections.
Another obstacle is that multiple emissions are treated very differently between vacuum-like showers and medium-induce showers.
For the vacuum evolution, the ``time'' variable is the virtuality scale with the space-time information integrated out, while the transport model evolves the systems in real-time, with virtuality integrated out below a specific in-medium scale.
Significant progress has been made in both theory and design of event-generators to solve this problem \cite{MehtarTani:2012cy,Mehtar-Tani:2017ypq,Cao:2017zih,Kauder:2018cdt,Putschke:2019yrg,PhysRevLett.120.232001,Caucal:2018ofz}.
In section \ref{section:match}, we discuss a possible prescription to interface the two types of showers in our simulation.
Section \ref{section:couple-to-hydro} contains details of coupling the transport model to a dynamically evolving medium with large longitudinal expansion.
The heavy flavor hadronization model and hadronic rescatterings are introduced in section \ref{section:hadronization}.
The hadronization routine applies a previously used implementation \cite{Cao:2013ita} of the high-$p_T$ fragmentation plus low-$p_T$ recombination model for heavy hadrons production \cite{Oh:2009zj}.
Finally, in section \ref{section:benchmark}, the model is benchmarked using a few values of fixed coupling constant and running coupling constants, before being systematically calibrated to data in the next chapter.

\section{Initial production of heavy flavor}
\label{section:hard}
\subsection{Factorization framework in proton-proton collisions}
\begin{figure}

\centering
\includegraphics[width=.8\textwidth]{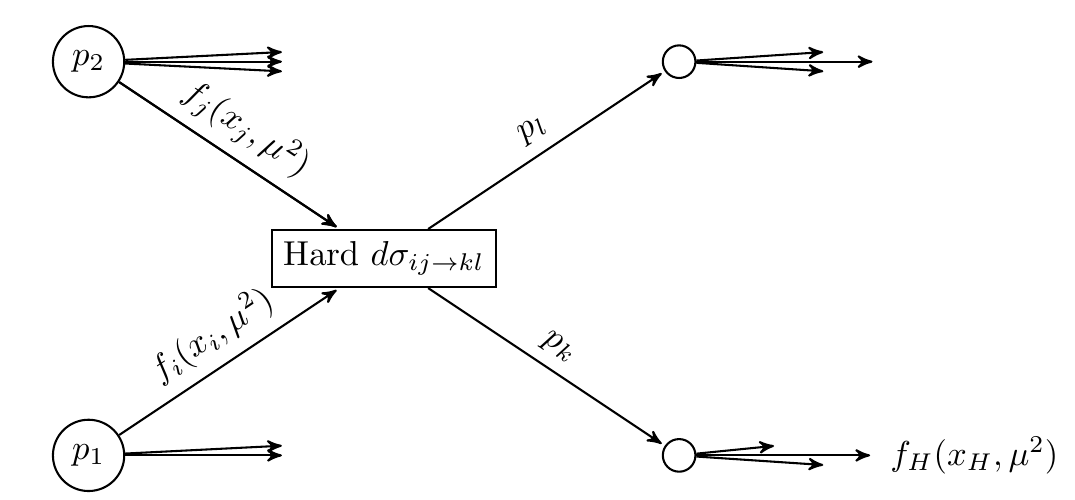}
\caption[A schematic demonstration of the factorization theorem. The]{A schematic demonstration of the factorization theorem. The momenta of the two incoming protons are $p_1$ and $p_2$. $f_{i,j}$ are the parton distribution function. $d\sigma$ is the perturbative matrix-element. $f_H$ is the fragmentation function.}
\label{fig:factorization}
\end{figure}

In the proton-proton collision, hard processes can be computed in the factorization framework using pQCD-based techniques as schematically demonstrated in \ref{fig:factorization}.
The incoming proton is a composite object and there is a certain ``probability'' of finding a parton $i(j)$ carrying $x_i(x_j)$ fraction of the momentum of the proton $p_1(p_2)$.
This ``probability'' is known as the parton distribution function (PDF) $f_i(x, Q^2)$.
It not only is a function of $x$, but also depends on the scale $Q^2$ at which the proton is probed.
The probing scale is required to be much larger than the non-perturbative scale $Q^2 \gg \Lambda^2$ such that $\alpha_s(Q^2)$ is small due to asymptotic freedom, and the process of partons $i$ and $j$ scattering into partons $k$ and $l$ is perturbatively computable.
The partonic final state eventually hadronizes via non-perturbative processes.
The parton fragmentation function is defined as the probability to find a certain hadron $H$ carrying a fraction $x_H$ of the parton's momentum.
Combing these pieces together, the cross-section for the inclusive production of the hadron $H$ can be written as \cite{Field:1989uq},
\begin{eqnarray}
\frac{d\sigma_{p+p\rightarrow H+X}}{dy d\mathbf{p}_T^2} = \frac{1}{\pi}\int dx_i dx_j f_i(x_i, Q^2) f_j(x_j, Q^2) \frac{d\sigma_{ij\rightarrow kl}}{d\hat{t}} \frac{1}{z_k}D^H(z_k, Q^2).
\end{eqnarray}
Although the parton distribution function $f$ and the parton fragmentation function $D$ are essentially non-perturbative objects, they parametrize universal long-distance physics and can be extracted from independent experiments at certain scales $Q_0^2$.
Moreover, the evolution from the ``definition'' scale $Q_0^2$ to the process scale $Q^2$ can be described by the Dokshitzer-Gribov-Lipatov-Altarelli-Parisi (DGLAP) evolution equations \cite{Gribov:1972ri,Altarelli:1977zs,Dokshitzer:1977sg} based on pQCD to increase the predictive power of the calculation.

\subsection{The DGLAP evolution equation and the vacuum parton shower} The DGLAP evolution takes into account that the initial high-virtuality parton $i$ (or $j$) could have come from a splitting process of a parton with lower virtuality parton $i'$ (or $j'$).
Similarly, the final state high virtuality parton $k$ (or $l$) could also split into a low virtuality parton $k'$ (or $l'$) before it turns into a hadron.
Though each splitting causes an additional power of $\alpha_s$, it is also magnified by a potentially large factor $\ln (Q^2/\mu^2)$ when $Q^2$ is much higher than the scale where the $f$, $D$ are defined.
The same argument also applies to partons $i', j', k', l'$. 
The DGLAP equations systematically resum contributions including an arbitrary number of parton splittings and evolve the scale from $\mu^2$ to the hard scale $Q^2$.
Take the evolution equation for quark distribution function as an example,
\begin{eqnarray}
\frac{\partial f_q(x, Q^2)}{\partial \ln Q^2} &=& \frac{\alpha_s C_F}{2\pi} \int_x^1 \frac{dz}{z} \left(\frac{1+z^2}{(1-z)_+} + \frac{3}{2}\delta(1-z) \right)  f_q\left(\frac{x}{z}, Q^2\right)\\\nonumber
&&+ \frac{\alpha_s}{2\pi} \int_x^1 \frac{dz}{z} \frac{z^2+(1-z)^2}{2}f_g\left(\frac{x}{z}, Q^2\right).
\end{eqnarray}
where $x$ is the momentum fraction carried by the parton. The ``+'' subscript on a function defines the operation,
\begin{eqnarray}
\int_0^1 \frac{f(x)}{(1-x)_+} = \int_0^1 dx \frac{f(x)-f(1)}{1-x}.
\end{eqnarray}
Using
\begin{eqnarray}
\int_0^1 dx \frac{1+x^2}{(1-x)_+} = -\frac{3}{2}, P_{qg}^{q(0)}(z) = C_F\frac{1+z^2}{1-z}, P_{q\bar{q}}^{g(0)}(z) = \frac{z^2+(1-z)^2}{2}, 
\end{eqnarray}
The equation can be cast into a form similar to the transport equation,
\begin{eqnarray}
\frac{\partial f_q(x, Q^2)}{\partial \tau} &=&  \int_x^1 \frac{dz}{z}\left\{ P_{qg}^{q(0)}(z) f_q\left(\frac{x}{z}, Q^2\right) 
+ P_{q\bar{q}}^{g(0)}(z) f_g\left(\frac{x}{z}, Q^2\right)\right\} \\\nonumber&& - f_q\left(x, Q^2\right) \int_0^1 dz P_{qg}^{q(0)}(z).
\end{eqnarray}
where $\tau = \alpha_s/2\pi \ln Q^2$ plays the role of a ``time'', and the right-hand side contains the gain-term (feeding from quark and gluon splittings) and loss-term (splitting of a quark).
This probabilistic interpretation is beneficial for building a phenomenological parton-shower picture: 
each hard parton has certain probabilities to splits into two or more partons within a ``time'' interval from $\tau$ to $\tau+\Delta \tau$.
The newly created partons can also split in the next ``time'' step.
In this way, one can mimic the production of the exclusive partonic final state from the sequence parton branchings using Monte Carlo techniques.

\subsection{Production in the nuclear environment}
The above framework explains very well the hard production process in proton-proton collisions.
In a nuclear environment, there are several differences.
First, the parton distribution functions inside a nucleus differ from a simple superposition of the nucleon PDFs.
The ratio between the nuclear PDF and proton PDF generally deviates from unity.
In particular, this ratio for small $x$ gluons is significantly below one, known as the ``nuclear shadowing'' effect. 
This ratio increases and becomes larger than one at larger $x$, termed as the ``anti-shadowing'' region.
The difference between the nuclear PDF and proton PDF belongs to the category of ``cold nuclear matter'' (CNM) effect, in contrary to the ``hot nuclear matter'' effect from the QGP medium.
The CNM effect has to be included to correctly interpret the experimental data, though the current level of uncertainty on the nuclear PDF is still significant.

\subsection{Inclusive calculation versus Monte-Carlo event generator}
In the course of my study, I have tried using both an inclusive cross-section calculation as well as a Monte-Carlo event generator to initialize the heavy quark production.
The inclusive calculation directly applies the factorization theorem and computes the inclusive spectra of heavy quark/hadron production spectrum; while the event generator used the probabilistic picture of the DGLAP evolution to build an exclusive final state.

\paragraph{Initialization from an inclusive calculation of heavy flavor production}
We use a  FONLL (Fixed-Order-Next-to-Leading-Log) calculation to generate the inclusive production cross-section of heavy flavors \cite{Cacciari:1998it}.
The FONLL program is a combination of fixed order (NLO) massive matrix-elements and a massless resummation program.
It computes the single inclusive differential cross-section of heavy quark/hadron production $d^2\sigma/dydp_T$ from which we sample the heavy quark's initial momentum.

This method has the advantage of being a first principle calculation when applied to proton-proton collisions. 
The main disadvantage is the lack of an exclusive partonic final state, causing several problems:
\begin{itemize}
\item[1.] Limitation to the study of open-heavy flavor.
For full jets, one needs the exclusive partonic final state. For quarkonia, the momentum correlations among the $Q$-$\bar{Q}$ pairs are important.
\item[2.] We cannot generate a space-time picture of the parton shower to implement medium modifications to parton evolution. 
Therefore, in this initialization routine, we have always assumed that the vacuum-like evolution is complete at time $\tau=0^{+}$.
\end{itemize}

\paragraph{Initialization from Monte-Carlo event generator}
We used Pythia (version 8.235) as the hard parton generator \cite{Sjostrand:2014zea,Sjostrand:2006za}.
Pythia implements leading order (LO) matrix-elements for hard QCD processes, including LO production of heavy flavor particles,
$g+g\rightarrow Q+\bar{Q}$ and $q+\bar{q}\rightarrow Q+\bar{Q}$.
A parton shower, including initial state radiation (ISR) and final-state radiation (FSR), is generated around the hard vertex.
At high energy, the LO production of heavy flavor is only a fraction of the total heavy flavor cross-section, the remainings are created in the parton showers via the so-called ``gluon splitting'' and ``flavor creation'' processes.
The former corresponds to a situation where the heavy flavor pair originates from a final state gluon splitting, and the latter produces the pair in initial state gluon splitting and is put-on shell by the hard scattering.
These contributions introduce certain non-back-to-back angular correlations.

This initialization method is not a first principle approach. 
Also, the generation of full parton showers at LHC energy can be slow, but the benefits are enormous,
\begin{itemize}
\item Though the parton shower in Pythia evolves as a function of virtuality $Q^2$; an approximate space-time picture can be reconstructed by defining the formation time to be $2x(1-x)E/Q^2$ for each branching. Then, it is easy to determine which splitting happened inside the medium and receives medium modifications.
\item It allows initialization of full jet and the study of quarkonia transport.
\end{itemize}

\begin{figure}

\centering
\includegraphics[width=.8\textwidth]{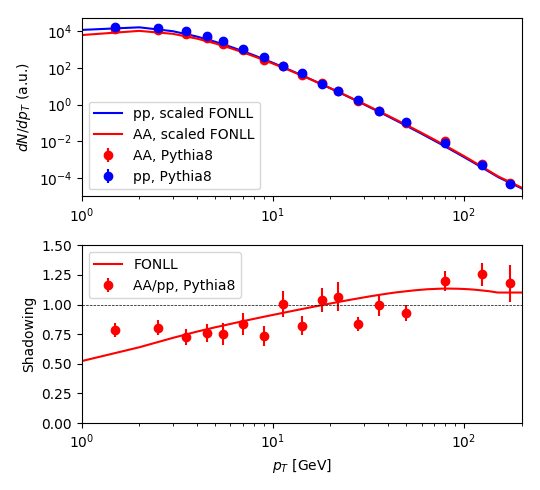}
\caption[Top plot: a comparison of $D$-meson production in proton-proton]{Top plot: a comparison of $D$-meson production in proton-proton collisions (blue) and in Pb-Pb collisions (red) with cold nuclear matter effect only. FONLL calculations are shown as lines, and Pythia8 simulations are shown as symbols. Bottom plot: the ratio between the production in Pb-Pb collision (cold nuclear matter effect only) to the proton-proton baseline shows the nuclear shadowing effect.}
\label{fig:pythia-fonll}
\end{figure}

\paragraph{A comparison of the proton-proton baseline and the CNM effect}
We checked whether the Pythia event generator predicts a similar proton-proton baseline compared to the first principle approach FONLL.
In the upper plot of figure \ref{fig:pythia-fonll}, we compare the $p_T$ differential cross-section of $p+p\rightarrow c$ from FONLL (lines) and Pythia simulations (symbols), and for Pb+Pb collision (red) and p+p collision (blue) at the LHC energy $\sqrt{s}=5.02$ TeV.
For proton-proton collisions, we use the CT10 parton distribution function \cite{Lai:2010vv}.
The nuclear PDF uses the EPS09 parametrization \cite{Eskola:2009uj}.

Though the absolute value of the cross-sections compared between FONLL and Pythia are different, the observables are usually presented as ratios between nuclear collisions and the proton-proton baseline where the normalization cancels, or other dimensional-less observables such as the momentum-space anisotropy of heavy mesons.
Therefore, we focus more on the shape of the spectra between the two calculation, which agree very well.
The ratio of initial charm spectra in Pb+Pb collisions and p+p collisions estimates the magnitude of the cold-nuclear matter effect on the nuclear modification factor $R_{AA}$ (without the hot QGP effect).
FONLL and Pythia simulations predict consistent modulation: the initial production AA spectra of charm quark at low-$p_T$ is suppressed compared to the pp spectra, due to the shadowing effect of the small-$x$ gluon. 
At higher $p_T$, the ratio increase and slightly shoots over unity, because partons from anti-shadowing contribute more at larger-$x$.

\section{Matching vacuum and medium-induced showers}
\subsection{Vacuum versus medium-induced shower phase-space}
\label{section:match}
The fate of vacuum-like showers in the hot-medium is complicated, and there have been studies for its phenomenological consequences \cite{Cao:2017zih,PhysRevLett.120.232001,Caucal:2018ofz}.
The prescription that we build in this section is by no means exact, but follow the reasoning from a recent work \cite{PhysRevLett.120.232001}.
The general idea is to identify different regions of phase-space of radiation and apply different means of computation (DGLAP / transport) to different regions based on how much medium-modification it would have received.

\begin{figure}

\includegraphics[width=.35\textwidth]{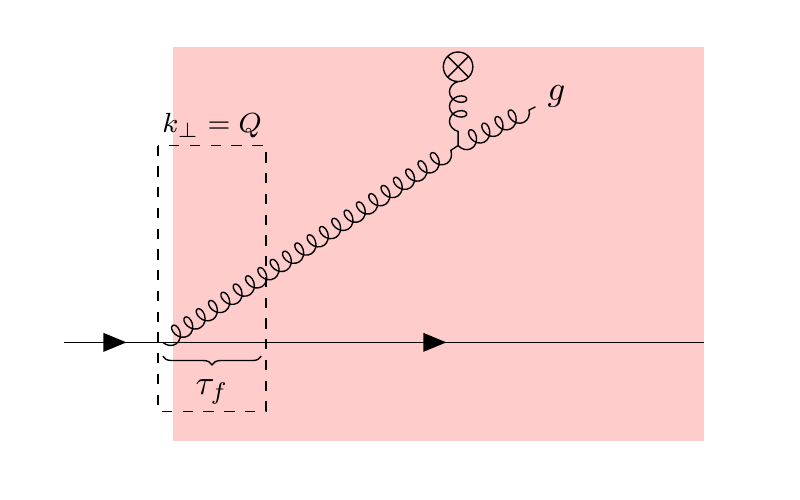}\includegraphics[width=.35\textwidth]{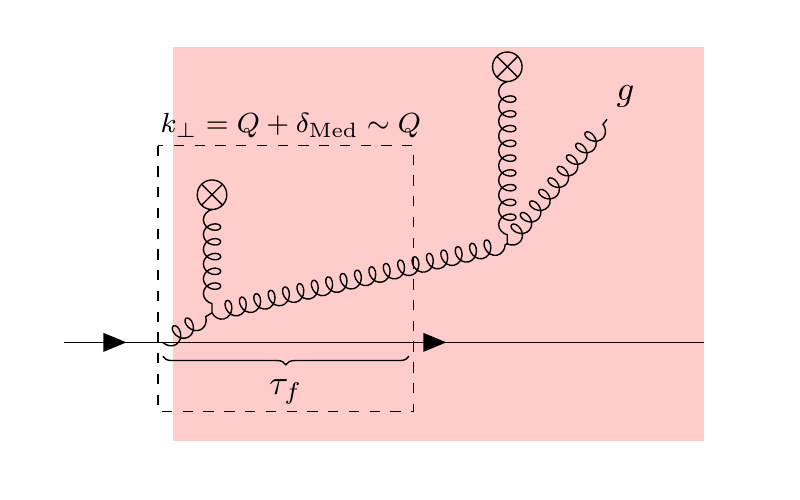}\includegraphics[width=.35\textwidth]{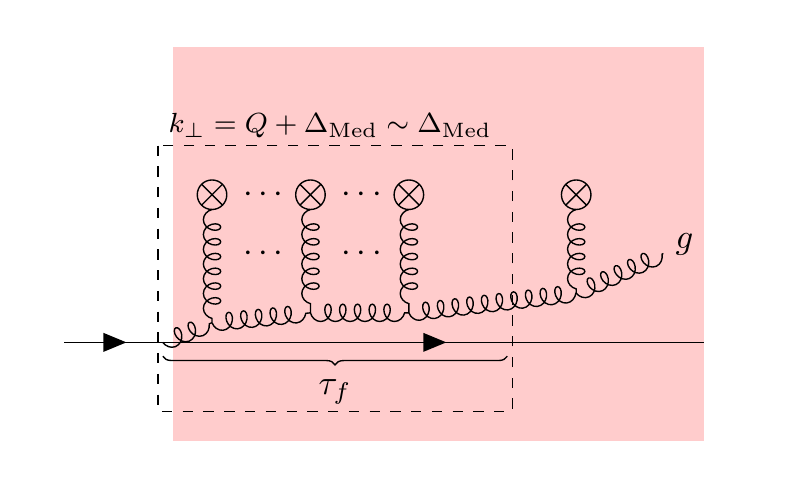}
\caption[Demonstration of the medium corrections to vacuum-like radiation]{Demonstration of the medium corrections to vacuum-like radiation with different formation time. The energy of the gluon is held fixed, while the virtuality is decreasing from left to right.}
\label{fig:vac-med-interface}
\end{figure}

Consider a vacuum splitting of a hard parton that enters the medium at $z=0$.
The vacuum splitting has a formation time of $\tau_f \sim 2x(1-x)E/k_\perp^2$.
Before it fragments in the vacuum, the system (quark plus gluon) likely interacts with one or more scattering centers (labeled by ``i'') in the medium at time $t_i$.
Whether these interactions contribute coherently to the ``vacuum-like'' splitting follows the same argument as before.
Scatterings that are well-separated from the formation processes $\tau_f \ll t_i$ are treated as independent; they only broaden the transverse momentum without changing the radiation probability.
For $t_i \lesssim \tau_f$, the branching probability of the vacuum-like radiation gets modified.
Now classify the radiations using the average ``number'' of scatterings  $N = \tau_f/\lambda$ (for the case of a static medium).
\begin{itemize}
\item For a branching with large virtuality (left of figure \ref{fig:vac-med-interface}) so that $N \ll 1$ or equivalently $Q^2 \gg  g^2 x(1-x)E T$. 
The chance for a medium modification of the vacuum branching probability is negligible. 
\item Hold the energy of the radiation and decrease its virtuality (middle of figure \ref{fig:vac-med-interface}) so that $N = \tau_f/\lambda \sim 1$ ($Q^2 \sim g^2 x(1-x)E T$). 
Now, there is an order one probability of scatterings within $\tau_f$, but the initial virtuality still dominates the transverse momentum of the gluon.
The probability for the branching should also be modified accordingly, for example, using the higher-twist formula that expands in terms of $1/Q^2$.
\item Further decrease the initial virtuality of the branching (right of figure \ref{fig:vac-med-interface}) until $N \gg 1$.
Eventually, the medium broadening of the transverse momentum is large compared to the parton's virtuality from initial production.
It is proper to associate this parton an in-medium virtuality $Q^2 \sim g^2\sqrt{x(1-x)E T^3}$. 
When this happens, the branching probability gets heavily modified by the medium and should be replaced by a medium-induced radiation calculation.
\end{itemize}
Summarizing the two extreme regions:
The unmodified DGLAP evolution applies to the high-virtuality part of the shower ($Q^2 \gg \alpha_s \omega T$), while medium-induced calculation, via a transport equation, applies to the low-virtuality shower $Q^2 \sim \sqrt{\hat{q}\omega}$ (relation obtained in a static medium).
It is therefore natural to use the comparative relation between the partons' initial virtuality $Q^2$ and the transverse momentum change contributed by medium broadening $\Delta k_\perp^2 = (\mathbf{k}_\perp(\tau=\tau_f) - \mathbf{k}_\perp(\tau=0))^2$ to separate the medium-induced radiation from the vacuum-like radiation.
The matching prescription is then to cut-out the vacuum branchings generated by Pythia in the region $\Delta k_\perp^2 \gtrsim Q^2$, the cut region is referred to as the ``vetoed'' region in the literature \cite{PhysRevLett.120.232001}).
For a dynamical and fluctuating medium, there is no simple relation as $\Delta k_\perp^2\sim \sqrt{\hat{q}\omega}$ in the static medium, but the ``preformed parton'' technique can be used to determine $\Delta k_\perp^2$ self-consistently for each vacuum branching (to be explained in the next paragraph).
In a finite medium, certain vacuum-like branchings may have a long formation time that they form outside of the medium.
Due to the uncertainty principle, these branchings do not resolve the details of the medium, and their branching probability remains unchanged in our model.
This separate treatment of different regions of phase-space still depends on the detailed choice of the separation scale, so in the future, it would be desirable to develop a unified theoretical treatment for both vacuum and medium-induced showers in the time evolution picture.

Focusing only on the vacuum-like radiation generated by heavy quarks, one traces back a heavy quark line in the Pythia event recorder to find all the gluons from its final state radiation (FSR) and the original four-momentum of the heavy quark at the initial production vertex.
These FSR gluons are first treated as ``unformed'' by the transport models, and they are allowed to undergo elastic broadening with the medium.
In this way, by the time these gluons reaches their formation times ($t-t_0>\tau_f$), one knows both the initial virtuality of the splitting $Q^2$, as well as how much medium broadening $\Delta k_\perp^2$ it has acquired.
Then, applying our previous approximation, vacuum-like branching with 
$\Delta k_\perp^2 < R_v Q^2$ remains unmodified, but $\Delta k_\perp^2 > R_v Q^2$ vacuum branchings are rejected because these contributions are already taken care by the medium-induced rate in the transport model.
The order one $R_v$ parameter is introduced to parametrize the uncertainty in this matching scale.

\subsection{Visualizing the matching on the Lund diagram}
The Lund diagram is a useful tool to visualize the phase-space for high energy parton splitting.
There are many different choice of kinematic variables, but here we choose the vertical axis to be $Y = \ln(1/x) = \ln(E/\omega)$, and the horizontal 
axis to be $X = \ln(1/\theta^2) = \ln(\omega^2/k_\perp^2)$.
Here $x$ is the energy fraction carried by the daughter parton in a particular splitting, and $\theta$ is the daughter's emission angle relative to the mother parton.
This arrangement is inspired by the soft and collinear limit of the QCD splitting function (for example $q\rightarrow q+g$),
\begin{eqnarray}
dP^{q}_{qg} \sim \frac{\alpha_s C_F}{\pi} \frac{dx}{x}\frac{d\theta^2}{\theta^2} = \frac{\alpha_s C_F}{\pi} d\ln\frac{1}{x} d\ln\frac{1}{\theta^2}.
\end{eqnarray}
Therefore, the probability distribution of a vacuum-like splitting vertex should be uniform, apart from the running coupling effect.
The closer a point lies towards the origin, the higher its virtuality.
The soft and collinear radiations reside at large $X$ and $Y$.
Also, constant-formation-time contours are simply straight lines $Y+X=\ln(E\tau_f/2)$.

\begin{figure}

\includegraphics[width=.5\textwidth]{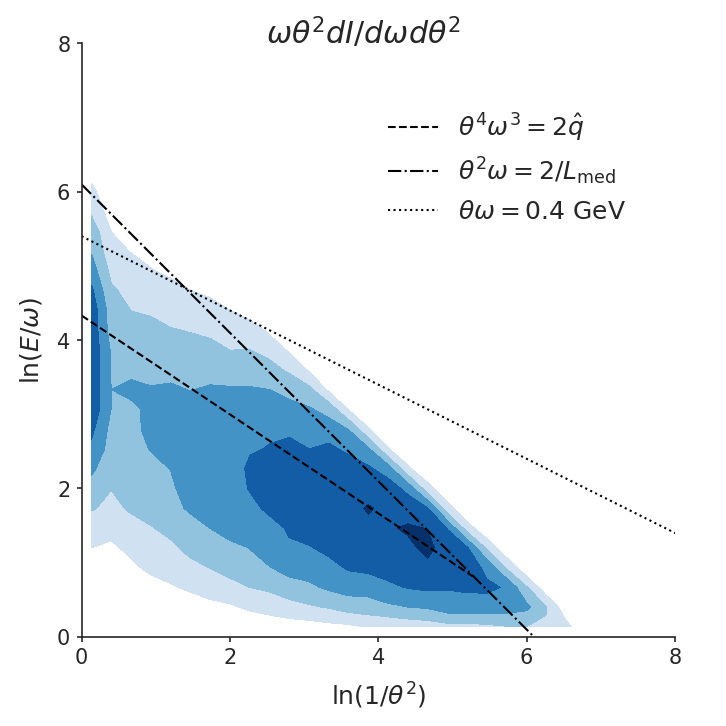}\includegraphics[width=.5\textwidth]{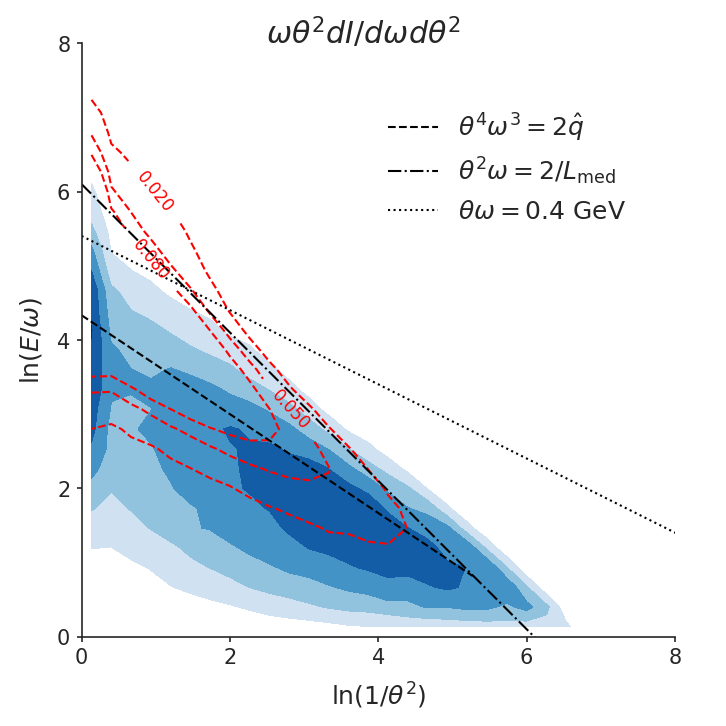}
\caption[Plotting the gluon radiations from a charm quark on the Lund]{Visualize the gluon radiations from a charm quark on the Lund diagram. The gluon emission has an energy $\omega$ with angle $\theta$ with respect to the heavy quark. The blue heat map in the left plot shows the distribution of vacuum-like emissions without medium effects. The heat map in the right plot shows the distribution of vacuum-like emissions with medium effects; while the red contour stands for the medium induced emissions. Emissions on the black dashed lines have $k_\perp^2 = \sqrt{2\omega \hat{q}}$. Emissions on the dashed-dotted lines have a formation time equals to the medium path-length. Finally, the dotted line is the non-perturbative cut-off in Pythia $k_\perp = 0.4$ GeV.}
\label{fig:lund}
\end{figure}

On the left of figure \ref{fig:lund}, we show the phase space occupied by the vacuum branching without a medium (left); on the right, it is medium-modified vacuum splitting (blue color map) and the medium induced radiation (red contour) from our simulation.
The simulation first identifies charm quarks with transverse momentum $90 < p_T <110$ GeV at the production vertex in Pythia, and then propagate them and their vacuum radiated gluons in a static medium with $T=0.3$ GeV with $\alpha_s = 0.3$ for a path length $L$.
We see that without the medium effect, the vacuum radiations fills the region bounded by the time-evolution limit $\tau_f < L$ (dash-dotted line) and the default non-perturbative bounds $k_\perp > 0.4$ GeV (dotted line) of Pythia. 
Inside the medium, the medium-induced radiations are distributed around the line $\tau_f\hat{q} = k_\perp^2$ which is $\theta^4\omega^3 = 2\hat{q}$ (dashed line) in the soft limit.  
However, this line is only an averaged estimation of the relation between $k_\perp, \omega$, and $\hat{q}$, since the actual outcome of the simulation strongly fluctuates.
The triangle area bounded by the line $\tau_f < L$ and the line $\theta^4\omega^3 = 2\hat{q}$ is where the vacuum-like radiation receives significant modification from medium interactions.
The rejection program introduced before suppresses the vacuum-like radiation in this region compared to the case without a medium.
Again, due to fluctuations, the triangular region is not entirely vetoed as the one demonstrated in \cite{PhysRevLett.120.232001}.

Concluding this section, the realm of the transport equation and the DGLAP evolution is separated when the parton virtuality is comparable to the acquired transverse momentum broadening within the formation time.
High virtuality evolution is approximated as unmodified, while low virtuality evolution is terminated and replaced by the medium-induced processes via the transport evolution. 
This procedure is, of course, only viable if we initialize the simulation with a parton shower event generator.
We are not able to do such a separation using heavy quark spectra obtained from FONLL.

\section{Particles coupled to an evolving medium}
\label{section:couple-to-hydro}
The coupling between hydrodynamics and hard parton transport often requires switching of the reference frame, as the velocity of the medium local-rest-frame relative to the lab frame is a function of space-time.

\paragraph{For diffusion dynamics} The diffusion equations are most easily written in the local-rest-frame of the medium.
Given a particle's four momentum in the lab frame ($p_{L}^\mu$), one first boost it into the medium local-rest-frame ($p_{M}^\mu$),
\begin{eqnarray}
p_{M}^\mu &=& L^\mu_\nu(\vec{\beta}) p_{L}^\nu\\
L^\mu_\nu &=& 
\begin{bmatrix}
\gamma & -\gamma\vec{\beta}\\
-\gamma\vec{\beta} & \mathbb{1} + \frac{\gamma^2}{\gamma+1}\vec{\beta}\vec{\beta}
\end{bmatrix}
\end{eqnarray}
where $\vec{\beta}$ is the velocity of the fluid cell relative to the lab frame, and $L^\mu_\nu$ is the Lorentz transformation.
One needs to be careful with that since the time step in the fluid rest frame $\Delta t_{M}$ is different from the one in the lab frame $\Delta t_{L}$.
Consider the particle trajectory $\Delta x_{L}^\mu$ within $\Delta t_{L}$ observed in the lab frame and boost it into the medium frame,
\begin{eqnarray}
\Delta x_{L}^\mu = \frac{p_{L}^\mu}{E_L} \Delta t_{L} \xrightarrow{\textrm{boost}} \Delta x_{M}^\mu = \frac{L^\mu_\nu(\mathbf{v}_{M}) p_{L}^\nu}{E_L} \Delta t_L = \frac{p_{M}^\mu}{E_L} \Delta t_L
\end{eqnarray}
Comparing the time-component of the equations, one finds the time step in the medium frame being related to the lab frame step by the ratio between the energy of the particle in the two reference frames,
\begin{eqnarray}
\Delta t_M = \frac{E_M}{E_L} \Delta t_L
\end{eqnarray}
Once the momentum is updated in the medium frame to $p'_M$, it is boosted back to the lab frame,
\begin{eqnarray}
x'^{\mu} &=& x^{\mu} + \frac{p_{L}^\mu}{E_L} \Delta t_{L} \\
p_{L}^{'\mu} &=& L^\mu_\nu(-\vec{\beta}) p_{M}^{'\nu}
\end{eqnarray}
where we have chosen to update position before the update of the momentum.

The choice of $\Delta t_L$ is also tricky. 
Because the relativistic hydrodynamics for heavy-ion collision is usually solved in $(\tau,x,y,\eta_s)$ coordinates, the hydrodynamic fields are propagated from one constant proper time $\tau = \sqrt{t^2 - z^2}$ to the next.
As a result, there are two consequences if we use a straightforward uniform time step  $\Delta t_L$ for all particles:
\begin{itemize}
\item[1.] Different particles will be at different proper times $\tau$ at a constant $t$. It requires the program to load the entire hydrodynamic temperature and velocity history into the memory, which can be a memory storage problem for 3+1 D hydro simulation (the memory consumption for boost-invariant hydrodynamics is not critical).
\item[2.] The time step in the medium-rest-frame for particles at large space-time rapidity would be too small.
\end{itemize}
For these practical reasons, we choose to propagate particles with a constant proper-time step $\Delta \tau$. 
It requires the time step in the lab frame is different for each particle, depending on its location and momentum. The time-step is solved by,
\begin{eqnarray}
\Delta \tau = \sqrt{(t+\Delta t_L)^2 - (z+v_z \Delta t_L)^2} - \sqrt{t^2 - z^2}.
\end{eqnarray}
This is (keeping the positive solution),
\begin{eqnarray}
\Delta t_L(p, x) = \frac{-(t-z v_z) + \sqrt{(t-z v_z)^2 - (1-v_z^2)(\Delta \tau^2 + 2\sqrt{t^2 - z^2}\Delta \tau )}}{2(1-v_z^2)}
\label{eq:dt-transformation}
\end{eqnarray}
This adaptive time step propagates a particle between constant proper-time hyper-surfaces, therefore only two steps of hydrodynamic information needs to be loaded into memory at any given time.
Also $\Delta t_L$ becomes larger for forward/backward particles.

\paragraph{For matrix-element scattering} Sampling matrix-element-based scattering is more complicated than solving the diffusion equation.
One can straightforwardly sample the initial state in the medium local-rest-frame, but the final state is most efficiently sampled in the center-of-mass frame of the few-body collisions.
The center-of-mass velocity relative to the local-rest-frame is,
\begin{eqnarray}
\vec{\beta}_{C} = \frac{\sum_{i\in \textrm{IS}} \vec{p}_i}{\sum_{i\in \textrm{IS}} E_i}
\end{eqnarray}
where ``IS'' stands for the initial state.
\begin{itemize}
\item[1.] For each hard parton, determine $\Delta t_L$ with equation \ref{eq:dt-transformation}.
\item[2.] Boost the particle to the medium rest-frame and sample the scattering rate $\Delta t_M R$ channel, and then sample the initial-state medium parton(s).
\item[3.] In the CoM frame of the initial state, sample the final state particles.
\item[4.] Boost back the final state particles to the medium rest frame.
\item[5.] Boost back to the lab frame.
\end{itemize}

\section{Heavy-flavor hadronization and hadronic stage}
\label{section:hadronization}
At a temperature near $T_c$, light hadrons can be sampled from the hydrodynamics energy-momentum tensor statistically.
For hard partons that may be off equilibrium, a microscopic hadronization model is in need.
The final hadronic system is also dense enough for the heavy hadron to interact.
Though the hadronic interactions are not analyzed as extensively as the QGP interaction, studies have shown hadronic rescatterings contribute to finite low-$p_T$ $v_2$ of D-mesons \cite{Cao:2015hia}.
Therefore we also include the afterburner stage for the heavy flavors.

\subsection{The instantaneous approximation of hadronization} 
The hadronization implementation is described in \cite{Cao:2013ita}.
It combines the fragmentation of heavy quarks at high momentum and the recombination with medium partons into hadrons at low momentum.
The hadronization is treated to be instantaneous on an isothermal hypersurface.
This instantaneous approximation has certain drawbacks.
First, hadronization is a long-distance process. 
In the rest frame of the heavy flavor, it takes time on a scale of $1/\Lambda_{QCD}$. 
With a large boost factor $E/M_H\sim E/M_Q$, the formation time of the heavy hadron can be comparable to macroscopic length scales.
For example, for a moderate $E=10$ GeV charm quark with $M_Q=1.3$ GeV, this time is estimated to be $8$ fm/$c$, which is certainly not instantaneous, considering the hydrodynamic stage only last for $O(10)$ fm/$c$.
Second, an instantaneous recombination process breaks energy conservation and the detailed balance.
To solve all of these problems, one may need to consider using a dynamical hadronization model \cite{He:2019vgs}.

\paragraph{Fragmentation} 
In high energy electron-positron collisions and proton-proton collisions, high momentum heavy quarks hadronize through the fragmentation mechanism.
The energetic heavy quark produces a bunch of hadrons with a heavy hadron that carries a certain fraction of the origin quark momentum $z = p_H/p_Q$.
The probability distribution of $z$ is known as the fragmentation function $D(z)$, and can be measured in, e.g., electron-positron colliders.
There are different parametrizations for $D(z)$ and the Peterson fragmentation function \cite{PhysRevD.27.105} is used in the present study,
\begin{eqnarray}
D(z) \propto \frac{1}{z(1-\frac{1}{z} - \frac{\epsilon}{1-z})^2}
\end{eqnarray}
where $\epsilon$ is a parameter that scales as $m_Q^{-2}$ ($\epsilon_c \approx 0.05, \epsilon_b \approx 0.006$).

\paragraph{Recombination}
In proton-proton collisions, heavy quarks can hadronize into mesons by the recombination with a light quark in the proton remnant \cite{Mehen:2003rf}.
In a heavy-ion collision, the recombination mechanism plays a far more essential role for low $p_T$ heavy flavors, given the abundance of thermal medium partons.
Early studies in nuclear collisions \cite{Oh:2009zj} assumed that the recombination probability can be computed from the wave function overlap between initial state partons and final state mesons or baryons, with the momentum of the medium parton integrated over the thermal distribution.
\begin{eqnarray}
\frac{dP_M(p', p)}{dp'^3} &=& \int dk^3 n_{\bar{q}}(k) W_{M}(p, k)\delta^{(3)}(\vec{p}'-\vec{p}-\vec{k}), \label{eq:meson_recombine}\\
\frac{dP_B(p', p)}{dp'^3} &=& \int dk_1^3 dk_2^3 n_{\bar{q}}(k_1)  n_{\bar{q}}(k_2) W_{B}(p, k_1, k_2)\delta^{(3)}(\vec{p}'-\vec{p}-\vec{k}_1 - \vec{k}_2), \label{eq:baryon_recombine}.
\end{eqnarray}
On the left are the differential probability for a heavy quark with momentum $p$ to hadronize into a heavy meson (first line) or a heavy baryon (second line) with momentum $p'$ through recombination.
They are equal to integration of light quark(s)/antiquark(s) momenta of the produced baryon/meson Wigner function $W$ times the thermal distribution function, subjected to three-momentum conservation.
The energy conservation is not imposed in the instantaneous $2\rightarrow 1$ coalescence approach.
The quark/antiquark distribution function is the Fermi-Dirac one, neglecting the chemical potential, 
\begin{eqnarray}
n = \frac{g_q V}{e^{\beta p\cdot u} + 1}
\end{eqnarray}
with $u$ the fluid velocity and $p$ the four momentum of the light quark / anti-quark.
$g$ is the degeneracy factor of the quark, and $V$ is a test volume that will eventually be canceled by the normalization factor in the Wigner function.
As a remark, we have assumed in the transport model that medium partons are massless because the thermal masses are higher-order effects for energy loss; but for recombination into bound states near $T_c$, it is crucial to use non-perturbative constituent masses of light quarks $m_u = m_d = 300$ MeV and $m_s = 475$ MeV.

Regarding the meson wave-function, there have been efforts using the Dirac equation to obtain a more realistic wave-function for different states of heavy mesons \cite{Zhao:2018jlw,PhysRevD.88.014021}.
The current model uses a parametrized Gaussian wave-function for simplicity,
\begin{eqnarray}
\phi_M(\vec{r}) &=& \left(\frac{1}{\pi \sigma^2}\right)^{3/4} e^{-\frac{r^2}{2\sigma^2}}.
\end{eqnarray}
$\sigma$ is related to the reduced mass of the two body system $\mu = m_1 m_2/(m_1+m_2)$ and the frequency of the  two-body potential $\omega$ by $\sigma = 1/\sqrt{\mu \omega}$.
These frequencies are estimated from the charge radius of different heavy mesons: $0.106$ GeV for charmed mesons and $0.059$ GeV for the bottom mesons.
The Wigner function is defined in terms of the relative distance $\vec{r}$ and relative momentum $\vec{q}$ between the quark and anti-quark,
\begin{eqnarray}
W_M(\vec{r}, q^2) &=& g_M \int d^3 \vec{a} e^{-i\vec{q}\cdot \vec{a}} \phi_M(\vec{r}+\vec{a}/2) \phi_M^*(\vec{r}-\vec{a}/2) \\
\vec{q} &=& \frac{E_2\vec{p}_1 - E_1\vec{p}_2}{E_1+E_2}.
\end{eqnarray} 
Averaging over the light quark's positions,
\begin{eqnarray}
W_M(q^2) &=& \frac{g_M}{V} (2\sqrt{\pi}\sigma)^3 e^{-\sigma^2 q^2},
\end{eqnarray}
which is the quantity needed in equation \ref{eq:meson_recombine},
\begin{eqnarray}
\frac{dP_M(p',p)}{dp'^3} &=& \int dk^3 \frac{g_q g_M}{e^{\beta p\cdot u} + 1} (2\sqrt{\pi}\sigma)^3 e^{-\sigma^2 q^2} \delta^{(3)}(\vec{p}'-\vec{p}-\vec{k}),
\end{eqnarray}
where the test volume in the distribution function has been canceled by the one in the Wigner function.

The same procedure applies to heavy baryons, with the three-body Wigner function in the Gaussian approximation as,
\begin{eqnarray}
f_B^W(q_1^2, q_2^2) = \frac{N g_B}{V^2} (2\sqrt{\pi\sigma_{1,2}\sigma_{12,3}})^6 e^{-q_{1,2}^2 \sigma_{1,2}^2 - q_{12,3}^2 \sigma_{12,3}^2}.
\end{eqnarray}
The relative momenta are defined as,
\begin{eqnarray}
\vec{q}_{1,2} &=& \frac{E_2 \vec{p}_1 -E_1\vec{p}_2}{E_1+E_2},\\
\vec{q}_{12,3} &=& \frac{E_3 (\vec{p}_1+\vec{p}_1) - (E_1+E_2)\vec{p}_3}{E_1+E_2 + E_3},
\end{eqnarray}
and the $\sigma$ related to the frequency and masses by,
\begin{eqnarray}
\sigma_{1,2}^{-1} &=& \sqrt{\omega \frac{m_1m_2}{m_1+m_2}}\\
\sigma_{12,3}^{-1} &=& \sqrt{\omega \frac{(m_1+m_2)m_3}{m_1+m_2+m_3}}
\end{eqnarray}

To synthesize these two competing mechanisms of hadronization, first, one samples the recombination probability in equations \ref{eq:meson_recombine} and \ref{eq:baryon_recombine} and determines whether the heavy quark coalesces with medium partons. 
If not, its hadronization will be handled by the Pythia fragmentation routine with the Peterson fragmentation function.

\subsection{Hadronic rescattering of heavy-meson in UrQMD}
Currently, UrQMD includes hadronic collisions between charmed mesons and $\pi$, $\rho$ mesons.
These cross-sections are obtained in \cite{Lin:2000jp}.
Hadronic cross-section of the charmed baryons and bottom hadrons are not included.

One modification made to the UrQMD heavy-flavor sector is that the kinematic effect of backreaction from heavy flavor mesons on the light sector is turned-off. 
It is achieved by resetting the light scattering partner's four-momentum back to its initial value.
This practice retains the same level of approximation of the linearized transport equation in the QGP phase and allows for an easy oversampling of the number of heavy flavor particles to obtain better statistics.

\section{Benchmark calculation of observables}
\label{section:benchmark}
In the last section of this chapter, we provide a benchmark calculation of the open-heavy flavor simulation framework by comparing to experimental data. 
A systematic calibration of model parameters and uncertainties will be discussed in the next two chapters.

\subsection{Open heavy flavor observables}
Experimentally, the ground states mesons $D^0, \bar{D}^0, D^{\pm}, B^{\pm}, D_s^{\pm}, B_s^{\pm}$ and the excited states $D^{*\pm}$ can be measured. 
Their nuclear modification factor and momentum anisotropy have been measured at both LHC and RHIC.
Currently, we focus on comparing to non-strange $D$ and $B$ mesons data.
Though strange heavy mesons $D_s, B_s$ are also very interested as they contain the strangeness enhancement information, the strangeness physics is not the main focus of this work.

The nuclear modification factor has already been introduced in chapter \ref{chapter:introduction}. 
Here we summarize how the momentum anisotropy observables are computed.
A list of the measurements and references can be find in table \ref{table:ALICE-obs} and table \ref{table:CMS-obs}.
\begin{center}
\begin{table}[h]
\caption{ALICE dataset}\label{table:ALICE-obs} 
\begin{tabularx}{\columnwidth}{XXX}
\hline 
 Observables & Centrality & Reference\\ 
\hline 
$D$-meson $v_2$ & 30-50\% & \cite{Acharya:2017qps}\\ 
\hline 
Event-engineered $D$-meson $v_2$ & 30-50\% & \cite{Grosa:2017zcz}\\ 
\hline 
$D$-meson $R_{AA}$ & 0-10, 30-50, 50-80\% & \cite{Acharya:2018hre}\\
\hline 
\end{tabularx}
\end{table}
\begin{table}[h]
\caption{CMS dataset}\label{table:CMS-obs} 
\begin{tabularx}{\columnwidth}{XXX}
\hline 
Observables & Centrality & Reference\\ 
\hline 
D${}^0$-meson $v_2$ & 0-10, 10-30, 30-50\% & \cite{Sirunyan:2017plt}\\ 
\hline 
D${}^0$-meson $R_{AA}$ & 0-10\%, 0-100\% & \cite{Sirunyan:2017xss}\\ 
\hline 
B${}^{\pm}$-meson $R_{AA}$ & 0-100\% & \cite{Sirunyan:2017oug}\\ 
\hline 
\end{tabularx}
\end{table}
\end{center}

\paragraph{Momentum anisotropy}
Heavy flavor momentum anisotropy at high-$p_T$ is thought to be the result of anisotropic energy loss because on average, hard partons emitted along the short axis lose less energy than those emitted along the long axis.
At low momentum, the momentum anisotropy is related to collective flow since the heavy quark interacts so frequently with the medium and tends to catch up with the flow velocity of the medium.
Both mechanisms produce $v_2$ relative to the common reference frame of the bulk geometry/bulk flow.
The $p_T$ differential $v_2$ is usually measured in a two-particle correlation approach,
\begin{eqnarray}
v_n\{2\}(p_T) = \frac{\mathfrak{Re}\langle d_n\{2\} \rangle}{\langle c_n\{2\}\rangle}.
\end{eqnarray}
$c_n$ is the event-wise two particle correlation of $N$ reference particles (REF, the bulk medium) within a certain kinematic range,
\begin{eqnarray}
c_n &=& \frac{|Q_n|^2 - N}{N(N-1)},  \\
Q_n &=& \sum_{i=1}^N e^{i n \phi},
\end{eqnarray}
and the event average ($\langle\cdots\rangle$) is weighted by $N(N-1)$.
$d_n$ is the correlation between the $M$ particles of interest (POI, in this case the heavy flavors) and the $N$ reference particles,
\begin{eqnarray}
d_n &=& \frac{q_n Q_n^* - m}{MN-m},  \\
q_n &=& \sum_{j=1}^M e^{i n \phi}
\end{eqnarray}
$m$ is the number of POI that is also counted as REF to subtract auto-correlations. 
The event average is weighted by the number of pairs $MN-m$.

\paragraph{Event-shape engineering on heavy-flavor $v_2$}
Event-shape engineering is a more recent idea to look at the detailed response of the hard sector to the medium geometry.
Experimentally, an ensemble of events belonging a certain centrality class is further classified according to its ``event shape'', measured by $q_2$,
\begin{eqnarray}
q_2 = \frac{|Q_2|}{\sqrt{N}}.
\end{eqnarray}
Due to event-by-event geometry fluctuations, the event shape in a given centrality class can vary dramatically.
The ALICE experiment then measures the D meson $v_2$ with events having the $20\%$ largest $q_2$ and events with the $60\%$ smallest $q_2$.
They found a large separation between the resulting $v_2$ of bias selected events compared to the $v_2$ calculated from unbiased events.
This measurement quantifies the response of the hard probe to the event geometry fluctuation while controlling multiplicity. 

\subsection{A first comparison to data}
We do not intend to optimize all the parameters in the model in this first comparison to data, but use a reasonable estimate of the parameters to understand the model.
The \trento\ parameters and the hydrodynamic transport coefficients are obtained from the high likelihood parameters in \cite{Bernhard:2018hnz}.
The heavy quarks start to lose energy from $0.6$ fm/$c$, and the matching condition between the vacuum-like radiation and the medium-induced radiation is $\Delta k_\perp^2 = R_v Q^2$ with $R_v = 1$.
We used only leading order contributions from the weakly coupled theory, and try both fixed coupling and running coupling.
The default switching scale between a large-$Q$ scattering small-$Q$ is $Q_{\textrm{cut}}^2 = 4 m_D^2$.

\begin{figure}
\centering
\includegraphics[width=\textwidth]{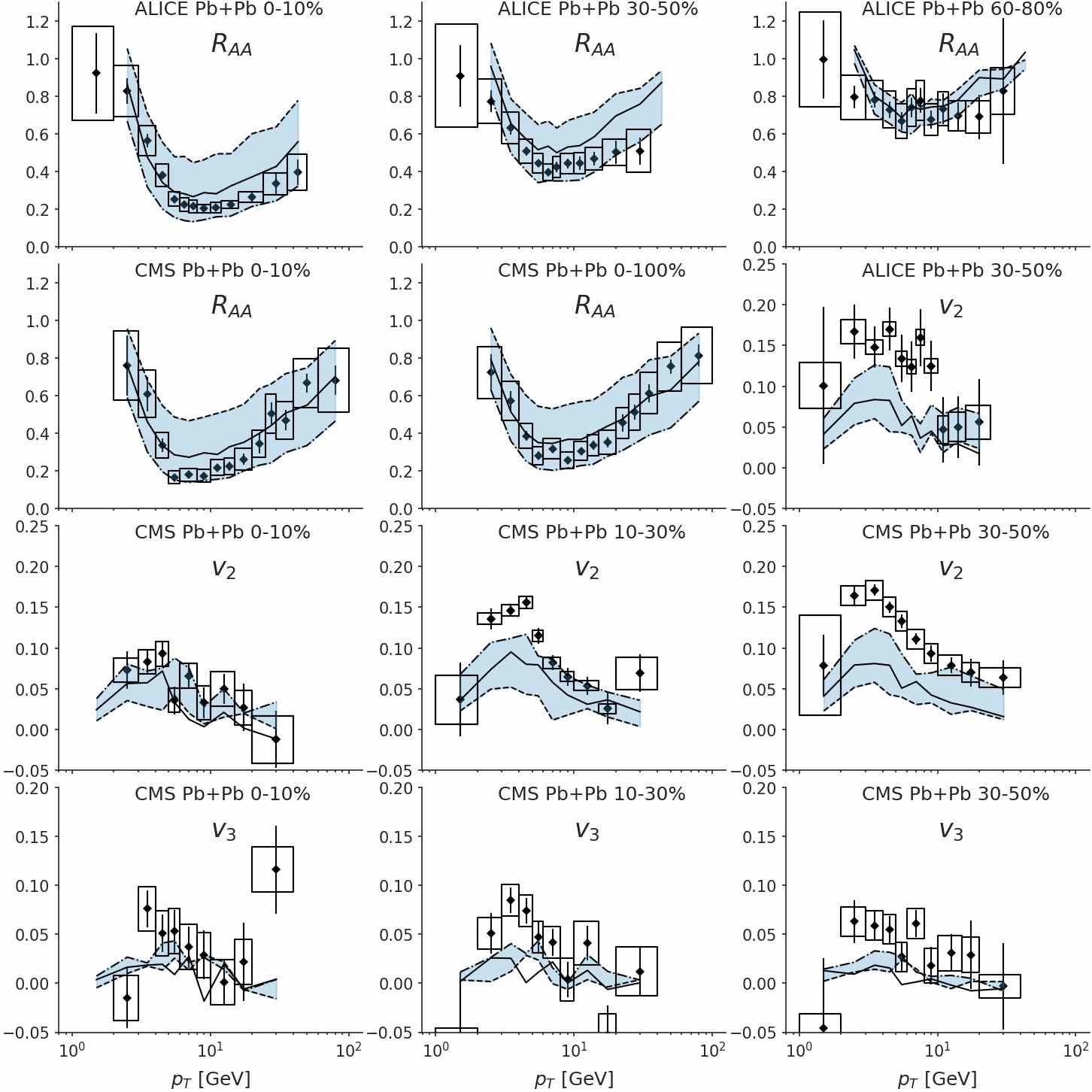}
\caption[Benchmark results using a fixed coupling constant $\alpha_s = 0.2$ (dashed),]{Benchmark results using a fixed coupling constant $\alpha_s = 0.2$ (dashed), $0.3$ (solid), and $0.4$ (dash-dotted). The blue bands fill between the results using $\alpha_s=0.2$ and $0.4$. They are compared to the experimental data (black symbols) obtained by the ALICE Collaboration \cite{Acharya:2017qps,Acharya:2018hre} and the CMS Collaboration \cite{Sirunyan:2017xss,Sirunyan:2017plt}.}
\label{fig:new:fix-a}
\end{figure}

\paragraph{Fixed coupling} First, we compute with a fixed coupling constant.
It is understood as an effective in-medium coupling for both elastic and radiative processes.
In figure \ref{fig:new:fix-a}, we present the results (lines and bands) with data points measured at $\sqrts{s} = 5.02$ TeV for D mesons (symbol with error bars and boxes).
Different line shapes corresponds to different coupling $\alpha=0.2$ (dashed), $0.3$ (solid), and $0.4$ (dash-dotted). 
The types of observables are shown within each subplot, indicating the experimental collaboration, the collision system, and the centrality.

Looking at the experimental measurements, $R_{AA}$ increases with the centrality classes and displays a minimum around $8 < p_T < 10$ GeV.
At high-$p_T$, the $R_{AA}$ increases slowly towards the baseline around one, noticing the log-scale of $p_T$.
At low-$p_T$, the $R_{AA}$ quickly rises.
There are many reasons for this, for example, the feed down from higher-$p_T$ particles due to energy loss;  the feeding from low-$p_T$ particles that are pushed outward by the strong medium radial flow.
Besides, the recombination hadronization mechanism also plays a part, as the D meson is gaining momentum (on average) in the recombination process.
Based on the comparison to $R_{AA}$, a phenomenological value for a fixed $\alpha_s$ is around $0.3$--$0.4$.
However, such values cannot explain the large momentum anisotropy in mid-central collisions, e.g., centrality 30-50\%.
This is usually referred as the $D$ meson $R_{AA}$--$v_2$ puzzle, which also appears for leading light hadrons.
There have been different solutions proposed to this problem, such as a sudden increase in the interaction strength near $T_c$, fine-tuning the general temperature-momentum dependence of the transport coefficients, et cetera. \cite{SCARDINA2016329,Xu:2017obm,Shi:2018vys}.
In the next two chapters, we will see if this discrepancy can be overcome by a fine-tuning of parameters in the current model.
A non-zero $v_3$ of $D$ mesons is evidence of heavy-flavor coupling to detailed event-by-event nuclear geometry fluctuations.
The calculation of $v_3$ is systematically below the data, despite the considerable statistical and systematic uncertainty.

\begin{figure}

\centering
\includegraphics[width=\textwidth]{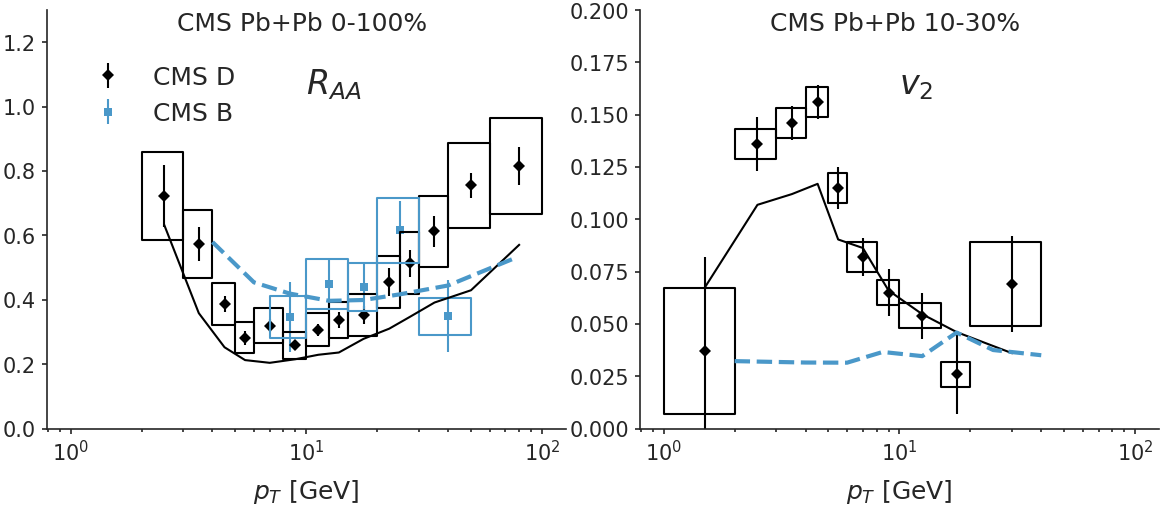}
\caption[Demonstrating the mass dependence of the observables using]{Demonstrating the mass dependence of the observables using $\alpha_s = 0.4$. 
The D meson and B meson results are labeled by black and blue, respectively.
Left plot: $R_{AA}$ for 0-100\% centrality. Right plot: $v_2$ for 10-30\% centrality.}
\label{fig:new:charm-bottom}
\end{figure}

In figure \ref{fig:new:charm-bottom}, we compare the calculation with $\alpha_s = 0.4$ for charmed meson $R_{AA}^D$ and bottom meson $R_{AA}^B$ at $0-100\%$ centrality and D meson and B meson flow at $10-30\%$ centrality.
The mass effect of bottom quarks is much stronger than for charm quarks; therefore, $R_{AA}^B$ at intermediate $p_T$ is higher than $R_{AA}^D$. 
At very high $p_T$, the ``dead cone'' of bottom quark becomes insignificant, and the $B$ and $D$ meson $R_{AA}$ converge.
Unlike the sudden increase of $v_2^D$ at low $p_T$, $v_2^B$ is always small, meaning that the bottom quarks do not catch up to the medium flow as the charm quarks do and remain far from equilibrium.

\begin{figure}

\centering
\includegraphics[width=\textwidth]{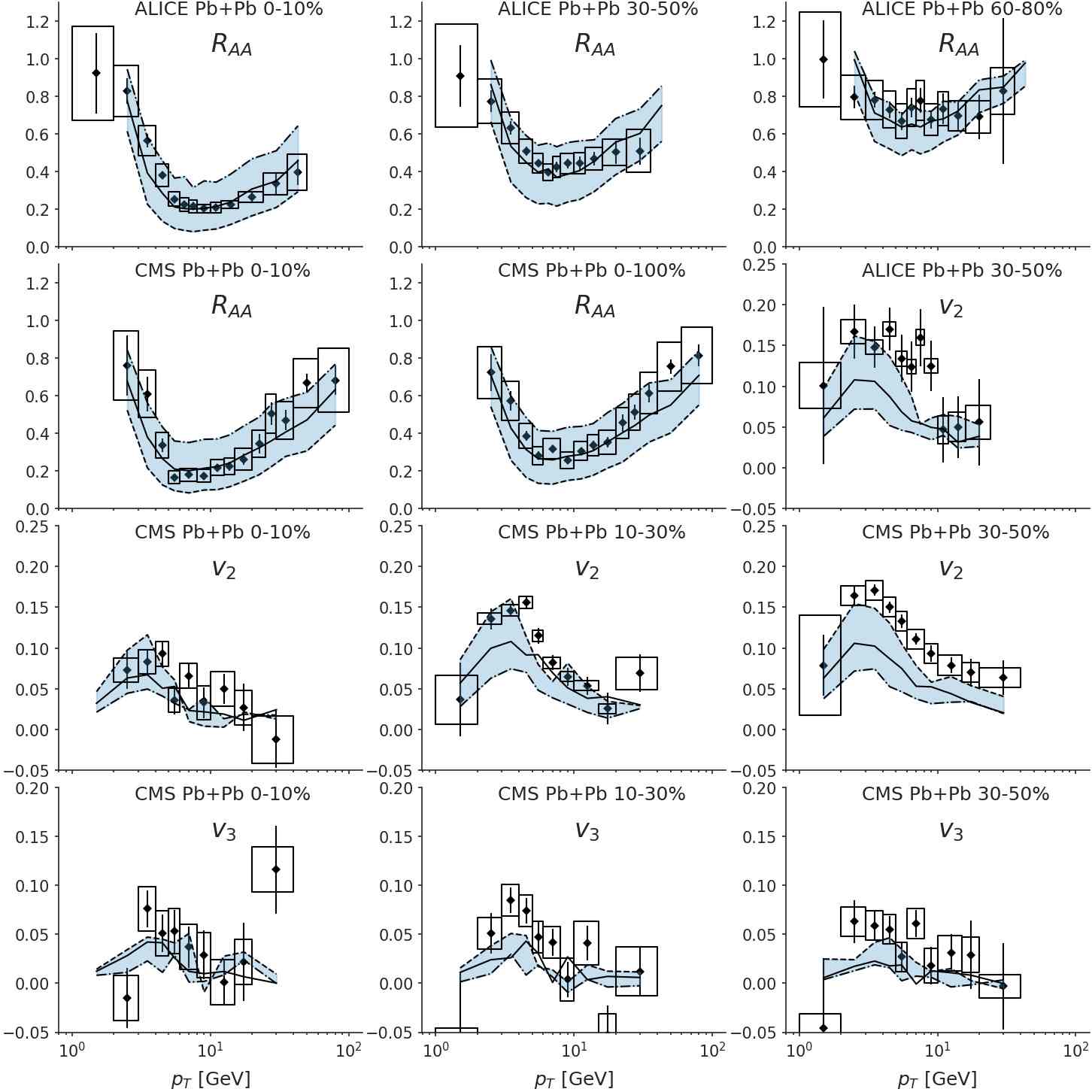}
\caption[Benchmark results using a running coupling constant. The]{Benchmark results using a running coupling constant. The medium scale that stops the low-$Q$ running is chosen at $Q_{\textrm{med}} = \mu\pi T = \pi T$ (dashed), $2\pi T$ (solid), and $4\pi T$ (dash-dotted). They are compared to the experimental data (black symbols) obtained by the ALICE Collaboration and the CMS Collaboration.}
\label{fig:new:run-a}
\end{figure}

\paragraph{Running coupling} Moving to a running coupling constant, the uncertainty of the in-medium coupling strength is transferred to the uncertainty of the medium scale in the running $\alpha_s$,
\begin{eqnarray}
\alpha_s(Q) = \frac{2\pi}{9}\frac{1}{\ln \left( \max\{Q, \mu\pi T\} / \Lambda\right)}.
\end{eqnarray}
Due to the running, heavy quark radiation at high energy will be reduced compared to low energy and the interaction strength with the medium is enhanced at low temperature relative to high temperature.

In the comparison shown in figure \ref{fig:new:run-a}, we choose $\mu = 1, 2, 4$, terminating the low-$Q$ running of $\alpha_s$ at $Q = \pi T$ (dashed), $2\pi T$ (solid), $4\pi T$ (dash-dotted).
We use $\pi T$ as a natural unit because it is the typical thermal scale in the finite-temperature field theory calculations.
Given that the entire heavy-flavor coupled-to hydrodynamic model is only an approximation, one should not think of the appearance of $\pi$ so seriously.
The $\mu=2\pi T$ choice explains the nuclear modification factor for all centralities very well but underestimates $v_2$ by 50\%.
The $\mu=\pi T$ case achieves a better agreement with $v_n$, but $R_{AA}$ is systematically off.
Therefore, going from fixed coupling to running coupling, the $v_2$ puzzle remains.

\begin{figure}

\centering
\includegraphics[width=\textwidth]{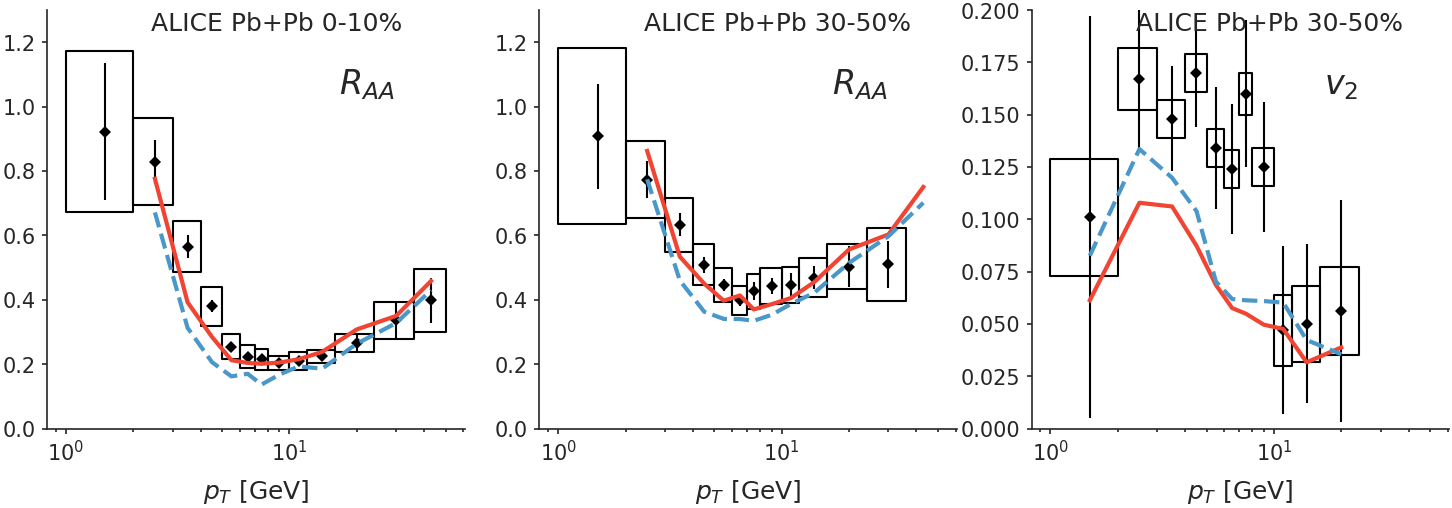}
\caption[Effect of changing the switch scale between small-$Q$ diffusion]{Effect of changing the switch scale between small-$Q$ diffusion modeling and large-$Q$ scattering modeling. $\mu=2$ is used. The red solid lines used a switching scale at $Q_{\textrm{switch}}^2 = 4 m_D^2$, and the blue dashed lines uses $16 m_D^2$.}
\label{fig:new:run-cut}
\end{figure}
\paragraph{Switching scale dependence} By construction, the energy loss should be insensitive to the switching scale between the small-$Q$ diffusion and the large-$Q$ scattering in the high energy, weakly coupled limit.
We check if this arguments holds for phenomenological application.
In figure \ref{fig:new:run-cut}, in addition to the default $Q_\textrm{cut}^2 = 4 m_D^2$ (red solid lines), we also use $Q_\textrm{cut}^2 = 16 m_D^2$ (blue dashed lines) to model an increase amount of probe-medium interaction by diffusion compared to scattering.
We find that the effect on high-$p_T$ observable is small.
Because the high-$p_T$ dynamics is dominated by the radiative energy loss, whose $Q_\textrm{cut}$ is indeed small as checked in chapter \ref{chapter:transport}.
Larger differences of $R_{AA}$ and $v_2$ is observed at low-$p_T$.
One reason for this is that the $Q_\textrm{cut}$ independence argument obtained for high energy partons does not work very well for low-velocity partons.
Another reason is that despite the scattering dynamics and the diffusion dynamics having a matched diffusion constant (second moment of the momentum transfer), they are differed in all other higher moments, in particular, the drag (first moment).
Remember that the drag coefficient in the diffusion dynamics is not a direct input from the weakly coupled theory, but is determined by the Einstein relation.
The Einstein relation only guarantees that the diffusion dynamics evolve the system to the same equilibrium as the scattering dynamics, but the non-equilibrium path it takes can be very different from that of the scattering dynamics.

This $Q_\textrm{cut}$ dependence may be undesirable at first sight, but one knows that the weakly coupled scattering picture does not necessarily work for the phenomenological coupling regime ($g\sim 2$), while the diffusion dynamics can be extended to the strongly coupled regime.
The $Q_\textrm{cut}$ parametrizes an important source of theoretical uncertainty in our modeling.

\begin{figure}

\centering
\includegraphics[width=\textwidth]{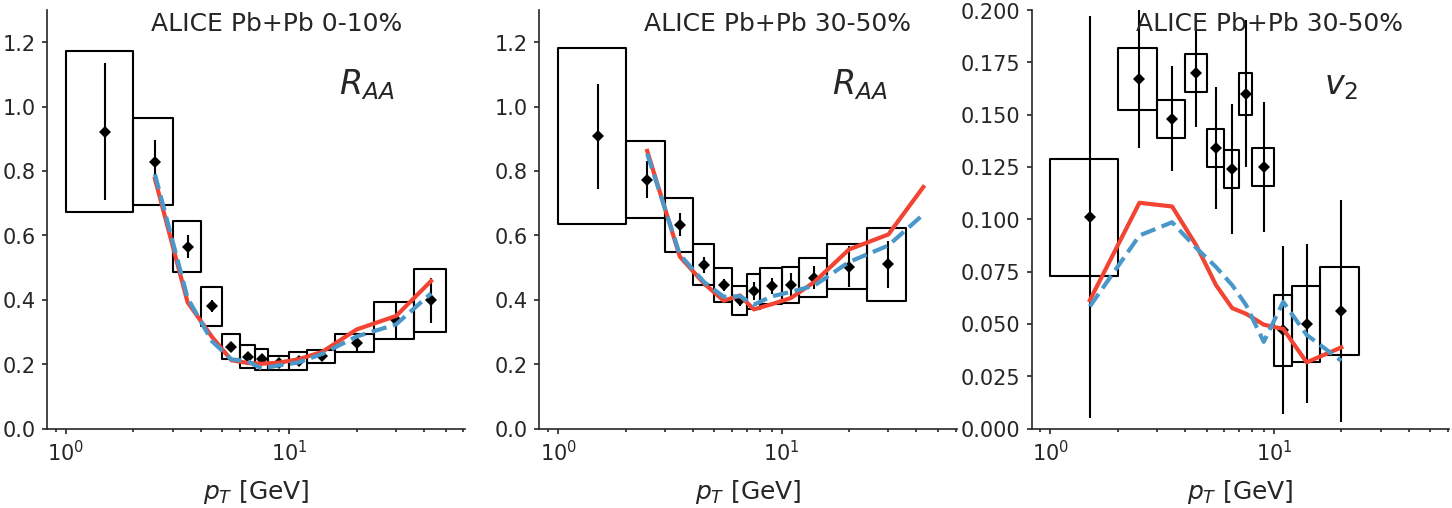}
\caption[Effect of changing the matching scale parameter $R_v$]{Effect of changing the matching scale parameter $R_v$ ($\Delta k_\perp^2 = R_v Q^2$) between the vacuum-like shower and the medium-induced shower. $\mu=2$ is used. The red solid lines use $R_v =1$, while the blue dashed lines use $R_v = 1000$, which leave the vacuum-like shower largely unmodified.}
\label{fig:new:run-match}
\end{figure}

\paragraph{Vacuum / medium-induced radiation matching scale dependence}
As explained, there are separated treatments of radiation in different regions of phase-space on the Lund-diagram.
Accordingly, we need to subtract the vacuum radiation that overlaps with the medium-induced region in the Pythia event generator.
In our earlier transport study of heavy flavor \cite{Ke:2018tsh}, this subtraction was not included; therefore, we would like to demonstrate the impact of this mistreatment here.

In figure \ref{fig:new:run-match}, two calculations are shown. 
The red dashed lines stand for the case where we removed vacuum-like radiations that satisfy $\Delta k_\perp^2 > Q^2$.
The solid blue lines are calculations without this subtraction.
The two calculations for $R_{AA}$ only differ for $p_T\gtrsim 20$ GeV, because only high-$p_T$ heavy quarks can undergo splittings that take long enough time to receive significant medium corrections.
Also, the difference is larger for central collisions than for peripheral collisions, because medium the effect for the latter is weaker.
No significant difference is observed for $v_n$.

\begin{figure}

\centering
\includegraphics[width=\textwidth]{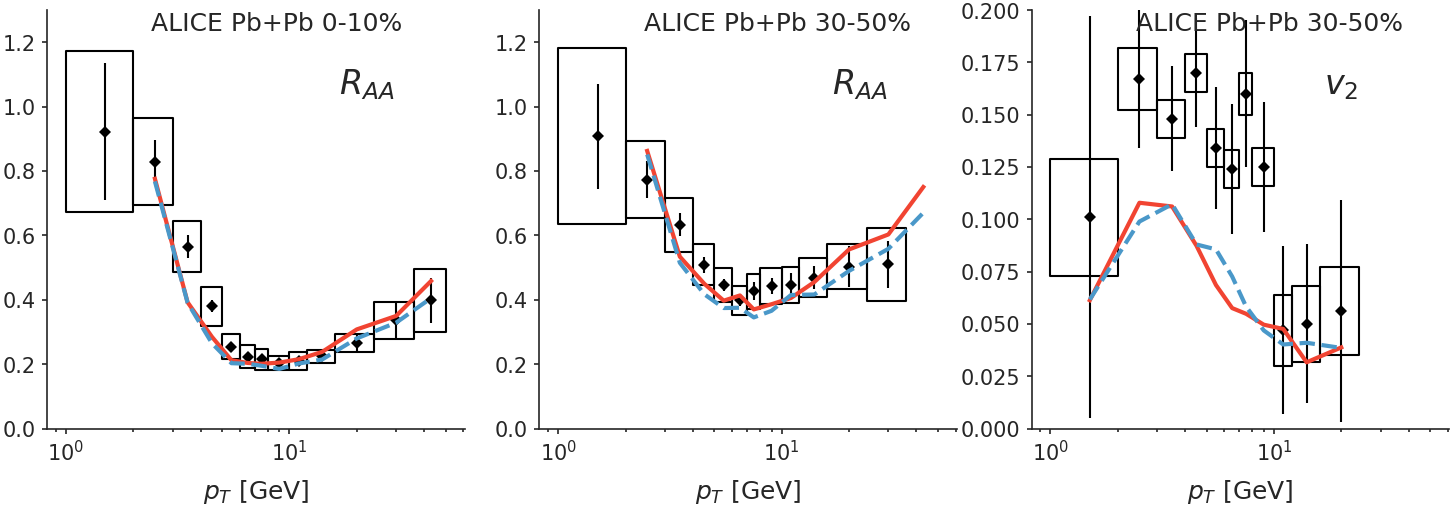}
\caption[The impact of using the ``local rate'' approximation. The red]{The impact of using the ``local rate'' approximation. The red solid lines use the default implementation; while the blue dashed lines perform the rescatterings in an imaginary infinite box using locally defined temperatures to mimic the ``local rate'' approximation.}
\label{fig:new:run-local}
\end{figure}

\paragraph{Performance of the ``local rate'' approximation}
Finally, it is interesting to examine the effect of an ``local rate'' approximation of the radiative processes on the observables.
It approximates the radiation probability in a medium with slowly varying temperature by the integration of radiation rates defined in an infinite static box with the local temperature at each point.
One can also refer to it as the ``adiabatic'' approximation because it assumes the temperature variation is slow compared to the formation time.

We know this approximation can be broken by the fast expansion of the QGP fireball and would like to quantify the impact.
It is easy to mimic the ``local'' approximation in our model; one can let the preformed-gluon rescattering procedure be done in an imaginary medium with the same temperature and flow velocity as those at the point of its production, instead of those in the evolving medium.
The resulting comparison is shown in figure \ref{fig:new:run-local}.
The local approximation is good except at very high-$p_T$ ($p_T > 30$ GeV).

\chapter{Bayesian model-to-data comparison}
\label{chapter:bayes}
We have discussed the modeling details of the heavy flavor transport in relativistic heavy-ion collisions, and have shown a comparison to data with rather a ``na\"ive'' guess of multiple parameters.
Till now, we have only varied a small subset of them to understand the model qualitatively.
In this section, we introduce the advanced statistical tool known as Bayesian analysis that can calibrate all parameters simultaneously to the experimental data.
For the full details of such an analysis, we refer the readers to this excellent dissertation on this subject \cite{Bernhard:2018hnz} in the context of heavy-ion collisions.

To facilitate the discussion, I define the problem for this chapter and introduce a few notations and terminologies.
We formulate the general task of a model-to-data comparison into the following form,
\begin{itemize}
\item A complex model $M$ with $n$ input parameters organized as a $n$-dimensional vector $\mathbf{p}$.
\item There exists a prior belief on the reasonable range of each parameter, known as the prior probability distribution, and for short ``$\mathrm{Prior}$''.
\item $n$ experimental measurements are organized as an observation vector $\mathbf{y}_{\exp}$ of dimension $m$, with given statistical and systematic uncertainties $\delta\mathbf{y}_{stat}$,$\delta\mathbf{y}_{sys}$.
\item The task is to infer the posterior probability distribution of $p$ ($\mathrm{Posterior}$), given the model $M$, the measurements $\delta\mathbf{y}_{\exp}\pm \delta \mathbf{y}$, and the $\mathrm{Prior}$.
\end{itemize}
The analysis proceeds in the following steps that are explained in each section.

\section{Model evaluation}
A prerequisite for this analysis is the ability to fast evaluate model $M$ at any point in the considered region of parameter space.
It is achieved by interpolating model calculations obtained at $N$ carefully designed parameter points.
This $N$ set of parameter vectors of length $n$ forms a so-called design matrix $\mathbf{D}$,
\begin{eqnarray}
\mathbf{D}_{N\times n} = 
\begin{bmatrix}
p_{11} & p_{12} & \cdots & p_{1n}\\
p_{21} & p_{22} & \cdots & p_{2n}\\
\vdots & \vdots & \ddots & \vdots \\
p_{N1} & p_{N2} & \cdots & p_{Nn}
\end{bmatrix}
\end{eqnarray}
where the first index is the label of different parameter set, and the second index labels different parameters.

We use an existing software \cite{lhs-r} of so-called Latin-Hyper-Cube design method \cite{MORRIS1995381} to determine the location of these points in parameter space.
It generates a semi-random design subject to the following constraints:
\begin{itemize}
\item The marginalized distribution on any parameter is a uniform distribution.
This is different from a grid design, where the marginalized distribution are spiky delta functions on the grid points.
\item The minimum distance between any two points in the parameter space is maximized.
This is different from a completely random design in which points may form tight clusters or leave sparsely occupied regions.
\end{itemize}
Usually, for a well-behaved model, the number of design points needed for a good interpolation increases linearly with the number of parameters $n$, in contrast to an exponential increasing with $n$ in a grid design.

The actually model evaluation on these points is the most time-consuming part of this analysis.
The outputs are organized into the observation matrix,
\begin{eqnarray}
\mathbf{Y}_{N\times m} = 
\begin{bmatrix}
y_{11} & y_{12} & \cdots & y_{1m}\\
y_{21} & y_{22} & \cdots & y_{2m}\\
\vdots & \vdots & \ddots & \vdots \\
y_{N1} & y_{N2} & \cdots & y_{Nm}
\end{bmatrix}
\end{eqnarray}
where the first index is the label of different parameter set, and the second index labels different observables.
The design matrix $\mathbf{D}$ and the observations matrix $\mathbf{Y}$ help to train a general interpolator to infer the calculated observables at any given parameter value.

\section{Data reduction}
The model $M$ is a mapping of an $n$-dimensional vector to an $m$-dimensional vector. 
One can certainly construct an array of independent $m$ scalar mappings, and interpolate each of them.
However, this na\"ive construction does not make use of the intrinsic correlations/structures in the training data, and can be very inefficient for practice usage.
Considering an observation with two values of $R_{AA}$ and $v_2$. Usually, the larger the $R_{AA}$ the model predicts, the smaller the $v_2$ is, and thus an anti-correlation is expected.
If one build interpolators for them independently, the interpolation uncertainties are also going to be independent, which does not reflect the correlation information.
However, if one interpolates the linear combinations $a R_{AA} \pm b v_2$; then a wise choice of $a, b$ significantly reduces the correlation between these two ``newly'' constructed observables.

The principal component analysis (PCA) is a systematic way to implement this idea.
The original vectors of observables are transformed into the principal-component (PC) space, with each PC a specific linear combination of the original observables, so that the covariances between the newly defined observables (the PCs) vanish.
Mathematically, this is the same as finding the singular value decomposition (SVD) of $\mathbf{\tilde{Y}}$. 
$\mathbf{\tilde{Y}}$ is the standardized observation matrix $\mathbf{Y}$,
\begin{eqnarray}
\tilde{y}_{ij} = \frac{y_{ij} - \mu_j}{\sigma_j}
\end{eqnarray}
with $\mu_j$ and $\sigma_j$ the mean and the standard deviation of column $j$.
Then the SVD proceeds as,
\begin{eqnarray}
\tilde{\mathbf{Y}}_{N\times m} = \mathbf{U}_{N\times N} \mathbf{\Sigma}_{N\times m} \mathbf{V}_{m\times m}.
\end{eqnarray}
Here $\mathbf{\Sigma}$ only contains the variance of each PCs on its diagonal.
The PCs are defined as the components after the $V$ transformation.
\begin{eqnarray}
z = \mathbf{V}y
\end{eqnarray}
It is evident that the covariance matrix of the $z$ observables is diagonalized,
\begin{eqnarray}
\mathrm{Var}(z_i, z_j) = \frac{1}{N}V_{ii'}\tilde{Y}_{ki'}V_{jj'}\tilde{Y}_{kj'} = \frac{1}{N}V\tilde{Y}^T\tilde{Y}V^T = \frac{1}{N}\mathbf{\Sigma}.
\end{eqnarray}
So different PCs are orthogonalized.

A data reduction is another benefit of using PCA.
Suppose we have sorted the variance in $\mathbf{\Sigma}$ from maximum to minimum.
For data with pronounced structures, often the first few PCs take into account the majority of the data variance.
Practically, a truncated set of PCs already gives a good representation of the original data, and this dramatically reduces the computations necessary for interpolating a large number of observables.
Finally, one can always go back from the PC space to the original space by the inverse transformation $y = V^{-1} z$.
The PCA software is provided by \cite{sklearn_api}.

\section{Model emulator}
With limited information on a finite number of design points contained in the matrices $D$ and $M$, the original mapping is approximated by a model emulator (a surrogate model) using a Gaussian Process (GP).
The Gaussian Process provides a non-parametric interpolation for scalar function with one or high dimensional input.
We shall let the readers refer to \cite{rasmussen2006gaussian} for the technical details and only summarize the basics of the Gaussian Process.

\paragraph{Gaussian Process} Take a uni-variate case as an example. Given an array of input and an array of output, polynomial interpolation is a common way to interpolate the data.
However, polynomial interpolation only uses local information of the grid, and its performance can be sensitive to the error of the output, e.g., statistical fluctuation in the simulation.
Moreover, it is hard to work with a Lain-hypercube design because the design points are not arranged on a regular grid.
In contrary, a GP does not make any assumption on the functional form of the interpolation but infers the output at a particular input based on how its output correlates with given outputs at other input points.
Mathematically, one assumes that elements of the predicted output $\mathbf{y}^*$ at input $\mathbf{x^*}$ and the known outputs $\mathbf{y}_{\textrm{train}}$ at the training points $\mathbf{x}_{\textrm{train}}$ form a multi-variate normal distribution,
\begin{eqnarray}
\begin{bmatrix}
\mathbf{y}^* \\
\mathbf{y}_{\textrm{train}}
\end{bmatrix}
\sim
\mathcal{N}\left(
\begin{bmatrix}
\mathbf{\mu}^* \\
\mathbf{\mu}_{\textrm{train}}
\end{bmatrix},
\begin{bmatrix}
\mathbf{\Sigma}(\mathbf{x}^*, \mathbf{x}^*)& \mathbf{\Sigma}(\mathbf{x}^*, \mathbf{x}_{\textrm{train}}) \\
\mathbf{\Sigma}(\mathbf{x}_{\textrm{train}}, \mathbf{x}^*)& \mathbf{\Sigma}(\mathbf{x}_{\textrm{train}}, \mathbf{x}_{\textrm{train}})
\end{bmatrix}
\right)
\end{eqnarray}
Without a loss of generality, one often standardizes the training data so that the mean values $\mathbf{\mu}^*$ and $\mathbf{\mu}_{\textrm{train}}$ are zero.
The $\mathbf{\Sigma}$s form the covariance matrix, and each of them has the same shape of the outer product of its two arguments.
Its matrix-element (the kernel function) are parametric, and one often takes a squared exponential form,
\begin{eqnarray}
\Sigma_{ij} = k(x_i, x_j) = \sigma^2 \exp\left(-\frac{(x_i-x_j)^2}{2l^2}\right).
\end{eqnarray}
$\sigma^2$ is the auto correlation and $l$ is the correlation length.
The covariance decays exponentially with the squared separation of the two input points.
In such a way, points that are close in inputs will also be close in outputs, and points that are far apart are effectively uncorrelated.
The squared exponential form is not the only possible kernel function; people have designed more sophisticated choices with more parameters for various problems. 

\paragraph{Conditioning a Gaussian Process} The outputs at training points are known.
Therefore, the probability distribution of $\mathbf{y}^*$ is obtained by conditioning the training outputs on their actual values,
\begin{eqnarray}
\mathbf{y}^* \sim &&\mathcal{N}\left(
\mathbf{\Sigma}(\mathbf{x}^*, \mathbf{x}_{\textrm{train}} )
\mathbf{\Sigma}^{-1}(\mathbf{x}_{\textrm{train}}, \mathbf{x}_{\textrm{train}} )\mathbf{y}_{\textrm{train}},\right.\\\nonumber
&&\left.
\mathbf{\Sigma}(\mathbf{x}^*, \mathbf{x}^*) - 
\mathbf{\Sigma}(\mathbf{x}^*, \mathbf{x}_{\textrm{train}} )
\mathbf{\Sigma}^{-1}(\mathbf{x}_{\textrm{train}}, \mathbf{x}_{\textrm{train}} )
\mathbf{\Sigma}(\mathbf{x}_{\textrm{train}},\mathbf{x}^*)
\right)
\end{eqnarray}
Note that the conditional multivariate normal distribution is still a normal distribution, with modified mean and covariance matrix.
One can check that if the predicted input approaches one of the training inputs, the distribution of the output approaches a $\delta$-function (as the limit of a narrow Gaussian) at the training output.

\paragraph{Hyperparameters and training} We have not discussed the parameters in the kernel function $k(x, x')$ too much yet.
For now, they are the auto-correlation $\sigma^2$ and the correlation length $l$. 
They are known as hyper-parameters (denoted as a vector $\mathbf{\theta}$), and should in principle, be treated as unknown parameters in the calibration.
But a common practice to reduce the complexity is to fix the hyper-parameters at a set of ``optimal values'' by minimizing the loss function $\mathcal{L}$,
\begin{eqnarray}
\mathcal{L} = -\ln p(\mathbf{y}|\mathbf{\theta}) = \frac{1}{2}\ln \det \mathbf{\Sigma}(\mathbf{\theta})  + \frac{1}{2}\mathbf{y}^T \mathbf{\Sigma}(\mathbf{\theta})^{-1} \mathbf{y} + \frac{N}{2}\ln(2\pi)
\end{eqnarray}
where $\mathbf{y}$ is the (PCA transformed) training data, and $N$ is the number of training points.
The minimization process is referred as ``training'' a Gaussian Process emulator.

\paragraph{Inference with uncertainty quantification} Unlike the polynomial interpolation, a GP does not provide a single estimation of the output but infers the probability distribution of the predicted outputs by predicting both the mean and the covariance matrix.
It is a huge advantage of the Gaussian Process to quantify its interpolation uncertainty.

\paragraph{Validation} Though the training process includes a penalty for over-fitting the data, whether the trained GP has an over-fitting problem can only be checked by validation.
In a validation procedure, one performs model calculations at novel points in the parameter space that is not used to train the GP; then, compare the GP's prediction $y_i \pm \sigma_i$ to the model calculation $y_{\textrm{validate}, i}$.
If an emulator is trained to work properly, then the standardized deviation $(y_i - y_{\textrm{validate}, i})/\sigma_i$ should follow approximately a standard normal distribution.

\paragraph{Multivariate inputs and outputs} The GP formulation can be easily generalized to higher-dimensional inputs by specifying a multidimensional kernel function.
For high dimensional outputs, one first applies the PCA analysis introduced in the previous section and the build individual GPs for each of the first $N_{PC}$ principal components that take most of the data variance.

\section{Bayes' theorem and Markov chain Monte Carlo}
With the model emulator $M$ (we are using the same symbol as the model, but one should always remember that the emulator is only a fast surrogate of the original model and comes with uncertainty), we apply Bayes' theorem, the essence of the statistical analysis.
Bayes' theorem provides a quantitative way to update the knowledge of model parameters with empirical observations,
\begin{eqnarray}
\mathrm{Posterior}(\mathbf{p}|M, \mathbf{y}_{\textrm{exp}}) \propto \mathrm{Likelihood}(\mathbf{y}_{\textrm{exp}}|M, \mathbf{p})\times\mathrm{Prior}(\mathbf{p}).
\end{eqnarray}
It states that the posterior probability distribution of parameters, given the model and experimental measurements, is proportional to the likelihood of describing the experiments with the model using this set of parameters, times the prior belief of the distribution of the parameters.
The likelihood function is often assumed to be a multivariate Gaussian,
\begin{eqnarray}
\mathrm{Likelihood}(\mathbf{p}) &=& (2\pi)^{-\frac{m}{2}} (\det|\Sigma|)^{-\frac{1}{2}} \exp\left\{-\frac{1}{2}\Delta \mathbf{y}^T \mathbf{\Sigma}^{-1} \Delta \mathbf{y}\right\}, \\ 
\Delta \mathbf{y} &=& \mathbf{y}(\mathbf{p}) - \mathbf{y}_{\textrm{exp}}
\end{eqnarray}
where the $\mathbf{y}(\mathbf{p})$ is the model emulators' prediction at parameter point $\mathbf{p}$, $m$ is the number of observables.
The prior distribution is often a multi-dimensional uniform distribution within a reasonable range. 
The covariance matrix contains various sources of uncertainties from both theory and experimental side.

\paragraph{A model dependent statement}One always defines a posterior with a given model; therefore, even the extraction of theoretically well-defined quantities can be affected by different dynamical modeling assumptions/approximations.
On the one hand, the ultimate solution is, of course, to improve the physical accuracy of the model.
On the other hand, one could use a flexible model or models with different (but reasonable) assumptions to extract the same quantity to establish a level of theoretical uncertainty.

\paragraph{The covariance matrix} covariance matrix is decomposed into different contributions,
\begin{eqnarray}
\mathbf{\Sigma} = \mathbf{\Sigma}_{\textrm{stat}} + \mathbf{\Sigma}_{\textrm{sys}} + \mathbf{\Sigma}_{\textrm{emulator}} + \mathbf{\Sigma}_{\textrm{truncation}} + \mathbf{\Sigma}_{\textrm{model, sys}}
\end{eqnarray}
\begin{itemize}
\item The statistical co-variance takes the diagonal form, $\mathbf{\Sigma}_{\textrm{stat}} = \delta_{ij}\delta\mathbf{y}_{\textrm{stat}, i}^2$. 
$\delta\mathbf{y}_{\textrm{stat}, i}$ is the experimental statistical uncertainty.
\item The experimental systematic uncertainties $\mathbf{\Sigma}_{\textrm{sys}}$ can be correlated for different observations, so generally its off-diagonal elements are non-zero,
\item The emulator covariance $\mathbf{\Sigma}_{\textrm{emulator}}$ is the prediction covariance of the GPs in the PC space and then transformed into the physical space.
\item The truncation covariance $\mathbf{\Sigma}_{\textrm{truncation}}$ take those less important principal components that are not being emulated by GPs into account. 
Its variance is first computed in the PC space and then transformed back to the physical space.
\item Finally, $\mathbf{\Sigma}_{\textrm{model, sys}}$ stands for the model uncertainty. 
It is always present but is hard to quantify using the model itself.
Therefore, the previous study \cite{Bernhard:2018hnz} assign a variable model systematic uncertainty parameter $\sigma$, and this parameter will be treated as uncertainty in the calibration as well.
The $\sigma$ stands for a uniform model uncertainty fraction on each principal component and is added to the emulator prediction covariance.
The $\sigma$ parameter is given an information prior distribution $P(\sigma) \propto \sigma^2 e^{-\sigma/0.05}$. Meaning an expectation of $15\%$ model uncertainty.
The exact origin of this model uncertainty is unknown, but it plays a row as a ``regulator'' in the fitting process to prevent the model trying to explain features that can never be described better than a $\sigma$ level precision.
\end{itemize}

\paragraph{Marginalize the posterior distribution} The resultant posterior distribution is a function of $n$ parameters.
To answer what is the probability distribution of one parameter folded with the uncertainty from other parameters, one looks at the marginalized distribution with the other $n-1$ parameters integrated out.
A Markov chain Monte Carlo (MCMC) sampling of the posterior function performs the marginalization.
The MCMC evolves an ensemble of $n$-dimensional walkers to thermalize to the target posterior distribution.
Then, one obtains the one-parameter marginalization by projecting the ensemble onto one dimension.
Similarly, a marginalization of the joint-distribution of two or more parameters can be obtained similarly.
The MCMC software is developed by \cite{emcee}.

\chapter{Results}
\label{chapter:results}
In this chapter, we perform Bayesian analysis on the heavy-flavor transport model and extract the heavy quark transport coefficients.
I want to present both our earlier extraction using older models and the present extract to emphasize the latest improvements.

\section{Lessons from earlier extractions of $\hat{q}_Q$}
In an earlier publication \cite{Ke:2018tsh}, we used a linearized Boltzmann model with the coherence factor approach to implement the LPM effect.
The coherence factor approach, described in section  \ref{section:compare_former} modifies the incoherent gluon radiation rate with an interference factor $2(1-\cos\Delta t/\tau_f)$. 
It also uses a multiple emission prescription by resetting $\Delta t=0$ after every emission.
We have commented on its advantages and disadvantages in \ref{section:compare_former}.

The heavy quark initial momentum distribution was obtained from the FONLL calculation.
We have already commented on the advantages and disadvantages of these choices.
Two different sets of nuclear PDF--EPPS16 \cite{Eskola:2016oht} and nCTEQ15\cite{Kovarik:2015cma}--were used to represent the uncertainty from the cold nuclear matter effect in the $\hat{q}$ extraction.

Regarding model parameters, the one parameter for the perturbative elastic and inelastic scatterings was controlled by $1/3 < \mu < 4$ in the running coupling. 
There was an additional pure diffusion process with a diffusion constant $\kappa_{NP}$ parametrized to peak at low temperature and low energy, in order to mimic the non-perturbative coupling between a low energy probe and the medium near $T_c$,
\begin{eqnarray}
\kappa_{NP} = T^3 \kappa_D \left(x_D + (1-x_D)\frac{1}{ET/\textrm{ GeV}{}^2}\right).
\end{eqnarray}
The $0<\kappa_D<8$ parameter was the overall strength of the diffusion, and the $0<x_D<1$ controlled the degree of energy-temperature dependence.
One can see that in the heavy quark limit $M\rightarrow \infty$, this parametrization becomes independent of mass.
An additional parameter was the in-medium energy loss starting time $\tau_0$ that was allowed to vary between $0.1$ fm/$c$ to $1.0$ fm/$c$ (before the onset of hydrodynamics).
The reason is our lack of a quantitative description of the production of color charge in the initial stages.
This starting time is a simple approximation that interactions starts after $\tau_0$ when the color carries approach a Boltzmann distribution.

The design of the four-dimensional parameter space  $(\tau_0, \mu, \kappa_D, x_D)$ had 80 design points.
The computation was carried on the distributed computing system Open Science Grid \cite{Pordes:2007zzb,Sfiligoi:2010zz} using about a million CPU hours.
The observables on which we calibrated are listed in tables \ref{table:ALICE-obs} and \ref{table:CMS-obs}. 
Including, $p_T$ dependent $D$-meson nuclear modification factor $R_{AA}$ and $p_T$ dependent (event-shape-engineered) azimuthal anisotropy $v_2$.
CMS measurements of the $B^{\pm}$-meson $R_{AA}$ were also included to constrain the mass dependence of the transport coefficients.

\begin{figure}

\includegraphics[width=\textwidth]{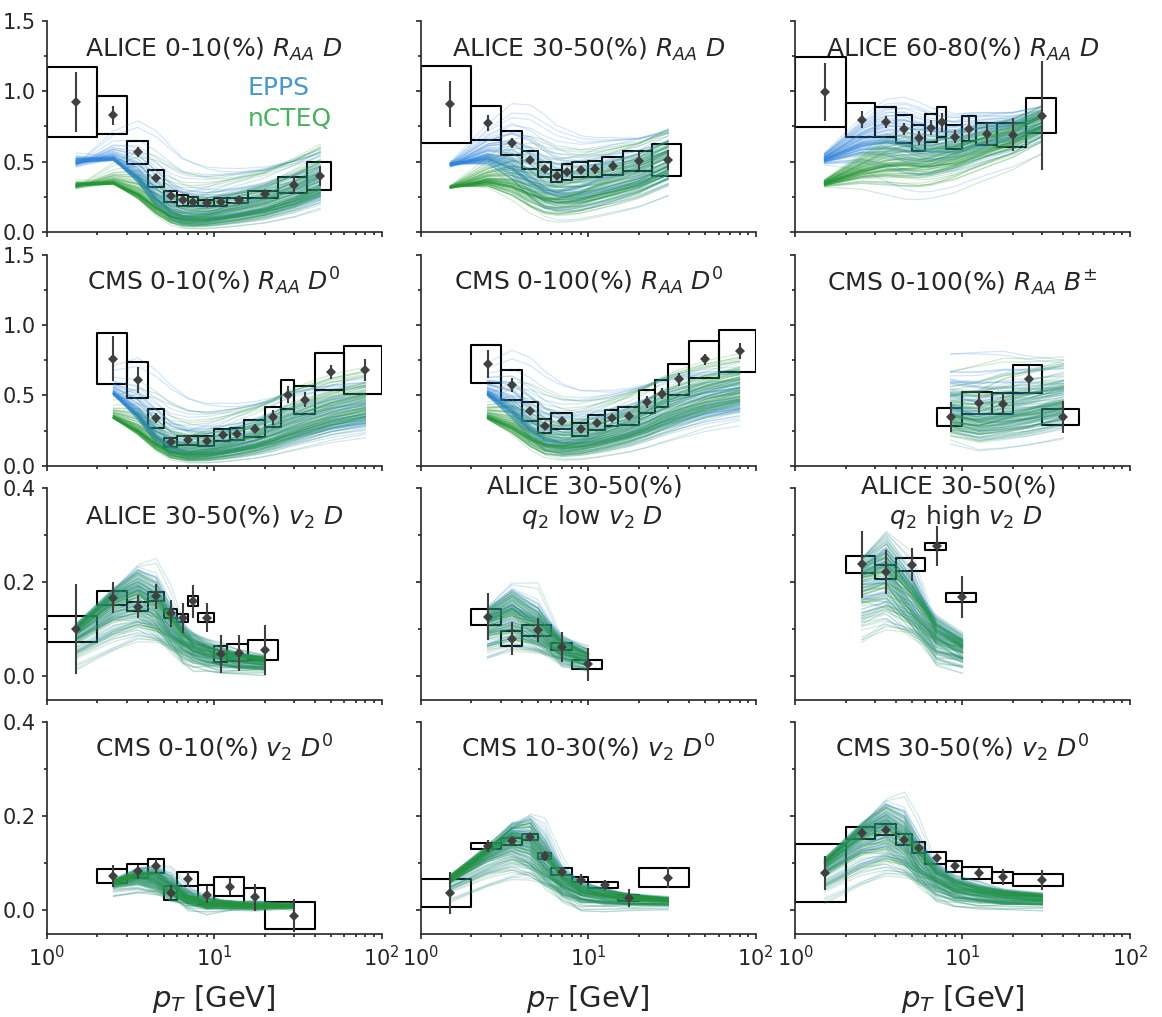}
\caption[The prior distribution of calculated observables compared to data.]{The prior distribution of calculated observables compared to data. The colors labeled the use of EPPS (blue) and nCTEQ15np (green) nuclear PDF.}
\label{fig:LBT:obs_prior}
\end{figure}

\begin{figure}

\includegraphics[width=\textwidth]{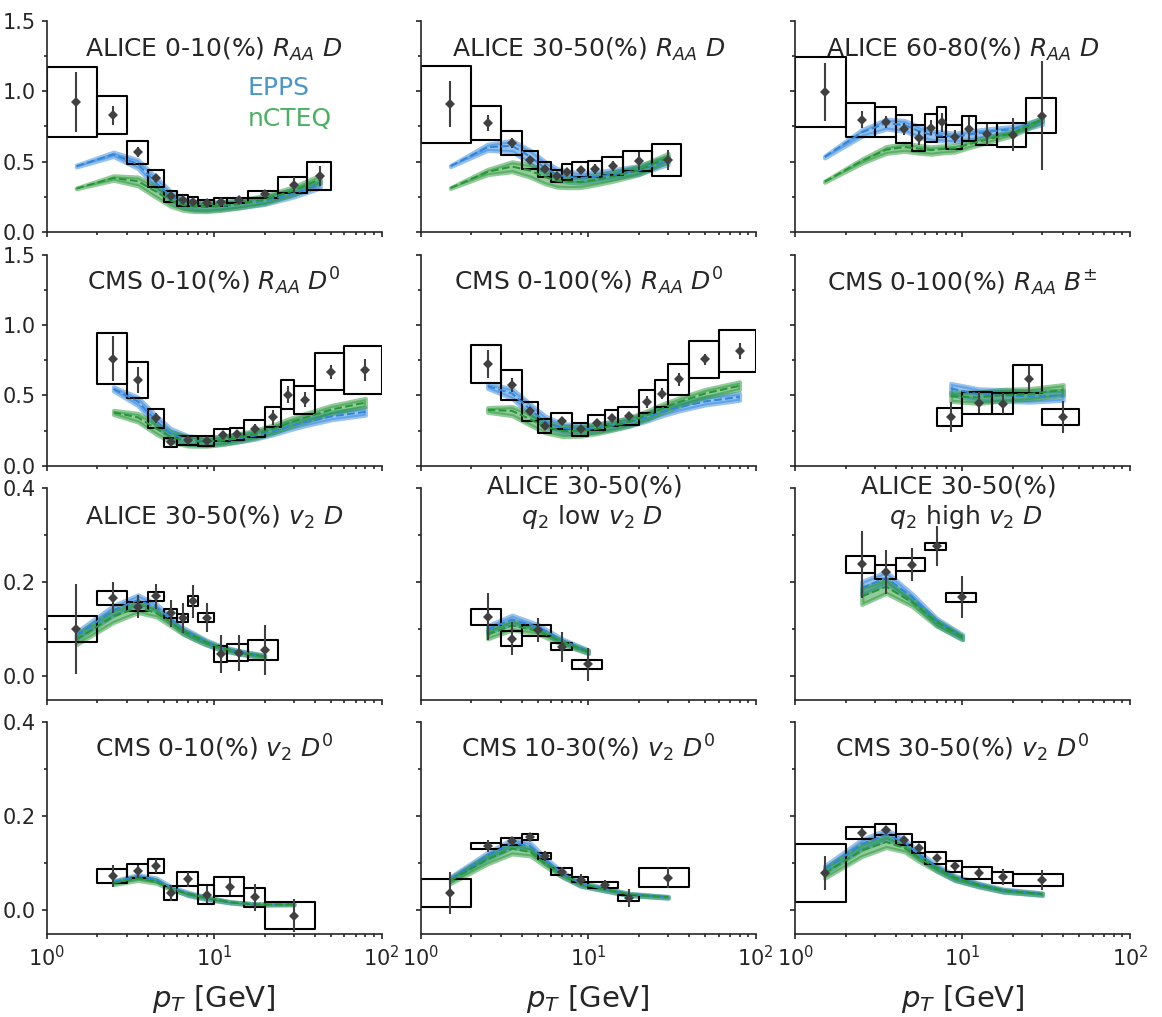}
\caption[The 90\% credible region of the posterior distribution of observables]{The 90\% credible region of the posterior distribution of observables compared to data. The colors labeled the use of EPPS (blue) and nCTEQ15np (green) nuclear PDF.}
\label{fig:LBT:obs_posterior}
\end{figure}

The prior and the posterior of the observables before and after the calibration is shown in figures \ref{fig:LBT:obs_prior} and \ref{fig:LBT:obs_posterior}.
Blue stands for using EPPS nuclear PDF and green stands for using the nCTEQnp nuclear PDF.
We found that the model after the calibration provided a good description of $R_{AA}$ and $v_2$ at the intermediate $p_T$ of the ALICE experiments.
However, it did not reproduce the fast uprising shape of $R_{AA}$ at high-$p_T$ of the CMS experiment.
Besides, the model seemed to underestimate the high-$p_T$ $v_2$ of the $30-50\%$ centrality bin measured by CMS.
The model can explain the correlation between the D-meson $v_2$ and the event-shape, though there is still significant fluctuation in the data.
The use of different nuclear PDFs had a negligible effect on $v_2$, but did affect the $R_{AA}$ at small and large $p_T$.
Another thing worth noting is that the $D$ and $B$ meson $R_{AA}$ were described at the same time.

\begin{figure}

\centering
\includegraphics[width=.6\textwidth]{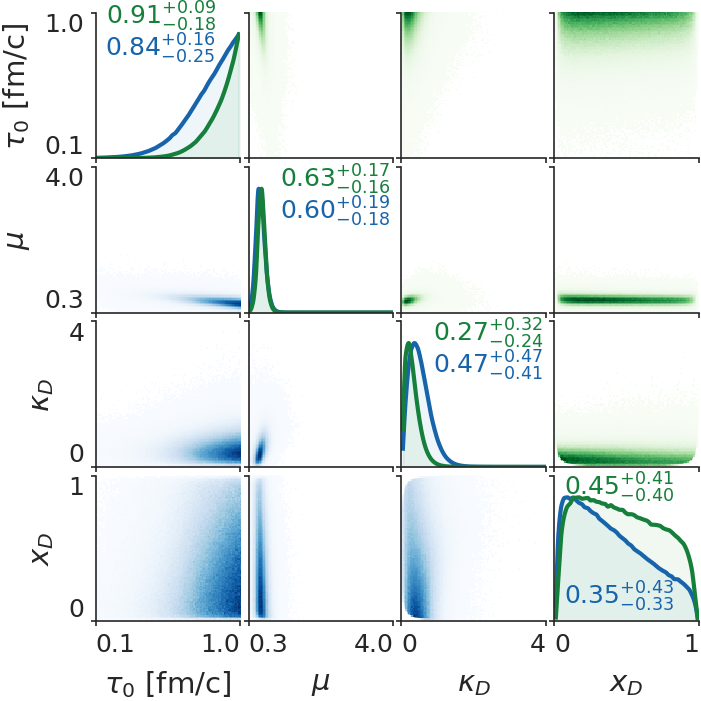}
\caption[Posteriors of single-parameter distributions (diagonal plots) and]{Posteriors of single-parameter distributions (diagonal plots) and two-parameter joint distributions (off-diagonal plots). The colors labeled the use of EPPS (blue) and nCTEQ15np (green) nuclear PDF.}
\label{fig:LBT:posterior}
\end{figure}

The inferred posterior probability distribution of the parameters is shown in figure \ref{fig:LBT:posterior}.
The diagonal plots show single parametrized distributions, and the off-diagonal ones display the two-parameter correlations.
We split the results that use different nuclear PDFs into the upper (EPPS, green heat map, and lines) and lower (nCTEQ15np, blue heat maps and lines) triangles.
One notices that the results from different nuclear PDF are consistent within the uncertainty; therefore, from now on I shall not stress on any differences between these two sets of results, but combine them into a single distribution to fold in the PDF uncertainty.
The favored parameters are $\mu \sim 0.6$ and $\kappa_D \sim 0.4$, indicating a large in-medium $\alpha_s$ and a small amount of additional diffusion.
The typical value of $\alpha_s$ is, in fact, so large that let one worry the use of a weakly-coupled approach in the first place.
For example, $\alpha_s(0.6\pi T)$ at $T=300$ MeV is 0.67, corresponding to $g \approx 3$. 
Moreover, the screening mass $m_D \sim 3.6 T$ is even larger than the average energy of the thermal partons $3T$. 
In the discussion of the next section, we will see that the extracted $\alpha_s$ is smaller once we use the improved implementation of the LPM effect developed in chapter \ref{chapter:transport}, though $g$ is still large.

\begin{figure}

\centering
\includegraphics[width=.8\columnwidth]{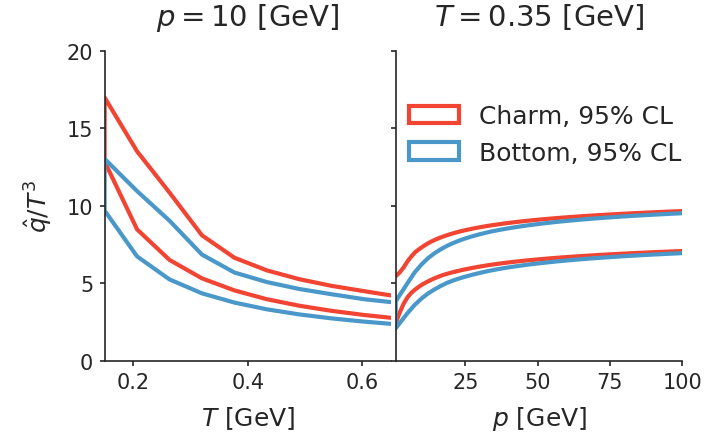}
\caption[The 95\% credible region of the heavy quark transport coefficient]{The 95\% credible region of the heavy quark transport coefficient $\hat{q}$ extracted using the model described in \cite{Ke:2018tsh}.
The charm quark results are shown in red, and bottom results in blue.
We have marginalized the uncertainty from using different nuclear PDFs.
Left plot: the temperature dependence at $p=10$ GeV. Right plot: the momentum dependence at $T=0.35$ GeV.
}\label{fig:LBT:posterior_qhat}
\end{figure}

\paragraph{Transport coefficients} In this analysis, the heavy quark transport coefficient $\hat{q}$ is computed by adding up the momentum broadening from both the scattering and the parametric diffusion,
\begin{eqnarray}\label{eq:qhat}
\hat{q} &=& 2T^3\kappa_D\left(x_D + (1-x_D)\frac{1}{ET/\textrm{GeV}{}^2}\right) + \hat{q}_{\textrm{el}}.
\end{eqnarray}
In a perturbative definition of the transport coefficients, the inelastic process does not contribute to heavy quark transport coefficient at leading order. 
In figure \ref{fig:LBT:posterior_qhat}, the 95\% credible region of $\hat{q}$ is shown as a function of temperature at fixed energy (left), and as a function of energy at fixed temperature (right).
Different colors label the results for charm (red) and bottom (blue) quarks.
The mass difference only causes a small difference in $\hat{q}$.

\begin{figure}

\centering
\includegraphics[width=.8\columnwidth]{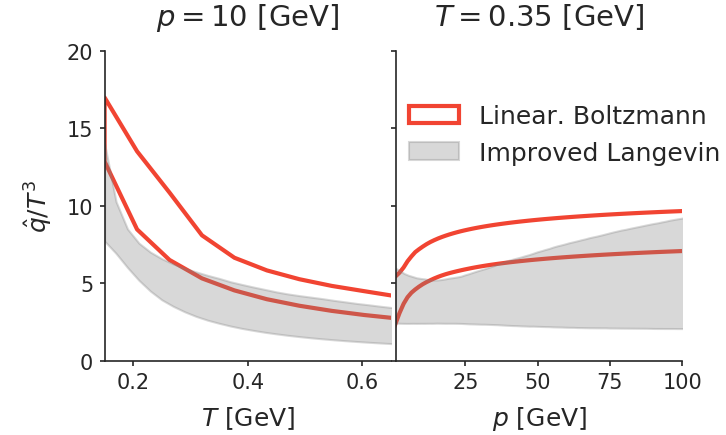}
\caption[Comparison of the 95\% credible region of the charm $\hat{q}$ using]{Comparison of the 95\% credible region of the charm $\hat{q}$ using different models. The red regions use the model described in \cite{Ke:2018tsh}, while the  shaded regions are obtained using  the improved-Langevin model \cite{Cao:2013ita}.}\label{fig:LBT:compare_qhat}
\end{figure}

\paragraph{Comparison to results from an improved-Langevin model}
The same transport coefficient is also extracted using the improved-Langevin model \cite{Cao:2013ita}.
It includes a diffusion modeling of the elastic interaction, a higher-twist single gluon emission rate, and a similar routine to implement multiple radiations.
This model is then coupled to the same medium as the one used here and compared to the same set of observables as this work does.
The resultant posterior (for charm quarks only) is shown as the shaded region in figure \ref{fig:LBT:compare_qhat}.
We see that the $\hat{q}$ extracted using the two models only overlap at the boundary of the credible region.
Their difference is comparable to the uncertainty band of either model, while both models provide a reasonable description of the data.
It suggests that the theoretical uncertainty that comes from the assumption made about the probe-medium is a significant one.
The ability to tune a switching scale parameter in the new model intends to include this type of theoretical uncertainty.

\section{Calibration using the improved transport model}
Finally, we apply the improved model to the extraction of the heavy quark transport coefficients.
Here we summarize the improvements:
\begin{itemize}
\item A more sophisticated implementation of the LPM effect to reduce modeling uncertainty of the radiative processes;
\item An interpolation between the diffusion picture and the scattering picture to take into account modeling uncertainty.
\item Separation of the high-virtuality evolution and the low-virtuality transport equation at a medium scale.
\end{itemize}

\begin{table}
\centering
\caption{Prior range of parameters}\label{table:new:prior}
\begin{tabular}{ccc}
\hline
Symbol & Description & Range \\
\hline
$\xi = \frac{\tau_0}{\tau_{\textrm{hydro}}}$ & Energy loss starting time & (.1, .9) \\
$c = \frac{Q_{\textrm{cut}}^2}{m_D^2}$ & Soft / hard switching scale & $(.1, 10.)$ \\
$R_v$ & Vacuum / Medium matching scale & $(0,7)$\\
$\mu$ & Running $\alpha_s$ stops at $Q = \mu\pi T$ & $(.6, 10)$ \\
$K$ & Magnitude of $\Delta \hat{q}/T^3$ & $(0, 15)$\\ 
$p$ & \multirow{2}{*}{$E$-dependence of $\Delta \hat{q}/T^3$} & $(-2, 2)$\\ 
$a$ &  & $(-1, 1)$\\ 
$q$ & \multirow{2}{*}{$T$-dependence of $\Delta \hat{q}/T^3$}  & $(-.5, 3)$\\ 
$b$ &   & $(-.5, 3)$\\ 
$\gamma$ & $\Delta \hat{q}_L = (E/M)^\gamma\Delta \hat{q}_L$  & (-1, 1)\\ 
\hline
\end{tabular}
\end{table}

\paragraph{Model parameters}
In the new analysis, we try to include as many theoretical uncertainties as possible, so we have more parameters than in the two previous studies.
They are listed in table \ref{table:new:prior}.
\begin{itemize}
\item The first parameter is again the energy loss starting time $\tau_i$.
In this analysis, we are comparing to data at two collision energies, and the hydrodynamic starting time $\tau_0$ varies from $1.2$ fm/$c$ to $0.6$ fm/$c$.
To account for this differences, we use the ratio $\xi = \tau_i/\tau_0$ as the single parameter for both energies.
It means that after $\xi$ fraction of the hydrodynamization time, the color density is assumed to be large enough to apply the linearized transport model.
\item The second parameter is switching scale parameter $1.0 < c < 10.0$ in $Q_{\textrm{cut}}^2 = c m_D^2$. For a typical coupling $g\sim 2$, $Q_{\textrm{cut}}$ is then varied from about $2T$ to $7T$.
\item The third parameter $0<R_v<7$ controls the matching condition between the vacuum-like radiation and the medium-induce radiation $\Delta k_\perp^2 = R_v Q^2$.
At $R_v = 0$, the vacuum-like radiation is completely forbidden once the daughter parton interacts with the medium; for $R_v \gg 1$, the vacuum-like radiation is effectively unmodified.
\item The $0.6 < \mu < 10$ parameter controls the in-medium strong coupling constant $\alpha_s(\max\{Q, \mu\pi T\})$.
\item The remaining six numbers $K,a,b,p,q, \gamma$ parametrize a correction to the weakly coupled transport coefficient $\hat{q} + \Delta\hat{q}$, $\hat{q}_L + \Delta\hat{q}_L$,
\begin{eqnarray}
\Delta\hat{q} &=& \frac{K T^3}{\left[1+\left(a\frac{T}{T_c}\right)^p\right]\left[1+\left(b\frac{E}{T}\right)^q\right]}, \\
\Delta\hat{q}_L &=& \left(\frac{E}{M}\right)^\gamma \frac{\Delta\hat{q}}{2}
\end{eqnarray}
$0 < K < 15$ is the overall magnitude of the correction. 
The deviation from the $T^3$ dependence and the energy dependence are parametrized using two dimensionless combinations $T/T_c$, and $E/T$.
The $\gamma$ parameter varies from $-1$ to $1$ allow the correction to be anisotropic.
Note that such a construction reverts to an isotropic diffusion when velocity approaches zero ($E\rightarrow M$).
\end{itemize}

\begin{figure}

\centering
\includegraphics[width=.5\textwidth]{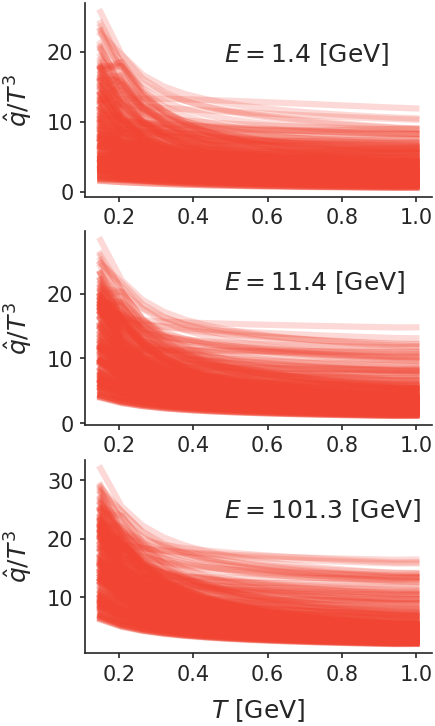}\includegraphics[width=.5\textwidth]{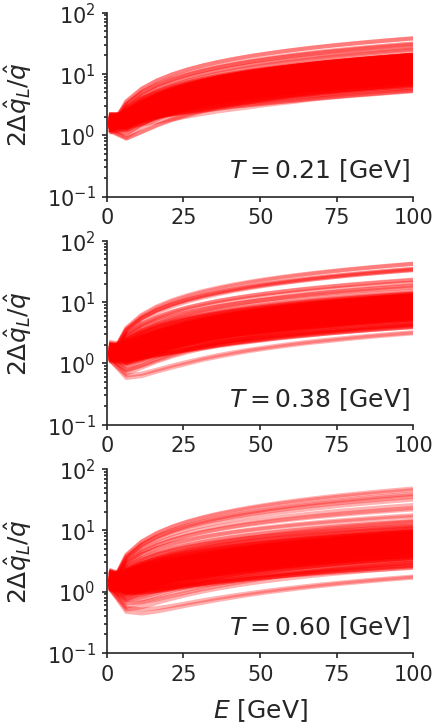}
\caption[Left: the prior range of the transport parameter $\hat{q}$ as function]{Left: the prior range of the transport parameter $\hat{q}$ as function of temperature at different heavy quark energy. Right: the prior range of the longitudinal diffusion parameter $\hat{q}_L$, plotted as ratio $2\hat{q}_L/\hat{q}$.}
\label{fig:new:design-qhat}
\end{figure}

\paragraph{Design and prior} 
We choose to give $\ln c, \ln R_v, \ln \mu, \ln a$, and $\ln b$ a uniform design and prior.
Therefore, the original parameter will have a non-uniform design and prior distribution.
The reason is that these parameters either cause a logarithmically slow change of the model prediction or has a large prior uncertainty that they are allowed to vary by orders of magnitude.
For example, the $\mu$ parameter enters the logarithmic running of $\alpha_s$ and we can rewrite the maximum possible $\alpha_s$ as,
\begin{eqnarray}
\alpha_{s,\max}(T) = \frac{2\pi}{9}\frac{1}{\ln(\mu) + \ln(\pi T/\Lambda_{\textrm{QCD}})}
\end{eqnarray}
Therefore, we assign a uniform prior to $\ln(\mu)$ so that $\alpha_s$ also varies notable within the prior range.
For the $c$ and $R_v$ parameter, we have seen in the previous benchmark calculation that the $R_{AA}$ and $v_2$ predictions depend somewhat weakly on the choice of these parameters; therefore they are also given a logarithm prior.
For the $a$ and $b$ parameters, one notices that asymptotic largeness or smallness of these numbers do not change the value of $\Delta \hat{q}$ notably.
By applying the logarithmic prior, we can explore both the large and small limits of these numbers while still having enough design points to control the interpolation uncertainty in the physically interesting regions ($a$ and $b$ are of order one). 

We sample 250 design points and 50 validation points. 
Combining $\mu, K, p, q, a, b$ and $\gamma$, the prior region of the heavy quark transport parameters are plotted as a function of temperature and energy in figure \ref{fig:new:design-qhat}. 
On the left, the 250 design $\hat{q}$'s as a function of temperatures are shown  (using charm mass for demonstration).
Each subplot shows quark energy at $1.4$ GeV, $11.4$ GeV and $101.3$ GeV.
The prior range of $\hat{q}$ varies over an order of magnitude.
On the right of the figure, we plot ratio $2\hat{q}_L/\hat{q}$ to indicate the degree of anisotropy of the transport parameters.

\begin{figure}

\centering
\includegraphics[width=\textwidth]{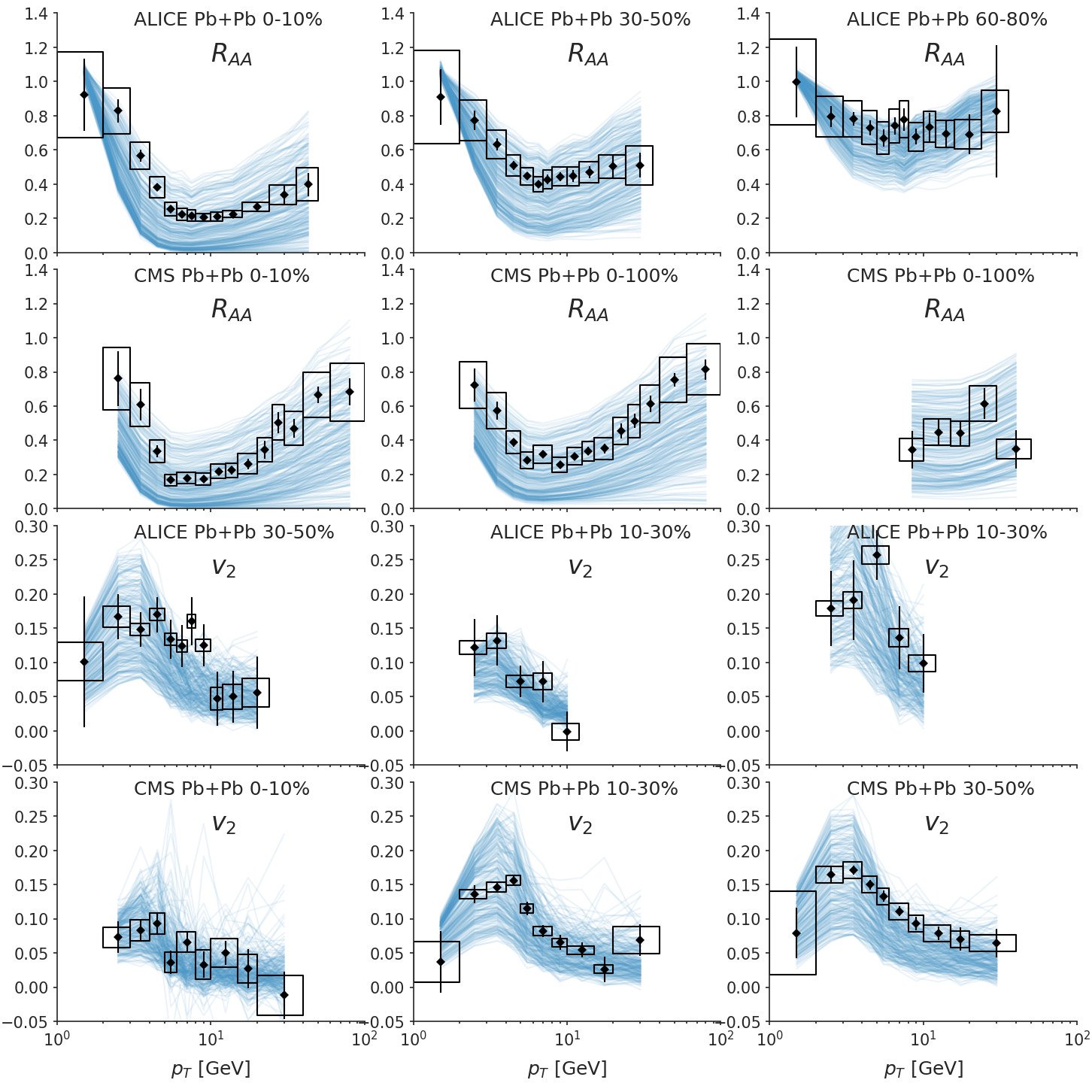}
\caption[The prior distribution of the calculated observables at LHC]{The prior distribution of the computed observables at LHC energy compared to data.}
\label{fig:new:obs_prior_LHC}
\end{figure}

\begin{figure}

\centering
\includegraphics[width=\textwidth]{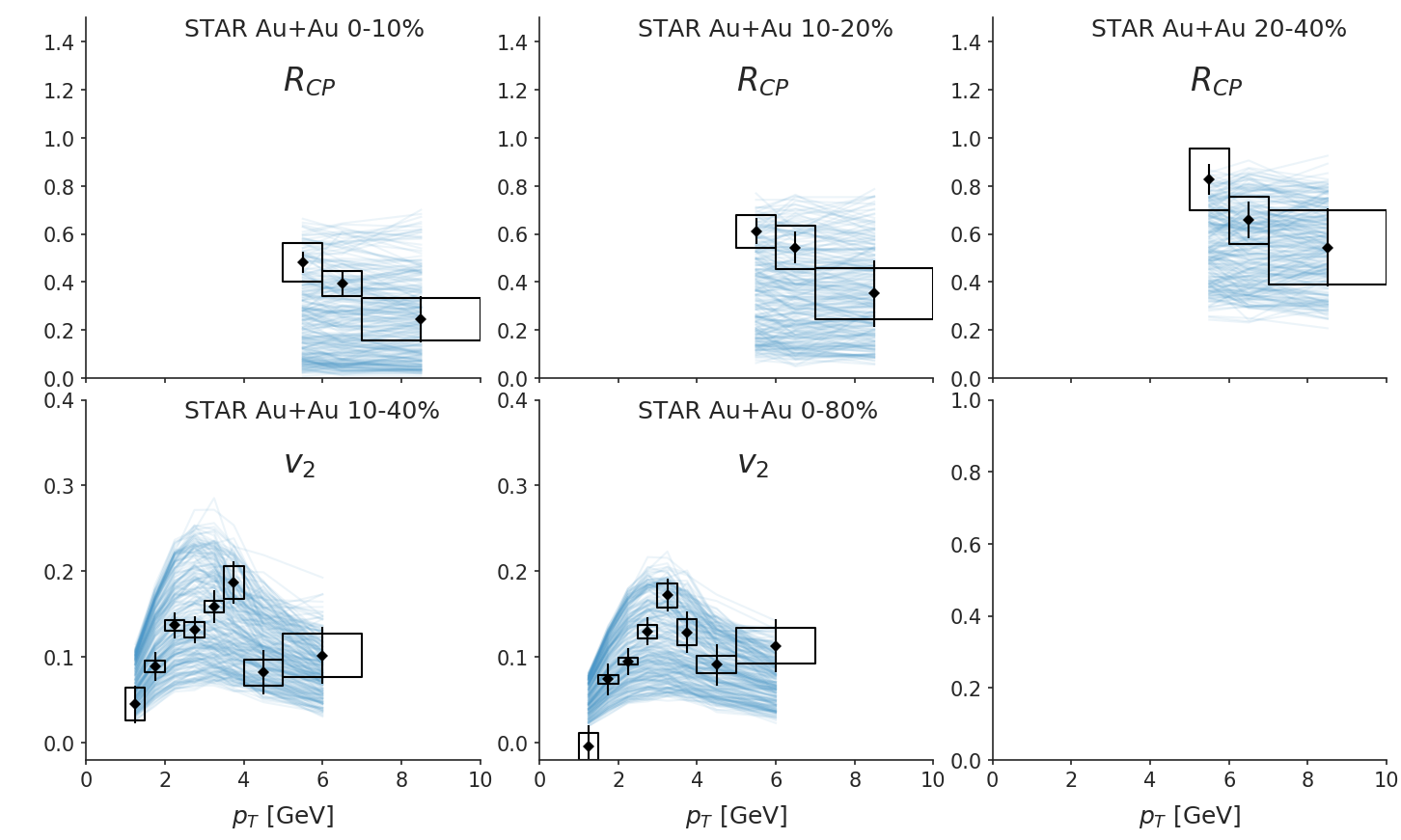}
\caption[The prior distribution of the calculated observables at RHIC]{The prior distribution of the computed observables at RHIC energy compared to data.}
\label{fig:new:obs_prior_RHIC}
\end{figure}

The computations of the model on both the design points and the validation points are performed on the NERSC super-computing platform using over two million CPU hours.
The observables calculated on the prior are shown in figure \ref{fig:new:obs_prior_LHC} at LHC energy $\sqrt{s}$ = 5.02 TeV and in figure \ref{fig:new:obs_prior_RHIC} at RHIC energy $\sqrt{s} = 200$ GeV.
In addition to the LHC dataset used in the last calibration, we also include a dataset at RHIC energy measured by the STAR Collaboration \cite{Adamczyk:2017xur,Adam:2018inb}.
We choose two observables at RHIC, namely D meson $R_{CP}$ and $v_2$. 
The new one, $R_{CP}$, is defined as the $N_{\textrm{bin}}$ normalized ratio between the D meson yield in a smaller centrality class $C$ to a larger centrality class $P$,
\begin{eqnarray}
R_{\textrm{CP}} = \frac{dN_\textrm{C}/dp_T N_{\textrm{bin,P}}}{dN_\textrm{P}/dp_T N_{\textrm{bin,C}}}.
\end{eqnarray}
Using the nuclear data as a reference has the advantage of canceling certain theoretical uncertainties, such as the nuclear PDF (if its impact-parameter dependence is neglected) and possible modifications to the initial production mechanism in the nuclear environment.
A problem we found at RHIC energy is that the very low-$p_T$ $R_{CP}$ is not well covered by the calculation, even exploring a broad parameter range.
It indicates one has to improve the model in this region of $p_T$, possibly by a more up-to-date dynamical hadronization model.
Our temporary solution is to only include the STAR $R_{CP}$ data above $5$ GeV in the calibration.

\begin{figure}

\centering
\includegraphics[width=.8\textwidth]{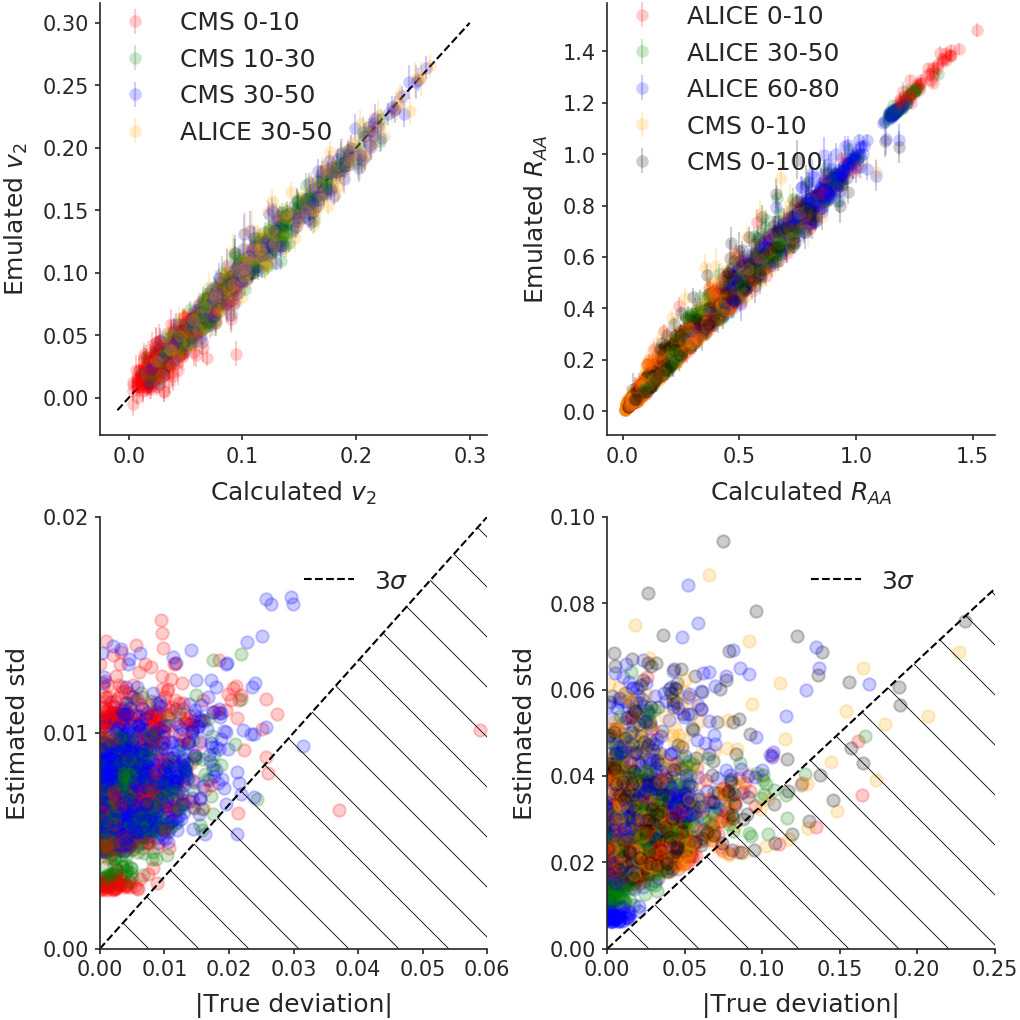}
\caption[Validation of the emulators performance. The top two plots shows]{Validation of the emulator's performance. The top two plots show the correlation between the calculated quantities ($x$ variable) versus the emulated quantities ($y$ variable). The bottom two plots compare the GP's estimated standard deviation $\sigma$ of the prediction to the actual deviation from the actual calculation. The dashed regions indicate where the actual deviation is outside the $3\sigma$ band of emulator's prediction.}
\label{fig:new:validation}
\end{figure}

\paragraph{Emulator validation} 
We perform the validation by comparing the emulator trained on the 250 design points to the actual calculation on the 50 validation points.
We visualize the validation in figure \ref{fig:new:validation}.
In the top row, the emulated $v_2$ (left) and $R_{AA}$ (right) are compared with the model calculations, and different colors label the data from different experiments and centralities.
The emulated values strongly correlate with the actual calculations around the $y=x$ lines.
Most points slightly miss the diagonal lines, meaning the emulator is not 100\% accurate.
To see if the emulator correctly accounts for the interpolation uncertainties, we plot the emulator's prediction uncertainty ($1\sigma$, $y$ axis) versus the absolute deviation between the prediction and the calculation (the $x$ axis).
The dashed line defines a shaded region where the actual deviation is greater than $\pm 3\sigma$ of emulator's estimated uncertainty.
We found that over $99\%$ of the prediction are within the $3\sigma$ region.
Therefore, the emulator correctly estimates its uncertainty and thus prevents over-fitting.

\paragraph{Covariance matrix} 
From chapter \ref{chapter:bayes}, the covariance matrix has the structure
\begin{eqnarray}
\Sigma = \Sigma_{\textrm{emulator}} + \Sigma_{\textrm{truncation}} + \Sigma_{\textrm{stat}} + \Sigma_{\textrm{sys}} + \Sigma_{\textrm{model, sys}} 
\end{eqnarray}
The construction of these terms is straight forward, except for the systematic covariance $\Sigma_{\textrm{sys}}$ of the experimental data.
Usually, experiments publish the marginalized uncertainty on each observable point $\delta y_{sys}$ (for example, $R_{AA}$ of a given centrality at a single $p_T$ bin), and may specify the nature of the uncertainty as ``correlated'' or ``uncorrelated''.
The correlation among uncertainties is crucial as it directly affects the interpretation of the quality of fit.
For instance,  if one assumes uncorrelated uncertainty, a prediction with $+5\%$ deviations on each of the $N$ data points is penalized by a factor $e^{-N(0.05 y)^2/\delta y^2}$; while the penalizing factor is only $e^{-(0.05 y)^2/\delta y^2}$,  assuming fully correlated uncertainty.
It is because fully correlated uncertainty allows the prediction to deviate from the data points with a systematic trend.

However, we lack the information to construct the full covariance matrix from $\delta y_{sys}$.
In this study, we simply parametrize the correlation as function of observables (labeled by $\alpha, \beta \in \{R_{AA}, v_2, R_{CP}\}$), centrality labeled by $m,n$ and transverse momentum (labeled by $i,j$),
\begin{eqnarray}
\mathbf{\Sigma}_{\textrm{sys}} = \delta_{\alpha\beta} C_{mn}  \exp\left\{-\frac{1}{2 L_{\textrm{corr}}^2} \left(\ln\frac{p_{T, i}}{p_{T, j}}\right)^2 \right\} \times \sigma^{\alpha m}_{\textrm{sys}, i}\sigma^{\beta n}_{\textrm{sys}, j}.
\end{eqnarray}
So, the covariance is zero if there are different observables or measurements from different experiments or different particle species.
The centrality correlation $C_{mn}$ is only applied to $R_{AA}$ and $R_{CP}$ as these quantities share the same baseline reference across different centrality, so a fraction of their uncertainty must be correlated across-centrality. 
By default, $C_{m=n}=1$ and $C_{m\neq n}=0.3$.
The correlation in the $p_T$ dimension is assumed to be a Gaussian in the $\ln p_T$ space with correlation length $L_{\textrm{corr}}$.
We use $\ln p_T$ based on the consideration that the original uncertainty should not be sensitive to the linear change of $p_T$ as there is no other scale present.
The default correlation length is $1$, meaning the uncertainty is effectively uncorrelated with measurements at a $p_T$ $2.7$ times larger or smaller.
Finally, this correlation modulation is applied to the completely correlation case of the systematic uncertainty $\sigma^{\alpha m}_{\textrm{sys}, i}\sigma^{\beta n}_{\textrm{sys}, j}$.
This construction is entirely parametric, except for the direct experimental inputs $\sigma^{\alpha m}_{\textrm{sys}, i}$.
We hope that future measurements will provide more information on the covariance structure of the published systematic uncertainties.

What we have done is to parametrize the unknown experimental covariance matrix by a reasonable ansatz using two parameters $C$ and $L_{\textrm{corr}}$.
One may try selecting different values $C$ and $L_{\textrm{corr}}$ to do a Bayesian analysis to investigate whether the calibrated parameters are sensitive to these choices.
However, due to the lack of knowledge, no value is superior to other choices.
It should be considered as another source of uncertainty in the model-to-data comparison.
In the Language of the Bayesian analysis, we treat $C$ and  $L_{\textrm{corr}}$ as hyperparameters that appear in the definition of the likelihood function and marginalize their distribution when focusing on other parameters.
It is given a uniform prior probability distribution within $0.2 < L_{\textrm{corr}} < 2.0$.
This range corresponds to $1\sigma$ reduction of the correlation once $p_T$ increases by a factor of $1.2$--$7.4$.
Meanwhile, the posterior distribution also infers the probability distribution of $L_{\textrm{corr}}$.
We can compare this inference to future experimental estimations of the uncertainty correlation for a consistency check.

\begin{figure}

\centering
\includegraphics[width=\textwidth]{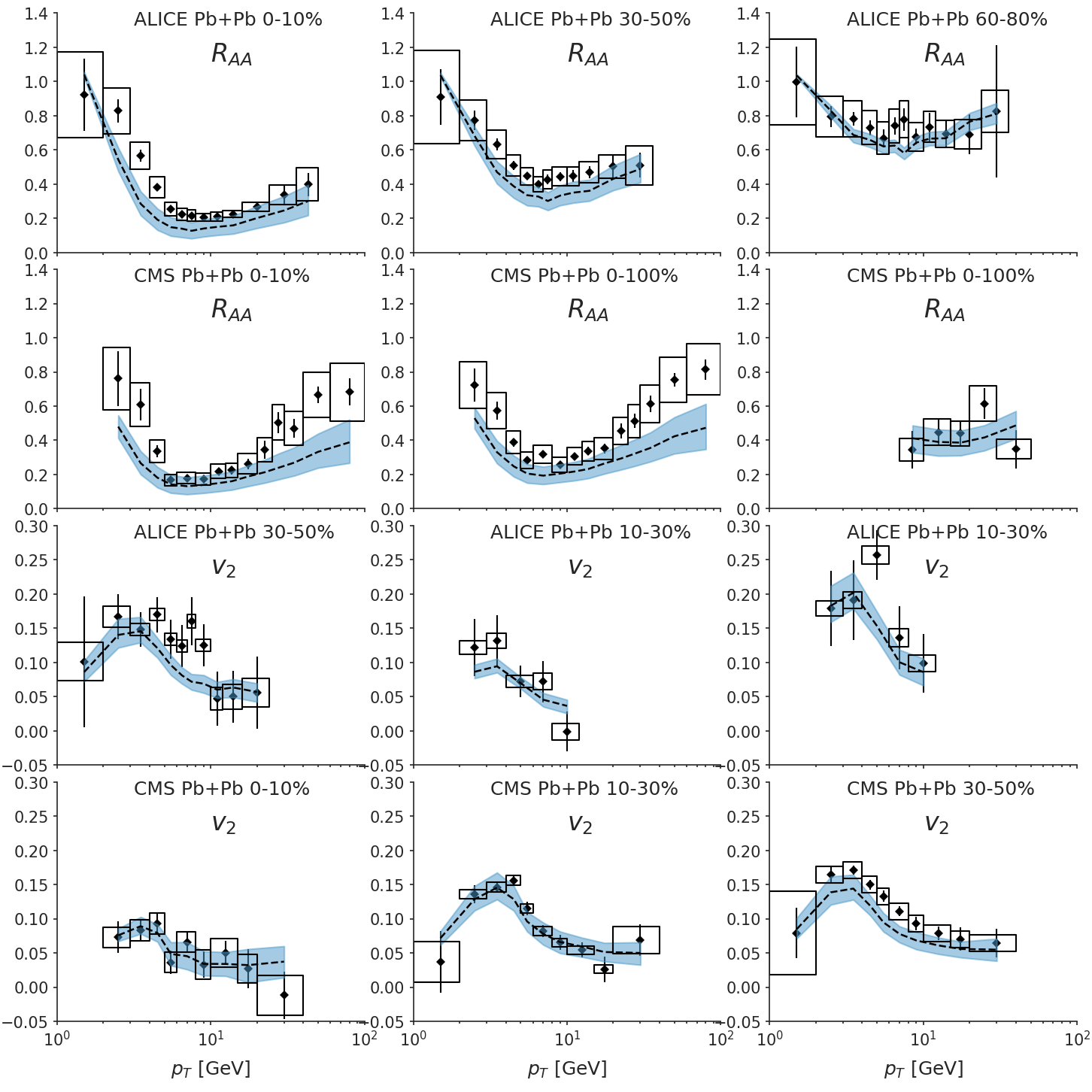}
\caption[The 90\% credible region (blue bands) of the posterior distribution]{The 90\% credible region (blue bands) of the posterior distribution of the observables at the LHC energy. Black dashed lines are the median prediction.}
\label{fig:new:obs_posterior_LHC}
\end{figure}

\begin{figure}

\centering
\includegraphics[width=\textwidth]{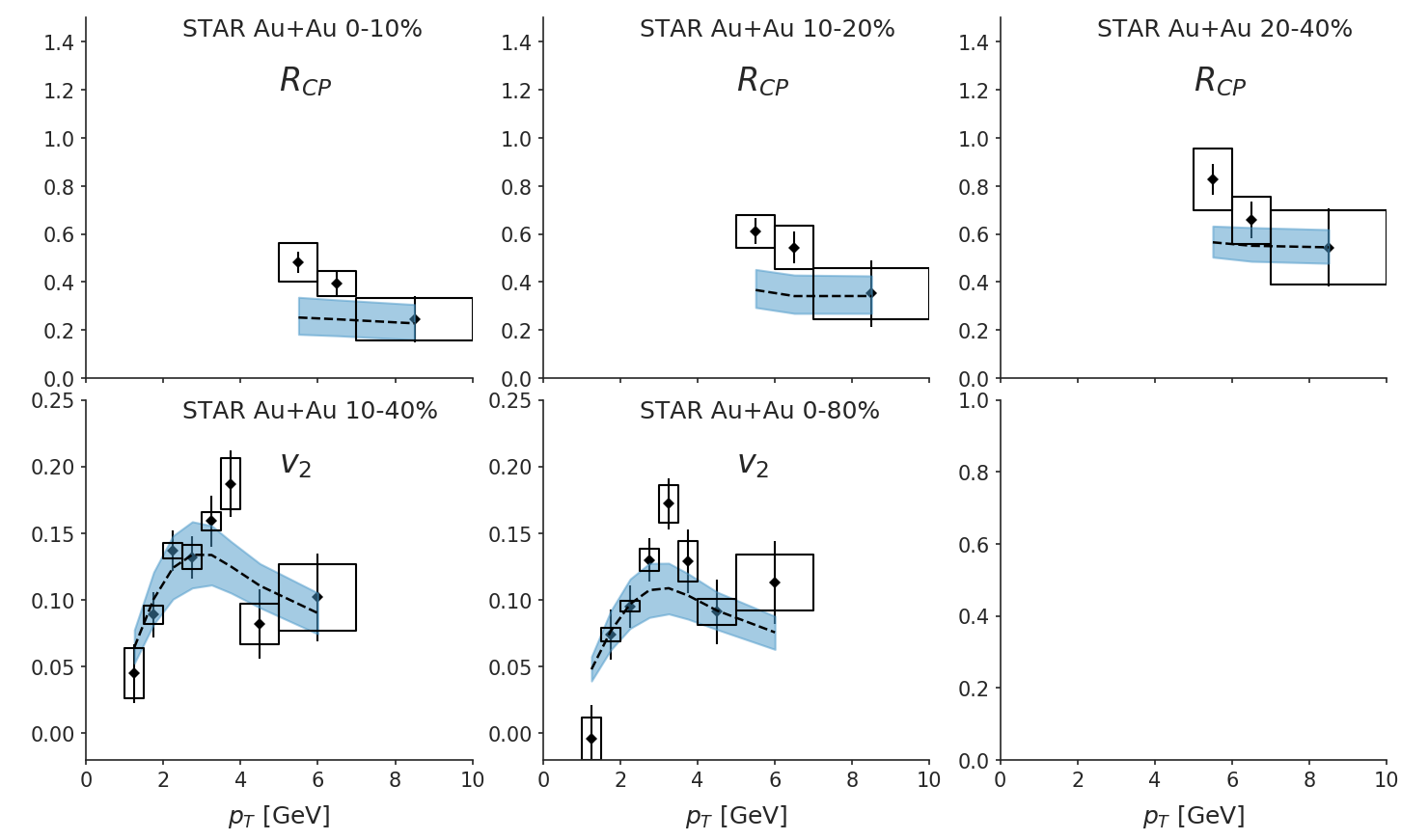}
\caption[The 90\% credible region (blue bands) of the posterior distribution]{The 90\% credible region (blue bands) of the posterior distribution of the observables at the RHIC energy. Black dashed lines are the median prediction.}
\label{fig:new:obs_posterior_RHIC}
\end{figure}

\paragraph{Posterior observables} The global level of agreement between the calibrated model and the data is shown in figure \ref{fig:new:obs_posterior_LHC} at the LHC energy, and figure \ref{fig:new:obs_posterior_RHIC} at the RHIC energy.
The black dashed lines show the median prediction, while the blue bands stand for $90\%$ credible region.
We remind the reader that because the model predicts anti-correlation between $R_{AA}$ and $v_2$, the lower and upper bounds of the uncertainty bands are also anti-correlated.
For example, prediction closer to the upper bounds of $R_{AA}$ likely hits the lower bounds of $v_2$.

The shape of the $D$-meson and the $B$-meson $R_{AA}$ and $D$-meson $v_2$ at the LHC energy are described by the calibrated model, while the absolute values of $R_{AA}$ are systematically below the data, so the $R_{AA}$-$v_2$ is not entirely solved in the current level of modeling.
A significant separation between the event-engineered $v_2$ is observed and is in a good agreement with data.
It means the model correctly accounts for the heavy-flavor response on the event-by-event geometry fluctuation of the medium.
At RHIC energy, $v_2$ is well described.  
The magnitude of $R_{CP}$ at $p_T> 5$ GeV \footnote{  Remember that the model is calibrated on the three data points above $p_T=5$ GeV} and its centrality dependencies are correctly reflected, though the $p_T$-shape is too flat compared to the data.

\begin{figure}

\centering
\includegraphics[width=\textwidth]{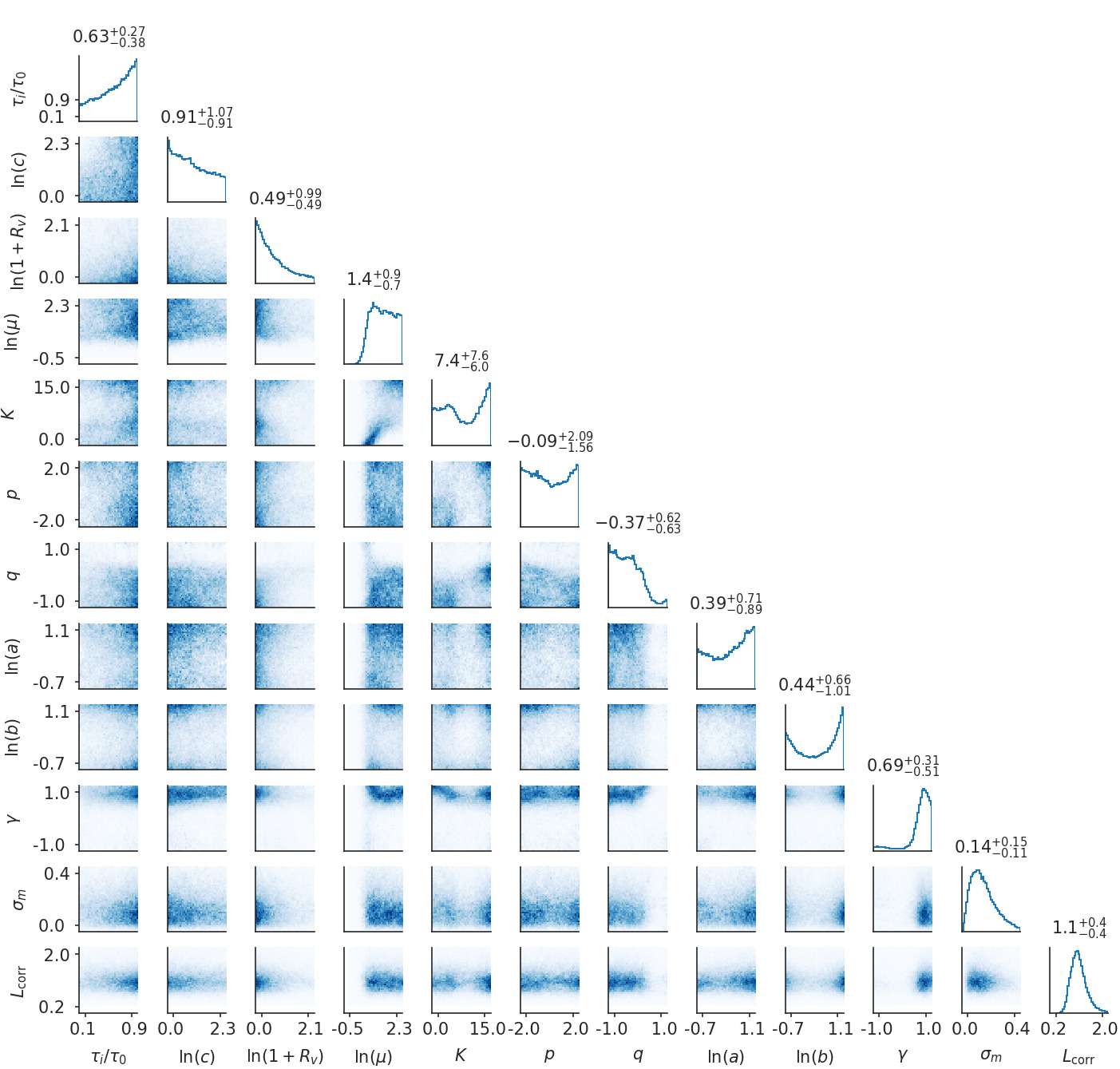}
\caption[The single-parameter posterior distributions (diagonal plots) and]{The single-parameter posterior distributions (diagonal plots) and two-parameter joint posterior distributions (off-diagonal plots).}
\label{fig:new:posterior}
\end{figure}

\paragraph{Posterior distribution of parameters} Figure \ref{fig:new:posterior} shows the single parameter posterior (diagonal plots) and two-parameter-joint posterior distributions (off-diagonal plots) of the 10 model parameters, plus the model systematic uncertainty parameter ($\sigma_m$).
Both the $\ln\mu$ parameter which controls the perturbative coupling and the $K$ parameter which controls the magnitude of parametric diffusion have rather broad distributions.
However, looking at the correlation between these two parameters, we find that the high-likelihood parameter values can either be around $\mu\sim 1.5, K\sim 0$, or around $\mu\sim 8, K\sim 15$.
So, a similar level agreement with data can either be achieved with a more perturbative-driven physics or a model with a large parametric diffusion constant.
Note that since the origin of this parametric diffusion can also come from a high-order correction to $\hat{q}$ from a weakly-coupled theory, we cannot immediately interpret a large $K$ value as a sizeable non-perturbative effect.

\begin{figure}

\centering
\includegraphics[width=.8\textwidth]{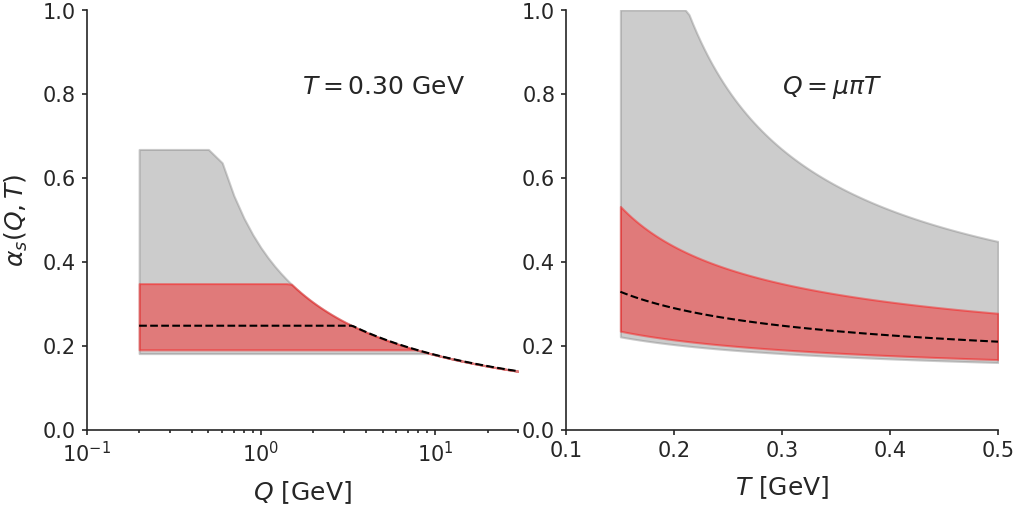}
\caption[Left plot: the scale dependence of the running coupling constant at]{Left plot: the scale dependence of the running coupling constant at $T=0.3$ GeV. Right plot: the temperature dependence of running coupling evaluated at the cut-off scale $Q=\mu\pi T$. The red bands are generated using 90\% credible region of the $\mu$ parameter; the gray bands are the prior distributions, and the black dashed lines are median predictions.}
\label{fig:new:posterior-alphas}
\end{figure}

The resulting posterior of $\alpha_s$ is plotted in figure \ref{fig:new:posterior-alphas}, the median value of $\alpha_s$ at $Q=\mu\pi T$ varies from 0.3 to 0.22 for the relevant temperature range $0.15 < T < 0.5$ GeV, corresponding to $g\sim 2$.
Compared to the previous extraction, the preferred in-medium coupling strength is smaller and is closer to the phenomenological values used by other studies \cite{Burke:2013yra}.
However, the coupling is still large compared to the weakly coupled assumption $g\ll 1$.
This $\alpha_s$ does not stand for the strength of all the probe-medium interaction, recalling that there is a significant parametric diffusion contribution to the elastic energy loss.
For radiative processes, though the $1\rightarrow 2$ splitting vertex explicitly uses this $\alpha_s$, the strength of the LPM effect is again controlled by the elastic broadening.

The calibration suggests a late energy loss starting time with a median value around 0.6 of the $\tau_{\textrm{hydro}}$.
The switching scale parameter does not have a strong preference as long as it is not too large, which is consistent with our model construction that physical processes should only weakly depend on this switching scale between diffusion and scattering modeling.
The posterior matching parameter $R_v$ tends a small value, suggesting a large region of phase-space of the vacuum-like radiation is removed.
\begin{figure}

\centering
\includegraphics[width=.5\textwidth]{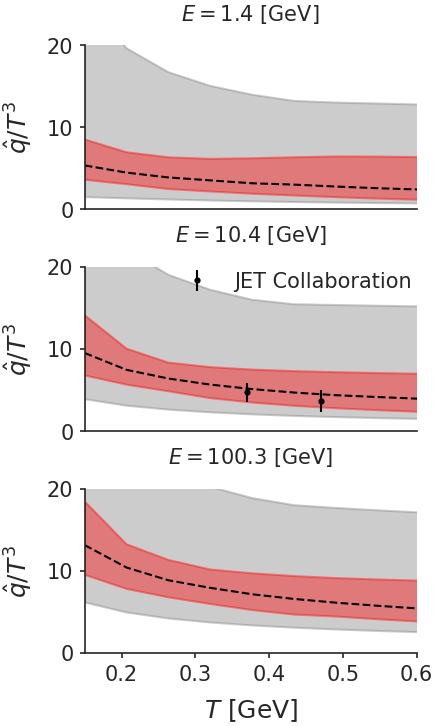}\includegraphics[width=.5\textwidth]{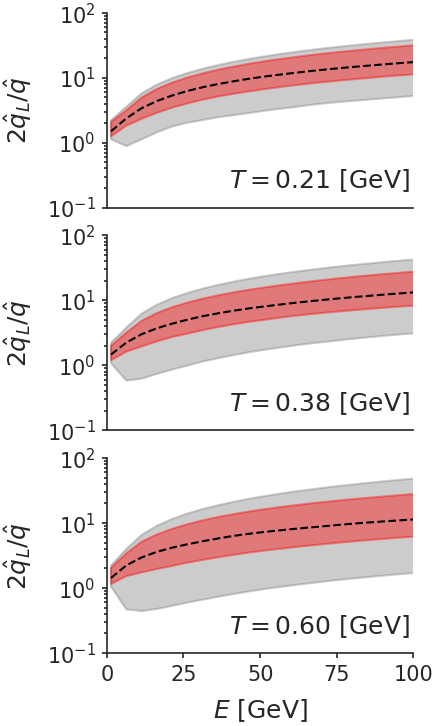}
\caption[The 90\% credible region of the charm quark transport coefficients]{The 90\% credible region of the charm quark transport coefficients (left plot) and of the longitudinal transport parameter (right plot) are shown in red. The prior range is shown in gray. We also compare the results to the previous JET Collaboration extraction of the light quark transport parameters at $p=10$ GeV (black symbols). }
\label{fig:new:posterior-qhat}
\end{figure}

It is hard to interpret the posteriors of $p, q, a, b, \gamma$ individually; instead, it is more instructive to directly look at the posteriors of the transport coefficients $\hat{q}, \hat{q}_L$.
We plot the 90\% credible range (red bands) of charm quark $\hat{q}, \hat{q}_L$ on top of their prior range (gray band) in figure \ref{fig:new:posterior-qhat}.
The values of $\hat{q}$ at $E=10.4$ GeV are found to be comparable to earlier extractions of the light quark $\hat{q}$ by the JET Collaboration \cite{Burke:2013yra}. 
Note that even though the JET Collaboration extracts the light quark $\hat{q}$, but at $p_T = 10$ GeV, the charm mass effect is small
We also present a first extraction of the longitudinal transport parameter $\hat{q}_L$. 
The longitudinal transport is quite anisotropic when compared to $\hat{q}$.

\begin{figure}

\centering
\includegraphics[width=.7\textwidth]{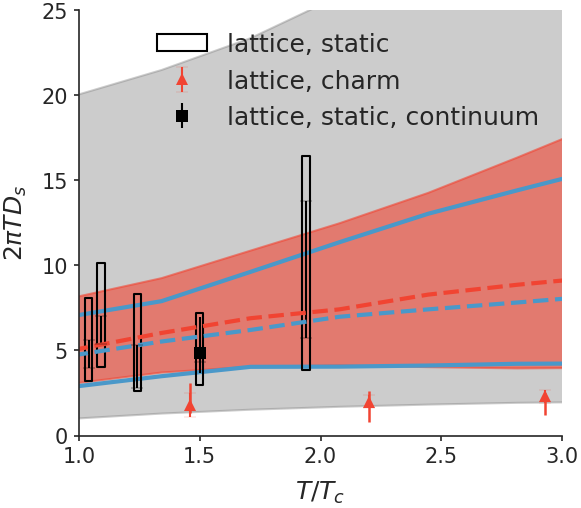}
\caption[The 90\% credible region of the spatial diffusion constant defined]{The 90\% credible region of the spatial diffusion constant defined in the zero momentum limit. The gray band indicates the prior range. The posterior of charm and bottom flavor are shown as a red band and blue boundary lines receptively. It is compared to lattice QCD evaluation in references \cite{Banerjee:2011ra} (black boxes), \cite{Francis:2015daa} (the black dot) and \cite{Ding:2012sp} (red triangles).}
\label{fig:new:posterior-Ds}
\end{figure}

The extracted heavy quark spatial diffusion constant is essentially the extrapolation of $\hat{q}$ down to zero momentum, which can be sensitive to the particular choice of the form of parametrization.
Nevertheless, the extraction (red band for 90\% credible region) is compared to various lattice calculations\footnote{ As a remark, it is not entirely rigorous to compare the phenomenological extracted $D_s$ with the lattice evaluation. Because in a transport model one separate the interactions into elastic and inelastic channels and the shown $D_s$ only contains elastic contributions; while lattice calculations do not make such distinctions.} \cite{Banerjee:2011ra,Ding:2012sp,Francis:2015daa} in figure \ref{fig:new:posterior-Ds}.
The charm and bottom quark spatial diffusion constants are similar and are consistent with lattice calculations in the static (infinitely heavy) limit of the heavy quark, while the lattice calculation using dynamical charm quark predicts a much lower value of $D_s$.
One may expect that our phenomenological extraction should give a similar separation between bottom and charm flavor because bottom quark is much closer to the ``heavy limit'' than charm.
However, we found that the mass dependence in the elastic part of our model is relatively weak. 
First, the mass only affects the phase-space integration of $t$-channel of the perturbative cross-sections.
Second, mass only enters the parametric diffusion part through a combination $E/T$, which is approximately $M/T$ at low momentum. Because charm and bottom masses are already much higher than the typical temperature, the parametric part only introduces a weak flavor dependence.
In the future, one may seek a more physically motivated flavor dependence parametrization of the transport parameters.

\chapter{Conclusion}
\label{chapter:conclusion}
In this dissertation, I have focused on understanding the transport properties of heavy flavor in the strongly coupled quark-gluon plasma applying model-to-data comparison methodology, aiming for model improvements and uncertainty quantification.

A prerequisite for the study is an ``accurate'' modeling of the physical ingredients to be tested.
It is not so trivial to model the heavy quark transport that is coupled to an event-by-event fluctuating and evolving medium.
On the one hand, this is because the finite medium-induced radiation formation time at high energy is much greater than the mean-free-path in semi-classical transport equations, and can be comparable to the medium evolution time scales.
On the other hand, there are two competing pictures regarding the heavy-quark-to-medium coupling: a weakly coupled picture modeled by scatterings, and a strongly coupled picture whose dynamics is often modeled by diffusion equations.
We developed a transport model for hard parton propagation in a near-equilibrium plasma. 
We implement an improved treatment of the LPM effect, and it is shown to reduce to theoretical baseline calculations in the idealized infinite static medium limit, and capture qualitative features in a finite and evolving medium.
The model also treats the large and small momentum transfer processes with different strategies of few-body scattering and diffusion (plus diffusion-induced radiation), which grants a flexible parametrization of diffusion-like deviations from the leading order weakly coupled approach.

The transport in a hot QGP stage is embedded in a more general ``transport'' picture including the initial production and high-virtuality evolution, hadronization near the transition temperature and hadronic dynamics and decay.
We identify a matching problem between the high-virtuality evolution and medium-induced evolution.
Currently, a unified formulation that smoothly connects the virtuality shower and the in-medium shower is still missing, and we terminate vacuum showers at a scale ($Q^2$) where they are likely to receive similar amounts of medium modification to the transverse momentum ($\Delta k_\perp^2 \sim Q^2$).
The exact location of the separation scale is then treated as an uncertainty of the model.

Finally, we apply Bayesian analysis to infer the model parameter distribution by comparing to heavy flavor measurements at both RHIC and the LHC.
The model parameters include uncertainties such as the in-medium coupling strength, energy loss starting time, matching scale between vacuum and medium-induced shower, diffusion versus scattering model, as well as parameterized deviations from weakly coupled calculations.

\begin{figure}
\centering
\includegraphics[width=.8\textwidth]{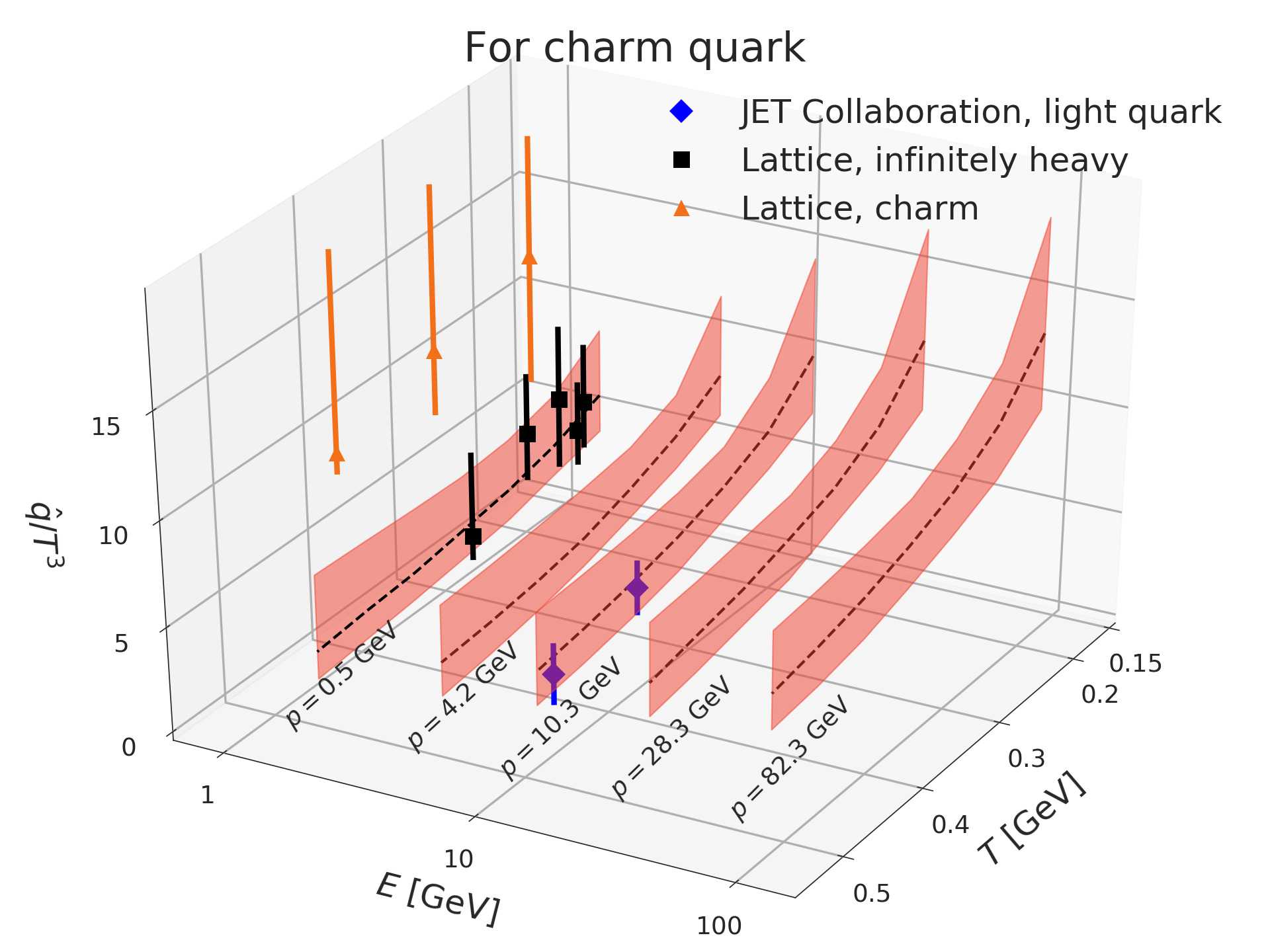}
\caption[This figure shows the main result of this dissertation. The 90\%]{This figure shows the main result of this dissertation. The 90\% credible transport coefficient $\hat{q}/T^3$ extracted for the charm flavor is displayed on the two-dimensional landscape of energy and temperature. The JET Collaboration extraction of light quark transport coefficients at $p=10$ GeV \cite{Burke:2013yra} (blue diamonds) and two lattice calculations of the momentum diffusion coefficient $\kappa$ ($\hat{q}=2\kappa$) \cite{Ding:2012sp,Banerjee:2011ra} are plotted for comparison.}
\label{fig:conclusion}
\end{figure}

We highlight the progress of this work in the conclusion figure \ref{fig:conclusion}.
It visualizes the $90\%$ credible region of the energy and momentum dependence of the heavy-quark momentum-diffusion transport parameter $\hat{q}$ scaled by $T^3$.
We found $\hat{q}/T^3$ gradually increases with $\ln E$ and displays an enhancement near the critical temperature.
Studying heavy flavor helps to connect the knowledge of in-medium transport properties at very high momentum (light quark limit) and very low momentum (static sources limit).
At relatively high momentum $p\sim 10$ GeV, it is consistent with the light quark transport parameter extracted by the JET Collaboration (blue).
At low momentum, it is consistent with lattice calculations in the heavy quark limit (black).
Future study with improved flavor dependence may be needed to understand the impact of using the ``heavy' limit in a dynamical model.
In the present calibration, the effective in-medium strong coupling constant is about $0.3$ and only contributes to a small fraction of the extracted $\hat{q}$ parameter.
The rest comes from the parametric contribution whose origin can be either perturbative or non-perturbative; either way, it suggests the necessity to model beyond leading order physics.

In conclusion, a transport model with perturbative parton evolution with a parametric probe-medium interaction term provides a reasonable description to the open-heavy flavor observables measured at RHIC and LHC, while the level of accuracy needs to be improved.
Extracted heavy quark transport coefficients as a function of energy and temperature are consistent with early phenomenological studies and lattice calculations.

The present model accuracy is still not enough to make the best use of future high-precision hard probe measurements in heavy-ion collisions.
We, therefore, list a few necessary points of improvements which may help to reduce or estimate the theoretical and modeling uncertainties.
\begin{itemize}
\item An interpolation formula between vacuum and medium-induced radiation: a calculation that connects virtuality evolution with in-medium time evolution will help to eliminate the matching scale uncertainty. Even though its effect is not strong for the present observables and $p_T$ range, it may impact more delicate jet observables.
\item Correlations among multiple emissions in the presence of a medium. We have been neglecting the correlation among subsequent emissions in the ``modified transport model''. In the infinite medium limit, this is because the probability of overlapping emissions scales as $\tau_{1,f} R(\omega_2) \sim \tau_{1,f} \alpha_s/\tau_{2,f}$ which is suppressed by $\alpha_s$. However, this higher-order effect can be important since the phenomenological $\alpha_s$ is not small. There are ongoing studies on this topic \cite{Arnold:2015qya,Arnold:2016kek,Arnold:2016mth,Arnold:2016jnq}.
\item Off equilibrium corrections to the linearized transport equation. One essential assumption in the linearized transport model is that medium partons follow local thermal distributions, even though the hydrodynamics used includes viscous corrections. 
In fact, the viscous correction and the momentum space anisotropy can be huge at early times of the hydrodynamic evolution. 
One needs to understand how these off-equilibrium effects change the interpretation of the transport coefficients one extracts assuming full thermal equilibrium of partons.
\item Dynamical hadronization model and improved treatment of energy loss in the hadronic stage.
Our current hadronization model has the problem of pinching long-distance physics into an instantaneous process. 
At low-$p_T$, the sudden recombination model breaks the detailed balance of the transport model and dynamically treating the recombination process would be desirable.
At high-$p_T$, the problem is more severe, as the large boost dilates the hadronization time. 
Moreover, the hadronic system near $T_c$ is still very dense, and it is inconsistent to apply the vacuum fragmentation function at $T\sim T_c$.
One possible solution for those high-$p_T$ heavy quarks (the recombination process is negligible) is to continue their partonic transport into the hadronic phase, and finally apply the vacuum fragmentation function when the system is dilute enough.
Meanwhile, one can also study the energy loss in the dense hadronic system to extend the extracted transport parameter to the region below $T_c$.
\item A calibration with the simultaneous tuning of the soft and hard sectors. Currently, a separate analysis calibrates the bulk medium evolution. With future high-precision hard-probe measurements, a simultaneous calibration of both soft and hard sector would be interesting.
For example, we find that the number of binary collision as a function of centrality is quite sensitive to the proton shape modeling in the Monte-Carlo Glauber model. 
The sensitivity of hard probe production to the number of binary collisions may help to improve the proton shape modeling in the soft sector. 
In turn, a better-calibrated medium may help to reduce the uncertainty in the hard parton energy loss study.
\end{itemize}

\begin{appendices}
\chapter{Few-body matrix-elements}
\label{app:ME}
This section provides the detailed $2\leftrightarrow 2$ and $2\leftrightarrow 3$ matrix-elements we used in the transport model.
The $2\leftrightarrow 2$ results are standard, and we do not re-derive them here.
A detailed derivation for the $2\leftrightarrow 3$ cross-sections is attached.

\subsection{$2\leftrightarrow 2$ processes}
The two-body scatterings between quarks, antiquarks, and gluons are standard, and we quote the results from existing references \cite{RevModPhys.59.465}.
For a light parton scattering, we keep only $\hat{t}$-channel contribution, the $\hat{s}$ and $\hat{u}$ channel contribution are suppressed at high energy.
\begin{eqnarray}
\overline{|M_{q_1q_2\rightarrow q_1q_2}|^2} &=& \frac{64\pi^2 \alpha_s^2}{9} \frac{s^2+u^2}{t^2} \\
\overline{|M_{gg\rightarrow gg}|^2} &\approx& 72\pi^2 \alpha_s^2 \frac{-su}{t^2}
 \\
\overline{|M_{qg\rightarrow qg}|^2} &\approx& 16\pi^2 \alpha_s^2 \frac{s^2+u^2}{t^2}
\end{eqnarray}
For the heavy quark, since we are interested in its diffusion dynamics at low $p_T$, we uses the exact leading order matrix-element in the vacuum.
\begin{eqnarray}
\overline{|M_{Qq\rightarrow Qq}|^2} &=& \frac{64\pi^2\alpha_s^2}{9} \frac{(M^2-u)^2 + (s-M^2)^2 + 2 M^2 t}{t^2}
\nonumber
\\
\overline{|M_{Qq\rightarrow Qq}|^2} &=& \pi^2 \left\{
32\alpha_s^2 \frac{(s-M^2)(M^2-u)}{t^2} \right.
\nonumber
\\
&+&\frac{64}{9}\alpha_s^2 \frac{(s-M^2)(M^2-u)+2M^2(s+M^2)}{(s-M^2)^2} \nonumber
\\
&+&\frac{64}{9}\alpha_s^2 \frac{(s-M^2)(M^2-u)+2M^2(u+M^2)}{(M^2-u)^2} \nonumber
\\
&+& \frac{16}{9}\alpha_s^2 \frac{M^2(4M^2 - t)}{(M^2-u)(s-M^2)} 
\nonumber
\\
&+& 16 \alpha_s^2 \frac{(s-M^2)(M^2-u)+M^2(s-u)}{t(s-M^2)}
\nonumber
\\
&-& \left. 16 \alpha_s^2 \frac{(s-M^2)(M^2-u)-M^2(s-u)}{t(M^2-u)}\right\}
\end{eqnarray}

\begin{figure}
\includegraphics[width=\textwidth]{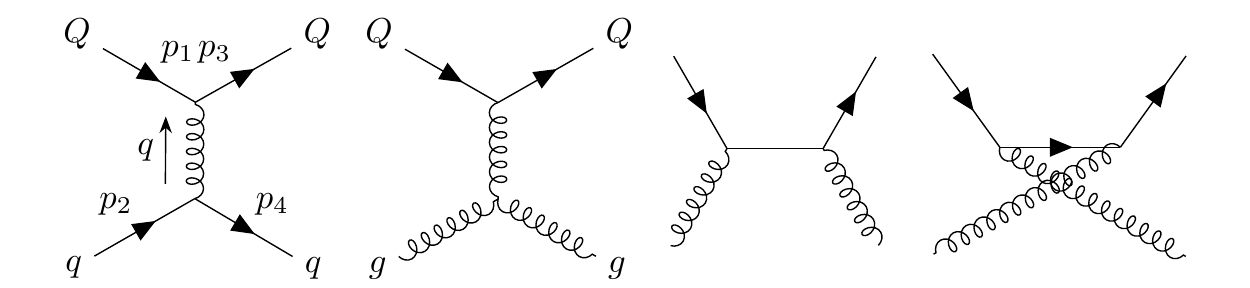}
\caption[Elastic processes: The first diagram corresponds to heavy quark]{Elastic processes: The first diagram corresponds to heavy quark ($Q$) - light quark ($q$, $\bar{q}$) scattering. The last three diagrams contribute to heavy quark ($Q$) - gluon ($g$) scattering.}\label{plots:feyn-elastic}
\end{figure}

\subsection{$2\rightarrow 3$ matrix-elements}
Large-Q $2\rightarrow 3$ inelastic processes are $g + i \rightarrow q+\bar{q} + i$, $q+i\rightarrow q+g+i$ and $g+i\rightarrow g+g+i$, where $i$ stands for a medium parton, and the other symbols stands for hard partons.
In the medium frame, the hard parton has an energy $E\gg T$, while the medium thermal parton has $E\sim T$, and the typical center-of-mass energy is $\sqrt{6ET}$.
We perform the calculation in the center-of-mass frame of the two incoming partons and let the hard parton move towards the $+z$ direction with momentum $p_1$, and the medium parton moving to the $-z$ direction with $p_2$.
The hard parton then splits into two daughter partons with momenta $k$ and $p_1 + q - k$.
The momentum transfer $q$ between the hard parton and the medium parton is thought to be large enough $|q| > Q_{\textrm{cut}}$ so we neglect the thermal correction to its propagator.

Our derivation largely follows the work of \cite{Fochler:2013epa} while relaxing the soft approximation $xq_\perp \ll k_\perp$ in \cite{Fochler:2013epa}, and we only use the collinear approximation $k_\perp^2, q_\perp^2 \ll x(1-x) \hat{s}$ with $x = k^+/\sqrt{s} = k_\perp e^y_k /\sqrt{s}$.
Also, we only include the contributions with a $\hat{t}$-channel momentum exchange between the medium and the hard partons.
The collinear approximation requires $y_k \gg \ln(k_\perp/\sqrt{s})$ so that $y_k$ cannot be arbitrarily small and $y_k>0>\gg -\ln(\sqrt{s}/k_\perp)$ is a reasonable range of application.
Because $\hat{s}\sim 6 ET$, we expect this approximation to break down when either the typical values of $q_\perp^2$ becomes comparable to $x(1-x)6ET$ or when $y_k<0$ ($x < k_\perp/\sqrt{s} \sim k_\perp/\sqrt{6ET}$).
We shall briefly mention the treatment of the $y_k<0$ region in the end.

The light-cone momentum for $p_1$ , $p_2$ and $k$ can written down directly using $\sqrt{s}$, $x$ and $k_\perp$, then applying the above collinear condition, the expression for $q$ (and therefore $p_3$ and $p_4$) is obtained by kinematic constraint up to corrections of order $\{k_\perp, q_\perp^2\}/x(1-x)\hat{s}$.
\begin{eqnarray}
p_1 &=& (\sqrt{s}, 0, \vec{0})\\
p_2 &=& (0, \sqrt{s}, \vec{0})\\
k &=& (x\sqrt{s}, \frac{k_\perp^2}{x\sqrt{s}}, \vec{k}_\perp)\\
q &\sim& (-\frac{q_\perp^2}{\sqrt{s}}, \frac{q_\perp^2 + k_\perp^2/x - 
2\vec{q}_\perp \cdot \vec{k}_\perp}{(1-x)\sqrt{s}}, \vec{k}_\perp)
\end{eqnarray}
Using the light-cone gauge with a light-like vector $n = (0, 1, 0)$, the gauge fixing condition $n\cdot A =0$ eliminates the ``+" component in the gluon (with momentum $p$) polarization vector, and is obtained by applying the transverse condition $\epsilon \cdot p = 0$ (up to a higher order correction to its normalization)
\begin{eqnarray}
\epsilon(p) &\sim& (0, \frac{2\vec{\epsilon}_\perp\cdot\vec{p}_\perp}{p^+}, \vec{\epsilon}_\perp).
\end{eqnarray}
With these preparations, the matrix-element is factorized into an amplitude for the splitting process (approximated in the collinear limit) times the amplitude for two-body collision with the medium parton.
We shall only derive explicitly the cases where the medium parton is a quark, for colliding with medium anti-quark and gluon, it is sufficient to replace the $H+q\xrightarrow{\hat{t}} H+q$ amplitude by $H+\bar{q}\xrightarrow{\hat{t}} H+\bar{q}$ and $H+g\xrightarrow{\hat{t}} H+g$.
In the end, we elucidate the connection of these results and the Bethe-Heitler limit of the solution to the AMY integral equation.

\paragraph*{Gluon splitting to quark-anti-quark pair}
\begin{figure}

\centering
\includegraphics[width=.5\textwidth]{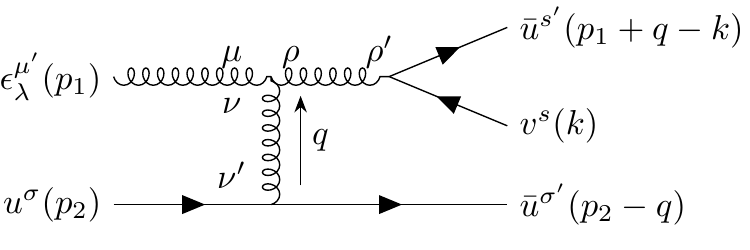}\\
\vspace{1em}
\includegraphics[width=.49\textwidth]{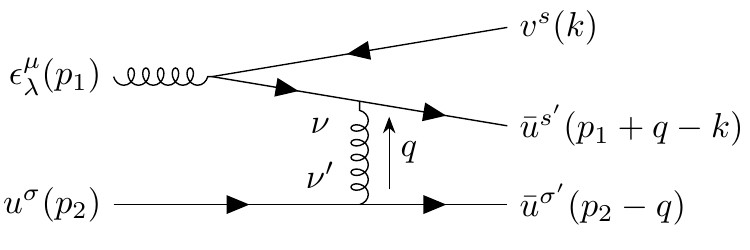}\hfill
\includegraphics[width=.49\textwidth]{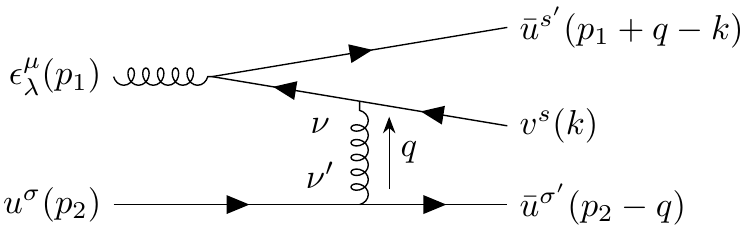}
\caption[Three diagrams $A$ (Top), $B$ (Bottom left), $C$ (Bottom right) that]{Three diagrams $A$ (Top), $B$ (Bottom left), $C$ (Bottom right) that contribute to the large angle scattering induced gluon splitting into quark-anti-quark pair in the forward region of the center-of-mass frame.}
\label{fig:feyn-g2qqbar}
\end{figure}

Three Feynman diagrams contribute to the kinematic region $y_k >0$ in the current approximation, as shown in figure \ref{fig:feyn-g2qqbar}.
We start from the amplitude for diagram $A$.
\begin{eqnarray}
i M_A &=& (-ig)^2(-g)f^{abc}(t^b)_{j'j}(t^c)_{i'i} \epsilon_\lambda^\mu(p_1) \\\nonumber
&&\frac{-i}{(p_1+q)^2}\left(g^{\rho\rho'}-\frac{n^{\rho}(p_1+q)^{\rho'}+n^{\rho'}(p_1+q)^\rho}{n\cdot (p_1+q)}\right) \bar{u}^s(p_1+q-k)\gamma_{\rho'}v^{s'}(k) \\ \nonumber
&&\frac{-i}{q^2}\left(g^{\nu\nu'}-\frac{n^{\nu}q^{\nu'}+n^{\nu'}q^\nu}{n\cdot q}\right) \bar{u}^{\sigma}(p_4)\gamma_{\nu'}u^{\sigma'}(p_2) \\ \nonumber
&& \left[g_{\mu\nu}(p_1-q)_\rho + g_{\nu\rho}(2q+p_1)_\rho + g_{\rho\mu}(-2p_1 -q)_\nu \right]
\end{eqnarray}
Next, express the projection matrix of the gluon propagator with momentum $p_1+q$ by the sum of tensor products of its polarization vectors, and identify the amplitude $iP_{A,\lambda'}^{ss'}$ for a gluon with polarization $\lambda'$ to split into the quark and anti-quark pair with spin $s$ and $s'$.
Also, use the high energy approximation to replace $\bar{u}^i(a)\gamma^\alpha u^j(b)$ by $(a+b)^\alpha \delta^{ij}$, then
\begin{eqnarray}
i M_A &\approx& -g^3 f^{abc}(t^b)_{j'j}(t^c)_{i'i} \delta^{\sigma\sigma'} \epsilon^\mu(p_1) \\\nonumber
&&\frac{1}{(p_1+q)^2} \sum_{\lambda'=\pm}\epsilon_{\lambda'}^{\rho}(p_1+q)\underbrace{\epsilon_{\lambda'}^{*,\rho'}(p_1+q) \bar{u}^s(p_1+q-k)\gamma_{\rho'}v^{s'}(k)}_{iP_{A,\lambda'}^{ss'}} \\ \nonumber
&&\frac{1}{q_\perp^2}\left(g^{\nu\nu'}-\frac{n^{\nu}q^{\nu'}+n^{\nu'}q^\nu}{n\cdot q}\right) (2p_2-q)_{\nu'} \\ \nonumber
&& \left[g_{\mu\nu}(p_1-q)_\rho + g_{\nu\rho}(2q+p_1)_\rho + g_{\rho\mu}(-2p_1 -q)_\nu \right] \\
&=& -g^3 f^{abc}(t^b)_{j'j}(t^c)_{i'i} \frac{1}{(p_1+q)^2}\frac{1}{q_\perp^2} \sum_{\lambda'=\pm}iP_{A,\lambda}^{ss'} \delta^{\sigma\sigma'}  \\ \nonumber
&& \epsilon_\lambda^\mu(p_1)2p_2^{\nu} \epsilon_{\lambda'}^{\rho}(p_1+q) \left[g_{\mu\nu}(p_1-q)_\rho + g_{\nu\rho}(2q+p_1)_\rho + g_{\rho\mu}(-2p_1 -q)_\nu \right].
\end{eqnarray}
Finally, we evaluate the contraction in the second line using the expression for $p_1, q$ and $\epsilon$, and keep only terms that are leading in $q_\perp^2/s$ to get,
\begin{eqnarray}
i M_A \approx -g^3 f^{abc}(t^b)_{j'j}(t^c)_{i'i}\delta^{\sigma\sigma'}\frac{2s}{q_\perp^2} \frac{x(1-x)}{(\vec{k}_\perp-x \vec{q}_\perp)^2} iP_{A,\lambda}^{ss'}.
\end{eqnarray}

Diagram B and C are similar, so we only write down diagram B in detail.
\begin{eqnarray}
i M_B &=& (-ig)^3 (t^bt^a)_{i'i}(t^b)_{j'j} \epsilon_\lambda^\mu(p_1) \\\nonumber
&&\frac{-i}{q^2}\left(g^{\nu\nu'}-\frac{n^{\nu}q^{\nu'}+n^{\nu'}q^\nu}{n\cdot q}\right) \\\nonumber
&&\bar{u}^s(p_1+q-k)\gamma_{\nu}\frac{i(\slashed{p_1}-\slashed{k})}{(p_1-k)^2}\gamma^{\mu}v^{s'}(k) \\ \nonumber
&&\bar{u}^{\sigma}(p_4)\gamma_{\nu'}u^{\sigma'}(p_2)
\end{eqnarray}
Again, we represent the tensor structure of the fermion propagator by the sum of tensor products of the spinors, identify the splitting amplitude $iP_{B,\lambda'}^{ss'}$ and use the high energy limit of the current,
\begin{eqnarray}
i M_B &\approx& ig^3 (t^bt^a)_{i'i}(t^b)_{j'j}  \\\nonumber
&&\frac{-i}{q_\perp^2}\left(g^{\nu\nu'}-\frac{n^{\nu}q^{\nu'}+n^{\nu'}q^\nu}{n\cdot q}\right) (2p_2-q)_\nu' \\\nonumber
&&\frac{1}{2p_1\cdot k} \sum_\sigma \bar{u}^s(p_1+q-k)\gamma_{\nu} u^{\sigma}(p_1-k) \underbrace{\epsilon_\lambda^\mu(p_1)\bar{u}^{\sigma}(p_1-k) \gamma^{\mu}v^{s'}(k)}_{iP_{B,\lambda}^{\sigma s'}}\\
&\approx& ig^3 (t^bt^a)_{i'i}(t^b)_{j'j} \frac{-i}{q_\perp^2}\frac{1}{2p_1\cdot k} iP_{B,\lambda}^{ss'}\\\nonumber
&&\left(g^{\nu\nu'}-\frac{n^{\nu}q^{\nu'}+n^{\nu'}q^\nu}{n\cdot q}\right) (2p_2-q)_{\nu'} (2p_1-q+2k)_\nu 
\end{eqnarray}
Note that $iP_{B}$ is different from $iP_{A}$ as the initial splitting parton has a different transverse momentum from diagram $A$.
Finally, we evaluate the contraction and get,
\begin{eqnarray}
i M_B &=& i g^3 (t^b t^a)){i'i} t^b{j'j} \delta^{\sigma\sigma'} \frac{2s}{q_\perp^2} \frac{x(1-x)}{k_\perp^2}  iP_{B,\lambda}^{ss'}
\end{eqnarray}
Diagram C can be obtained similarly,
\begin{eqnarray}
i M_C &=& -i g^3 (t^a t^b)){i'i} t^b{j'j} \delta^{\sigma\sigma'} \frac{2s}{q_\perp^2} \frac{x(1-x)}{(\vec{k}_\perp-\vec{q}_\perp)^2}  iP_{C,\lambda}^{ss'} 
\end{eqnarray}
To sum the contributions from all three diagrams, apply $f^{abc}t^c = -i[t^a, t^b]$ to $iM_A$, and the result is,
\begin{eqnarray}
i (M_A+M_B+M_C) &=& ig^3 \frac{2s}{q_\perp^2} (t^b)_{j'j} x(1-x)\\\nonumber
&&\left\{(t^a t^b)_{i'i} \left(\frac{iP_{A,\lambda}^{ss'} }{(\vec{k}_\perp-x \vec{q}_\perp)^2} - \frac{iP_{C,\lambda}^{ss'}}{(\vec{k}_\perp-\vec{q}_\perp)^2}\right) \right. \\\nonumber
&&\left.-(t^a t^b)_{i'i}\left(\frac{iP_{A,\lambda}^{ss'} }{(\vec{k}_\perp-x \vec{q}_\perp)^2} - \frac{iP_{B,\lambda}^{ss'}}{k_\perp^2}\right) \right\}
\end{eqnarray}

Now we have to address what those splitting amplitudes are.
Label the four momenta as $p_g = c$, $p_q = a$, $p_{\bar{q}} = b$, then use the following representation for the spinors,
\begin{eqnarray}
u^s(p) = (\sqrt{p\cdot \sigma} \xi^s, \sqrt{p\cdot \bar{\sigma}} \xi^s)^T
v^s(p) = (\sqrt{p\cdot \sigma} \eta^s, -\sqrt{p\cdot \bar{\sigma}} \eta^s)^T.
\end{eqnarray}
$\sigma_{i=\{1,2,3\}}$ are Pauli matrices, $\sigma = (1_{2\times 2}, \vec{\sigma})$, and $\bar{\sigma} = (1_{2\times 2}, -\vec{\sigma})$.
The square root of the matrix is,
\begin{eqnarray}
\sqrt{p\cdot \sigma} =
\left.
\begin{bmatrix}
p^- & -p_L^\perp \\
-p_R^\perp & p^+
\end{bmatrix}\right.^{1/2} 
= \frac{1}{\sqrt{2(E\pm M)}}(p\cdot\sigma \pm \mathbf{1}M)\\
\sqrt{p\cdot \bar{\sigma}} =
\left.
\begin{bmatrix}
p^+ & p_\perp^- \\
p_R^\perp & p^-
\end{bmatrix}\right.^{1/2} 
= \frac{1}{\sqrt{2(E\pm M)}}(p\cdot\bar{\sigma} \pm \mathbf{1}M)\\
\end{eqnarray}
where $M$ is the mass of the particle, $p^\pm = E\pm p_z$, and $p_{R,L}^\perp = p_x \pm  i p_y$.
Currently, we only consider the massless case, and the splitting amplitude is,
\begin{eqnarray}
&&\epsilon_{\lambda, \mu}(c) \bar{u}_s(a)\gamma^\mu v_{s'}(b)\\
&=&\frac{1}{\sqrt{2a}\sqrt{2b}}(\xi^T_s a\cdot\sigma, \xi^T_{s} a\cdot \bar{\sigma})
\begin{bmatrix}
\epsilon\cdot\bar{\sigma} & 0 \\
0 & \epsilon\cdot\sigma
\end{bmatrix}
\begin{bmatrix}
b\cdot\sigma \eta_{s'}\\
b\cdot\bar{\sigma} \eta_{s'}
\end{bmatrix}
\\
&=&\frac{1}{2\sqrt{ab}}
\xi_s^T
\begin{bmatrix}
a^- & -a^\perp_L \\
-a^\perp_R & a^+
\end{bmatrix}
\begin{bmatrix}
0 & \sqrt{2}\delta_{\lambda R}\\
\sqrt{2}\delta_{\lambda L} & \frac{\sqrt{2}c^\perp_\lambda}{c^+}
\end{bmatrix}
\begin{bmatrix}
b^- & -b^\perp_L \\
-b^\perp_R & b^-
\end{bmatrix}
\eta_{s'}\\\nonumber
&-&
\frac{1}{2\sqrt{ab}}
\xi_s^T
\begin{bmatrix}
a^+ & a^\perp_L \\
a^\perp_R & a^-
\end{bmatrix}
\begin{bmatrix}
\frac{\sqrt{2}c^\perp_\lambda}{c^+} & -\sqrt{2}\delta_{\lambda R}\\
-\sqrt{2}\delta_{\lambda L} & 0
\end{bmatrix}
\begin{bmatrix}
b^+ & b^\perp_L \\
b^\perp_R & b^-
\end{bmatrix}
\eta_{s'}
\\
&=&\frac{1}{\sqrt{2ab}}
\xi_s^T
\begin{bmatrix}
-a^\perp_L b^- \delta_{\lambda L} - a^- b^\perp_L \delta_{\lambda R} + a^\perp_L b^\perp_R\frac{c^\perp_\lambda}{c^+} &
a^\perp_L b^\perp_L \delta_{\lambda L} + a^- b^+ \delta_{\lambda R} - a^\perp_L b^+\frac{c^\perp_\lambda}{c^+}
\\
a^+ b^- \delta_{\lambda L} + a^\perp_R b^\perp_R \delta_{\lambda R} - a^+ b^\perp_R\frac{c^\perp_\lambda}{c^+} &
-a^+ b^\perp_L \delta_{\lambda L} - a^\perp_R b^+ \delta_{\lambda R} + a^+ b^+\frac{c^\perp_\lambda}{c^+}
\end{bmatrix}
\eta_{s'}\\\nonumber
&-&\frac{1}{\sqrt{2ab}}
\xi_s^T
\begin{bmatrix}
-a^\perp_L b^+ \delta_{\lambda L} - a^+ b^\perp_R \delta_{\lambda R} + a^+ b^+\frac{c^\perp_\lambda}{c^+} &
-a^\perp_L b^\perp_L \delta_{\lambda L} - a^+ b^- \delta_{\lambda R} + a^+ b^\perp_L\frac{c^\perp_\lambda}{c^+}
\\
-a^- b^+ \delta_{\lambda L} - a^\perp_R b^\perp_R \delta_{\lambda R} + a^\perp_+ b^+\frac{c^\perp_\lambda}{c^+} &
-a^- b^\perp_L \delta_{\lambda L} - a^\perp_R b^- \delta_{\lambda R} + a^\perp_R b^\perp_L\frac{c^\perp_\lambda}{c^+}
\end{bmatrix}
\eta_{s'}
\end{eqnarray}
Keep the leading terms in the collinear limit which are products of $(+)(+)$ or $(+)(\perp)$ components of the momenta, and drop terms that are of order $(+)(-)$, $(\perp)(\perp)$ and $(\perp)(-)$,
\begin{eqnarray}
&&\epsilon_{\lambda, \mu} \bar{u}_s(a)\gamma^\mu v_{s'}(b)\\
&=& \frac{1}{\sqrt{2ab}}
\xi_s^T
\begin{bmatrix}
a^\perp_L b^+ \delta_{\lambda L} + a^+ b^\perp_R \delta_{\lambda R} - a^+ b^+\frac{c^\perp_\lambda}{c^+} & 0\\
0 & -a^+ b^\perp_L \delta_{\lambda L} - a^\perp_R b^+ \delta_{\lambda R} + a^+ b^+\frac{c^\perp_\lambda}{c^+}
\end{bmatrix}
\eta_{s'}
\end{eqnarray}
There are four combinations for the possible initial state polarization and final state spins:
\begin{eqnarray}
\epsilon_{\lambda, \mu} \bar{u}_s(a)\gamma^\mu v_{s'}(b) = \frac{x\vec{a} - (1-x)\vec{b}}{\sqrt{2x(1-x)}}
\begin{cases}
x, \hfill \lambda=L, s=\uparrow\\
-(1-x), \hfill \lambda=L, s=\downarrow\\
(1-x), \hfill \lambda=R, s=\uparrow\\
-x, \hfill \lambda=R, s=\downarrow\\
\end{cases}
\end{eqnarray}
Where we have used $a^+ = (1-x)c^+, b^+ = xc^+$ and $c_\perp = a_\perp+b_\perp$.
Sum over the spins and average over polarization for the squared amplitude,
\begin{eqnarray}
\frac{1}{2}\sum_\pm |P|^2 = \frac{2(x^2 + (1-x)^2)}{x(1-x)} \left((1-x)\vec{a}_\perp-x\vec{b}_\perp\right)^2.
\end{eqnarray}
This result goes back to the standard splitting function if we compute it in the frame where $a_\perp = -b_\perp$. 
However, there is no such frame that $a_\perp = -b_\perp$ satisfies simultaneously for the splitting in diagram A, B, and C.
Therefore different amplitude needs to be inserted for each diagram, and we find
\begin{eqnarray}
\label{eq:gq2qqbarq}
\frac{\sum_{\lambda, s, s', \sigma, \sigma', a, b}|M^2|_{g+q\rightarrow q+\bar{q}+q}}{2d_F 2d_A} &=& g^4 \frac{2C_F}{d_A}\frac{4s^2 x(1-x)}{q_\perp^4}  \\\nonumber
&\times& g^2\frac{(x^2+(1-x)^2)}{2} \left(C_F \vec{A}^2 + C_F \vec{B}^2 - (2C_F- C_A)\vec{A}\cdot\vec{B}\right).
\end{eqnarray}
The vectors $\vec{A}$ and $\vec{B}$ are,
\begin{eqnarray}
\vec{A} &=& \frac{\vec{k}_\perp - x\vec{q}_\perp}{(\vec{k}_\perp - x\vec{q}_\perp)^2} -  \frac{\vec{k}_\perp - \vec{q}_\perp}{(\vec{k}_\perp - \vec{q}_\perp)^2}, \\
\vec{B} &=& \frac{\vec{k}_\perp - x\vec{q}_\perp}{(\vec{k}_\perp - x\vec{q}_\perp)^2} -  \frac{\vec{k}_\perp}{\vec{k}_\perp^2}.
\end{eqnarray}
The final squared matrix-element has been factorized into the two-body scattering part (first line) and the collinear splitting part (second line) with the desired leading order QCD splitting function. 

\paragraph*{Quark splits to quark and gluon}
\begin{figure}

\centering
\includegraphics[width=.5\textwidth]{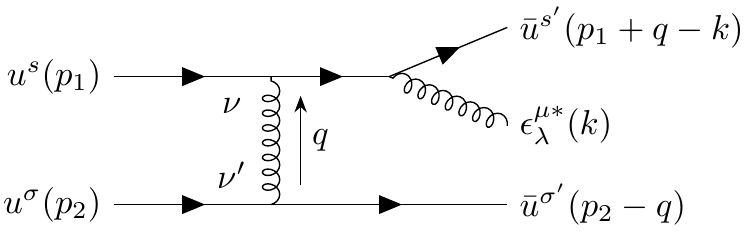}\\
\vspace{1em}
\includegraphics[width=.49\textwidth]{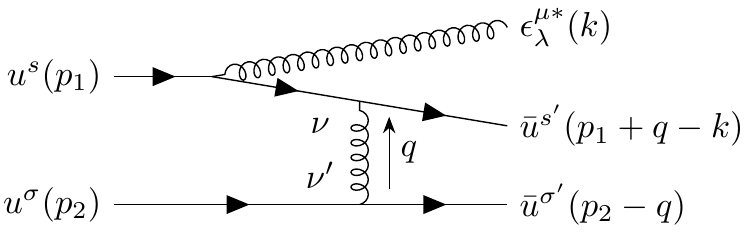}\hfill
\includegraphics[width=.49\textwidth]{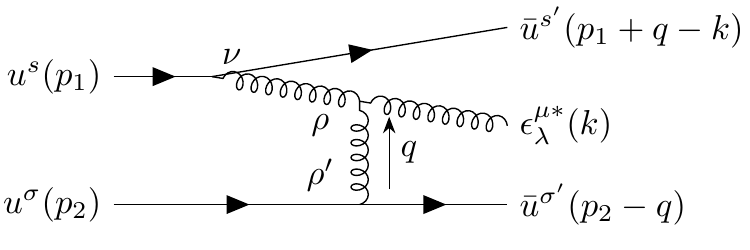}
\caption[Three diagrams $A$ (Top), $B$ (Bottom left), $C$ (Bottom right) that]{Three diagrams $A$ (Top), $B$ (Bottom left), $C$ (Bottom right) that contribute to the large angle scattering induced a quark splitting into a quark and a gluon in the forward region of the center-of-mass frame.}
\label{fig:feyn-q2qg}
\end{figure}

The Feynman diagrams to be included for $q+q\rightarrow q+g+q$ are shown in Figure \ref{fig:feyn-q2qg}.
The calculation uses precisely the same technique we used for the gluon splitting channel, and we present the result directly,
\begin{eqnarray}
\label{eq:qq2qgq}
\overline{|M^2|}_{g+q\rightarrow g+g+q} &=& 
 g^4 \frac{C_F}{d_F}\frac{4s^2}{q_\perp^4}x(1-x) \\\nonumber
&\times&g^2\frac{1+(1-x)^2}{x}  
\left(C_F\vec{A}^2 + C_F\vec{B}^2 - \left(2C_F-C_A\right)\vec{A}\cdot\vec{B}\right)\\
\vec{A} &=& \frac{\vec{k}_\perp - \vec{q}_\perp}{(\vec{k}_\perp - \vec{q}_\perp)^2} -  \frac{\vec{k}_\perp - x\vec{q}_\perp}{(\vec{k}_\perp - x\vec{q}_\perp)^2} \\
\vec{B} &=& \frac{\vec{k}_\perp - \vec{q}_\perp}{(\vec{k}_\perp - \vec{q}_\perp)^2} -  \frac{\vec{k}_\perp}{\vec{k}_\perp^2}
\end{eqnarray}

\paragraph*{Gluon splitting to two gluons}
\begin{figure}

\centering
\includegraphics[width=.5\textwidth]{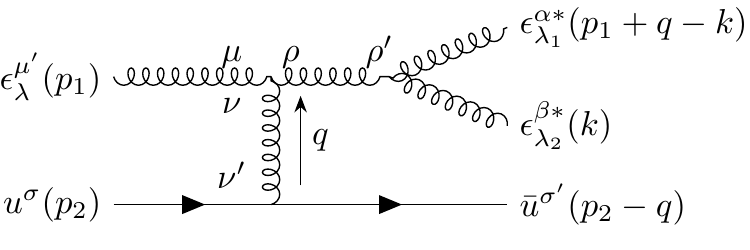}\\
\vspace{1em}
\includegraphics[width=.49\textwidth]{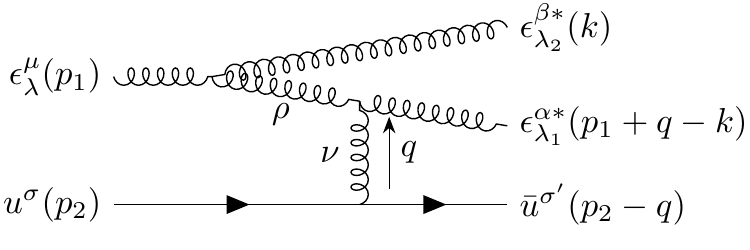}\hfill
\includegraphics[width=.49\textwidth]{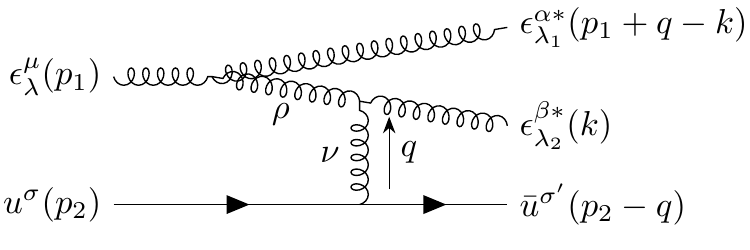}
\caption[Three diagrams $A$ (Top), $B$ (Bottom left), $C$ (Bottom right) that]{Three diagrams $A$ (Top), $B$ (Bottom left), $C$ (Bottom right) that contribute to the large angle scattering induced gluon splitting into two gluons in the forward region of the center-of-mass frame.}
\label{fig:feyn-g2gg}
\end{figure}

Finally, for $g+q\rightarrow g+q+g$, the Feynman diagrams are shown in Figure \ref{fig:feyn-g2gg}. 
The simplification of the two-body collision amplitude can be done similarly as the previous two channels. 
We only write down the splitting amplitude $g\rightarrow g+ g$ in detail.
Suppressing the color index, we label the initial gluon with $\epsilon_1^\mu(p)$, and the two daughter gluons with $\epsilon_2^\nu(k)$ and $\epsilon_3^\rho(q)$.
The splitting amplitudes are then (omitting the factor $-gf^{abc}$)
\begin{eqnarray}
iP &=& \epsilon^\mu_1\epsilon^\nu_2\epsilon^\rho_3
\left[
g_{\mu\nu} (p+k)_{\rho} +  g_{\nu\rho} (p+k)_{\mu} + g_{\rho\mu} (-q-p)_{\nu}
\right]\\
&=& -\vec{\epsilon}_{1,\perp}\cdot \vec{\epsilon}_{2,\perp} \left[(p+k)^+\frac{\vec{\epsilon}_{3,\perp}\cdot \vec{q}_\perp}{q^+} - \vec{\epsilon}_{3,\perp}\cdot (\vec{p}_\perp+\vec{k}_\perp)\right] \\\nonumber
&&-\vec{\epsilon}_{2,\perp}\cdot \vec{\epsilon}_{3,\perp} \left[(-k+q)^+\frac{\vec{\epsilon}_{1,\perp}\cdot \vec{p}_\perp}{p^+} - \vec{\epsilon}_{1,\perp}\cdot (-\vec{k}_\perp+\vec{q}_\perp)\right]
\\\nonumber
&&-\vec{\epsilon}_{3,\perp}\cdot \vec{\epsilon}_{1,\perp} \left[(-q-p)^+\frac{\vec{\epsilon}_{2,\perp}\cdot \vec{k}_\perp}{k^+} - \vec{\epsilon}_{2,\perp}\cdot (-\vec{q}_\perp-\vec{p}_\perp)\right]
\end{eqnarray}
There are four possible combinations of the polarization vectors, and their respective amplitudes are,
\begin{eqnarray}
iP = \sqrt{2}\left[x\vec{q}_\perp - (1-x)\vec{k}_\perp\right]\times 
\begin{cases}
\frac{1-x+x^2}{x(1-x)}, \hfill \lambda_1=\lambda_2=\lambda_3\\
-1, \hfill \lambda_1\neq\lambda_2=\lambda_3 \\
\frac{1}{x}, \hfill \lambda_1=\lambda_3\neq\lambda_2\\
\frac{1}{1-x}, \hfill \lambda_1=\lambda_2\neq\lambda_3
\end{cases}
\end{eqnarray}
Summing over the squared amplitude of all four cases and averaging over the initial gluon polarization, one gets the desired leading order QCD splitting function,
\begin{eqnarray}
2\frac{1+x^2+(1-x)^4}{x^2(1-x)^2} \left[x\vec{q}_\perp - (1-x)\vec{k}_\perp\right]^2.
\end{eqnarray}
Substituting the amplitude in each diagram, the final squared matrix-element is 
\begin{eqnarray}
\label{eq:gq2ggq}
\overline{|M^2|}_{g+q\rightarrow g+g+q} &=&
g^4 \frac{C_A}{d_F}\frac{4s^2x(1-x)}{q_\perp^4} \\\nonumber
&\times&g^2\frac{1+x^4+(1-x)^4}{x(1-x)}   
\left(C_A\vec{A}^2 + C_A\vec{B}^2 - C_A\vec{A}\cdot\vec{B}\right)\\
\vec{A} &=& \frac{\vec{k}_\perp - x\vec{q}_\perp}{(\vec{k}_\perp - x\vec{q}_\perp)^2} -  \frac{\vec{k}_\perp - \vec{q}_\perp}{(\vec{k}_\perp - \vec{q}_\perp)^2} \\
\vec{B} &=& \frac{\vec{k}_\perp - x\vec{q}_\perp}{(\vec{k}_\perp - x\vec{q}_\perp)^2} -  \frac{\vec{k}_\perp}{\vec{k}_\perp^2}
\end{eqnarray}

\paragraph{Regulating the $2\rightarrow 3$ squared matrix-elements}
The requirement that the few-body matrix-elements only apply to processes with $q>Q_{\textrm{cut}}$ removes the divergence in the $q$ integration.
The collinear divergence when $k$ approaches $q$ or $xq$ is regulated by including a gluon thermal mass. 
In practice, the collinear divergence is further regulated by the LPM effect.
The cross-section is obtained by integrating over the final-state phase-space, parameterized by $k_\perp^2$, the rapidity of $k$ in the center-of-mass frame $y_k$, and the solid angle of the recoil medium particle.

\paragraph{Soft limit: the Gunion-Bertsch approximation}
The result we obtained for the $g\rightarrow g+g$ and $q\rightarrow q+g$ channel have a soft limit that goes back to the well known Gunion-Bertsch form. 
In the soft limit, we require the radiated gluon energy to be small enough such that $xq_\perp \ll k_\perp$.
Then, the splitting amplitudes for both $g\rightarrow g+g$ and  $q\rightarrow q+g$ are simplified into the same form,
\begin{eqnarray}
\overline{|M|}^2_{22} x(1-x)g^2 \frac{2(1-x+O(x^2))}{x} C_A \left(\frac{\vec{k}_\perp}{k_\perp^2}-\frac{\vec{k}_\perp-\vec{q}_\perp}{(\vec{k}_\perp-\vec{q}_\perp)^2}\right)^2
\end{eqnarray}
Neglecting the $O(x^2)$ terms in the splitting function, the result is the same as the improved version of the Gunion-Bertsch cross-section \cite{Fochler:2013epa} used in the full Boltzmann partonic transport model BAMPS \cite{Xu:2004mz},
\begin{eqnarray}
\overline{|M|}^2_{22} 8\pi C_A\alpha_s (1-x)^2 \left(\frac{\vec{k}_\perp}{k_\perp^2}-\frac{\vec{k}_\perp-\vec{q}_\perp}{(\vec{k}_\perp-\vec{q}_\perp)^2}\right)^2
\end{eqnarray}

\paragraph{The backward ($y_k < 0$) region}
We have mentioned at the beginning of the derivation that the condition $k_\perp^2 < x(1-x)\hat{s}$ restricts the splitting to happen only for the parton moving in the $+z$ direction in the center-of-mass frame ($y_k > 0$).
For splittings that happen in the backward region, another set of diagrams contribute, where the splitting comes from the parton that moves in the $-z$ direction in the center-of-mass frame.
Also one needs a different gauge $A^- = 0$.
The derivation is similar to the previous ones, but with the definition of $x$ and $q$ changed to $x = k^-/\sqrt{s}$, and $q = p_1-p_3$.

To combine the results that are obtained in different regions of phase space ($y_k > 0$ and $y_k < 0$), we follow \cite{Fochler:2013epa} and defines,
\begin{eqnarray}
\bar{x} &=& \frac{(k + |k_z|)}{\sqrt{s}} = \frac{k_\perp e^{|y_k|}}{\sqrt{s}}\\ 
\bar{q} &=& \Theta(y_k)(p_2-p_4) + \Theta(-y_k)(p_1-p_3)
\end{eqnarray}
which replaces the original $x$ and $q$ in our formula, and the resultant matrix-elements can be used for both forward and backward regions.

\paragraph{Relation to the Bethe-Heitler limit of the AMY formalism}
Now we show the connection between the $2\rightarrow 3$ cross-section obtained here and the Bethe-Heitler limit of the AMY equation.
In the Bethe-Heitler limit, the AMY integral equation can be solved approximately by treating $1/\tau_f$ as the leading factor. 
One obtains the splitting rate for each different channels (denoting $\vec{a}/a^2$ as $\vec{\phi}_{a}$), 
\begin{eqnarray}
R_{q\rightarrow q+g}^{BH} &\propto& g^2 P_{qg}^{q(0)}(x) \int d k^2 d q^2 \mathcal{A}(q^2) \left\{
C_A\vec{\phi}_k\cdot\left(\vec{\phi}_k-\vec{\phi}_{k-q}\right) \right.\\\nonumber
&&+\left. (2C_F-C_A) \vec{\phi}_k\cdot\left(\vec{\phi}_k-\vec{\phi}_{k+xq}\right)
+ C_A \vec{\phi}_k\cdot\left(\vec{\phi}_k - \vec{\phi}_{k+(1-x)q}\right)
\right\}
\\
R_{g\rightarrow g+g}^{BH} &\propto& g^2 P_{gg}^{g(0)}(x) \int d k^2 d q^2 \mathcal{A}(q^2) \left\{
C_A\vec{\phi}_k\cdot\left(\vec{\phi}_k-\vec{\phi}_{k-q}\right) \right.\\\nonumber
&&+\left. C_A \vec{\phi}_k\cdot\left(\vec{\phi}_k-\vec{\phi}_{k+xq}\right)
+ C_A \vec{\phi}_k\cdot\left(\vec{\phi}_k - \vec{\phi}_{k+(1-x)q}\right)
\right\}
\\
R_{g\rightarrow q+\bar{q}}^{BH} &\propto& g^2 P_{q\bar{q}}^{g(0)}(x) \int d k^2  d q^2 \mathcal{A}(q^2) \left\{
(2C_F-C_A)\vec{\phi}_k\cdot\left(\vec{\phi}_k-\vec{\phi}_{k-q}\right) \right.\\\nonumber
&&+\left. C_A \vec{\phi}_k\cdot\left(\vec{\phi}_k-\vec{\phi}_{k+xq}\right)
+ C_A \vec{\phi}_k\cdot\left(\vec{\phi}_k - \vec{\phi}_{k+(1-x)q}\right)
\right\}
\end{eqnarray}
with the collision kernel $\mathcal{A} = g^2 T m_D^2/q^2(q^2+m_D^2)$. These expressions look different from the incoherent rate computed using the cross-section derived in the previous section; however, we would like to show that they are equivalent once integration over $dk^2$ is performed.
Therefore, the incoherent rate we used in the Boltzmann equation indeed recovers the Bethe-Heitler limit of the AMY integral equation.

To show this, we start from the $2\rightarrow 3$ rate formula using the matrix-elements from equations \ref{eq:gq2qqbarq}, \ref{eq:qq2qgq} and \ref{eq:gq2ggq}. 
For the $q\rightarrow q+g$ channel, the rate in the Boltzmann equation is,
\begin{eqnarray}
R_{q\rightarrow q+g} &\propto& g^2 P_{qg}^{q(0)}(x) \int  \frac{f(p_2)dp_2^3}{2E_2(2\pi)^3} d q^2 \frac{g^4}{q^4}\\\nonumber
&&  \int d k^2\left\{
C_F\left( \vec{\phi}_{k-q}-\vec{\phi}_{k-xq} \right)^2
+ C_F\left( \vec{\phi}_{k-q}-\vec{\phi}_{k} \right)^2\right.\\\nonumber
&&\left.
- (2C_F-C_A)\left( \vec{\phi}_{k-q}-\vec{\phi}_{k-xq} \right)\cdot \left( \vec{\phi}_{k-q}-\vec{\phi}_{k} \right)
\right\}
\end{eqnarray}
Focusing on the three products (squares) of $\vec{\phi}$s under the $dk^2$ integration, we are going to expand the first term in each product and then shift the argument of the first $\vec{\phi}$ to $k$, 
\begin{eqnarray}
R_{q\rightarrow q+g} &\propto& g^2 P_{qg}^{q(0)}(x) \int  \frac{f(p_2)dp_2^3}{2E_2(2\pi)^3} d q^2 \frac{g^4}{q^4}\\\nonumber
&&  \int d k^2\left\{
C_F\vec{\phi}_{k}\left( \vec{\phi}_{k}-\vec{\phi}_{k+(1-x)q} \right)
- C_F\vec{\phi}_{k}\left( \vec{\phi}_{k-(1-x)q}-\vec{\phi}_{k} \right)\right.
\\\nonumber
&&+ C_F\vec{\phi}_{k}\left( \vec{\phi}_{k}-\vec{\phi}_{k+q} \right)
- C_F\vec{\phi}_{k}\left( \vec{\phi}_{k-q}-\vec{\phi}_{k} \right)
\\\nonumber
&&\left.
- (2C_F-C_A)\vec{\phi}_{k}\cdot \left( \vec{\phi}_{k}-\vec{\phi}_{k+q} \right)
+(2C_F-C_A)\vec{\phi}_{k} \cdot \left( \vec{\phi}_{k-(1-x)q}-\vec{\phi}_{k+xq} \right)
\right\}
\end{eqnarray}
Next, flip the sign of $q$ under the integration.
Meanwhile, insert a $-\vec{\phi}_k +\vec{\phi}_k$ in the brackets of the last term,
\begin{eqnarray}
R_{q\rightarrow q+g} &\propto& g^2 P_{qg}^{q(0)}(x) \int  \frac{f(p_2)dp_2^3}{2E_2(2\pi)^3} d q^2 \frac{g^4}{q^4}\\\nonumber
&&  \int d k^2\left\{
2C_F\vec{\phi}_{k}\left( \vec{\phi}_{k}-\vec{\phi}_{k+(1-x)q} \right)
+ 2C_F\vec{\phi}_{k}\left( \vec{\phi}_{k}-\vec{\phi}_{k+q} \right)
\right.
\\\nonumber
&&
- (2C_F-C_A)\vec{\phi}_{k}\cdot \left( \vec{\phi}_{k}-\vec{\phi}_{k+q} \right)
+(2C_F-C_A)\vec{\phi}_{k} \cdot \left( \vec{\phi}_{k+(1-x)q} -\vec{\phi}_k \right) \\\nonumber
&&\left.+(2C_F-C_A)\vec{\phi}_{k} \cdot \left(\vec{\phi}_k-\vec{\phi}_{k+xq} \right)
\right\}
\end{eqnarray}
After this manipulation, the first (second) term cancels the $C_F$ part of the fourth (third) term, 
\begin{eqnarray}
R_{q\rightarrow q+g} &\propto& g^2 P_{qg}^{q(0)}(x) \int  \frac{f(p_2)dp_2^3}{2E_2(2\pi)^3} d q^2 \frac{g^4}{q^4}\\\nonumber
&&  \int d k^2\left\{
C_A\vec{\phi}_{k}\cdot \left( \vec{\phi}_{k}-\vec{\phi}_{k+q} \right)
+C_A\vec{\phi}_{k} \cdot \left( \vec{\phi}_k - \vec{\phi}_{k+(1-x)q}\right) \right.\\\nonumber
&&\left.+(2C_F-C_A)\vec{\phi}_{k} \cdot \left(\vec{\phi}_k-\vec{\phi}_{k+xq} \right)
\right\}
\end{eqnarray}
which is the same integration as the one obtained from the Bethe-Heitler limit of the AMY equation (neglecting the screen mass in $\mathcal{A}$ when $q^2 \gg m_D^2$)
Similarly, the equivalence also exists for the $g\rightarrow g+g$ channel and the $g\rightarrow q+\bar{q}$ channel.

\end{appendices}

\backmatter

\bibliography{dissertation}{}
\bibliographystyle{JHEP}

\end{document}